\documentclass[pdftex,11pt]{article}
\usepackage[pdftex]{graphicx}
\usepackage[pdftex,                
bookmarks=true,
bookmarksnumbered=true,
hypertexnames=false,
breaklinks=true,
linkbordercolor={0 0 1}]{hyperref} 
\usepackage{tabularx}    
\usepackage{import}      
\usepackage{amsmath, amssymb} 
\usepackage{pdfpages}
\usepackage[24hr,mmddyyyy]{datetime}
\usepackage{hyphenat}			
\usepackage{lineno}
\usepackage{booktabs}			
\usepackage{textcmds}	
\usepackage{xspace}
\usepackage{longtable}	
\usepackage{bm}		
\usepackage{afterpage}

\usepackage[margin=0pt,font=small,labelfont=bf,labelsep=colon]{caption}

\usepackage[backend=bibtex8,doi=false,style=numeric,sorting=none]{biblatex}

\newcommand{\citep}[1]{\cite{#1}}

\def\bfseries{\fontseries \bfdefault \selectfont \boldmath}

\usepackage[margin=0.25in]{caption}

    \usepackage[outermarks]{titlesec}
    \titleformat{\section}
      {\normalfont\Large\bfseries}{\thesection}{1em}{}
    \titleformat{\subsection}
      {\normalfont\large\bfseries}{\thesubsection}{1em}{}
    \titleformat{\subsubsection}
      {\normalfont\normalsize\bfseries}{\thesubsubsection}{1em}{}
    \titleformat{\paragraph}
      {\normalfont\normalsize\bfseries}{\theparagraph}{1em}{}
    \titleformat{\subparagraph}
      {\normalfont\normalsize\bfseries}{\thesubparagraph}{1em}{}

    \titlespacing*{\section}      {0pt}{3.5ex plus 1ex minus .2ex} {2.3ex plus .2ex}
    \titlespacing*{\subsection}   {0pt}{3.25ex plus 1ex minus .2ex}{1.5ex plus .2ex}
    \titlespacing*{\subsubsection}{0pt}{3.25ex plus 1ex minus .2ex}{1.5ex plus .2ex}
    \titlespacing*{\paragraph}    {0pt}{3.25ex plus 1ex minus .2ex}{1.5ex plus .2ex}
    \titlespacing*{\subparagraph} {0pt}{3.25ex plus 1ex minus .2ex}{1.5ex plus .2ex}

\newcommand{\impc}{{\ensuremath{{\rm\,Mpc}^{-1}}}\xspace}

\newcommand{\ihmpc}{{\ensuremath{h{\rm\,Mpc}^{-1}}}\xspace}

\newcommand{\kms}{{\ensuremath{\rm\ km\ s^{-1}}}\xspace}

\newcommand{\lya}{{Ly-$\alpha$}\xspace}
\newcommand{\simlt}{\lower.5ex\hbox{$\; \buildrel < \over \sim \;$}}
\newcommand{\simgt}{\lower.5ex\hbox{$\; \buildrel > \over \sim \;$}}

\newcommand{\iMpc}{\impc}

\newcommand{\ihMpc}{\ihmpc}

\newcommand{\micron}{{\mbox{$\mu$m}}\xspace}
\newcommand{\arcsec}{{\mbox{$^{\prime\prime}$}}\xspace}

\renewcommand\thetable{\arabic{section}.\arabic{subsection}.\arabic{table}}
\renewcommand\theequation{\arabic{section}.\arabic{subsection}.\arabic{equation}}

\newcommand{\Nnueff}{{N_{\nu,{\rm eff}}}\xspace}
\newcommand{\summnu}{{\Sigma m_\nu}\xspace}

\newcommand{\lyaf}{{Ly-$\alpha$\ forest}\xspace}
\newcommand{\lyb}{{Ly-$\beta$}\xspace}

\newcommand{\otwo}{{[O\,II]}\xspace}
\newcommand{\othree}{{[O\,III]}\xspace}
\newcommand{\htwo}{{H\,II}\xspace}

\newcommand{\kmaxeff}{{k_{\rm max}}\xspace}
\newcommand{\fnl}{{f_{\rm NL}}\xspace}
\newcommand{\fnlk}{{f_{\rm NL}(k)}\xspace}
\newcommand{\nf}{{{n_{f_{\rm NL}}}}\xspace}
\newcommand{\fnlstar}{{f^*_{\rm NL}}\xspace}
\def\pvm#1{
}

\begin{document}

\renewcommand{\thefigure}{\thesection .\arabic{figure}}
\renewcommand{\thetable}{\thesection .\arabic{table}}
\renewcommand{\theequation}{\thesection .\arabic{equation}}




\begin{titlepage}

\begin{center}
\begin{Large}
{The DESI Experiment Part I:
Science,Targeting, and Survey Design\\}
\end{Large}
\end{center}

\begin{center}
DESI Collaboration: 
Amir Aghamousa$^{73}$,
Jessica Aguilar$^{76}$,
Steve Ahlen$^{85}$,
Shadab Alam$^{41,59}$,
Lori E. Allen$^{81}$,
Carlos Allende Prieto$^{64}$,
James Annis$^{52}$,
Stephen Bailey$^{76}$,
Christophe Balland$^{88}$,
Otger Ballester$^{57}$
% ...

\end{center}

\begin{center}
{\it (Affiliations can be found after the references)}
\end{center}

\begin{abstract}

DESI (Dark Energy Spectroscopic Instrument) is a Stage IV
ground-based dark energy experiment that will study
baryon acoustic oscillations (BAO) and the growth of structure through
redshift-space distortions with a wide-area galaxy and quasar
redshift survey.  To trace the underlying dark matter
distribution, spectroscopic targets will be selected in four classes
from imaging data.  We will measure luminous red galaxies up to
$z=1.0$.  To probe the Universe out to even higher redshift, DESI will
target bright \otwo emission line galaxies up to $z=1.7$.
Quasars will be targeted both as direct tracers of the underlying dark
matter distribution and, at higher redshifts ($ 2.1 < z < 3.5$), for
the \lyaf~absorption features in their spectra, which will be used to
trace the distribution of neutral hydrogen.  When moonlight prevents
efficient observations of the faint targets of the baseline survey,
DESI will conduct a magnitude-limited Bright Galaxy Survey 
comprising approximately 10 million galaxies with a median $z\approx
0.2$.  In total, more than 30 million galaxy and quasar redshifts will
be obtained to measure the BAO feature and determine the matter power
spectrum, including redshift space distortions.

\end{abstract}

\end{titlepage}

\cleardoublepage


\setcounter{secnumdepth}{3}
\setcounter{tocdepth}{3}	

\pagenumbering{roman}		
\tableofcontents
\cleardoublepage

\pagestyle{headings}		
\pagenumbering{arabic}		


\section{Overview}\label{s1:Exec}
\setcounter{equation}{0}\setcounter{figure}{0}\setcounter{table}{0}	
DESI is a Stage IV ground-based dark energy experiment that will study
baryon acoustic oscillations (BAO) and the growth of structure through
redshift-space distortions (RSD) with a wide-area galaxy and quasar
redshift survey.  DESI is the successor to the successful Stage-III
BOSS redshift survey and complements imaging surveys such as the
Stage-III Dark Energy Survey (DES, operating 2013--2018) and the
Stage-IV Large Synoptic Survey Telescope (LSST, planned start early in
the next decade).  DESI is an important component of the DOE Cosmic
Frontier program, meeting the need for a wide-field spectroscopic
survey identified in the 2011 ``Rocky-III'' dark energy community
planning report.  In addition to providing Stage IV constraints on
dark energy, DESI will provide new measurements that can constrain
theories of modified gravity and inflation, and that will measure the sum
of neutrino masses.
 
The DESI instrument is a robotically-actuated, fiber-fed spectrograph
capable of taking up to 5,000 simultaneous spectra over a wavelength
range from 360 nm to 980 nm.  The fibers feed ten three-arm
spectrographs with resolution $R= \lambda/\Delta\lambda$ between 2000 and 5500,
depending on wavelength.  This powerful instrument will be installed
at prime focus on the 4-m Mayall telescope in Kitt Peak, Arizona,
along with a new optical corrector, which will provide a three-degree
diameter field of view.  The DESI collaboration will also deliver a
spectroscopic pipeline and data management system to reduce and
archive all data for eventual public use.

The DESI instrument will be used to conduct a five-year survey
designed to cover 14,000 deg$^2$.  To trace the underlying dark matter
distribution, spectroscopic targets will be selected in four classes
from imaging data.  We will measure luminous red galaxies (LRGs) up to
$z=1.0$, extending the BOSS LRG survey in both redshift and survey
area.  To probe the Universe out to even higher redshift, DESI will
target bright \otwo emission line galaxies (ELGs) up to $z=1.7$.
Quasars will be targeted both as direct tracers of the underlying dark
matter distribution and, at higher redshifts ($ 2.1 < z < 3.5$), for
the \lyaf~absorption features in their spectra, which will be used to
trace the distribution of neutral hydrogen.  When moonlight prevents
efficient observations of the faint targets of the baseline survey,
DESI will conduct a magnitude-limited Bright Galaxy Survey (BGS)
comprising approximately 10 million galaxies with a median $z\approx
0.2$.  In total, more than 30 million galaxy and quasar redshifts will
be obtained to measure the BAO feature and determine the matter power
spectrum, including redshift space distortions.

In the following document, we primarily refer to this baseline survey,
which would span 14,000 deg$^2$.  We also calculate numbers for a
minimum survey spanning 9,000 deg$^2$, which is still sufficient to
meet the requirements of a Stage-IV project.
  
DESI provides at least an order of magnitude improvement over BOSS
both in the comoving volume it probes and the number of galaxies it
will map.  This will significantly advance our understanding of the
expansion history of the Universe, providing more than thirty
sub-percent-accuracy distance measurements.  Precision on the
expansion history of the Universe is a powerful probe of the nature of
dark energy.  This can be quantified with the Dark Energy Task Force
figure of merit (DETF FoM), which measures the combined precision on
the dark energy equation of state today, $w_0$, and its evolution with
redshift $w_a$.  DESI galaxy BAO measurements achieve a DETF FoM of
133, more than a factor of three better than the DETF FoM of all
Stage-III galaxy BAO measurements combined.  The FoM increases to 169
with the inclusion of \lyaf BAO, and 332 including galaxy broadband
power spectrum to $k=0.1$ \ihMpc.  
DESI clearly satisfies the DETF criteria for a Stage-IV experiment.
Moreover, the FoM grows to 704 when the galaxy broadband power
spectrum data out to $k<0.2$ \ihMpc are included.  

In addition, DESI will measure the sum of neutrino masses with an
uncertainty of 0.020 eV (for $k_{\rm max} < 0.2$ \ihMpc), sufficient
to make the first direct detection of the sum of the neutrino masses
at 3-$\sigma$ significance and rule out the the inverted mass
hierarchy at 99\% CL, if the hierarchy is normal and the masses are
minimal.  DESI will also place significant constraints on theories of
modified gravity and of inflation by measuring the spectral index
$n_s$ and its running with wavenumber, $\alpha_s$.  The BGS will
enable the best ever measurements of low redshift BAO and RSD,
including the use of multiple-tracer methods that exploit galaxy
populations with different clustering properties, and it will yield
novel tests of modified gravity theories using the velocity fields of
cluster infall regions.  Because the nearby galaxies of the BGS are
too clustered to fill all of the targets, in parallel with the BGS,
DESI will conduct a survey of Milky Way stars, that can be used to
trace the dark matter halo of the Milky Way and probe the small-scale
structure of $\Lambda$CDM.

DESI will provide an unprecedented multi-object spectroscopic
capability for the U.S. through an existing NSF telescope
facility. Many other science objectives can be addressed with the DESI
wide field survey dataset and through bright time and piggy-back
observation programs.  Much as with SDSS, a rich variety of projects
will flow from the legacy data from the DESI survey.

DESI will overlap with the DES and LSST survey areas, which are
primarily in the Southern hemisphere but which will have equatorial
and northern ecliptic regions.  DESI will be a pathfinder instrument
for the massive spectroscopic follow-up required for future large area
imaging surveys such as LSST.

This portion of the Final Design Report summarizes the
DESI scientific goals, the target selection, and survey design.  The
accompanying instrument portion of the FDR describes the instrument and optical
design, integration and test plan, and the data management system.
The companion Science Requirements Document provides information that
guides the design.  The DESI construction management plan is presented
in the accompanying Project Execution Plan.  Likewise, project cost
and schedule are available in appropriate Project Office documents.


\clearpage

\section{Science Motivation and Requirements}\label{s2:Science}
\setcounter{equation}{0}\setcounter{figure}{0}\setcounter{table}{0}

\newcommand{\fixme}{{\bf FIXME???:}}
\newcommand{\dragan}[1]{\textcolor{red}{Dragan: #1}}


\subsection{Introduction}

DESI will explore some of the most fundamental questions in physical science:
what is the composition of the Universe at large and what is the
nature of space-time?  These questions are now open to exploration
because of recent discoveries. We summarize here the framework used to
express these questions and the parameters used to quantify our
understanding.


There are several pillars of the cosmological model that are now well
established: 1) a period of rapid acceleration --- {\it inflation} or
a similar process --- occurred in the early Universe, generating the
primordial fluctuations, which seeded large scale structures, galaxies
and galaxy clusters, which grew during the decelerating, matter
dominated era 2) gravitational instabilities produced acoustic
oscillations in the plasma, which were imprinted about
400,000 years after this inflation period, when photons decoupled
from atoms and produced the Cosmic Microwave Background 3) this was
followed by a period of matter domination, when small density
fluctuations grew into large-scale structure, 4) comparatively
recently, there was a transition to accelerated
expansion driven by either a modification to General Relativity or a
new form of energy -- {\it dark energy} -- not due to any particles
known or unknown, and which contributes about 68\% of the Universe's
energy density, and 5) about 27\% of the energy density
today is due to matter outside the Standard Model of particle physics
-- {\it dark matter} -- which is responsible for large-scale structure
formation and accounts for galaxy rotation curves and the motions of
galaxies in clusters.  

That the Universe is expanding more and more rapidly was first
revealed through measurements of Type Ia supernovae \cite{perlmutter,
  riess}, and subsequently confirmed using other techniques.  Within
General Relativity, accelerated expansion requires $\rho + 3p < 0$,
where $\rho$ is the total energy density and $p$ is the total pressure of the
matter, radiation, and other ingredients.  
The total equation of state $w=p/\rho$ must be less than $-1/3$ for accelerated expansion.  The equation of state need not be a constant; in general it depends on time, or equivalently the scale size of the universe $a=1/(1+z)$. From now on,
we let $w$ denote the equation of state of the dark energy component
alone.  


For ordinary non-relativistic matter, the pressure is negligible
compared to the energy due to the rest mass and thus $w=0$.  For photons
and other massless particles, $w=1/3$. The cosmological constant term
is equivalent to dark energy with $w=-1$. Generally, energy with an
equation of state $w(a)$ evolves as $\rho(a)=\rho(a=1)F(a)$, where
$F(a)=1$ for a cosmological constant and for a general equation of
state $w(a)$ is
\begin{equation}
F(a)\equiv \exp\left[{3\int_a^1\frac{da'}{a'}(1+w(a'))}\right].\label{eq:rho}
\end{equation}
It is standard to parameterize the equation of state as
\begin{equation}
w(a)=w_0+(1-a)w_a,
\end{equation}
which accurately reproduces distances for a wide range of models.

The contributions to the energy density of the Universe are
conventionally expressed relative to the critical density
\begin{equation}
\rho_{\rm crit}=\frac{3H_0^2}{8\pi G},
\end{equation}
which would be just sufficient to slow the expansion ultimately to
zero in the absence of a dark energy component $\Omega_\Lambda$.  We write
\begin{equation}
\Omega_m=\frac{\rho_m}{\rho_{\rm crit}}.
\end{equation} 
We define $\Omega_{\rm r}$ for radiation and $\Omega_{DE}$
for dark energy analogously.  The curvature term $\Omega_k=-k/H_0^2$
is defined so that General Relativity requires 
\begin{equation}
\Omega_{\rm r}+\Omega_{m}+\Omega_{k}+\Omega_{DE}=1
\end{equation}
for a Universe with spatial curvature $k$.
The expansion rate of the Universe is given by
\begin{equation}
H(a)\equiv\frac{\dot a}{a}=H_0\left[\Omega_{\rm r}a^{-4}+\Omega_m a^{-3}+\Omega_k a^{-2}+\Omega_{DE}F(a)\right]^{1/2}.
\end{equation}
The contribution from radiation, $\Omega_{\rm r}$ is
negligible today and inflation predicts that the curvature is
zero. The Hubble constant today is $H_0=h\times 100$ km/s/Mpc$
\approx$ 70 km/s/Mpc.

\begin{figure}[!t]
\begin{center}
\includegraphics[height=3.5in]{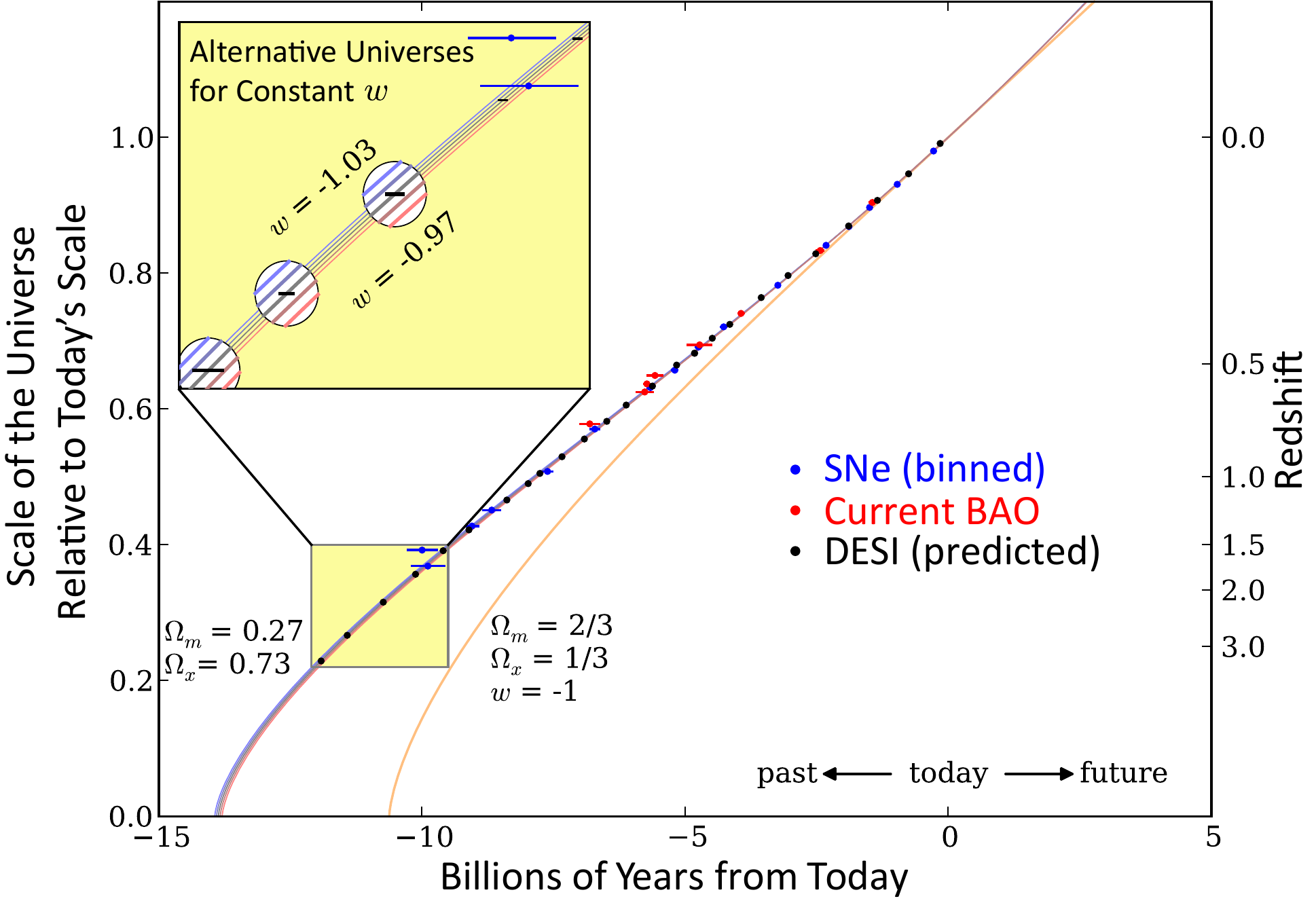}
\caption{\label{fig:rubin} 
The expansion history of the Universe for different models of dark energy,
holding the present-day Hubble constant fixed.
The inset shows the spacing between five models with
constant $w$ ranging from $-0.97$ to $-1.03$, showing the 
exquisite precision required to distinguish these.  
Overlaid are measurements of the distance-redshift relation, translated
into errors on lookback time at each redshift. 
Measurements from current supernovae, binned in redshift, are shown in blue;
current BAO measurements from BOSS DR9, WiggleZ, and 6dF are shown in red;
projections for DESI are shown in black.
DESI measurements have the ability to make very tight constraints
on dark energy, although we caution that this figure shows variations in
only one cosmological parameter.  Full forecasts, such as 
those presented in \S~\ref{sec:scienceForecastResults}, 
must marginalize over other cosmological
parameters such as $\Omega_m$ and $H_0$.
  }
\end{center}
\end{figure}

We have three possible explanations for the accelerating expansion of
the Universe:  a cosmological constant, equivalent to static dark
energy with $w=-1$; a dynamical dark energy with $w(a)\ne -1$;
or a failure of General Relativity. DESI is designed to address this fundamental question about the nature of the Universe. The challenge of distinguishing the cosmological constant solution from dark energy with $w$ near $-1$ is displayed in Figure~\ref{fig:rubin}.

The Dark Energy Spectroscopic Instrument (DESI) \citep{Levi2013gra}
will provide precise spectroscopic redshifts of more than thirty
million objects. From these will come three-dimen\-sional maps of the
distribution of matter covering unprecedented volume.  DESI will 
survey an enormous volume at $0.4<z<3.5$ using luminous red galaxies,
emission line galaxies, and quasars, producing tight constraints
on the large-scale clustering of the Universe.  
In addition, DESI
will perform a Bright Galaxy Survey (BGS) of the $z<0.4$ Universe, allowing
the study of cosmic structure in the dark-energy-dominated epoch
with much denser sampling.
These data will help
establish whether cosmic acceleration is due to a mysterious component
of the Universe or a cosmic-scale modification of GR, and will
constrain models of primordial inflation.

DESI will have a dramatic impact on our understanding of dark energy
through its primary measurement, that of baryon acoustic oscillations.
Waves that propagated in the electron-photon-baryon plasma before
recombination imprint a feature at a known comoving physical scale
(150 Mpc or $4.6 \times 10^{24}$ m) in the distribution of separations
between pairs of galaxies.  Localizing this baryon acoustic oscillation
(BAO) feature and comparing its apparent size to the known physical
scale provides a measurement of the distance to the galaxy sample and
thus the expansion history of the Universe.  The BAO measurement was
singled out by the Dark Energy Task Force \citep{2006astro.ph..9591A}
as having the fewest experimental uncertainties among the techniques
for measuring dark energy; it simply depends on the galaxy locations, rather
than their shapes or brightnesses. DESI's two-point correlation
measurements will also detect the anisotropies in galaxy clustering ---
redshift space distortions (RSD) --- due to the peculiar velocities of
galaxies generated by density perturbations. This gives a direct
measurement of the properties of gravity at each redshift, through its
effect on galaxies' motions.

In addition to the constraints on dark energy, the galaxy and
Ly-$\alpha$ flux power spectra will reflect signatures of neutrino
mass, 
scale dependence of the primordial density fluctuations from
inflation, and possible indications of modified gravity. To realize
the potential of these techniques requires an enormous number of
redshifts over a deep, wide volume and thus a substantial investment
in a new instrument with capabilities well beyond existing facilities 
and for which we can utilize a substantial portion of the observing time.

The DESI survey will have considerable impact beyond these cosmological
highlights on the study of galaxies, quasars, and stars.  Spectroscopy
is a core tool of astrophysics, and the ability to combine many millions
of spectra with modern wide-field, multi-wavelength imaging surveys will
yield rich opportunities.  While the DESI collaboration includes members
planning to work on these topics, we do not discuss these in this
design report, as they are not driving requirements.  We make one
brief exception for the Milky Way Survey (\S~\ref{sec:mws}), as it 
will involve a substantial number of targets that piggyback on the 
Bright Galaxy Survey, using fibers that have no suitable galaxy 
available within their patrol radius.


\clearpage
\subsection{Measuring Distances with Baryon Acoustic Oscillations}

DESI will measure the expansion of the Universe by observing the
imprint of baryon acoustic oscillations set down in the first 380,000
years of its existence. This pattern has the same source as the
pattern seen in the cosmic microwave background, but DESI will map it
as a function of cosmic time, while the CMB can see it only at one
instant.  The pattern is imprinted on all matter at large scales and
can be viewed by observing galaxies of various kinds or by observing
the distribution of neutral hydrogen across the cosmos, and shows up
as excess correlations at the characteristic distance of the sound
horizon at decoupling.

\subsubsection{Theory}


Initial fluctuations in density and pressure provided sources for sound
waves that propagated in the photon-electron-baryon plasma of the
early Universe (see, for example, \cite{Eisenstein:1997ik}). These
sound waves propagated with a speed approximately $c/\sqrt{3}$ until the
Universe cooled sufficiently for electrons and ions to recombine to neutral atoms, causing the
sound speed to drop dramatically.  An excess of matter was left both at the
source of the wave and at the surface where these waves terminated.  
The matter excesses at these locations left their imprint  on the large-scale structure of galaxies and
hydrogen gas.
Before a wave stopped, it traveled a co-moving distance $s
\approx 150$ Mpc, which can be computed to precision 0.3\% from
cosmological parameters extremely well measured in CMB.




Viewed transversely, the 150-Mpc ruler subtends an angle $\theta$ such that 
\begin{equation}
s=(1+z)D_A(z) \theta= \theta\int_0^z\frac{c\,dz'}{H(z')}\
\end{equation}
where $D_A(z)$ is the angular-diameter distance to an object at
redshift $z$. The final equality holds only if the curvature is zero.

While the CMB gives us a purely angular correlation function, the
characteristic scale is present in the three-dimensional distribution
of large-scale structure. Viewed along the line of sight, correlations
are enhanced for galaxy pairs separated by $\Delta z$ such that
\begin{equation}
  \frac{c \Delta z}{H(z)}\approx s
\end{equation}
This latter measurement requires a spectroscopic survey to resolve the
full three-dimensional density distribution of galaxies.

The observation of the peak in the two-point correlation function thus
provides a means of measuring both the angular diameter distance,
$D_A(z)$ and the Hubble expansion rate, $H(z)$.  The ability of the
BAO method to directly probe $H(z)$ is unique among dark
energy probes. This becomes progressively more important at higher
redshifts since $H$ measures the instantaneous expansion rate (and
through it, the total energy density of the Universe) while $D_A$
measures the integrated expansion history. Measuring both
improves our ability to distinguish between different cosmological
models.


\subsubsection{BAO in Galaxies}

\begin{figure}
\begin{center}
\includegraphics[width=3.5in]{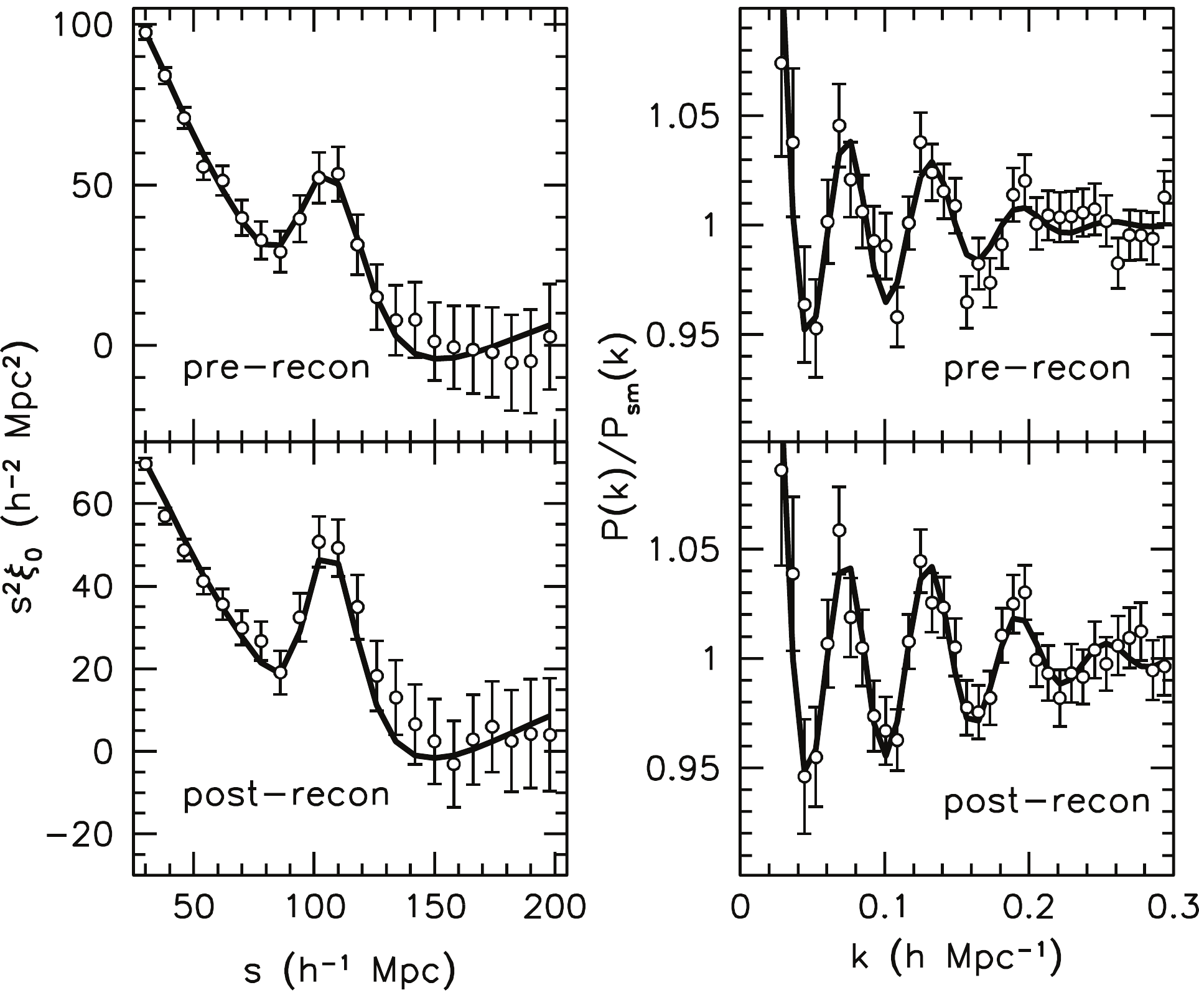}
\caption{The angle-averaged correlation functions [left] and power
  spectra [right], before [top] and after [bottom] reconstruction
  measured using the BOSS DR11 CMASS galaxy sample
  \citep{2014MNRAS.441...24A}.  The BAO feature is clearly detected at
  over $7\sigma$ as a peak in the correlation function and a
  corresponding set of oscillations in the power spectrum.}
\label{fig:BOSSBAOAnderson}
\end{center}
\end{figure}

The best developed application of the BAO technique uses galaxies as
tracers of the matter distribution; the BAO feature appears in the
two-point correlation function of galaxies, the probability, in excess
 of random, that two galaxies are separated by a distance $r$.   This has been achieved
with high statistical significance in several measurements spanning
the redshift range from $z=0$ to $z=1$. The highest significance
detection ($> 7 \sigma$) is currently that of the Baryon Oscillation
Spectroscopic Survey (BOSS) using the $z>0.45$ sample
\citep{2014MNRAS.441...24A,2016MNRAS.457.1770C}. We show representative data in
Figure~\ref{fig:BOSSBAOAnderson}. These data measured the
distance-like quantity $D_V(z) \equiv ((1+z) D_A)^{2/3}
(cz/H(z))^{1/3}$ to a redshift of 0.57 to 1.0\%, the most precise
measurement using the BAO technique. The lower redshift $z < 0.45$
sample in BOSS constrained the same combination of distances to $2\%$.
At still lower redshifts, the 6-Degree Field Galaxy Redshift Survey
\citep{2011MNRAS.416.3017B} measured the distance to $z=0.106$ with
4.5\% accuracy At a somewhat higher redshift, the WiggleZ galaxy
survey measured the distance to a redshift of $0.7$ to 4\%
\citep{2012MNRAS.425..405B}. This combination of these measurements
has for the first time enabled mapping the distance-redshift relation
purely from BAO measurements.

Most of these measurements used the galaxy correlation function
averaged over the orientation of the pair to the line of sight to
measure $D_V$, a combination of $D_A$ and $H$. More recent work has also
measured the correlation functions transverse and parallel to the line
of sight, allowing one to break the degeneracy between $D_A$ and $H$
that exists in purely angle averaged measurements.

The current generation of surveys is an excellent proving ground for analysis techniques. For instance, the BOSS
experiment compared analyses done in Fourier and in configuration space and used different algorithms
for estimating distances from the resulting two-point functions. All these yielded consistent distance measurements,
given the statistical precision of the measurements. While the level of consistency is not at the level required
by DESI, ongoing surveys provide a clear roadmap for developing and validating improvements to these analysis
techniques. The current measurements provide an important validation of our forecasts for DESI presented below.

The non-linear evolution of the matter density field broadens the acoustic peak,
potentially decreasing the precision on the distance measurement, and causes a small
shift in the peak location, thereby biasing the distance.
Ref. \citep{2007ApJ...664..675E} pointed out that because this broadening
is caused by the large-scale velocity flows resulting from gravitational 
forces, the effect may be substantially reversed by estimating the 
velocity fields from the large-scale structure map and
moving the galaxies back to their initial positions.  In addition
to a notable improvement in the recovered statistical errors, 
this reconstruction also mitigates the
shifts in the distance scale due to nonlinear evolution, with numerical tests
showing suppression to below
$0.1\%$.  Reconstruction was first applied to
the SDSS-II galaxy survey \citep{2012MNRAS.427.2132P}, improving the
statistical precision by a factor of $1.7$. Galaxy samples from more recent SDSS results, DR11, 
yield similar improvements after reconstruction. See Figure~\ref{fig:BOSSBAOAnderson}.
As with the other analysis methods, we expect improvements to
reconstruction algorithms before the DESI measurements become
available.
We however choose to be conservative and assume a
reconstruction performance similar to what has already been
demonstrated with current data.

\paragraph{Observational Systematics} The BAO method is simple in principle --- all one requires are the
three-dimensional positions of galaxies. The need to preserve the BAO
feature along the line of sight sets the requirement on redshift
precision. This precision, as stated in the Level 2 Survey Data Set
Requirements is $\sigma_z/(1+z) \sim 0.0005$ per galaxy, 
which is easily within the state-of-the-art and achieved throughout our wavelength range in the spectrograph design.


The angular and radial selection
functions of the survey can induce systematic uncertainties. The angular selection function is determined
by the imaging survey used for targeting, and may be spuriously
modulated by photometric calibrations, seeing and extinction
variations, and image deblending. All of these effects are
intrinsically angular effects and therefore may be separated from the
BAO feature, which is a feature in three-dimensional physical space
(not isolated to the angular degrees of freedom).  A similar
separation is possible for systematics in the radial selection
function of the survey.  The impact of these is therefore expected to
be small. In addition, there has been considerable work 
\citep{Ross12,2014MNRAS.444....2L}
developing techniques to further mitigate these effects. 

The ongoing BAO surveys provide the opportunity to
identify and quantify observational systematics. 
DESI will benefit greatly from this work, but it 
also faces some unique challenges.  The most important of these
arise from the fiber positioning
system and from the forest of sky lines, which impinge on the radial 
selection function.
The limited patrol radius of the fiber
positioners causes the highest density regions to be sampled less
completely than lower density regions.  This particularly affects the observer's line of sight and can skew the anisotropic
correlation pattern.  High sky brightness at 
certain wavelengths makes it difficult to find \otwo\ emission lines, thereby reducing 
the spectroscopic completeness at specific redshifts.
Initial studies have shown that these survey artifacts can influence 
the measured clustering, but we expect both to be correctable to good
accuracy, as the source of the variations can be tracked with high
fidelity.  
Finding the optimal method to achieve the full statistical precision
inherent in the data is an ongoing project of the science team.

\paragraph{Theoretical Systematics} 

The robustness and accuracy of the BAO method derive from the
simplicity of the early Universe and the precision with which we know
the speed and time of propagation of sound waves in the primordial
plasma. The evolution of density fluctuations in the Universe is very
well described by linear perturbation theory and is now exquisitely
tested by the recent measurements of temperature fluctuations in the
Cosmic Microwave Background radiation by the {\it Planck}
satellite \citep{2013arXiv1303.5076P,2015arXiv150201582P,2015arXiv150201589P}. 
The current CMB measurements constrain the size of the BAO
standard ruler to $0.3\%$. 
This uncertainty
is folded into our forecasts for DESI. Furthermore, any
miscalibrations in the acoustic scale would affect principally the
determination of the Hubble constant, not the dark energy constraints
\citep{2004PhRvD..70j3523E}.

The sound waves travel a comoving distance of 150 Mpc, setting the BAO
scale to be much larger than the scale of gravitational collapse even
in the present Universe (about 10 Mpc).  Analytical calculations,
verified by direct numerical simulations, have found the nonlinear
evolution of the density field alters the BAO scale by 
$0.3\%$ at the present epoch, and even less at the higher redshifts
probed by DESI.

Galaxy formation may result in an additional shift in the BAO
scale due to mismatched weighting of high and low density
regions. Initial perturbative and numerical studies 
\citep{2008ApJ...686...13S,2008PhRvD..77b3533C,SMSCSH08,2009PhRvD..80f3508P,2010ApJ...720.1650S,
2011ApJ...734...94M,2012PhRvD..85j3523S}
also find these shifts to be small, with the
most extreme shifts of order $0.5\%$. 
As mentioned above, density-field reconstruction applied to simulations
reduces these shifts to the 0.1\% level without the need for further 
modeling.  We expect that further modeling from theory and simulations 
will allow us to robustly limit these uncertainties to well below the 
DESI statistical limits.
In addition, the DESI target samples are designed to overlap in multiple redshift
ranges, allowing empirical tests of the robustness of the BAO measurements to
different tracer populations.

A recently discovered astrophysical effect that could affect the BAO
feature arises from the relative velocities of the baryons and the
dark matter at the recombination epoch
\citep{2010PhRvD..82h3520T,2015arXiv151003554B}. This modulates the formation of the
earliest protogalaxies and potentially could persist to their
descendants (some of which would be measured by DESI).  This
modulation is due to the same pressure forces that create the BAO, and
the impact could shift the measured acoustic scale.  While this effect is expected
to be negligible for the galaxies probed by DESI, the possibility of a 
systematic bias in the inferred distance scale can not
be ruled out on theoretical grounds.  Fortunately,
\cite{2011JCAP...07..018Y} demonstrate that this effect would also
create a distinctive three-point function signal measurable in DESI that
would diagnose any contamination from this effect (also \cite{2015MNRAS.448....9S}).

All of the above strongly argue that the theoretical systematic
effects associated with the BAO-scale measurements are either
intrinsically or correctable to below the $0.1\%$ level required by
DESI.

\subsubsection{BAO in the \texorpdfstring{\lya}{Ly-alpha} Forest}

\newcommand{\vx}{\mathbf{v}}
\newcommand{\vk}{\mathbf{k}}

Measuring BAO with galaxies as tracers is a mature method
\cite{2012arXiv1203.6594A,2012MNRAS.425..405B}.  Such measurements
become much more difficult for $z\gtrsim 2.0$ where galaxy redshifts
are harder to get.  However, measuring dark energy properties at this
high redshift allows us to probe the Universe well before the advent
of accelerated expansion.  An interesting possibility is that dark
energy density does not become completely negligible at high redshift,
as predicted by the cosmological constant or other models with
$w\simeq -1$, but rather remains at a level predicted by some
particle-physics models and detectable by future surveys
\cite{2012ApJ...749L...9R, 2010ApJ...714.1460A, 2009JCAP...04..002X,
  2008JCAP...06..004L, 2006JCAP...06..026D,
  2001PhRvD..64l3520Dt}. Such a component can only be measured or
excluded by a technique sensitive to the expansion history at high
redshift.

The \lyaf~provides the means to measure BAO at redshifts larger than
$2$. The forest is a collection of absorption features in the spectra
of distant quasars blue-ward of the \lya emission line
\cite{1971ApJ...164L..73L}. These features arise because the light
from a quasar is absorbed by neutral hydrogen in the intergalactic
medium. Since the quasar light is constantly red-shifting, hydrogen at
different redshifts absorbs at different observed wavelengths in the
quasar spectrum. The amount of absorption reflects the local density
of neutral hydrogen, which in turn traces the dark matter field on
sufficiently large scales.
Numerical simulations and analytical work show that for plausible
scenarios, the \lyaf is well within the linear biasing regime of
scales relevant for BAO \cite{MCDON03, 2009JCAP...10..019S,
  2012JCAP...03..004S}. Therefore, measuring three-dimensional
correlations in the flux fluctuations of the \lyaf provides an
accurate method for detecting BAO correlations
\cite{MCDON03,2003dmci.confE..18W,2007PhRvD..76f3009M,2011MNRAS.415.2257M}

\begin{figure*}[!tb]
	\centering
	\includegraphics[height=1.5in]{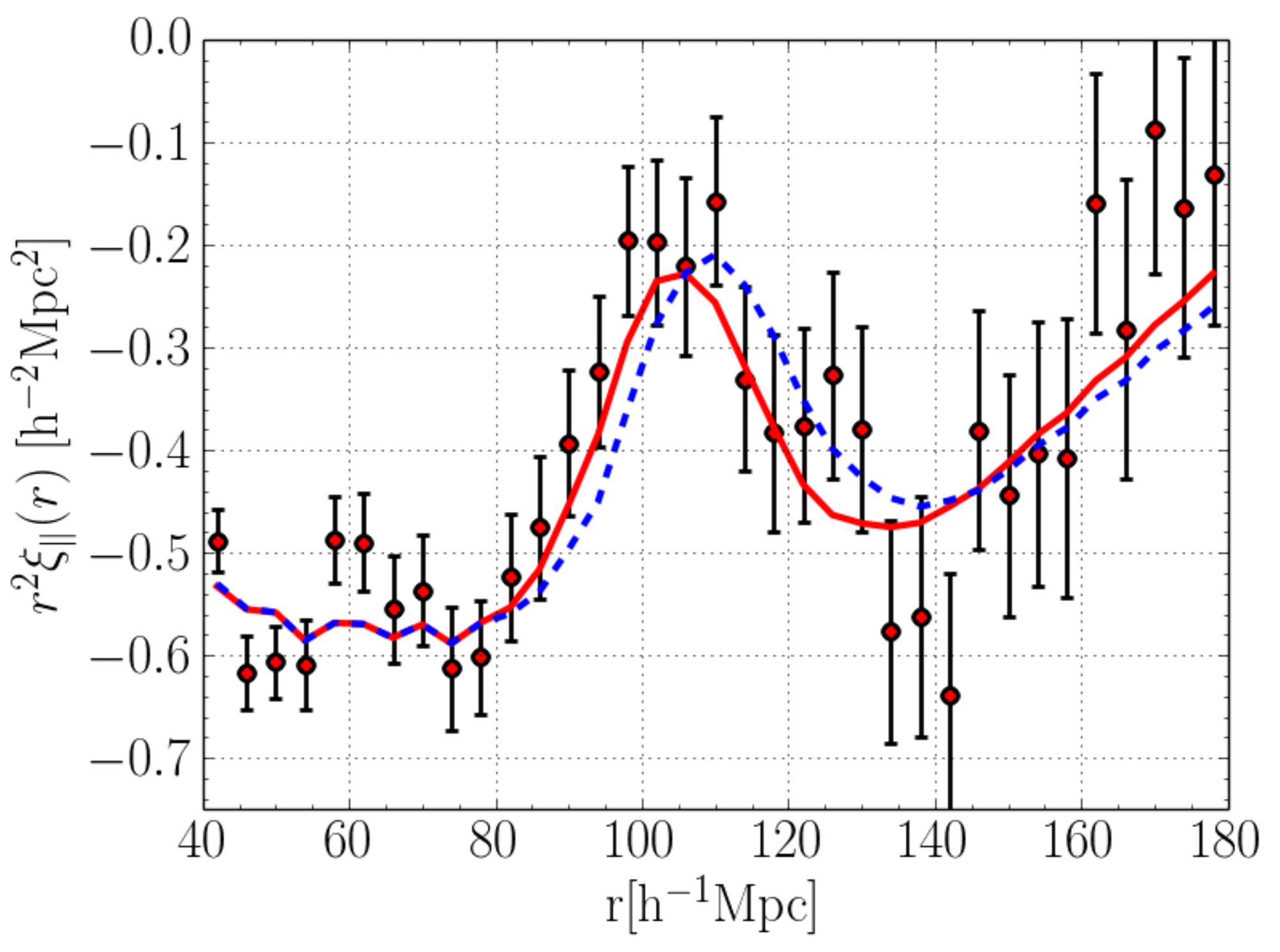}
	\includegraphics[height=1.5in]{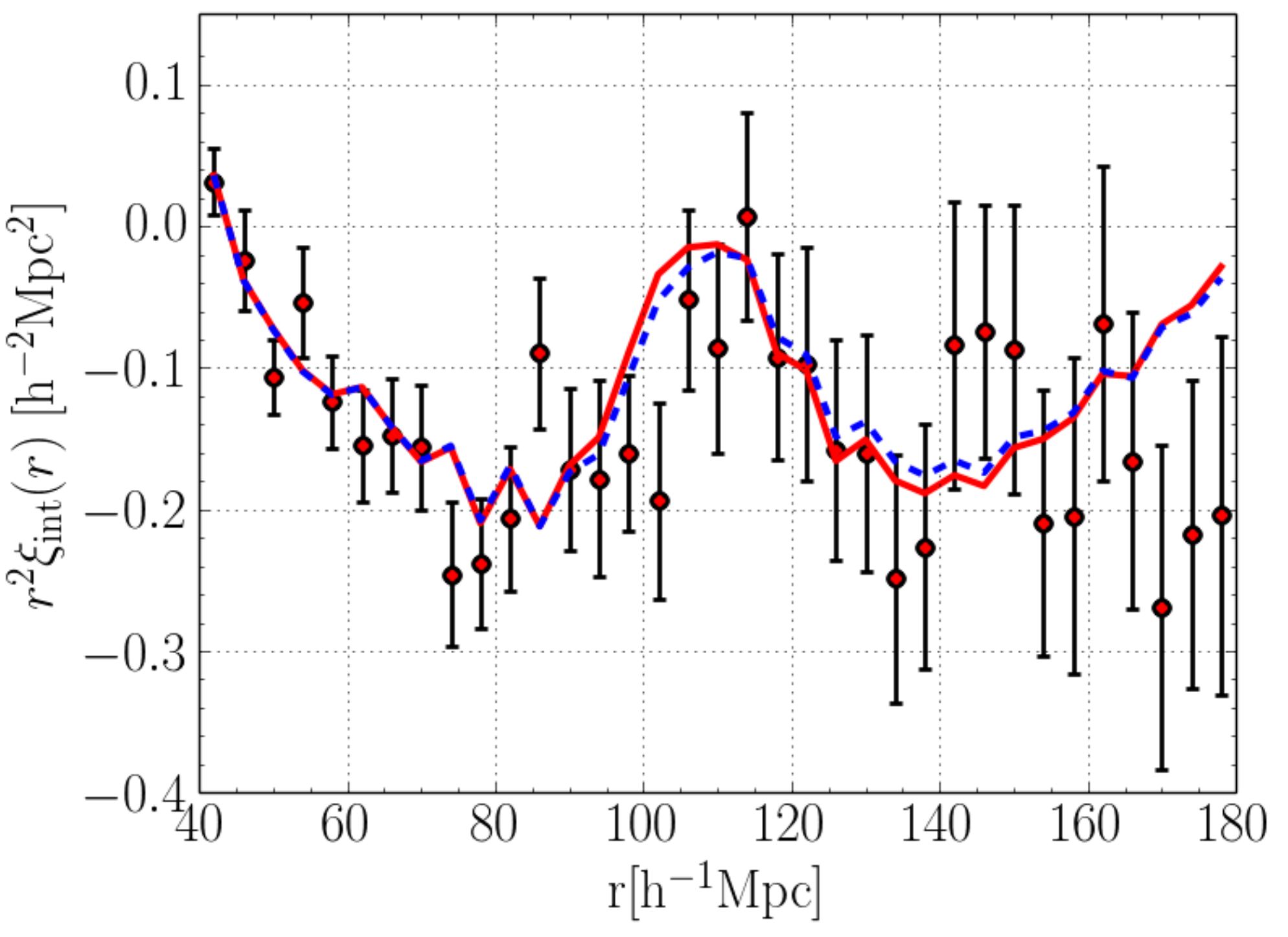}
	\includegraphics[height=1.5in]{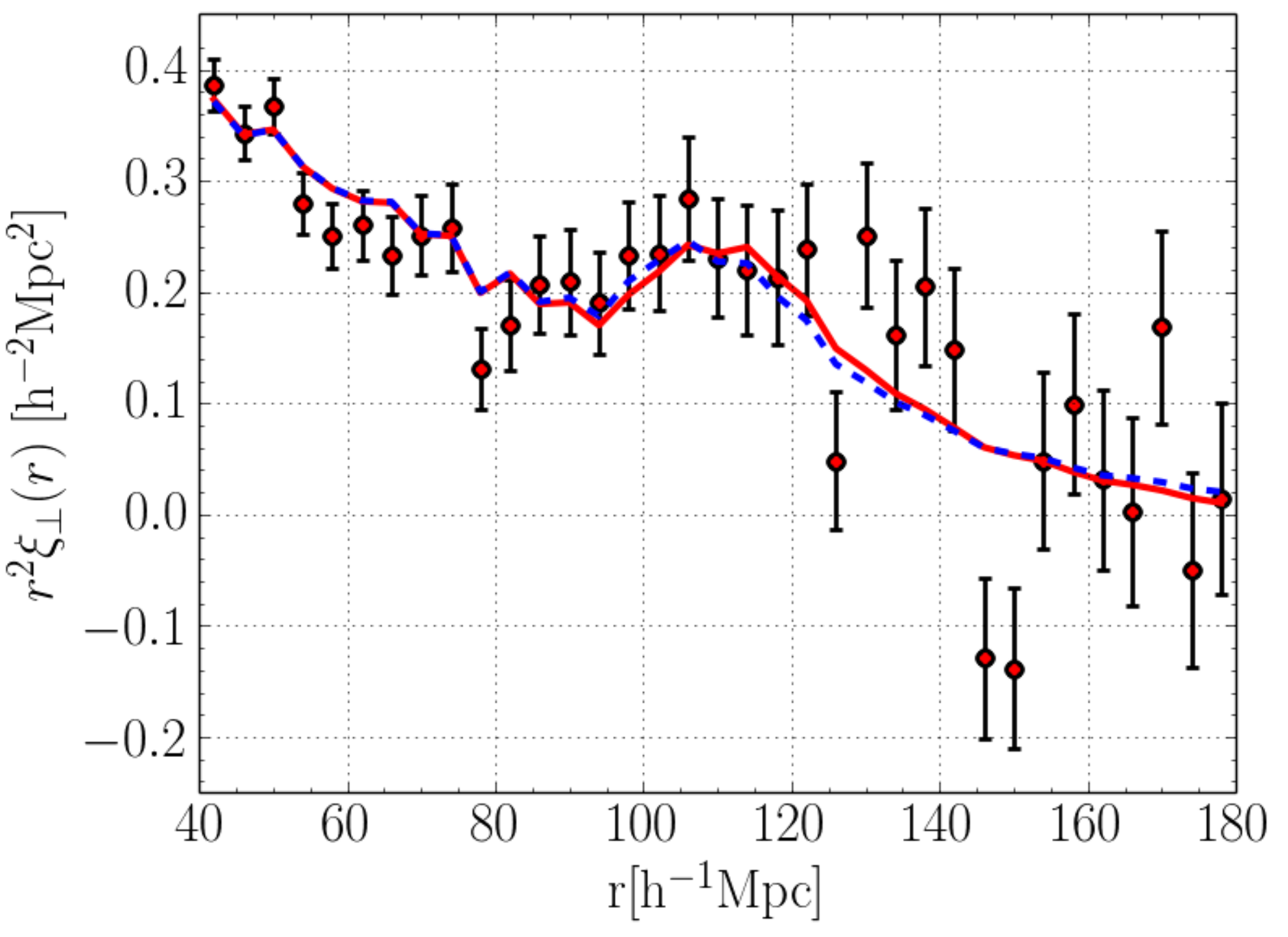}
	\caption{Correlation functions of \lyaf flux fluctuations
		based on the BOSS DR11 quasars \citep{2014arXiv1404.1801D}, 
		binned in the cosine of the angle to the line of sight, $\mu$
	    ($\mu=1$ is along the line of sight, $\mu=0$ is perpendicular
		to the line of sight). From left to right, the bins are
		$\mu > 0.8$, $0.5 < \mu < 0.8$ and $\mu < 0.5$. The points are the
		measured correlation function, the solid line is the best fit model, 
		while the dashed line is the best fit assuming a fiducial cosmology.
		These results measure the optimal combination $D_A^{0.3} H^{-0.7}$ to 
		$2\%$.
	}
\label{fig:lyabao}
\end{figure*}

Using the \lyaf to measure the three-dimensional structure of the
Universe became possible with the advent of BOSS, which was the first
survey to have a sufficiently high density of quasars to measure
correlations on truly cosmological scales. This was done in 2011
\cite{2011JCAP...09..001S}. At the beginning of 2013, the first
detection of BAO in the \lyaf was published in a series of papers
\cite{2013A&A...552A..96B,2013JCAP...04..026S,2013JCAP...03..024K}.
These were recently updated to the almost complete BOSS sample in
\cite{2014arXiv1404.1801D} (Figure~\ref{fig:lyabao}) yielding a
$5\sigma$ detection of the BAO feature.

The redshift-space distortions in the \lyaf are larger than in
galaxy-based measurements \cite{2011JCAP...09..001S,MCDON03}.  Thus
the signal-to-noise for the radial modes is considerably higher
than for transverse modes. Consequently, in contrast to the galaxy
measurements, the \lyaf BAO measurements measure the Hubble parameter
$H(z)$ with greater precision than the angular diameter distance
$D_A(z)$. For instance, \cite{2014arXiv1404.1801D} find that the 
combination $D_A^{0.3} H^{-0.7}$ is optimally constrained to $\sim 2\%$.



\paragraph{Systematics} 

Inevitably, there will be systematic effects that could distort the
\lya measurements, but these should produce broadband contamination
and would not affect our ability to measure an isolated feature in the
data, such as the BAO peak.  However, unless carefully accounted for,
these systematics could contaminate secondary science, such as \lya
broadband power measurements, neutrino masses and warm dark matter
constraints.

Astrophysical contaminants include sources of non-gravitational large
scale fluctuations, such as He II reionization and fluctuations in
the photo-ionization background 
\cite{2010ApJ...713..383W,2011MNRAS.415..977M,2014MNRAS.440.2406M,2014PhRvD..89h3010P,2014MNRAS.442..187G}.
There are also targeting systematics
-- quasars with significant absorption in the forest region are
considerably easier to target, since they are easier to distinguish
from stars. As a result, observed \lyaf regions are not sampling the
Universe randomly, but prefer overdense lines of
sight. Back-of-the-envelope calculations show that this effect is
small, although more work should be done to confirm
this\footnote{There is an additional effect because \lya quasar lines
of sight terminate in quasars, which are themselves tracers of the
underlying structure, but this can be explicitly shown to be a small
effect.}.  Finally, there are metal contaminations. For example, Si
III that tracks the hydrogen fluctuations produces a line that contaminates the \lyaf\
flux measurements at separation of 2271 km/s. The cross-correlation
between \lyaf\ absorption and Si III absorption, if misinterpreted as
Ly-$\alpha$-to-\lya correlations could bias the BAO
measurements \cite{2011JCAP...09..001S,2014arXiv1404.1801D}.  Further
contamination arises where the metal absorption traces large scale
structures at a significantly different redshift. For example C IV
traces structure at $z=1.7$ at wavelengths which probe the \lyaf\ at
$z=2.4$ \cite{2014arXiv1404.4569P}.  For BAO measurements these can be
reliably corrected by including them as a part of the model. For other
uses, such as broadband power spectrum measurements, a combination of
nuisance modeling, accurate mock spectra and numerical simulations
should remove any potential biases associated with these
complications.

Perhaps the most important systematic effects will come from
imperfections in the instrument and data reduction.  For example,
artificial features in the mean transmission at the position of
galactic Balmer transitions were noticed in BOSS data
\cite{2013A&A...552A..96B}. These were tracked down to the imperfect
interpolation in calibration vectors when these features were masked
in calibration stars. Although such effects are on average
calibrated out, they can in principle produce sharp features in
correlation at certain pairs of wavelengths that could potentially
contaminate the BAO measurements.  Other effects include noise
calibration and its Poisson nature, imperfect sky subtraction,
etc. Fortunately, there are no fundamental obstacles to modeling the
listed systematics with a carefully executed pipeline. The sheer
amount of data that will be available and the relatively high
signal-to-noise of true small scale fluctuations in the forest will
allow us to check the data in many different ways and validate the data reduction pipeline.


\clearpage
\subsection{Measuring Growth of Structure with Redshift Space Distortions}

DESI will observe redshifts, which reflect the velocities due to
expansion, but also the peculiar velocities due to gravitational
attraction by large scale structure.  Peculiar velocities are
observable in redshift surveys because they alter the correlations between
galaxies along the line of sight, resulting in an anisotropy
in the observed clustering.
Comparing the expansion history and the growth of large scale
structure from redshift space distortions will allow DESI to test
General Relativity.

\label{s2:rsd}

\subsubsection{Theory}


Galaxies and quasars are point tracers of the underlying cosmic
structure. The physics of how they trace the dark matter fluctuations
is well understood based on arguments about locality of galaxy
formation \cite{2009JCAP...08..020M,2012PhRvD..86h3540B,
2006PhRvD..74j3512M}.  On
very large scales bias is scale independent and redshift-space distortions are
described by linear perturbation theory.
Beyond-linear perturbative
corrections can be used on intermediate scales before perturbation theory
breaks down entirely on small scales
\cite{2011JCAP...11..039S,
2012JCAP...11..009V,2012JCAP...11..014O}.


The measurement of the growth of structure relies on redshift-space distortions seen in galaxy surveys.  Even though we expect the clustering of galaxies in real space to have no preferred direction, galaxy maps produced by estimating distances from redshifts obtained in spectroscopic surveys reveal an anisotropic galaxy distribution. The anisotropies arise because galaxy redshifts, from which distances are inferred, include components from both the Hubble flow and peculiar velocities driven by the clustering of matter. Measurements of the anisotropies allow constraints to be placed on the rate of growth of clustering~\citep{2010Natur.464..256R,2011PhRvD..84h3523S}.

On large scales, the observed large-scale structure is basically described by a small fractional perturbation $\delta(\bf x)=\delta\rho(\bf x)/{\overline \rho}=(\rho(\bf x)-{\overline \rho})/{\overline \rho}$   
 to the uniform density. Ignoring the higher-order contributions, the perturbation in redshift space ($\delta_{s}$) is related to the real space perturbation at  
directional cosine $\mu$ between line-of-sight direction and the wave-number ${\bf k}$, by the Kaiser relation ~\citep{1987MNRAS.227....1K},
\begin{equation}
\delta_s({\bf k})=\delta({\bf k})(1+\beta\mu^2)
\label{eq:Kaiser}
\end{equation}
Here $\beta=f/b$, where $b$ is the galaxy bias and $f$ is related to the linear growth function $D(a)$ by
\begin{equation}
f=\frac {d\ln D(a)}{d\ln a} \,.
\end{equation}
In the linear regime, density perturbations grow proportional to $D(z)$ which increases with 
decreasing $z$. 

In GR, $D(z)$ is completely specified by the expansion history even in the presence of dark energy; this is no longer generically true in alternative theories
of gravity. The behavior of $f$ in GR is given, to a good approximation, by
\begin{equation}
\label{e.growthindex}
f \simeq \Omega_m(z)^\gamma,
\end{equation}
where $\gamma$ is the growth index, approximately 0.55 in GR, 
and where $\Omega_m(z)$ is the fraction of the total energy density 
in the form of matter at redshift $z$.  
In alternative gravity theories, a common simple parameterization of the modified
growth rate is to alter the growth index $\gamma$.
\cite{2010MNRAS.404..239S}
demonstrated that a DESI-like survey could constrain $\gamma$ to
0.04 ($7\%$). More general modifications might involve modifying (in
a time- and scale-dependent manner) the potentials that enter the
metric. Precise growth measurements over a wide range of redshifts
and scales, combined with constraints from overlapping CMB and weak
lensing surveys, make large galaxy surveys like DESI excellent probes of gravity (see
\cite{2013arXiv1309.5385H} for a recent review).  Here, we focus on 
scale-independent growth rates for large-scale structure, but the 
DESI data set will allow more complicated investigations.

As an important example of extensions, we highlight the Bright Galaxy
Survey, where we will be mapping a smaller volume ($z<0.4$) at
substantially higher number density and with more diversity of
galaxies.  This redshift range is crucial because it is when dark
energy dominates and any associated modifications of gravity would be
expected to be strongest.  Getting the best precision out of this
limited volume requires spectroscopy to produce a 3-D map of the
density field.  The BGS will test for modifications of gravity
directly via the redshift-distortion method, including the novel
methods of using multiple tracers in order to suppress sample variance
\cite{2012PhRvD..86j3513H}.  But the search can be extended via
spectroscopic detection of clusters and groups, along with galaxy
halo occupation modeling, to measure the amplitude of clustering by
halo abundances \cite{2013MNRAS.430..725V, 2012ApJ...745...16T}.  The
maps can also be correlated with weak lensing maps (e.g., from DES,
LSST, Euclid, or CMB-S4) to measure the amplitude of clustering
\cite{2012PhRvD..86h3504Y,2013MNRAS.432.1544M}. Comparing the observed
velocity field to the expected velocity field sourced from the lensing
matter overdensities enables further tests of modified gravity models
of cosmic acceleration \cite{2010Natur.464..256R}.  Finally, the more
detailed map will allow tests of screening theories on smaller scales
\cite{2013arXiv1309.5389J,2013ApJ...779...39J}, in which one considers
the response of individual galaxies to the predicted gravitational
field.

In the Kaiser approximation, the redshift space power spectrum, $P_{s}$, is given by
\begin{equation}
P_s({\bf k})= (b+f\mu^2)^2 P_m({ k})
\end{equation}
where $P_{m}$ is the linear theory mass power spectrum. In principle,
this prescribed anisotropy provides a means of measuring $f$, and
through it the growth of gravitational structures. However, in the
above, the measurements of $f$ are degenerate with the amplitude
of the matter power spectrum. Therefore the combination $f(z)
\sigma_8(z)$ is the actual observable, where the normalization of
the power spectrum $P(k)$ is proportional to $\sigma_8^2(z)$
\footnote{$\sigma_8^2$ is defined to be the variance of the matter
density field averaged in spheres of 8 $h^{-1} {\rm Mpc}$ and
traditionally used to parametrize the amplitude of the power spectrum.
}.

\subsubsection{Systematics}

Galaxies are expected to follow the same gravitational
potential as the dark matter and hence have the same velocities. 
The main theoretical systematic uncertainty in RSD is that nonlinear velocity effects extend to rather large scales and
give rise to a scale-dependent and angle-dependent clustering signal. It is easy to see these effects in any
real redshift survey: one sees elongated features along the line of sight,
called the Fingers of God (FoG).
The FoG are caused by random velocities inside virialized objects such as
clusters, which scatter galaxies
along the radial direction in redshift space, even if they have a localized
spatial position in real space.
This is just an extreme example and other related effects, such as nonlinear infall streaming
motions, also cause nonlinear corrections. In addition, RSD measure velocities as sampled at the 
galaxy positions. One is thus probing not the velocity field, but rather the momentum density field. 
Galaxies are a biased tracer of the dark matter and this introduces scale dependent effects into 
RSD statistics even if galaxies are simply a linear tracer of the dark matter. 

There are a plethora of approaches 
\cite{2011MNRAS.417.1913R, 2013MNRAS.429.1674C,2013PhRvD..87l3510S,
2011JCAP...11..039S,
2012JCAP...11..009V,2012JCAP...11..014O}
to modeling redshift space
distortions in the literature, and the analyses in Table
\ref{tab:fs8constraints} make use of many of them.  It has been firmly
established that the Kaiser formula is inadequate to recover
information faithfully on the quasilinear scales of interest, and so
most analyses now adopt some form of perturbative corrections.
However, because these corrections depend strongly on the halo bias
\cite{ReiWhi11,Vlah13}, methods calibrated on purely the dark matter power 
spectrum are of limited utility.
Moreover, the details of the mapping between galaxies and dark matter
halos also strongly modify the correlation function, mostly through
FoG effects.
All of these effects can induce 10\% effects on RSD at $k \sim 0.1$~$h$/Mpc.  Current 
models of RSD are able to reproduce these nonlinear effects at the percent level for $k<$~0.05--0.1~$h$/Mpc.  Extending 
this to smaller scales would increase the power of the DESI RSD survey. This will require us to improve our bias models and the realism of our simulations.

Most of the observational systematics examined in detail in the SDSS-III BOSS 
\cite[see][]{Ross12} primarily affect clustering on the largest scales;
currently these are of little concern for RSD measurements, for which the signal
comes primarily from the smallest scales included in the measurements.  The most
important systematic effect is the estimate of a survey's radial selection
function \cite{SamPerRac11,Ross12}.  Since the redshift distribution of targets
cannot be predicted precisely a priori, it must be measured directly from the
observed galaxies' redshift distribution.  Doing so removes some cosmological
radial modes from the observed galaxy overdensity field, resulting in a bias in
the monopole-quadrupole amplitudes at the $<0.2 \sigma$ level.  The ratio of
systematic to statistical uncertainty should remain relatively constant with
survey area for a given redshift distribution, since the statistical errors on
the correlation function and $n(z)$ shrink at the same rate.

\subsubsection{Current Status of RSD Measurements}

\begin{figure}[!htb]
\begin{center}
\includegraphics[height=2.5in]{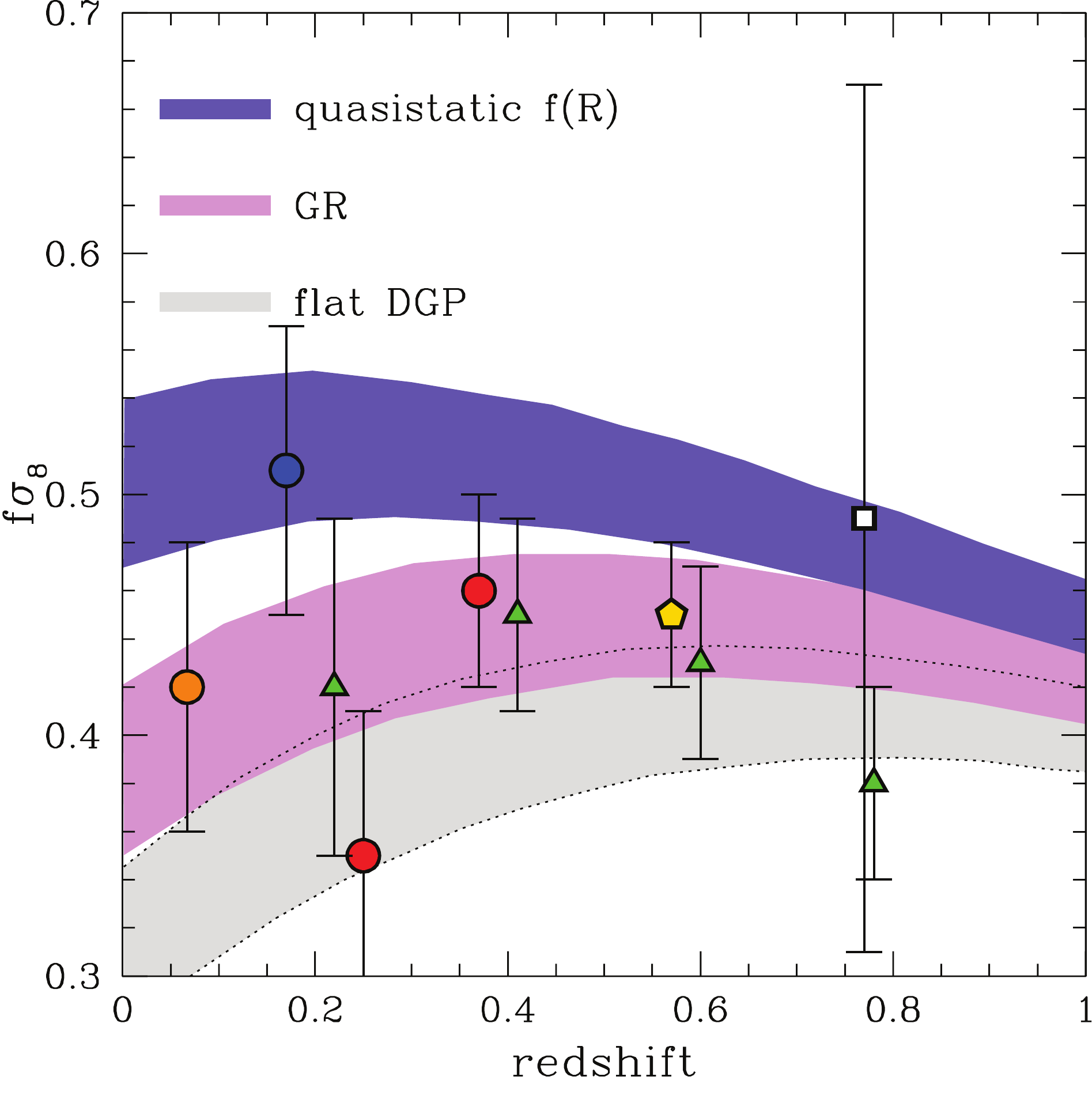}
\includegraphics[height=2.5in]{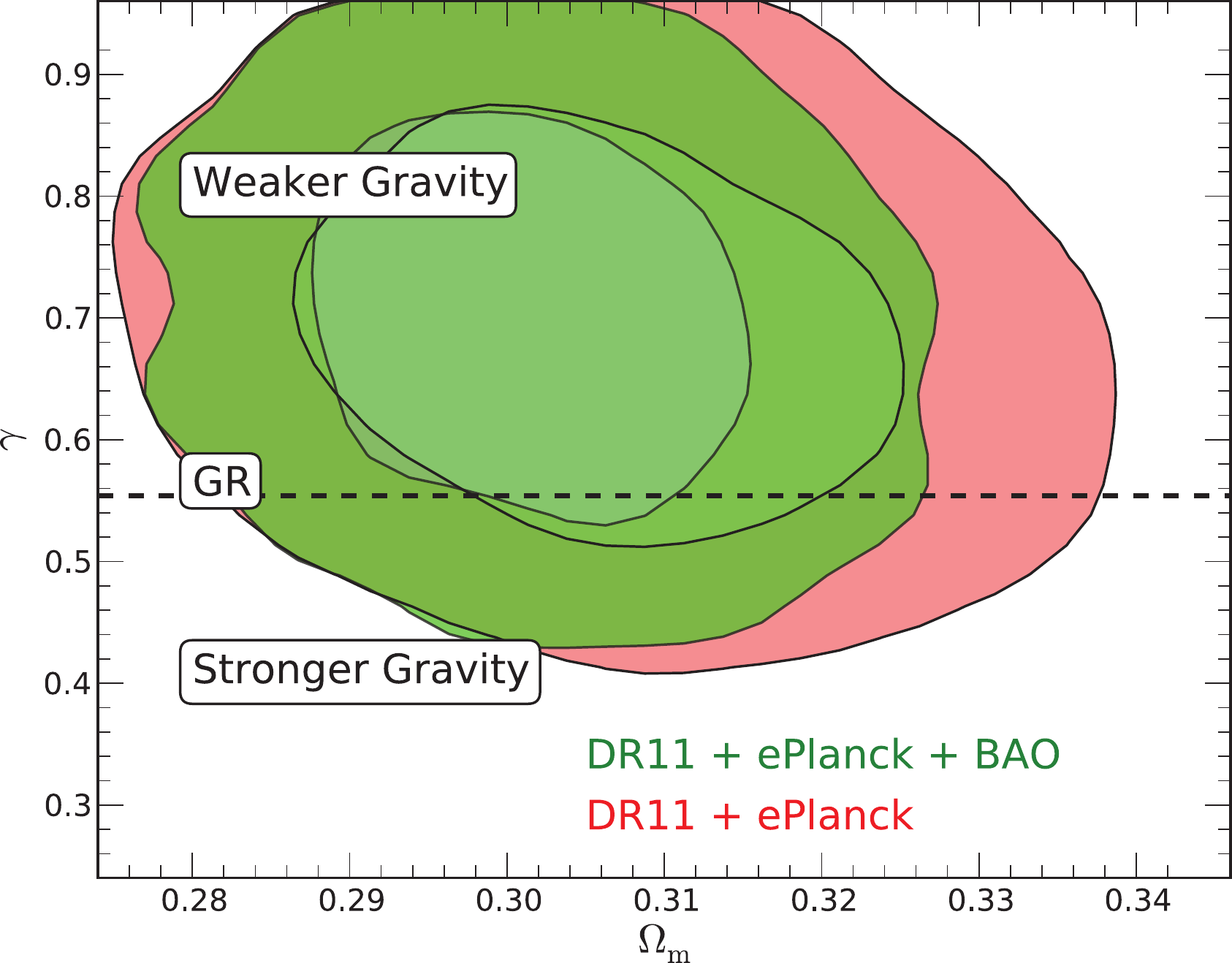}
\caption{{\em Left:} The data points show the CMASS DR11 measurement
  of $f\sigma_8$ (gold pentagon; \cite{Samushia14})
along with similar, low redshift, measurements and $1\sigma$ error bars
as presented in Table~\ref{tab:fs8constraints}. The three stripes show
theoretical predictions for different gravity models allowing for uncertainty in
the background cosmological parameters, constrained using only the
WMAP 7 data \cite{Komatsu11}.  Figure adapted from \cite{Samushia13}.
{\em Right:} Joint constraints in the $\Omega_m$-$\gamma$ plane from
BOSS DR11, where $\gamma$ is the growth index of structure, as defined
in Eq. (\ref{e.growthindex}). Figure taken from \cite{Samushia14}.
}
\label{fig:measfs8}
\end{center}
\end{figure}
\begin{figure}[!hbt]
\begin{center}
\includegraphics[height=2.5in]{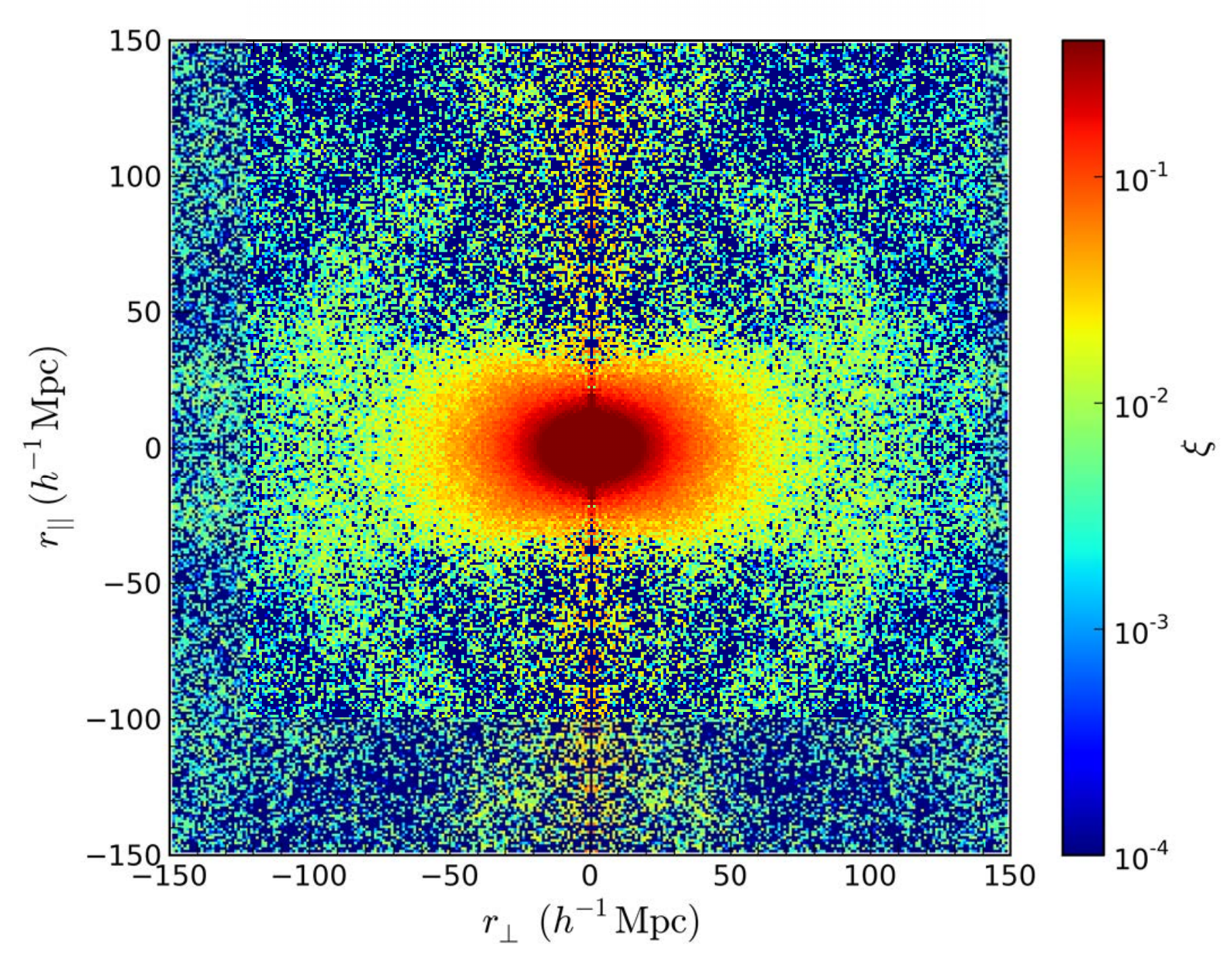}
\caption{The two-dimensional correlation function of the BOSS 
DR11 CMASS galaxies, measured perpendicular (x-axis) and parallel (y-axis)
to the line of sight. The BAO ring, distorted by redshift space distortions
is clearly visible, as is the characteristic squashing of the correlation 
function on large scales.}
\label{fig:butterfly}
\end{center}
\end{figure}

Redshift-space-distortion measurements have now been performed on a
host of surveys, which we summarize in Table \ref{tab:fs8constraints}
and show in the left panel of Figure~\ref{fig:measfs8}; taken
together, these surveys provide a measure of the growth rate of cosmic
structure good to about $ 3\%$ in the low redshift Universe.  Almost
all of these measurements of $f\sigma_8$ are derived from the
anisotropy in the two-point correlations of the observed galaxy
density field.
The anisotropic correlation from SDSS-III BOSS DR11 CMASS sample is
shown in Figure~\ref{fig:butterfly}.  While there have been some
analyses directly on the two-dimensional correlation function
$\xi(r_p, r_\pi)$ \cite[e.g.,][]{Okumura08,Song11,Beutler12,2014arXiv1407.2257S}, 
most
authors further compress the data into multipoles
\cite[e.g.,][]{Bla11a,SamPerRac11,Reid12,Beutler13,Samushia14} or wedges \cite{Kazin13,Sanchez14}.
Efficient information compression is necessary when the covariance
matrix of the observables 
are estimated from a finite number of mock surveys
\cite{2014MNRAS.439.2531P}.

\begin{table}[!t]
\begin{center}
\caption{Compilation of RSD-based $f\sigma_8$ measurements from 
\cite{Samushia13}.  For the BOSS DR11 galaxy sample we cite the
measurement of \cite{Samushia14}. Other analyses of DR11 find
consistent results
\cite{Sanchez14, Beutler13}} \label{tab:fs8constraints}
\begin{tabular}{llll} 
\hline
z & $f\sigma_8$ & survey &
reference\\ \hline 
0.067 & $0.42 \pm 0.06$ & 6dFGRS & \protect\cite{Beutler12}\\
0.17 & $0.51 \pm 0.06$ & 2dFGRS & \protect\cite{Per04}\\ 
0.22 & $ 0.42 \pm 0.07$ & WiggleZ & \protect\cite{Bla11a}\\ 
0.25 & $0.35 \pm 0.06$ & SDSS LRG & \protect\cite{SamPerRac11}\\ 
0.37 & $0.46 \pm 0.04$ & SDSS LRG & \protect\cite{SamPerRac11}\\ 
0.41 & $ 0.45 \pm 0.04 $ & WiggleZ & \protect\cite{Bla11a}\\ 
0.57 & $0.45 \pm 0.03$ & BOSS CMASS & \protect\cite{Samushia14}\\ 
0.6 & $ 0.43 \pm 0.04$ & WiggleZ & \protect\cite{Bla11a}\\ 
0.77 & $0.49 \pm 0.18$ & VVDS & \protect\cite{Guz08}\\
0.78 & $ 0.38 \pm 0.04$ & WiggleZ & \protect\cite{Bla11a}\\
0.80 & $0.47 \pm 0.08$ & VIPERS &   \protect\cite{2013AA...557A..54D}\\
1.4  & $0.48 \pm 0.12$ & FastSound & \protect\cite{2015arXiv151108083O} \\
\hline
\end{tabular}
\end{center}
\end{table}

Most of these measurements assume a flat $\Lambda$CDM cosmology to model
the redshift-distance relation (see \cite{2014arXiv1407.2257S} for an
exception);
dropping this assumption degrades the measurement of
$f\sigma_8$.  However, the combination of geometric and dynamical
constraints available from the analysis of anisotropic galaxy
clustering is quite complementary to isotropic BAO measurements for
constraining dark energy. For instance, in the case of SDSS-III BOSS DR11 for a flat
$w$CDM cosmology, the combination of Planck and the BOSS BAO measurements
constrain $w = -1.01 \pm 0.08$
\cite{2014MNRAS.441...24A}, while including the geometric and
dynamical information in the quadrupole correlation function (term
proportional to $\mu^2$) yields $w=-0.993 \pm 0.056$
\citep{Samushia14}.

Considering instead tests of gravity given a ``known'' expansion
history, Figure~\ref{fig:measfs8} shows that for a flat $\Lambda$CDM
cosmology in general relativity, the predicted redshift evolution of
the observable $f\sigma_8$ is quite mild in the redshift range that
has been studied observationally.  These observations can begin to
distinguish between gravity models ($f(R)$ and $DGP$ are shown),
though there is still substantial uncertainty in the theoretical
predictions simply due to uncertainties in both the matter density
$\Omega_m$ and overall matter power spectrum normalization,
$\sigma_8$. The right-hand side of Figure \ref{fig:measfs8} shows
constraints in the $\Omega_m$-$\gamma$ plane from BOSS DR11
\cite{Samushia14}. These data yield a 16\% constraint on the growth
index. DESI will improve on the precision of the growth constraint
from all previous measurements by a factor of $\sim$4--10
\cite{2014JCAP...05..023F}, depending on advances in analysis and
theoretical modeling.  In addition, it will provide measurements to
significantly higher redshifts.


  Two surveys in particular are pathfinders for DESI
targets: WiggleZ \cite{Drinkwater10} analyzed emission line galaxies with bias $b$ near 1,
while SDSS-II and SDSS-III BOSS study  luminous red galaxies (LRGs) with a bias near 2.  
WiggleZ included much smaller scales in
their RSD analysis, which led to impressive constraints given the number of
galaxies in the survey.  However, they were not able to generate easily a large
$N$-body simulation volume capable of resolving the halos expected to host
emission line galaxies, and so their theoretical modeling is necessarily less
well-tested.  By comparison, LRGs are hosted by massive halos that can easily be 
simulated.  The perturbation-based model of \cite{Reid12} was carefully
calibrated against $N$-body-based mock-galaxy catalog and included realistic
effects like the ``Fingers-of-God'' (the elongated structure in the right panel of
Figure~\ref{fig:butterfly}).  However, because these effects are so strong, their
analysis was restricted to relatively large scales.  

Ongoing
progress in combining the perturbative analytic results with those of
N-body simulations should pave way for the increased theoretical
prediction accuracy necessary to extract RSD information at small
spatial scales


\clearpage
\subsection{Distance, Growth, Dark Energy, and Curvature Constraint Forecasts}

\label{sec:scienceForecastIntro}
DESI's observational program defined in the Requirements Document and described in this Report specifies the numbers of galaxies and \lyaf sources and their distribution that will be measured.  Using the specified quality of those observations, we can predict the precision with which cosmological parameters will be determined by DESI.  Thanks to the unprecedented scope of DESI's spectroscopic measurements, these measurements will take us to a new level --- Stage-IV --- in cosmological exploration.



\subsubsection{Forecasting Overview}

\label{sec:scienceForecastMethod}
\newcommand{\allzgoalR}{0.17}
\newcommand{\allzreqR}{0.21}
\newcommand{\lowestzgoalR}{0.22}
\newcommand{\lowestzreqR}{0.28}
\newcommand{\midzgoalR}{0.31}
\newcommand{\midzreqR}{0.39}
\newcommand{\elgzgoalR}{0.37}
\newcommand{\elgzreqR}{0.46}
\newcommand{\hzgoalH}{0.84}
\newcommand{\hzreqH}{1.05}
\newcommand{\lzreqH}{0.58}
\newcommand{\lzreqD}{0.35}
\newcommand{\sysreqH}{0.26}
\newcommand{\sysreqD}{0.16}
\newcommand{\peskgoalRSD}{0.74}
\newcommand{\opkgoalRSD}{0.38}

We use the Fisher matrix formalism to estimate the parameter constraining
power of the finished survey, largely following \cite{2014JCAP...05..023F}.
Our baseline cosmological model is flat $\Lambda$CDM.  This
model is specified by seven parameters, which are listed together with
their fiducial values in Table \ref{tab:fid}.   Parameter symbols have their
conventional meanings.
%
Our standard fiducial parameter values follow the {\it Planck} 2013
results, specifically the P\-+WP\-+highL\-+BAO (P from {\it Planck}, WP from {\it WMAP}, highL from high resolution CMB experiments like ACT and SPT) column of Table 5 of
\citep{2013arXiv1303.5076P}.  The difference from the {\it Planck} 2015
is negligible for these purposes.  In addition to the
conventional six parameters of the minimal cosmological model, we also
always vary the amount of tensor modes; however
this is largely irrelevant because the
T/S measurement is completely dominated by $Planck$ and essentially
uncorrelated with other parameters.

\begin{table}[hbt]
  \caption{Parameterization of the cosmological model and
    parameter values for the fiducial model.  The seven parameters in the upper part of the table are always free. Parameters
    in the second half of the table are extensions of the simplest model
    discussed below.
      }

  \label{tab:fid}

  \centering
  \begin{tabular}{ccp{9.5cm}}
\hline
    Parameter & Value & Description \\
\hline
$\omega_b$ & $0.02214$ & Physical baryon density $\omega_b=\Omega_b h^2$ \\
& & $h =H_0/(100~{\rm km~s^{-1} Mpc^{-1}})$)\\
$\omega_m$ & $0.1414$  & Physical matter density $\omega_m=\Omega_m h^2$
(including neutrinos which are non-relativistic at $z=0$)\\
$\theta_s$ & $0.59680$ degrees & Angular size of sound horizon at the
surface of last scattering acting as a proxy for Hubble's constant
 \\
$A_s$ & $2.198\times 10^{-9}$ & Amplitude of the primordial power
spectrum at $k=0.05 \iMpc$ (for the numerical Fisher matrix we actually
use $\log_{10}A_s$)
\\
$n_s$ &  $0.9608$ & Spectral index of primordial matter fluctuations
with $P(k) \propto k^{n_s}$\\
$\tau$ & $0.092$ & Optical depth to the last scattering surface
assuming instantaneous reionization.\\
$T/S$ & $0$ & Ratio of tensor to scalar perturbations (we assume
inflationary tensor fluctuation's spectral index $n_t=-\frac{1}{8}T/S$) \\
\hline
$w_0$ & $-1$ & Equation of state of dark-energy $p=w\rho$ \\
$w_a$ & $0$ & Variable equation of state of dark energy of the form
$w=w_0+(1-a)w_a$ \\
$\Omega_k$ & 0 & Curvature of the homogeneous model \\
$\alpha_s$ & 0 & Running of the spectral index $\alpha_s=d \log n_s/d \log
k$ with pivot scale $k=0.05 \iMpc $ \\
$\summnu$ & $0.06$ eV & Sum of neutrino masses (we assume they are degenerate)
\\
$\Nnueff$ & 3.04 & Effective number of neutrino species
($\Nnueff>3.04 \rightarrow$ dark radiation). \\
\hline
  \end{tabular}
\end{table}

Isolating the BAO feature gives the most robust, but also most
pessimistic, view of the information that one can recover from galaxy
clustering measurements, since BAO can be measured even in the
presence of large unknown systematic effects (very generally, these
will not change the BAO scale \cite{2010ApJ...720.1650S}).
We  quote errors on the transverse and radial BAO scales as errors on
$D_A(z)/s$ and $H(z) s$, respectively, where $s$ is the BAO length scale. For
galaxy and quasar clustering, these measurements are correlated at each redshift
with a correlation coefficient of 0.4.

We also quote errors on an isotropic dilation factor $R/s$, defined as
the error one would measure on a single parameter that rescales radial
and transverse directions by equal amounts.
In this case, for a small change in $R$, the corresponding variations in the model values of $D_A$ and $H$ are
\begin{equation}
D_A = \left(1+\frac{\delta R}{R_{\rm fid}}\right) D_{A,{\rm fid}}
\end{equation}
and
\begin{equation}
H = \left(1+\frac{\delta R}{R_{\rm fid}}\right)^{-1} H_{\rm fid}
\end{equation}
where $D_{A,{\rm fid}}(z)$ and $H_{\rm fid}(z)$ are the angular diameter
distance and Hubble parameter in a fiducial Universe.
An explicit definition of $R$ in terms of
the measured $H$ and $D_A$ is generally not needed and depends on the experimental
scenario.  The simplest cases are easy to understand: for a purely
transverse measurement (e.g., photometric survey) $R=D_A$, while for a
purely radial measurement (e.g., something closer to the \lyaf, although it is
not purely radial) $R=H^{-1}$ (or
$R=H^{-1} H_{\rm fid} D_{A,{\rm fid}}$, 
if one is concerned about inequivalent units). 
For intermediate cases like
typical galaxy clustering, the appropriate combination of $H$ and $D_A$ can
always be determined given the covariance matrix between them.  For example,
it is approximately proportional to $D_V(z) \equiv ((1+z) D_A)^{2/3}
(cz/H(z))^{1/3}$ in analyses of spherically averaged
clustering, such as from 6dF, BOSS, and WiggleZ.


Going beyond BAO, we use ``broadband'' galaxy power, i.e. measurements
of the power spectrum as a function of redshift, wavenumber and angle
with respect to the line of sight.  This treatment automatically
recovers all available information from the two-point clustering, i.e. not just the shape of the
isotropic power spectrum, but also redshift-space distortions,
Alcock-Paczynski \cite{1979Natur.281..358A}, and the BAO information.

The broadband Fisher matrix is calculated by combining the inverse variance
of the power spectrum $P({\bf k})$ of each Fourier mode with the derivative of power
in each mode with respect to set of cosmological parameters.  We divide the survey
into a set of redshift slices and coadd the resulting matrices.
The model for the three-dimensional power spectrum of the galaxy or Ly-$\alpha$
distribution is
\begin{equation}
\tilde{P}(k,\mu,z) = b(z)^2 (1+\beta(z) \mu^2)^2 P_{mass}(k,z) D(k,\mu,z) ~,
\label{eq:Pspec}
\end{equation}
where $\mu$ is the angle of the wavevector to the line of sight,
$k$ is the wavenumber,
$b$ is the linear bias parameter, $\beta$ the redshift space
distortion parameter and $D(k,\mu,z)$ is a non-linear correction
calibrated from simulations (for the \lyaf\ this is given by 
\cite{2003ApJ...585...34M} and 
for galaxies it is based on the information damping factors
of \cite{2007ApJ...665...14S}).  
The Fisher matrix calculation will integrate over all $\mu$ and a suitable 
range of $k$.
The inverse variance of the power spectrum of each mode gets contributions from 
both the intrinsic sample variance and the shot noise.
This results in an effective volume
$V_{\rm eff}(\tilde{P})$ of each redshift slice that is given by $V_{\rm
  eff}(\tilde{P})=[1+1/(n\tilde{P})]^{-2}V_{\rm survey}$ 
\cite{1994ApJ...426...23F}.
The value
$nP$ represents the ratio of true clustering power to that from shot noise.
Alternatively, it can be seen as the signal-to-noise ratio per mode (redshift,
wavenumber, and orientation slice): if $nP>1$ then roughly the signal
exceeds the sample variance uncertainty for that mode.

For the galaxy survey, we use large-scale broadband power up to some quoted $\kmaxeff$.
At small scales, $k>\kmaxeff$, we continue to use BAO information.
We use two simple choices of $\kmaxeff$: 0.1~$\ihMpc$ and 0.2~$\ihMpc$.
These cutoffs are intended to indicate sensitivity of results to the
effective scale where information is recovered after making
corrections for non-linearity, after marginalization over
suitable non-linear bias parameters.
It will be a major program of the next
decade to figure out exactly how to do this fitting in practice for a
high precision survey like DESI; how well we can do this will
determine how well we can measure parameters.
As discussed in \cite{2014JCAP...05..023F}, $\kmaxeff\sim 0.1~\ihMpc$ 
corresponds roughly to the performance of current analyses, while 
$\kmaxeff\sim 0.2~\ihMpc$ is more of a stretch goal for the DESI era 
(some improvement over current analysis can be expected simply by going to 
higher redshift where the non-linear scale is smaller).

The redshift-space distortions can effectively constrain two parameter
combinations, $b(z) \sigma(z)$ and $f(z) \sigma(z)$, where
$\sigma(z)\propto P_{\rm mass}^{1/2}(z,k)$ is the RMS normalization of
the linear mass density fluctuations as a function of $z$.  In Table
\ref{tab:scienceBaseline}, we quote projected constraints on $f
\sigma$ for different maximum $k$ assumptions e.g., $f \sigma_{0.1}$
means the error calculation included information up to
$\kmaxeff=0.1~\ihMpc$. These fractional errors are equivalent to what
one usually sees quoted as an error on ``$f \sigma_8$''.
The $f \sigma_k$ precision we project for DESI, aggregated over all redshifts, 
is 
$\sim$\peskgoalRSD\% for $k_{\rm max}=0.1 \ihMpc$, or $\sim$\opkgoalRSD\% for
$k_{\rm max}=0.2\ihMpc$.

\subsubsection{Baseline Survey}


\label{sec:scienceForecastBaseline}

Our baseline assumption for science projections is that DESI runs over
an approximately five-year period 
covering 14,000 deg$^2$ in area.  DESI will target four types of
objects: Bright Galaxies (BGS), Luminous Red Galaxies (LRGs), Emission Line Galaxies (ELGs)
\cite{2013ApJ...767...89M}, and quasars.  Details on how these objects
are targeted can be found in Section~\ref{s4:TargetSelection}.  In
what follows, most calculations are done for this baseline survey.  We
additionally provide several relevant calculations for the required
minimum survey with the same target number densities over 9,000
instead of 14,000 deg$^2$ in area.

\begin{figure}[!p]
\centering
\includegraphics[height=2.4in]{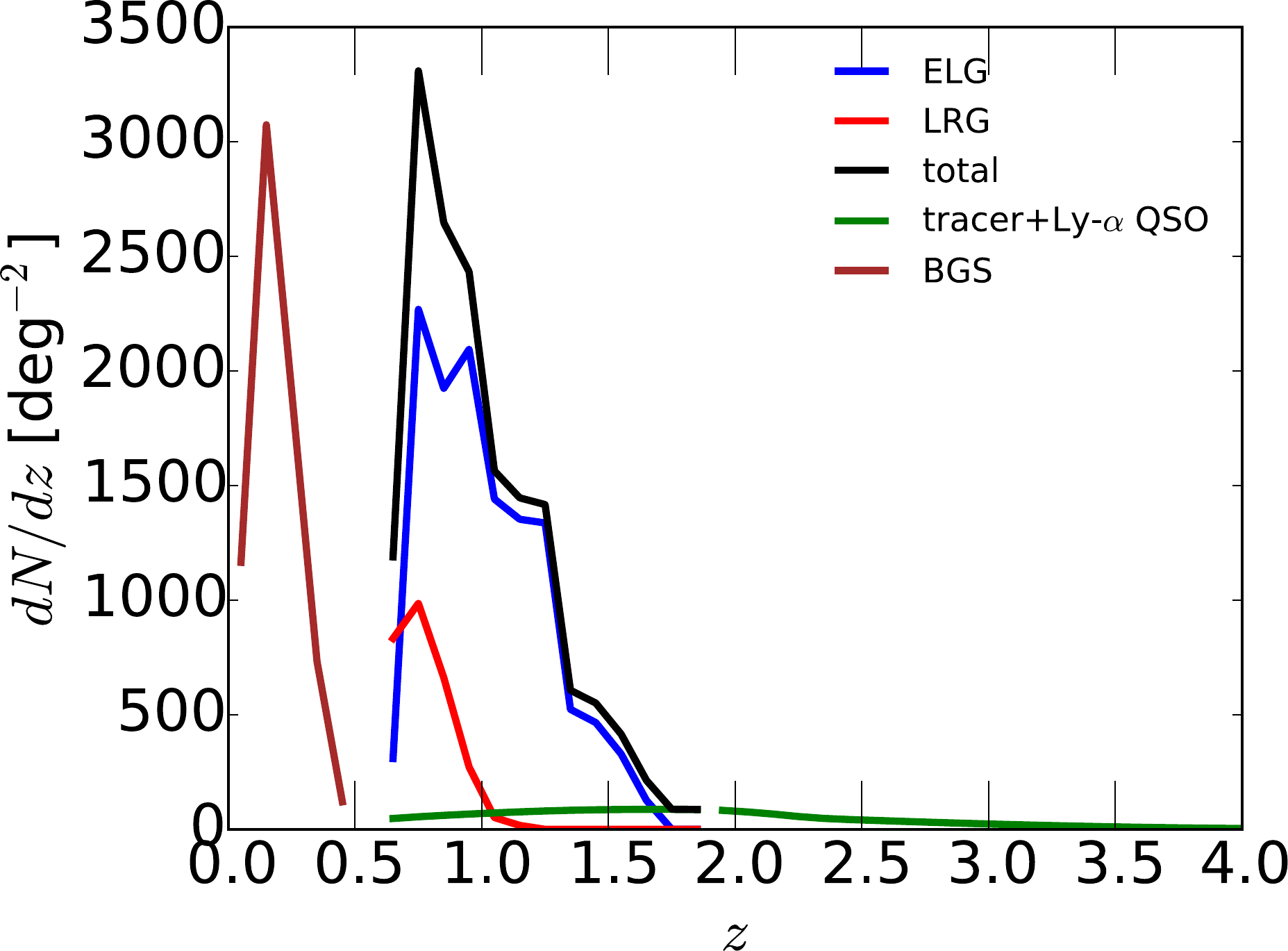}
\caption{
DESI number densities, per unit $z$, per square degree, used in cosmology
projections (Table \ref{tab:scienceBaseline} and \ref{tab:lyafnumbers}).
}
\label{fig:dNdz}
\centering
\includegraphics[height=2.4in]{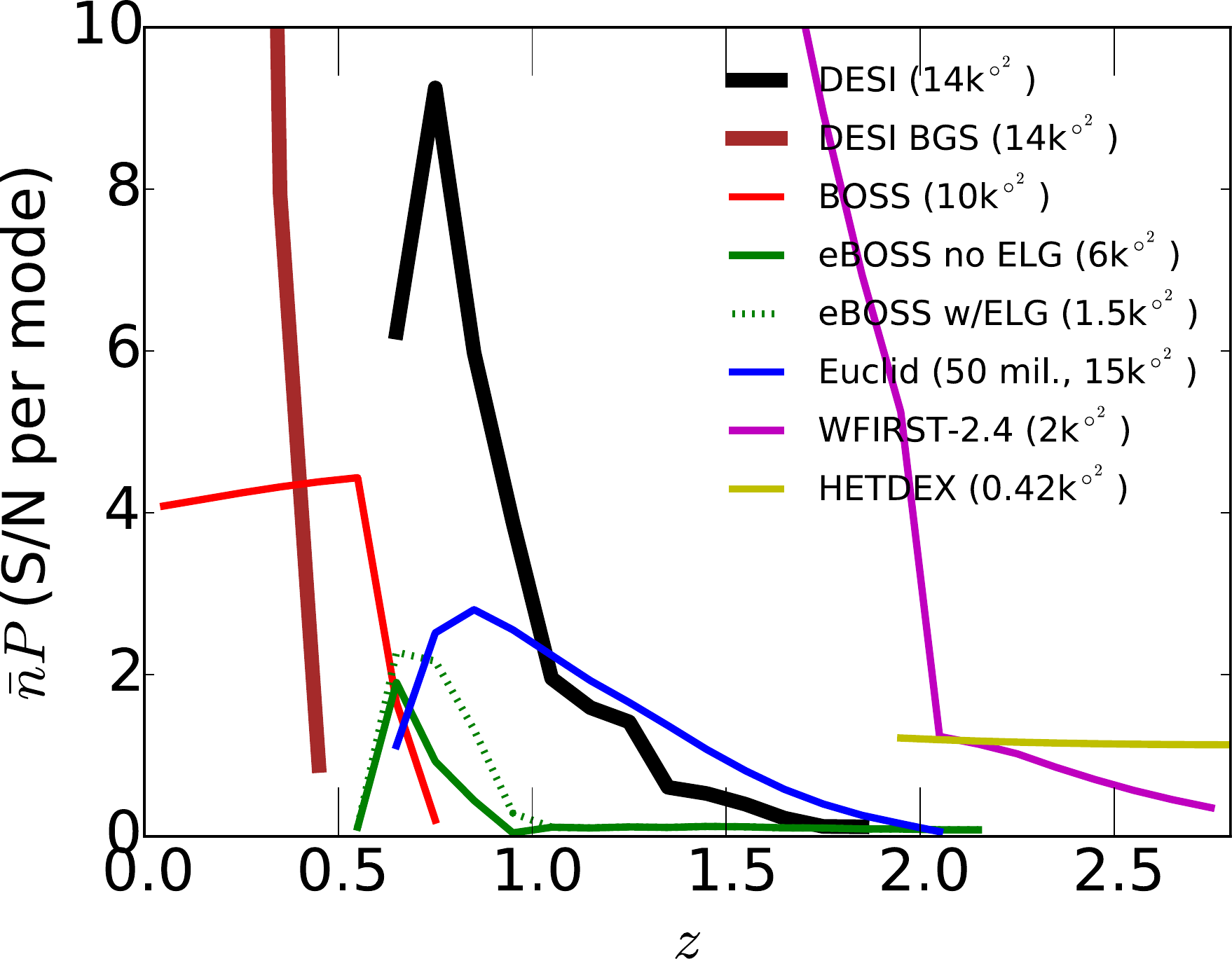}
\caption{
Signal to noise comparison of the DESI galaxy survey against other precursor 
(Stage II and Stage III) and upcoming (Stage IV) spectroscopic surveys. Shown 
is $\bar{n} P(k=0.14~\ihMpc, \mu=0.6)$. The DESI forecasts do not include the
\lyaf contribution. Including this would give an effective $\bar{n}P\sim 0.3$
at $z\sim 2.5$. Note that the large area covered by DESI provides an advantage 
reflected in Figure~\ref{fig:binnedR}.
}
\label{fig:nP}
\centering
\includegraphics[height=2.4in]{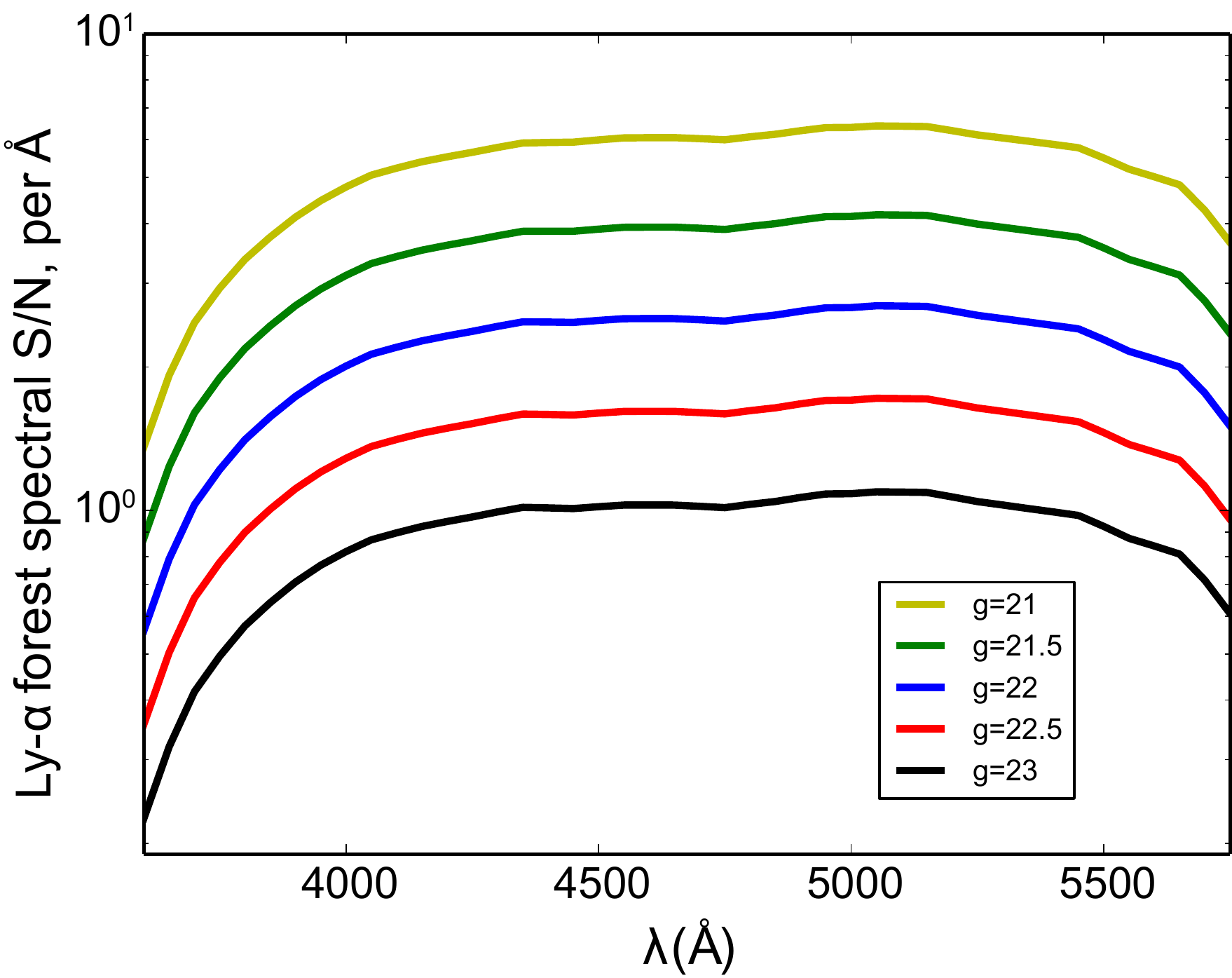}
\caption{
Signal-to-noise ratio per \AA\ used for DESI quasar spectra (detector noise,
not absorption noise), for different
$g$ magnitudes, accounting for mean \lyaf absorption.
}
\label{fig:qSN}
\end{figure}

The number densities used here, plotted in Figure~\ref{fig:dNdz}, are
based on the selection criteria for each object type described in the following chapter.

We assume fiducial biases follow constant $b(z) D(z)$, where $D(z)$ is
the linear growth factor normalized by $D(z=0) \equiv 1$.  For LRGs we
use $b_{\rm LRG}(z) D(z) = 1.7$.  For ELGs we use $b_{\rm ELG}(z) D(z)
= 0.84$ \cite{2013ApJ...767...89M}.  For quasars we use $b_{\rm
  QSO}(z) D(z) = 1.2$ (loosely based on \cite{2009ApJ...697.1634R}).
For the BGS, we use $b_{\rm BGS}(z) D(z) = 1.34$, but the results are 
insensitive to this value because of the much higher number density 
in most of the BGS volume.
Note that these forms keep the observed clustering amplitude of each
individual tracer constant with redshift, in agreement with observations
(more detailed references for bias evolution are given below, in 
sections \ref{sec:bgs}, \ref{sec:lrg}, \ref{sec:ELGtargprop}, and 
\ref{sec:qsotargets} for
BGS, LRGs, ELGs, and QSOs, respectively).

The signal-to-noise for typical BAO-scale modes in redshift space is shown
in Figure~\ref{fig:nP}, along with the same quantity computed for several
other experiments for comparison \cite{2014JCAP...05..023F}.

We evaluate $\bar{n}P$ at $k=0.14~\ihMpc$, $\mu=0.6$, an approximate
center-of-weight point for BAO measurements.  We chose these values by
looking for the point where $\bar{n}P=1$ corresponded to the optimum
in a trade-off between area and number density at fixed total number
of objects (specifically, for the full range of parameters covered by
DESI LRGs and ELGs). This definition reflects the origin of the idea
that $\bar{n}P=1$ is a special point, but it should be kept in mind
that achieving $\bar{n}P$ by this definition does leave a survey
significantly farther away from the sample variance limit than the
traditional definition $k=0.2~\ihMpc$, $\mu=0$.

The spectral signal-to-noise ratio that we use, computed using the
{\tt bbspecsim} code \cite{2012SPIE.8446E..0QM}, is shown in
Figure \ref{fig:qSN}. \pvm{This is out of date but changes make almost no
difference to results - Kirkby was supposed to be making a new verison of 
this figure.}


\subsubsection{Summary of Forecasts}

\label{sec:scienceForecastResults}

%

\begin{figure}[tb]
\centering
\includegraphics[height=3in]{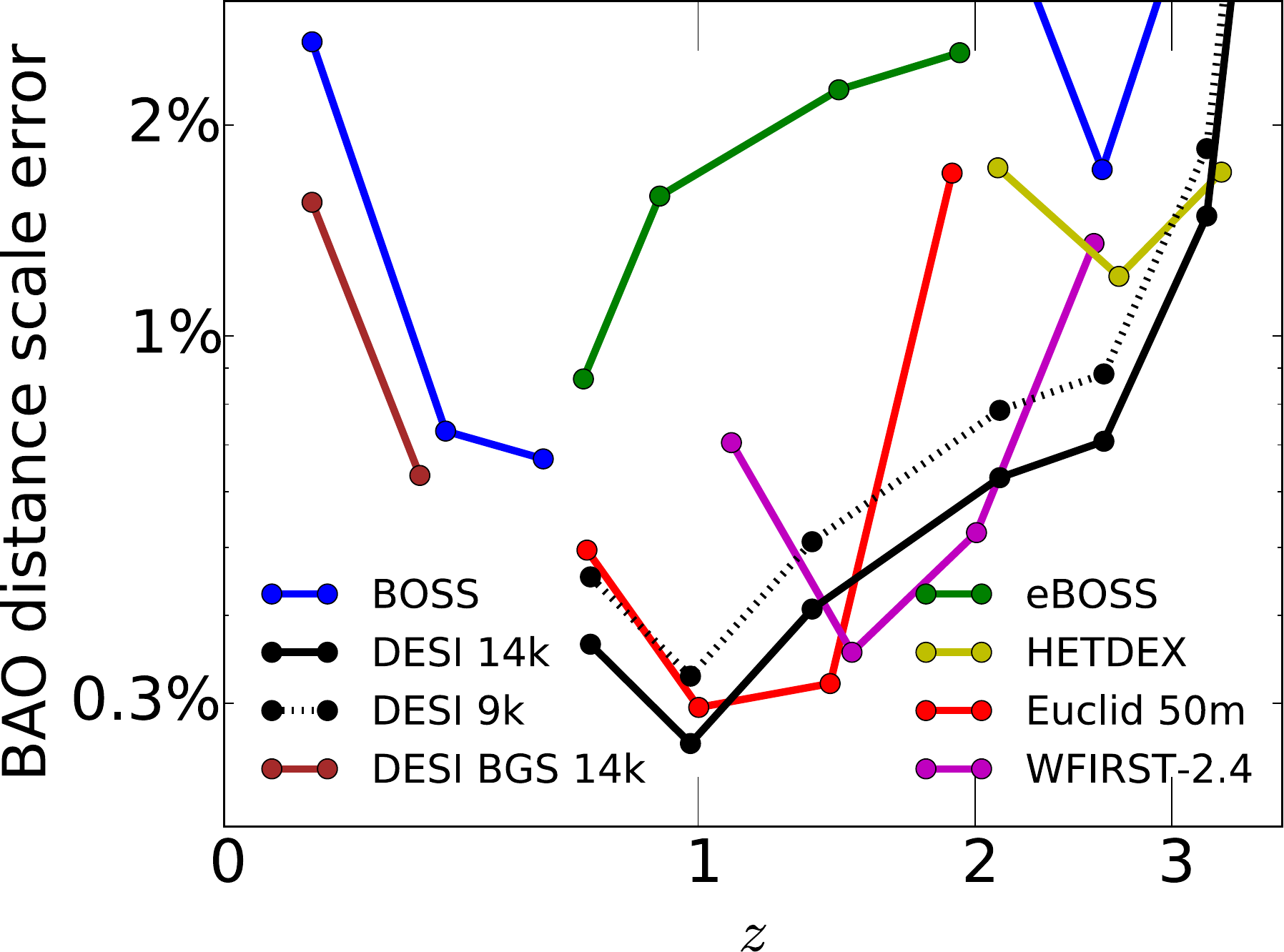}
\caption{
The fractional error on the dilation factor, $R$, as a function of redshift 
presented in comparable bins for DESI, BOSS, {\it Euclid}, {\it WFIRST},
HETDEX, and eBOSS.  This gives an 
indicative error on distance measurements to each redshift.  
The forecasts for a 14,000 deg$^2$ DESI Bright Galaxy Survey (BGS) are also shown.
DESI will provide 
the best measurements over much of the region and is competitive with 
space-based missions, which will come later.  We use 50 million total
galaxies for Euclid, following their Definition Study Report
\cite{2011arXiv1110.3193L}, although it has been suggested that this
may be optimistic \cite{2013arXiv1305.5422S}.  
}
\label{fig:binnedR}
\end{figure}

Table \ref{tab:scienceBaseline} lists the basic galaxy and quasar BAO
distance measurement projections, and RSD $f(z) \sigma_8(z)$ error
projections for two different $\kmaxeff$ values for our baseline 14K
survey.  
We provide the same set of calculations in Table
\ref{tab:scienceBaseline9K} for our threshold 9K survey.
Tables \ref{tab:scienceBaselineBG} and 
\ref{tab:scienceBaselineBG9k} 
shows the projections for the 
Bright Galaxy Survey for 14K and 9K square degrees, respectively.
Table~\ref{tab:lyafnumbers} lists the \lyaf BAO distance measurement
projections, including cross-correlations with quasars in the same
redshift range for a $z>1.9$ \lyaf survey; Table~\ref{tab:lyafnumbers9K}
presents the same calculations for the threshold 9K survey.  The BAO
errors are also shown in Figure~\ref{fig:binnedR}, along with those
from other experiments for comparison (see \cite{2014JCAP...05..023F}
for a description of the other experiments).

\begin{table}[!p]
\centering
\caption{
Summary of forecasted constraints achievable by DESI, covering 14,000 deg$^2$.
Indications of signal to noise, $nP$, are given at two values of $k,\mu$ = $\{0.2~\ihMpc,0\}$ and $\{1.4~\ihMpc,0.6\}$.
The fractional error
 on the normalization of
$f(z) P^{1/2}(k,z)$ is  ${\sigma_{f\sigma_k }}/{f \sigma_k}$, assuming known shape of the power spectrum and
known geometry, using $\kmaxeff=k~$\ihMpc.
The dilation factor $R$ is defined to be a parameter
rescaling the radial and transverse distances
by equal factors.
}
\label{tab:scienceBaseline}
\newcolumntype{C}{>{\centering\arraybackslash}X}
\footnotesize
\begin{tabularx}{\textwidth}{CCCCCCCCCCCC}
\midrule
$z$ &  
$\frac{\sigma_{R/s}}{R/s}$ &
$ \frac{\sigma_{D_A/s}}{D_A/s}$ & 
$\frac{\sigma_{Hs}}{Hs}$ &
$\bar{n}P_{0.2,0}$ & $\bar{n}P_{0.14,0.6}$ & $V$ &
$\frac{dN_{ELG}}{dz~ d{\rm deg}^2}$ & $\frac{dN_{LRG}}{dz ~d{\rm deg}^2}$ &
$\frac{dN_{QSO}}{dz~ d{\rm deg}^2}$ &
$\frac{\sigma_{f\sigma_{0.1} }}{f \sigma_{0.1}}$
& $\frac{\sigma_{f \sigma_{0.2}}}{f \sigma_{0.2}}$ \\
 & \% & \% & \% &
 &  & [$h^{-1}{\rm Gpc}^3$] & & & & \% & \%  \\
\midrule
0.65 & 0.57 & 0.82 & 1.50 & 2.59 & 6.23 & 2.63 & 309 & 832 & 47 & 3.31 & 1.57 \\ 
0.75 & 0.48 & 0.69 & 1.27 & 3.63 & 9.25 & 3.15 & 2269 & 986 & 55 & 2.10 & 1.01 \\ 
0.85 & 0.47 & 0.69 & 1.22 & 2.33 & 5.98 & 3.65 & 1923 & 662 & 61 & 2.12 & 1.01 \\ 
0.95 & 0.49 & 0.73 & 1.22 & 1.45 & 3.88 & 4.10 & 2094 & 272 & 67 & 2.09 & 0.99 \\ 
1.05 & 0.58 & 0.89 & 1.37 & 0.71 & 1.95 & 4.52 & 1441 & 51 & 72 & 2.23 & 1.11 \\ 
1.15 & 0.60 & 0.94 & 1.39 & 0.58 & 1.59 & 4.89 & 1353 & 17 & 76 & 2.25 & 1.14 \\ 
1.25 & 0.61 & 0.96 & 1.39 & 0.51 & 1.41 & 5.22 & 1337 & 0 & 80 & 2.25 & 1.16 \\ 
1.35 & 0.92 & 1.50 & 2.02 & 0.22 & 0.61 & 5.50 & 523 & 0 & 83 & 2.90 & 1.73 \\ 
1.45 & 0.98 & 1.59 & 2.13 & 0.20 & 0.53 & 5.75 & 466 & 0 & 85 & 3.06 & 1.87 \\ 
1.55 & 1.16 & 1.90 & 2.52 & 0.15 & 0.40 & 5.97 & 329 & 0 & 87 & 3.53 & 2.27 \\ 
1.65 & 1.76 & 2.88 & 3.80 & 0.09 & 0.22 & 6.15 & 126 & 0 & 87 & 5.10 & 3.61 \\ 
1.75 & 2.88 & 4.64 & 6.30 & 0.05 & 0.12 & 6.30 & 0 & 0 & 87 & 8.91 & 6.81 \\ 
1.85 & 2.92 & 4.71 & 6.39 & 0.05 & 0.12 & 6.43 & 0 & 0 & 86 & 9.25 & 7.07 \\ 

\hline
\end{tabularx}

\vspace{0.1in}

\centering
\caption{
Like Table \ref{tab:scienceBaseline}, except with DESI covering only 
9,000 deg$^2$.
}
\label{tab:scienceBaseline9K}
\newcolumntype{C}{>{\centering\arraybackslash}X}
\footnotesize
\begin{tabularx}{\textwidth}{CCCCCCCCCCCC}
\midrule
$z$ & 
$\frac{\sigma_{R/s}}{R/s}$ &
$ \frac{\sigma_{D_A/s}}{D_A/s}$ & 
$\frac{\sigma_{Hs}}{Hs}$ &
$\bar{n}P_{0.2,0}$ & $\bar{n}P_{0.14,0.6}$ & $V$ &
$\frac{dN_{ELG}}{dz~ d{\rm deg}^2}$ & $\frac{dN_{LRG}}{dz ~d{\rm deg}^2}$ &
$\frac{dN_{QSO}}{dz~ d{\rm deg}^2}$ &
$\frac{\sigma_{f\sigma_{0.1} }}{f \sigma_{0.1}}$
& $\frac{\sigma_{f \sigma_{0.2}}}{f \sigma_{0.2}}$ \\
 & \% & \% & \% &
 &  & $h^{-1}{\rm Gpc}^3$ & & & & \% & \%  \\
\midrule
0.65 & 0.71 & 1.02 & 1.87 & 2.59 & 6.23 & 1.69 & 309 & 832 & 47 & 4.12 & 1.96 \\ 
0.75 & 0.59 & 0.86 & 1.58 & 3.63 & 9.25 & 2.03 & 2269 & 986 & 55 & 2.62 & 1.26 \\ 
0.85 & 0.59 & 0.86 & 1.53 & 2.33 & 5.98 & 2.34 & 1923 & 662 & 61 & 2.64 & 1.26 \\ 
0.95 & 0.61 & 0.91 & 1.52 & 1.45 & 3.88 & 2.64 & 2094 & 272 & 67 & 2.61 & 1.24 \\ 
1.05 & 0.72 & 1.12 & 1.70 & 0.71 & 1.95 & 2.90 & 1441 & 51 & 72 & 2.79 & 1.39 \\ 
1.15 & 0.75 & 1.17 & 1.74 & 0.58 & 1.59 & 3.14 & 1353 & 17 & 76 & 2.80 & 1.42 \\ 
1.25 & 0.76 & 1.19 & 1.74 & 0.51 & 1.41 & 3.35 & 1337 & 0 & 80 & 2.81 & 1.44 \\ 
1.35 & 1.15 & 1.87 & 2.52 & 0.22 & 0.61 & 3.54 & 523 & 0 & 83 & 3.62 & 2.16 \\ 
1.45 & 1.22 & 1.99 & 2.66 & 0.20 & 0.53 & 3.70 & 466 & 0 & 85 & 3.82 & 2.34 \\ 
1.55 & 1.45 & 2.37 & 3.14 & 0.15 & 0.40 & 3.84 & 329 & 0 & 87 & 4.40 & 2.84 \\ 
1.65 & 2.20 & 3.59 & 4.74 & 0.09 & 0.22 & 3.95 & 126 & 0 & 87 & 6.36 & 4.50 \\ 
1.75 & 3.59 & 5.79 & 7.86 & 0.05 & 0.12 & 4.05 & 0 & 0 & 87 & 11.11 & 8.49 \\ 
1.85 & 3.64 & 5.87 & 7.97 & 0.05 & 0.12 & 4.13 & 0 & 0 & 86 & 11.53 & 8.82 \\ 

\hline
\end{tabularx}

\vspace{0.1in}

\centering
\caption{
Like Table \ref{tab:scienceBaseline}, except for the DESI Bright Galaxy Survey,
covering 14,000 deg$^2$.
}
\label{tab:scienceBaselineBG}
\newcolumntype{C}{>{\centering\arraybackslash}X}
\footnotesize
\begin{tabularx}{\textwidth}{CCCCCCCCCC}
\midrule
$z$ &  
$\frac{\sigma_{R/s}}{R/s}$ &
$ \frac{\sigma_{D_A/s}}{D_A/s}$ & 
$\frac{\sigma_{Hs}}{Hs}$ &
$\bar{n}P_{0.2,0}$ & $\bar{n}P_{0.14,0.6}$ & $V$ &
$\frac{dN_{BGS}}{dz~ d{\rm deg}^2}$ &
$\frac{\sigma_{f\sigma_{0.1} }}{f \sigma_{0.1}}$
& $\frac{\sigma_{f \sigma_{0.2}}}{f \sigma_{0.2}}$ \\
 & \% & \% & \% &
 &  & [$h^{-1}{\rm Gpc}^3$] & & \% & \%  \\
\midrule
0.05 & 4.33 & 6.12 & 12.10 & 146.60 & 352.91 & 0.04 & 1165 & 33.24 & 14.08 \\ 
0.15 & 1.66 & 2.35 & 4.66 & 59.47 & 144.69 & 0.23 & 3074 & 12.47 & 5.25 \\ 
0.25 & 1.07 & 1.51 & 2.97 & 14.84 & 36.43 & 0.58 & 1909 & 7.69 & 3.25 \\ 
0.35 & 0.91 & 1.32 & 2.44 & 3.21 & 7.94 & 1.04 & 732 & 5.83 & 2.60 \\ 
0.45 & 1.56 & 2.39 & 3.69 & 0.35 & 0.87 & 1.55 & 120 & 6.35 & 3.77 \\ 

\hline
\end{tabularx}
\end{table}

\begin{table}[!p]
\centering
\caption{
Like Table \ref{tab:scienceBaselineBG}, but for a
9,000 deg$^2$ Bright Galaxy Survey.
}
\label{tab:scienceBaselineBG9k}
\newcolumntype{C}{>{\centering\arraybackslash}X}
\footnotesize
\begin{tabularx}{\textwidth}{CCCCCCCCCC}
\midrule
$z$ &  
$\frac{\sigma_{R/s}}{R/s}$ &
$ \frac{\sigma_{D_A/s}}{D_A/s}$ & 
$\frac{\sigma_{Hs}}{Hs}$ &
$\bar{n}P_{0.2,0}$ & $\bar{n}P_{0.14,0.6}$ & $V$ &
$\frac{dN_{BGS}}{dz~ d{\rm deg}^2}$ &
$\frac{\sigma_{f\sigma_{0.1} }}{f \sigma_{0.1}}$
& $\frac{\sigma_{f \sigma_{0.2}}}{f \sigma_{0.2}}$ \\
 & \% & \% & \% &
 &  & [$h^{-1}{\rm Gpc}^3$] & & \% & \%  \\
\midrule
0.05 & 5.39 & 7.63 & 15.09 & 146.60 & 352.91 & 0.02 & 1165 & 41.46 & 17.56 \\ 
0.15 & 2.07 & 2.93 & 5.81 & 59.47 & 144.69 & 0.15 & 3074 & 15.55 & 6.54 \\ 
0.25 & 1.33 & 1.89 & 3.70 & 14.84 & 36.43 & 0.38 & 1909 & 9.59 & 4.05 \\ 
0.35 & 1.14 & 1.64 & 3.04 & 3.21 & 7.94 & 0.67 & 732 & 7.27 & 3.24 \\ 
0.45 & 1.94 & 2.98 & 4.60 & 0.35 & 0.87 & 1.00 & 120 & 7.92 & 4.71 \\ 

\hline
\end{tabularx}

\vspace{0.1in}

\centering
\caption{$z>1.9$ \lyaf quasar survey, over 14000 sq. deg.
Parameter errors are in percent relative to the BAO scale, $s$.}
\begin{tabular}{lcccc}
\midrule
$z$ & $\frac{\sigma_{R/s}}{R/s}$ (\%) & $ \frac{\sigma_{D_A/s}}{D_A/s}$ (\%) &
    $\frac{\sigma_{Hs}}{Hs}$ (\%)& $\frac{dN_{QSO}}{dz~ d{\rm deg}^2}$ \\ 
\hline
1.96 & 1.43 & 2.69 & 2.74 & 82 \\ 
2.12 & 1.02 & 1.95 & 1.99 & 69 \\ 
2.28 & 1.09 & 2.18 & 2.11 & 53 \\ 
2.43 & 1.20 & 2.46 & 2.26 & 43 \\ 
2.59 & 1.34 & 2.86 & 2.47 & 37 \\ 
2.75 & 1.53 & 3.40 & 2.76 & 31 \\ 
2.91 & 1.81 & 4.21 & 3.18 & 26 \\ 
3.07 & 2.16 & 5.29 & 3.70 & 21 \\ 
3.23 & 2.75 & 7.10 & 4.57 & 16 \\ 
3.39 & 3.86 & 10.46 & 6.19 & 13 \\ 
3.55 & 5.72 & 15.91 & 8.89 & 9 \\ 
3.70 & - & - & - & 7 \\ 
3.86 & - & - & - & 5 \\ 
4.02 & - & - & - & 3 \\ 

\midrule
\end{tabular}
\label{tab:lyafnumbers}

\vspace{0.1in}

\centering
\caption{Like Table \ref{tab:lyafnumbers}, except with DESI covering only 
9,000 deg$^2$.}
\begin{tabular}{lcccc}
\midrule
$z$ & $\frac{\sigma_{R/s}}{R/s}$ (\%) & $ \frac{\sigma_{D_A/s}}{D_A/s}$ (\%) &
    $\frac{\sigma_{Hs}}{Hs}$ (\%)& $\frac{dN_{QSO}}{dz~ d{\rm deg}^2}$ \\ \hline
1.96 & 1.78 & 3.35 & 3.42 & 82 \\ 
2.12 & 1.27 & 2.43 & 2.48 & 69 \\ 
2.28 & 1.37 & 2.72 & 2.63 & 53 \\ 
2.43 & 1.49 & 3.07 & 2.82 & 43 \\ 
2.59 & 1.67 & 3.57 & 3.08 & 37 \\ 
2.75 & 1.91 & 4.24 & 3.44 & 31 \\ 
2.91 & 2.25 & 5.26 & 3.96 & 26 \\ 
3.07 & 2.69 & 6.60 & 4.62 & 21 \\ 
3.23 & 3.43 & 8.86 & 5.70 & 16 \\ 
3.39 & 4.81 & 13.05 & 7.72 & 13 \\ 
3.55 & 7.14 & 19.85 & 11.09 & 9 \\ 
3.70 & - & - & - & 7 \\ 
3.86 & - & - & - & 5 \\ 
4.02 & - & - & - & 3 \\ 

\midrule
\end{tabular}
\label{tab:lyafnumbers9K}
\end{table}


\begin{figure}[!p]
\centering
\includegraphics[height=2.4in]{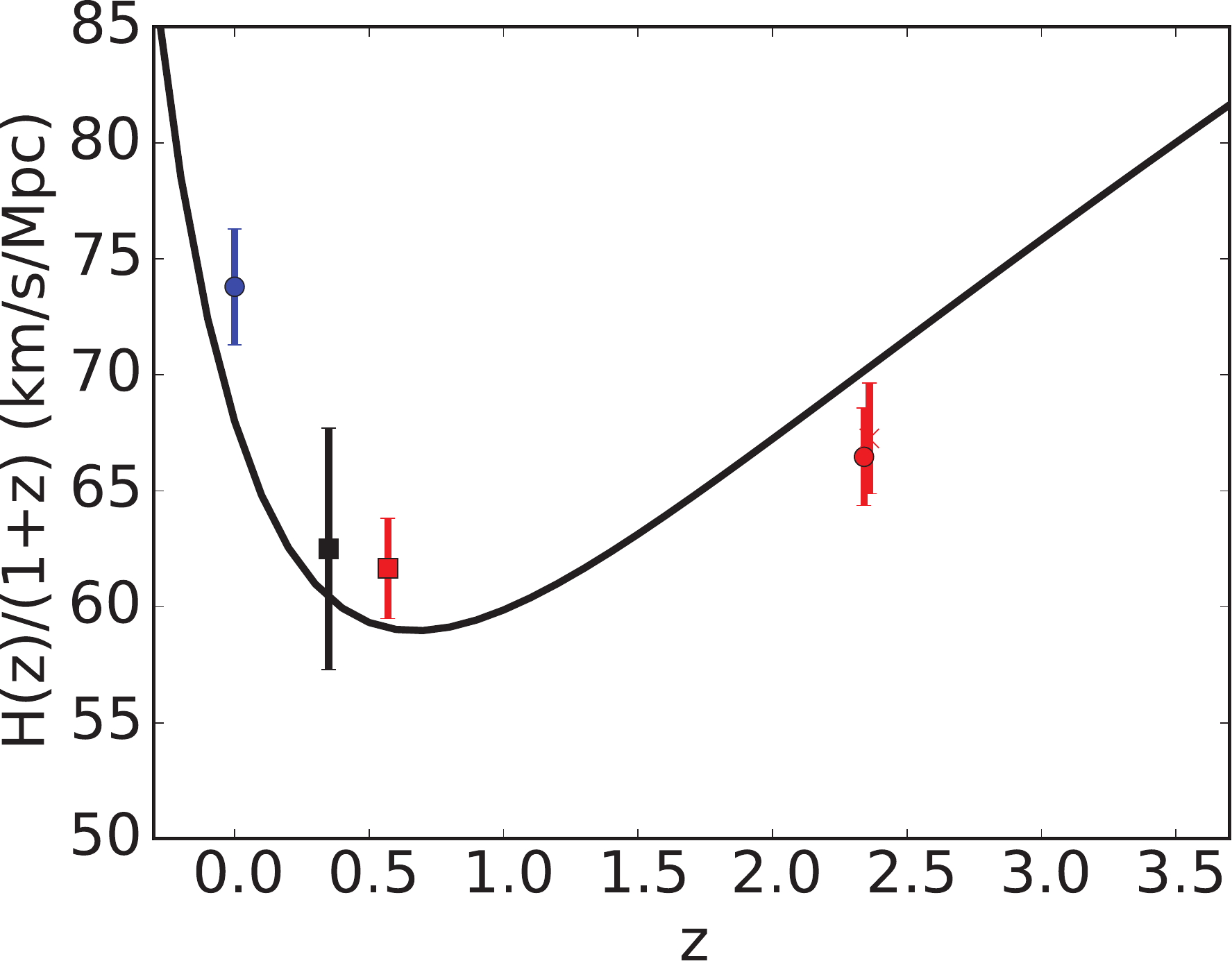}
\vskip 0.05in
\includegraphics[height=2.4in]{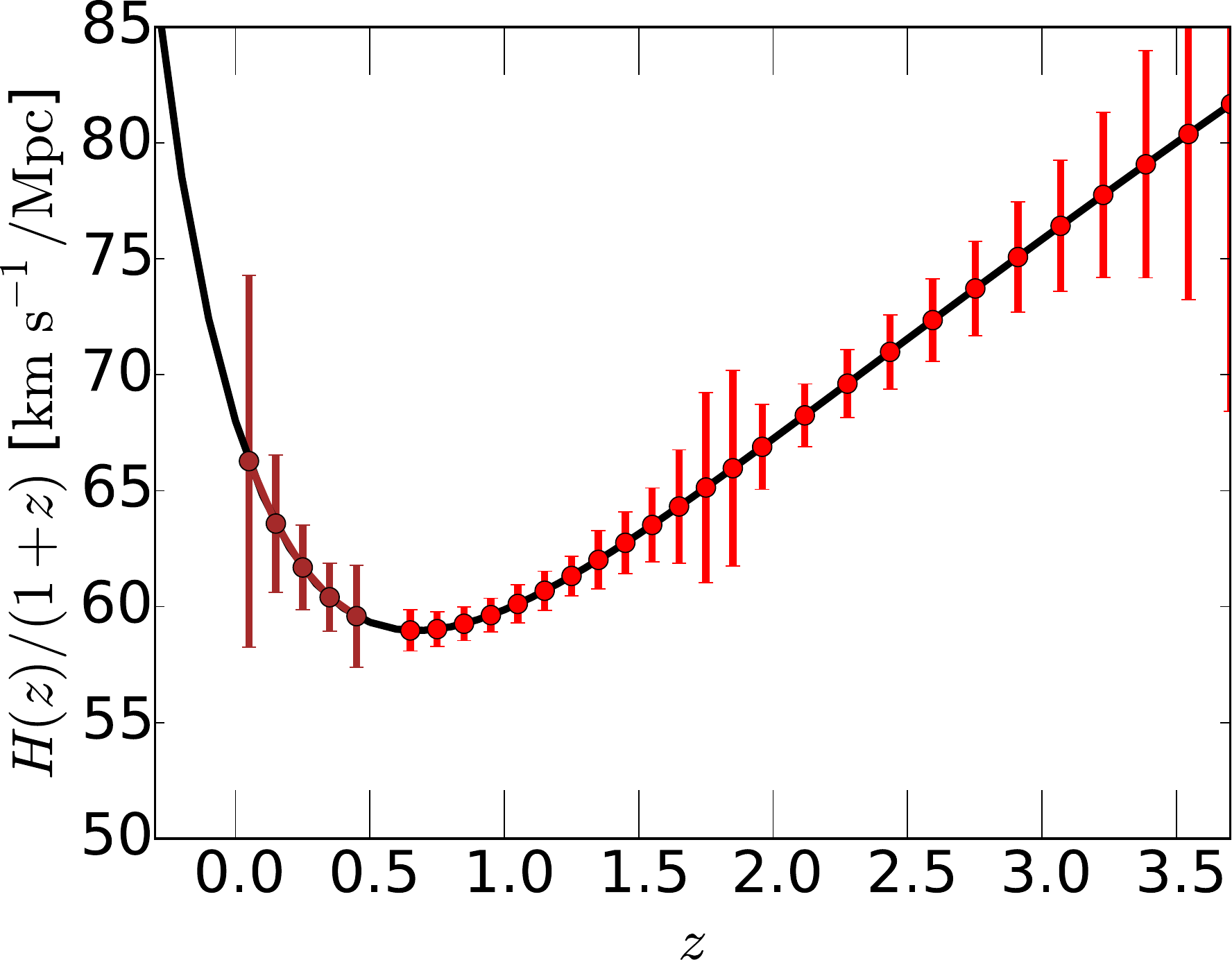}
\caption{
Expansion rate of the Universe as a function of redshift.
In the upper plot, the filled blue circle is the $H_0$ measurement of
\cite{2011ApJ...730..119R},
the solid black square shows the SDSS BAO
measurement of \cite{2013MNRAS.431.2834X},
the red square shows the BOSS galaxy BAO measurement
of \cite{2014MNRAS.441...24A}, the red circle shows the BOSS \lyaf\ BAO
measurement of \cite{2014arXiv1404.1801D}, and the red x shows the
BOSS \lyaf\ BAO-quasar cross-correlation measurement of
\cite{2014JCAP...05..027F}.
The lower plot shows projected DESI points.
}
\label{fig:expansion}

\begin{center}
\includegraphics[height=2.4in]{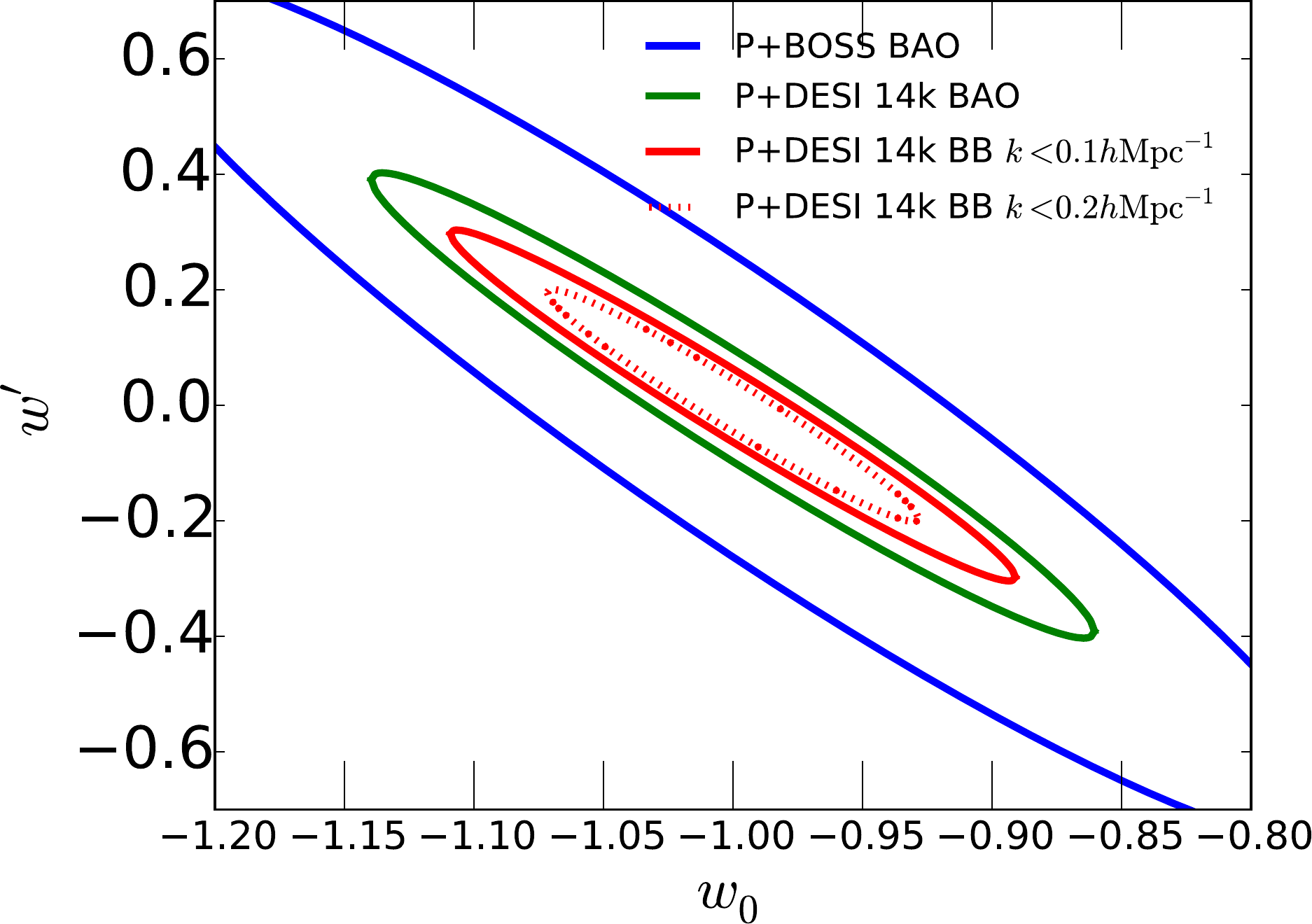}
\caption{
The $w_0-w_a$ plane showing projected limits (68\%) from DESI using 
just BAO and using the broadband (BB) power spectrum. 
Also shown is the limit from 
BOSS BAO.  {\it Planck} priors are included in all cases, and DESI includes 
the BGS and non-redundant part of BOSS. The figure of merit 
of the surveys is inversely proportional to the areas of the error ellipses.
}
\label{fig:w0wa}
\end{center}
\end{figure}

DESI will provide high precision measurements of the Universe's
expansion rate over billions of years.  Using the \lyaf technique,
coverage will include the early times when the expansion rate was
decreasing (when the matter density, not the dark energy density, was
controlling the rate).  In Figure~\ref{fig:expansion} we show how DESI
will improve these measurements over those existing today.

Table \ref{tab:FoM} shows Dark Energy Task Force (DETF) Figures of
Merit (FoMs) \cite{2006astro.ph..9591A}. For the common normalization
convention that we follow, the FoM is simply $\left(\sigma_{w_p}
  \sigma_{w^\prime}\right)^{-1}$ where $w(z) = w_p + (a_p-a)w^\prime$
and $a_p$ is chosen to make the errors on $w_p$ and $w^\prime$
independent.  Because the DETF FoM model is defined to include the
possibility of curvature, we include curvature projections in Table
\ref{tab:FoM}. 
The figure of merit results are reflected in
Figure~\ref{fig:w0wa}.

Importantly, Table \ref{tab:FoM} shows that these surveys exceed the
Stage IV FoM threshold.  We take this to be a value of 110, based on a
10-fold improvement of the value of 11 from
\cite{2011ApJ...737..102S}.  This is the same Stage IV definition that
LSST used in their Conceptual Design Report.  The 9,000 square degrees
DESI survey achieves 121 with galaxies and Ly-$\alpha$ forest BAO.  We
note that these computations include only BAO and CMB, without even
the Stage II Supernovae Ia results from \cite{2011ApJ...737..102S}.
Including DESI galaxy broadband clustering or other dark energy probes
boost the Figure of Merit well above 110.

As this 9000 square degree survey forecast meets the Stage IV threshold and hence the
Mission Need, we have adopted it as the quantitative basis for the
Level 1 Science Requirement for the DESI project.  We aggregate the
BAO performance into three redshift ranges, $R$ in $0.0<z<1.1$ and
$1.1<z<1.9$ and $H$ in $1.9<z<3.7$, for the L1 requirements, so as to
leave flexibility in the exact redshift distribution of targets.  An
extensive discussion of how the FOM depends on variation in survey
parameters was presented in the DESI Conceptual Design Review.

\begin{table}
\centering
\caption{DETF Figures of Merit and uncertainties $\sigma_{w_p}$ and $\sigma_{\Omega_k}$.  
$\sigma_{w_p}$ is the error on $w$ at the pivot redshift, which also equal to the
error on a constant $w$ holding $w_a=0$.
$\sigma_{\Omega_k}$ is the error on the curvature of the Universe, $\Omega_k$.
All DESI lines contain the BGS, and BOSS in the range $0.45<z<0.6$ that does
not substantially overlap with DESI.
All cases include $Planck$ CMB constraints.
The pivot point, where $w(a)$ has minimal uncertainty is indicated by $a_p$.
We note that a FoM of 110 is 10 times the Stage II level of \cite{2011ApJ...737..102S}, 
which we take to be the definition of Stage IV.  DESI BAO galaxy exceeds
this threshold even with a 9,000 square degree survey.}
\label{tab:FoM}
\begin{tabular}{lcccc}
\hline
Surveys & FoM & $a_p$ & $\sigma_{w_p}$ & $\sigma_{\Omega_k}$ \\
\hline
BOSS BAO & 37 & 0.65 & 0.055 & 0.0026 \\
DESI 14k galaxy BAO & 133 & 0.69 & 0.023 & 0.0013 \\
DESI 14k galaxy and Ly-$\alpha$ forest BAO & 169 & 0.71 & 0.022 & 0.0011 \\
DESI 14k BAO + gal. broadband to $k<0.1~h~{\rm Mpc}^{-1}$ & 332 & 0.74 & 0.015 & 0.0009 \\
DESI 14k BAO + gal. broadband to $k<0.2~h~{\rm Mpc}^{-1}$ & 704 & 0.73 & 0.011 & 0.0007 \\
DESI 9k galaxy BAO & 95 & 0.69 & 0.027 & 0.0015 \\
DESI 9k galaxy and Ly-$\alpha$ forest BAO & 121 & 0.71 & 0.026 & 0.0012 \\
DESI 9k BAO + gal. broadband to $k<0.1~h~{\rm Mpc}^{-1}$ & 229 & 0.73 & 0.018 & 0.0011 \\
DESI 9k BAO + gal. broadband to $k<0.2~h~{\rm Mpc}^{-1}$ & 502 & 0.73 & 0.013 & 0.0009 \\

\hline
\end{tabular}
\end{table}

\begin{figure}[!h]
\begin{center}
\includegraphics[height=2.5in]{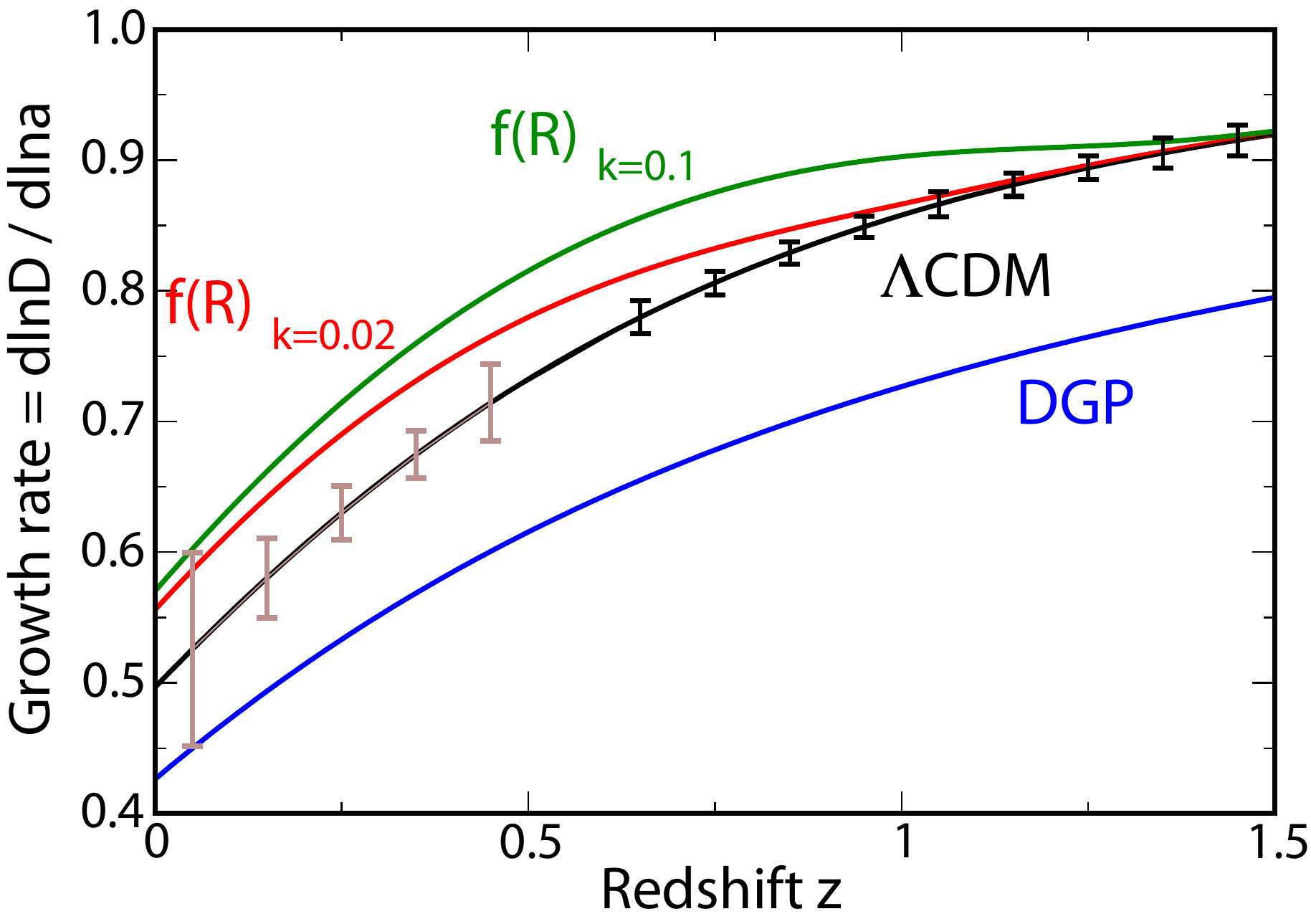}
\caption{Growth of structure, $f$, as a function of redshift, showing
projected DESI measurements and their ability to discriminate against
alternative gravity models, $f(R)$ (whose scale-dependent growth we show
evaluated at two different scales) and DGP. The brown (light) error bars at $z<0.5$
correspond to DESI Bright Galaxy Survey; these are expected
to improve when information from the multiple tracers in the BGS is included.
Adopted from the Snowmass report on the growth of cosmic structure \cite{2013arXiv1309.5385H}.
\label{fig:testGR}}
\end{center}
\end{figure}

The measurements of $f\sigma_8$ from redshift-space distortion provide the
means for testing General Relativity.  Figure~\ref{fig:testGR} shows the rate
of growth of structure, $f$, as a function of the redshift.  Forecasted DESI
errors, assuming information at $k<0.2~\ihMpc$, are shown on the $\Lambda$CDM
curve. Alternative gravity models generically predict scale-dependent growth,
and here we show theoretical expectations for the $f(R)$ modified theory of
gravity evaluated at two scales (two values of $k$), as well as predictions
for the DGP braneworld theory. DESI can clearly distinguish between these models.

\subsubsection{Forecasting Details}

\label{sec:scienceForecastDetails}

\newcommand{\vtheta}{\mathbf{\theta}}
\newcommand{\vC}{\mathbf{C}}

\paragraph{Galaxy and Quasar Clustering}

Our treatment of isolated galaxy BAO follows
\cite{2007ApJ...665...14S}, assuming 50\% reconstruction, i.e.,
reduction of the BAO damping scale of
\cite{2007ApJ...665...14S} by a factor 0.5, except at very low number
density, where we degrade reconstruction based on
\cite{2010arXiv1004.0250W}.


Bias uncertainty is modeled by a free parameter in
each redshift bin, generally of width $\Delta z=0.1$, for each type of galaxy.
Our results are not sensitive to the redshift bin width
\cite{2014JCAP...05..023F}.
For the broadband signal, we use the same information damping factors
from \cite{2007ApJ...665...14S} as we use for BAO.
This is well-motivated from a theoretical point of view
as the non-linear clustering suppresses
all linear theory information, not just BAO \cite{2008PhRvD..77b3533C}.
We also include the reconstruction factor
(50\% reduction in damping length),
assuming that reconstruction will recover non-BAO information as well.
See \cite{2014JCAP...05..023F} for more discussion.

\paragraph{\texorpdfstring{\lya}{Ly-alpha} Forest}

DESI will also probe large-scale structure using the \lyaf
\cite{2006ApJS..163...80M,2011JCAP...09..001S},
i.e., the \lya~absorption by
neutral gas in the intergalactic medium in the spectra of high redshift quasars
(it may be possible to do even better at faint magnitudes using
Lyman-break galaxies \cite{2007PhRvD..76f3009M}). The distribution of intergalactic gas can be used as a complementary tracer to galaxies of the underlying matter distribution for BAO and broadband power spectrum characteristics.

The constraints from the \lyaf are difficult to predict accurately, because
they require careful simulation
\cite{2005ApJ...635..761M,2005MNRAS.360.1471M}.
The forecasts described below we believe are a conservative assessment. We 
limit the application of  \lyb forest data to BAO only (see below), 
and do not include 
cross-correlations with quasar density, nor  statistics beyond the power
spectrum, such as the bispectrum, which are known to be powerful for breaking IGM model degeneracies
(e.g., \cite{2003MNRAS.344..776M}). Finally, we only use the redshift
range $z=2-2.7$. 

We model the three dimensional power spectrum of \lya using Eq. (\ref{eq:Pspec})
 and, except as otherwise noted, we use the method of
\cite{2007PhRvD..76f3009M} to estimate the errors obtainable by
DESI. We use Table I of \cite{MCDON03} to model the dependence of $b$,
$\beta$, and fitting parameters of $D$. While these are primarily
valid near $z\approx 2.25$, for BAO the model dependence is not
significant. For broadband spectra constraints the bias and damping
parameters depend on the amplitude and slope of the linear power
spectrum, temperature-density relation \cite{2001ApJ...562...52M}, and
mean level of absorption \cite{2000ApJ...543....1M}, all of which are
varied in our Fisher matrix calculations.  To help constrain these
parameters, we include the one-dimensional power spectrum, which could
be measured from the hundreds of existing high resolution
spectra \cite{2000ApJ...543....1M,2004MNRAS.347..355K}.





While past projections used the rest wavelength range
$1041<\lambda<1185$~\AA\ (following \cite{2006ApJS..163...80M}), for the BAO constraints only, we expand the
range to include the \lyb forest and
move slightly closer to the quasar,
$985<\lambda<1200$~\AA, reflecting our increasing confidence that we understand
the relevant issues well enough to measure BAO across this range
\cite{2013arXiv1307.3403I}. (The \lyb forest is the wavelength range 
$\sim 973-1026$\AA\ where there is \lyb absorption on top of the \lya 
absorption. This \lyb absorption corresponds to the same gas we see in the
standard \lya forest and should provide some extra information, but we simply
assume it can be mostly removed as a source of noise and the underlying \lya 
used to measure BAO to shorter wavelengths in each quasar spectrum.) 
Gains from this enhancement of effective
number density (and cross-correlations with quasars) are substantial
because the measurement is quite sparse, i.e.,
in what for galaxies we would call the shot-noise limited regime.


The cross-correlation of quasars with the \lyaf
\cite{2013JCAP...05..018F} provides a complementary
measurement of BAO at high redshift. We combine the two probes of structure in
the same volume as described in \cite{2014JCAP...05..023F}.
The correlation of \lya absorption in quasar spectra can also provide
other cosmological information, beyond BAO: cosmological parameter
constraints from the line of sight power spectrum
\cite{2006ApJS..163...80M,2013arXiv1306.5896P, 2006JCAP...10..014S,
2005PhRvD..71j3515S}, and from the full shape of the three-dimensional
clustering \cite{MCDON03}. In the projections below we distinguish
between \lyaf BAO measurements and broadband measurements that include
the one-dimensional power spectrum measurement.

\clearpage
\subsection{Cosmology Beyond Dark Energy}

\label{sec:scienceBeyond}

While the fundamental goal of DESI is the measurement of the expansion
rate of the Universe through BAO and RSD, the enormous spectroscopic
survey will measure the two-point correlation function and
power-spectrum over a broad range of scales and redshifts.  These data
will open up broader investigations into cosmology and particle
physics.

The broadband power spectrum will provide tests of inflation through
its scale dependence. Inflation can also be tested through the scale
dependence of the bias of dark matter halos, which constrains the
primordial non-Gaussianity.  The power spectrum will also reflect the
damping of structure by free-streaming neutrinos and thereby give a
measure of the sum of the neutrino masses, and possibly reveal
previously unknown nearly massless species.

\subsubsection{Inflation}

The inflationary paradigm is the leading explanation for the origin of
the fluctuations of primordial density, which in turn seeded the
large-scale structure we observe today. In its simplest formulation,
inflation predicts perturbations in the initial distribution that are
very nearly scale-independent and Gaussian-distributed about the mean.
Inflation has been tested primarily with the CMB observations ---
starting with {\it COBE} measurements on large scales in the early
1990s and continuing with the increasingly precise {\it WMAP} and {\it
  Planck} measurements in this millennium. However the CMB temperature
measurements are not expected to improve greatly after {\it Planck}
(though CMB polarization has a lot to offer, in particular in testing
for signatures of inflationary gravity waves). Large-scale structure
measurements have become increasingly precise thanks to 2dF, SDSS, and
WiggleZ. These complement the CMB measurements in temporal and spatial
scales.  The next frontier for tests of inflation is large-scale
structure.  DESI's unparalleled three-dimensional picture of the
evolution of structure will contribute powerfully.

\paragraph{Spectral Index and Its Running}

Inflation predicts that the primordial spectrum of density fluctuations is
nearly a power law in wavenumber $k$. The power law is specified by the
spectral index defined as
\begin{equation}
n_{\rm s}(k_0)=\left .\frac{d \ln P}{d \ln k}\right |_{k_0}
\label{eq:primpower}
\end{equation}
where $k_0$ is some reference scale, typically chosen to be $k_0=0.05~{\rm
Mpc}^{-1}$. A perfect power law would correspond to a constant $n_s$; in
reality, inflation also predicts a small ``running'' with wavenumber
parameterized with the parameter $\alpha=dn_{\rm s}/d\ln k$, again defined at $k_0$.
 The primordial power spectrum can therefore be written as \cite{Kosowsky_Turner}
\begin{equation}
P(k)=P(k_0)(k/k_0)^{n_{\rm S}(k_0)+\frac 12\alpha \ln(k/k_0)}.
\label{eq:primpowerexpand}
\end{equation}

The exact Harrison-Zel'dovich primordial spectrum has $n_{\rm s}=1$, while
inflation predicts slight deviations from unity.  Ruling out $n_{\rm s}=1$ at
a significant level of confidence would strengthen the case for
inflation \cite{Dodelson_book}. Recent {\it Planck} data currently favor $n_{\rm s}<1$ at
5$\sigma$;
$n_{\rm s}=0.968\pm 0.006$ \citep{Ade:2015lrj}. The current
limit on running of the spectral index obtained by {\it Planck} is $d
n_{\rm s} / d \ln k =-0.003\pm 0.007$ (95\% CL). Because it is in the
regime of linearity for a wide range of $k$, the \lyaf is an excellent
complementary probe of $\alpha_s$. 

In Table.~\ref{tab:FisherInflation} we present forecasts on inflationary
observables obtained with the Fisher-matrix formalism described in
Section~\ref{sec:scienceForecastMethod}, applied to the power
spectrum obtained from DESI galaxies, quasars, and \lyaf, combined with CMB data from the
{\it Planck} satellite.
\begin{table}
\begin{center}
\caption{Projected constraints on inflationary observables obtained by DESI.
In all cases, we include constraints from the {\it Planck} satellite and BAO
information from DESI galaxies, quasars and the \lyaf. We show the result of including information
from the broadband galaxy power spectrum (``Gal'') out to $\kmaxeff=0.1$ and $0.2~\ihMpc$,
and from the \lyaf. The numbers in parentheses show the relative improvement over
{\it Planck}.
Broadband \lyaf constraints include $\sim 100$ existing
high resolution spectra to constrain the IGM model.  $n_s$ constraints
assume fixed $\alpha_s$.  Both constraints are marginalized over $\summnu$,
and the fiducial values are $n_s=0.963$, $\alpha_s=0$.
\label{tab:FisherInflation}}
\begin{tabular}{lcccc}
\hline
Data & $\sigma_{n_{\rm s}}$ & $\sigma_{\alpha_{\rm s}}$ \\
\hline
Gal ($k_{\rm max}=0.1 \ihMpc$) & 0.0025 (1.3) & 0.005 (1) \\ 
Gal ($k_{\rm max}=0.2 \ihMpc$) & 0.0022 (1.5) & 0.004 (1.3) \\ 
Ly-$\alpha$ forest & 0.0029 (1.1) & 0.0027 (1.9) \\ 
Ly-$\alpha$ forest + Gal ($k_{\rm max}=0.2$) & 0.0019 (1.7) & 0.0019 (2.7) \\ 

\hline
\end{tabular}
\end{center}
\end{table}
%
The table shows strong constraints on $n_{\rm s}$,
and improvements up to a factor of three over $Planck$ alone, under the assumption that there is no significant running in the spectral index.  
Achieving these constraints will require excellent control of broad-band 
systematics in the \lyaf and galaxy analyses.  But the effort is worthwhile,
as these measurements
can have far-reaching implications on our understanding of the very early
Universe, as we now describe.

For the spectral index, the increased accuracy implies much better
constraints on models of inflation. With the DESI+$Planck$ constraints,
excellent constraints on the spectral index will effectively reduce
the allowed region in the plane of $n_{\rm s}$ and $r$, the ratio of
tensor to scalar modes, to a vertical line pinned at the measured
value of $n_{\rm s}$. Combining these results with better measurements of the
$r$ from the small-scale CMB experiments will lead to much better
constraints on inflationary models. Even without the accompanying $r$
measurements, better determination of the spectral index is important:
for example, for inflationary potentials $V(\phi)\propto
\phi^m$, where $\phi$ is the inflaton field, the spectral index and
the total number of e-folds of inflation $N$ are related via $1-n_{\rm
  s}=(m+2)/(2N)$ \cite{Liddle_Lyth}. Hence, for this class of models
the duration of the inflationary phase would be determined by DESI very
precisely.

Implications of the precise measurements of the running of the spectral index
$\alpha_{\rm s}$ are even more impressive.  In standard single-field
slow-rolling inflationary models, the running of the spectral index is of the
order $O((1-n_s)^2)\sim 1 \times 10^{-3}$ if $n_s\sim 0.96$. This means that
DESI will start to approach the region of expected detection in minimal
inflationary models. More importantly, a detection of running {\it larger}
than the slow-roll prediction would imply either that inflation involves
multiple fields, or a breakdown of the slow roll
approximation \cite{Easther_Peiris}, or else that a non-canonical kinetic term
is controlling inflationary dynamics \cite{Chung_Shiu_Trodden}.  {\it Any}
detection of the running of the spectral index would represent a
significant advance in our understanding of the physics of inflation.

\paragraph{Primordial non-Gaussianity}

One of the fundamental predictions of the simplest inflationary models
is that the density fluctuations in the early Universe that seeded
large-scale structure were nearly Gaussian distributed.  A single
field slow-roll inflation with canonical kinetic energy and adiabatic
vacuum predicts very small amount of non-Gaussianity. A violation of
any of these conditions, however, may lead to large non-Gaussianity.
A simple, frequently studied model is that of non-Gaussianity of the
local type, $\Phi=\phi_G+\fnl (\phi_G^2-\langle\phi_G^2\rangle)$,
where $\Phi$ is the primordial curvature fluctuation and $\phi_G$ is a
Gaussian random field.  A detection of nonzero $\fnl$ would rule out
the simplest model of inflation, while a non-detection at a level of
$\fnl<O(1)$ would rule out many of its alternatives.

The tightest existing upper limits on non-Gaussianity have been
obtained from observations of the cosmic microwave background by the
{\it Planck} experiment\citep{Ade:2015ava}. Recently, a number
of inflationary models have been proposed which predict a potentially
observable level of non-Gaussianity, these include those from
fast-roll inflation
\citep{Chen:2006nt,Khoury:2008wj,Noller:2011hd,Ribeiro:2012ar,Noller:2012ed},
quasi-single field inflation \citep{Chen:2009we,Chen:2009zp}, warm
inflation \citep{Gupta:2002kn,Moss:2007cv}, and non-Bunch-Davies or
excited initial states
\citep{Chen:2006nt,Holman:2007na,Meerburg:2009ys,Agarwal:2012mq}.
There are also hybrids of multi-field and non-slow-roll models
\citep{Langlois:2008qf,Arroja:2008yy,RenauxPetel:2009sj}, and the
inclusion of isocurvature modes in the non-Gaussian correlations
\citep{Langlois:2011zz,Langlois:2011hn,Langlois:2012tm}.  Improved
limits on non-Gaussianity would rule out some of these models.
Conversely, a robust detection of primordial non-Gaussianity would
dramatically overturn the simplest model of inflationary cosmology,
and provide information that would help us significantly improve our
understanding of the nature of physical processes in the early
Universe.

Until recently, the most powerful methods to place limits on $\fnl$
were based on the bispectrum of the CMB. The constraints from CMB data
have improved starting from $\sigma(\fnl)\simeq 3000$ with {\it COBE}
\cite{Komatsu_thesis} to $\sigma(\fnl)\simeq 20$ with {\it WMAP}
\cite{WMAP9_final_results}, to the tight constraint of
$\sigma(\fnl)\simeq 5.8$ with {\it Planck}'s first year data
\citep{2013arXiv1303.5084P} and finally to $\sigma(\fnl)\simeq 5.0$ with the 2015 data from Planck \citep{Ade:2015ava}. It is therefore impressive and maybe even surprising
that a powerful LSS survey such as DESI can provide comparable but
highly complementary constraints to {\it Planck}. Moreover, as we now
show, DESI and {\it Planck} in combination can provide very tight
constraints on distinct {\it classes} of physically motivated
inflationary models.

Powerful constraints on non-Gaussianity can come from the effect that
it has on the clustering of dense regions on very large scales
\citep{2008PhRvD..77l3514D}.
Essentially, the bias of dark matter halos assumes a unique, scale-dependent
form at large spatial scales in the presence of primordial non-Gaussianity of
local type
\begin{equation}
b(k)\equiv b_0+\Delta b (k) =b_0 + \fnl(b_0-1)\delta_c\, \frac{3\Omega_MH_0^2}{a\,g(a) T(k)c^2 k^2},
\label{eq:bias}
\end{equation}
where $b_0$ is the usual Gaussian bias (on large scales, where it is
constant), $\fnl$ is the parameter that indicates departures from
Gaussianity (when $\fnl\neq 0$), $\delta_c\approx 1.686$ is the
collapse threshold, $T(k)$ is the transfer function and $g(a)$ is the
growth suppression factor.  Notice the unique $k^{-2}$ scale
dependence in the presence of primordial non-Gaussianity.  Since the
bias $b(k)$ is readily measured from the correlation function of
galaxies or quasars, classes of inflationary models can be tightly
constrained.  A first application of this method has been presented
using the large-scale clustering of quasar and luminous red galaxies
(LRG) galaxy data from the Sloan Digital Sky Survey (SDSS)
\citep{2008JCAP...08..031S}.  The result, a non-detection with one
sigma error $\sigma(\fnl)\simeq 25$, was (at the time) comparable to
the CMB constraints from {\it WMAP}. DESI will provide constraints
competitive, and very complementary, to those from {\it Planck},
provided that we have systematics under control
\citep{2013arXiv1311.2597H, 2014PhRvD..89b3511G, Agarwal_sys}

Forecasts for DESI indicate that the 1$\sigma$ error on the local model from
DESI alone will be $\sigma(\fnl)\simeq 5$, and about a factor of two better
when combined with the final $Planck$ temperature and polarization data. From
the fundamental physics point of view, these constraints are very exciting, as
they probe not only primordial non-Gaussianity but are likely to detect the
additional non-Gaussian signal due to late-time nonlinear interactions of the
photon-baryon fluid with gravity (with $\fnl^{\rm late}\simeq$
few \cite{Pitrou_fNL_nonlin,Bartolo_AdvAstro}), and thus provide an additional
test of cosmology.

\begin{figure}[tb]
\centering
\includegraphics[height=3in]{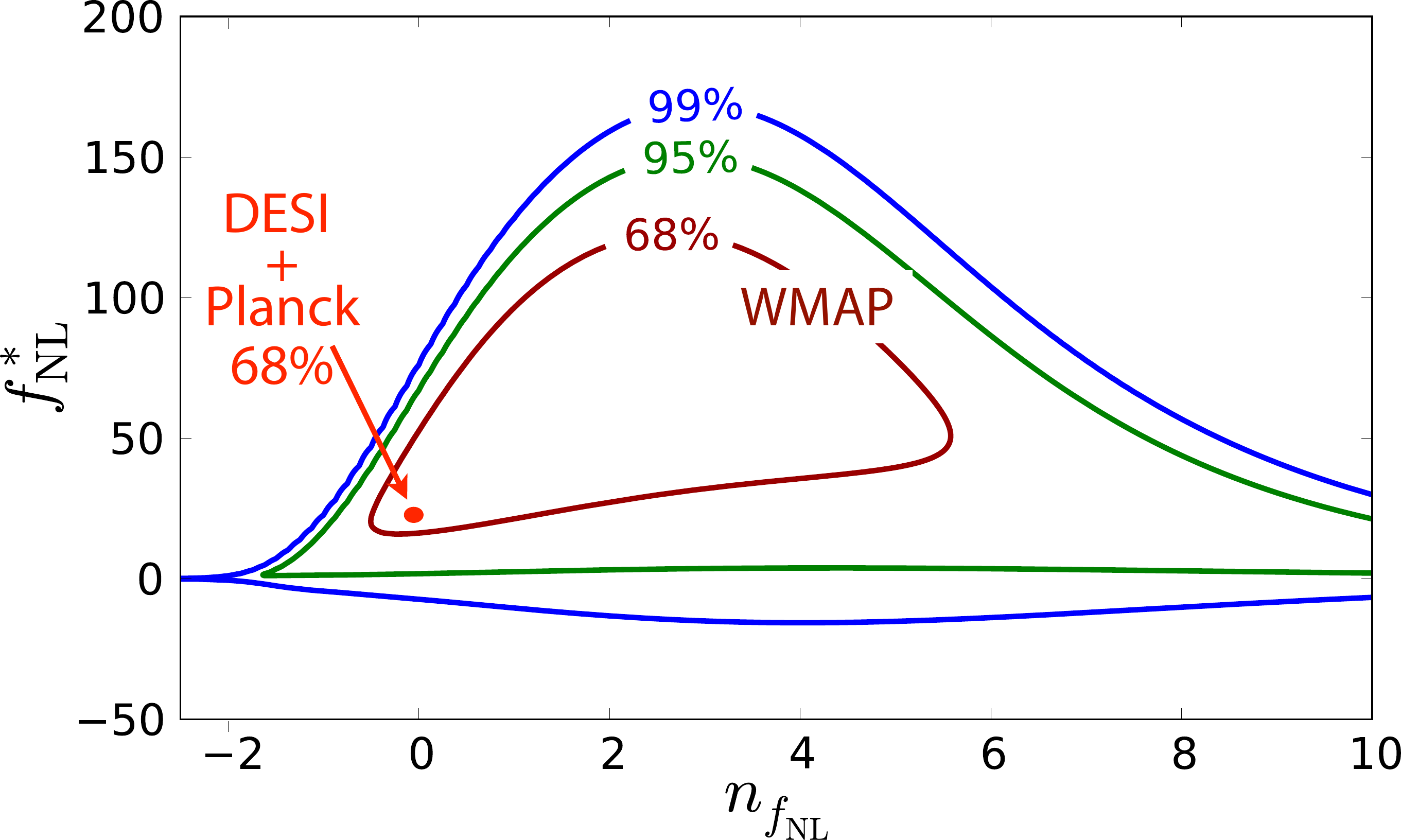}
\caption{Constraints on the models of primordial non-Gaussianity with
``running'', where the usual parameter $\fnl$ is promoted to a power-law
function of wavenumber, $\fnlk=\fnlstar (k/k_*)^\nf$. The larger contours show
constraints on $\fnlstar$ and $\nf$ from a first analysis that was applied to
WMAP 7 data \cite{Becker_Huterer_nfnl}. The size of the red dot shows the 68\%
C.L.\ forecast on the joint constraint expected from the combination of the
DESI and full {\it Planck} data sets, based on projections in
Ref.~\cite{Becker_Huterer_Kadota}.}
\label{fig:fnlk}
\end{figure}

More generally, inflationary models predict a range of possibilities for the
scaling of the bias $\Delta b\propto k^{-m}$. For example, $m=2$ for
the local model parameterized by $\fnl$ as in Eq.~(\ref{eq:bias}); multi-field
inflationary models generically produce $0<m\lesssim 2$, and models with
modifications to the initial quantum state can produce an even stronger
scaling with $m=3$.  Because many of these models therefore leave a
strong imprint in the clustering of galaxies and quasars, DESI will be able to
strongly constrain whole classes of inflationary models. We show an illustration
in Figure~\ref{fig:fnlk}, where we present  constraints on the models with
``running'' of non-Gaussianity, where the usual parameter $\fnl$ now runs with
wavenumber, $\fnlk=\fnlstar (k/k_*)^\nf$. The larger contours show constraints
on $\fnlstar$ and $\nf$ from a first analysis that was applied to WMAP 7
data \cite{Becker_Huterer_nfnl}, while the small, red contour shows the 68\%
C.L.\ forecast on the joint constraint expected from the combination of the
DESI and full Planck data sets, based on projections in
Ref.~\cite{Becker_Huterer_Kadota}.  The latter constraint will shrink the area
in the $\fnlstar-\nf$ plane by about a factor of 100.

To achieve such excellent constraints, the galaxies measured in DESI must have
sufficiently large bias, since only for biased tracers is the non-Gaussian
scale-dependent clustering revealed.  One way to further improve the errors
is by combining two tracers of LSS, one with a high bias and one with a low
bias. In this case it may possible to cancel sampling variance, which is the
dominant source of error on large
scales \cite{Seljak_var,Seljak_McDonald_var}, but due to low number density
this will have to include an additional tracer of structure, potentially
combining with the LSST and DES data.

More detailed studies of halo mass distribution of BOSS galaxies,
combined with numerical simulations of non-Gaussian models
\cite{desjac09} as well as studies of how to mitigate the large-angle
systematic errors \cite{Huterer_sys,Leistedt_sys,Agarwal_sys} are needed to
provide a better definition of the ultimate reach of DESI for
non-Gaussianity studies. However it seems certain that DESI constraints will
be at least comparable to the best limits from CMB and that they will provide
an excellent temporal and spatial complement to the latter.

\subsubsection{Neutrinos}


The effects of neutrinos in cosmology are well understood (for a review,
see \cite{Lesgourgues_neutrino_book}). They decouple from the cosmic plasma
when the temperature of the Universe is about $1$ MeV, just before
electron-positron annihilation. While ultra-relativistic, they behave as extra
radiation (albeit not electromagnetically coupled) with a temperature equal to
$(4/11)^{1/3}$ of the temperature of the cosmic microwave background. As the
Universe expands and cools, they become non-relativistic and ultimately behave
as additional dark matter.

\paragraph{Neutrino Mass}

The mass of neutrinos has two important effects in the
Universe \cite{Lesgourgues_neutrino_book}.  First, as the neutrinos become
non-relativistic after the time of CMB decoupling they contribute to the
background evolution in the same way as baryons or dark matter, instead of
becoming completely negligible as they would if massless (like photons). This
affects anything sensitive to the background expansion rate, e.g., BAO
distance measurements.  Second, the process of neutrinos becoming
non-relativistic imprints a characteristic scale in the power spectra of
fluctuations.  This is termed the `free-streaming scale' and is roughly equal
to the distance a typical neutrino has traveled while it is
relativistic. Fluctuations on smaller scales are suppressed by a
non-negligible amount, of the order of a few percent. This allows us to put
limits on the neutrino masses.

\begin{figure}
\centering
\includegraphics[height=3in]{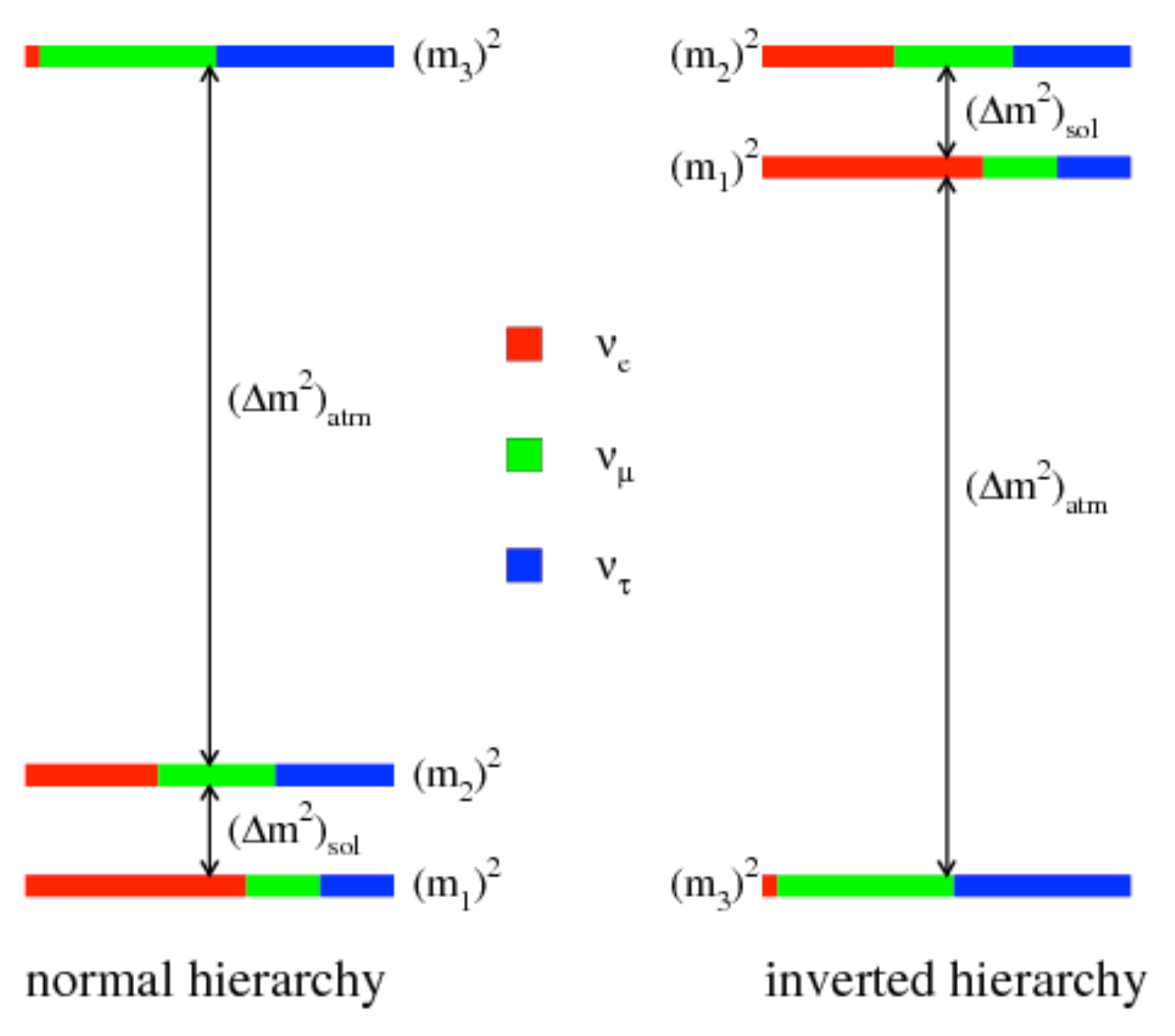}
\caption{The two possible neutrino mass hierarchies.  Also shown is
  what fraction of each mass eigenstate corresponds to a neutrino
  flavor eigenstate.  DESI will be sensitive to the sum of the
  neutrino masses and possibly to the mass hierarchy.}
\label{fig:nuhierarchy}
\end{figure}

From neutrino mixing experiments we know the differences of the
squares of masses of the neutrino mass eigenstates.  The splitting
between the two states with similar masses is $\Delta
m^2_{21}=(7.50\pm 0.20)\times 10^{-5}{\rm~eV}^2$, while the splitting
between the highest and lowest masses squared is $\Delta
m^2_{32}=2.32^{+0.12}_{0.08}\times 10^{-3}{\rm~eV}^2$.  Two things are
not known: the absolute mass scale, and whether the two states close
together are more or less massive than the third state.  In what is
called the normal hierarchy, the close states are less massive.  In
this configuration, the lowest possible masses in eV are 0, 0.009, and
0.048, so the minimal sum of neutrino masses is 0.057 eV.  In the
inverted hierarchy, the minimal masses are 0, 0.048, and 0.049 eV, for
a total of 0.097 eV.  This is shown in Figure~\ref{fig:nuhierarchy}.

Table \ref{tab:FisherNeutrinos} shows our projected $\summnu$ constraints,
obtained through Fisher matrix calculations as discussed above and in 
\cite{2014JCAP...05..023F}.
\begin{table}
\begin{center}
\caption{Constraints on the sum of neutrino masses 
from DESI forecasts in combination with constraints from the {\it Planck}
satellite. The experiment combinations are identified as described in the
caption of Table \ref{tab:FisherInflation}.  The last four cases include the
information from {\it Planck} and DESI BAO measurements.  Fiducial values are
$\summnu=0.06$ eV, $\Nnueff=3.04$.  $\summnu$ constraints assume fixed
$N_\nu$, while $N_\nu$ is marginalized over $\summnu$.
\label{tab:FisherNeutrinos}}
\begin{tabular}{lcl}
\hline
Data & $\sigma_{\summnu}$ [eV] & $\sigma_{\Nnueff}$ \\
\hline
Planck & 0.56 & 0.19 \\ 
Planck + BAO & 0.087 & 0.18 \\ 
\hline
Gal ($k_{\rm max}=0.1 \ihMpc$) & 0.030 & 0.13 \\ 
Gal ($k_{\rm max}=0.2 \ihMpc$) & 0.021 & 0.083 \\ 
Ly-$\alpha$ forest & 0.041 & 0.11 \\ 
Ly-$\alpha$ forest + Gal ($k_{\rm max}=0.2$) & 0.020 & 0.062 \\ 

\hline
\end{tabular}
\end{center}
\end{table}

With a projected resolution of 0.020 eV, DESI will make a precision measurement of the sum of the neutrino masses
independent of the hierarchy and therefore determine the absolute mass scale for neutrinos, a
measurement that is otherwise very challenging. Furthermore, if the masses
were minimal and the hierarchy normal, DESI would be able to exclude the inverted hierarchy at 2$\sigma$.


\paragraph{Dark Radiation (e.g., sterile neutrinos)} 

The other parameter relevant for neutrino physics is the effective
number of neutrino species $\Nnueff$, which parameterizes the energy
density attributed to any non-electromagnetically interacting
ultrarelativistic species (including e.g. axions) in units of the
equivalent of one neutrino species that fully decouples before
electron-positron annihilation. Extra radiation shifts the redshift of
matter radiation equality and changes the expansion rate during the
CMB epoch, although it does not significantly affect the Universe at
the epoch probed by DESI.  The value for the standard cosmological
model is $\Nnueff=3.04$\footnote{The small increase with respect to
  $N_\nu=3$ is due to the fact that some neutrinos are still coupled
  at the onset of electron-positron annihilation.}
\citep{Mangano:2005cc}.  The detection of any discrepancy from the
expected value would be a truly major result, as it would indicate a
sterile neutrino \cite{Hamann}, a decaying particle
\cite{Fischler_Meyers}, a nonstandard thermal history
\cite{Bashinsky_Seljak}, or perhaps that dark energy does not fade
away to $\sim 10^{-9}$ at the time of recombination as expected for
the cosmological-constant model \cite{Calabrese}.  All of these
possibilities represent important extensions of the standard
cosmological model, and uncovering them would present a major advance
of our understanding of the Universe.  Our forecasts for this
parameter are also shown in Table \ref{tab:FisherNeutrinos}. Again we
see that the effective number of neutrino species will be measured to
$\sim 10\%$ or better, providing strong constraints on the alternative
models involving extra sterile neutrinos, axions or partly thermalized
species.

\begin{figure}[!t]
\begin{center}
\includegraphics[height=5in]{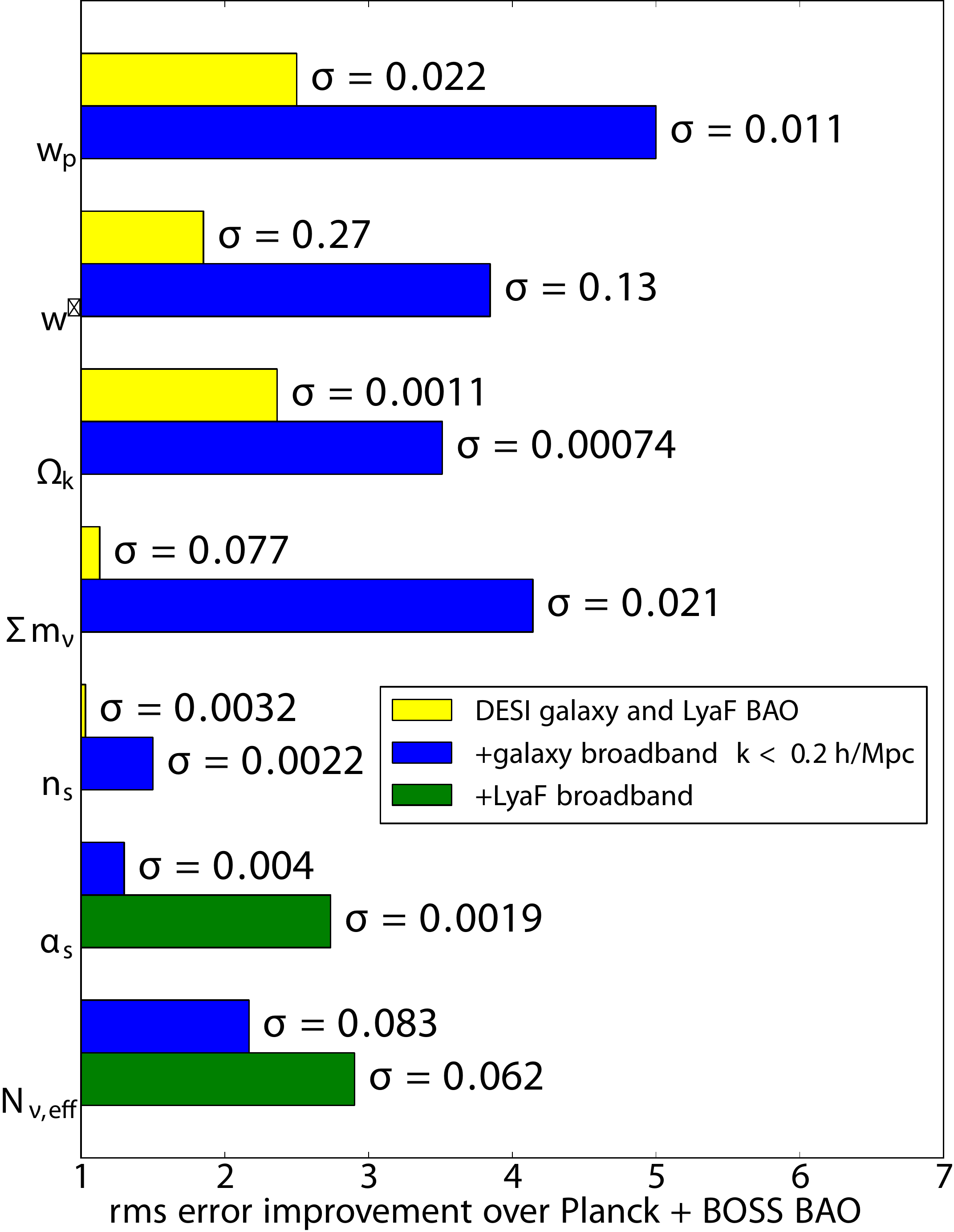}
\caption{Improvement in the measurements of $w_p$, $w'=w_a$, 
$\Omega_k$, $\sum m_{\nu}$ the sum of the neutrino masses, 
$n_s$ the spectral index, $\alpha_s$ the running of the spectral index, and 
$N_{\nu,{\rm eff}}$ the number of neutrino-like (relativistic) species.
\label{fig:bars}}
\end{center}
\end{figure}

In Figure~\ref{fig:bars} we show the improvement in the measurement of
several fundamental parameters from cosmology and neutrino physics.
The standard is taken to be the results from BOSS together with
$Planck$.  Displayed is the ratio of the uncertainty from BOSS over
the uncertainty from DESI, with $Planck$ always included.

\clearpage
\subsection{The Milky Way Survey: Near-Field Cosmology from Stellar Spectroscopy}
\label{sec:mws}

During conditions unusable for faint galaxy work, DESI will pursue
the Bright Galaxy Survey, mapping 10 million galaxies to $z\sim0.4$
in pursuit of the clustering analyses, such as from BAO and RSD,
as described earlier in this chapter.
As detailed in section \ref{sec:bgs},
the areal density of these bright galaxies is comparable to the
fiber density of DESI.  Achieving a high completeness in the
face of clustering and Poisson fluctuations requires multiple visits,
leading to an excess of fibers compared to targets.
Indeed, some fibers will be unable to reach a viable galaxy target even on
the first pass, and this fraction increases on subsequent passes.

Bright stars are the natural secondary target, and we expect that
any bright galaxy survey with the DESI fiber positioner will produce
a very large sample of stars as a by-product.  This sample is also
of high science interest, leading to the definition of the Milky
Way Survey. 
At 17th magnitude, even a short (8-10 min) DESI exposure measures
an excellent spectrum with $S/N = 25$ per pixel, which will yield
the radial velocity to a few km/s precision and the metallicity.
We expect the BGS to generate at least 10 million such spectra.
Spectroscopy of individual stars provides radial velocity, effective
temperature, surface gravity, chemical abundance distribution, and
approximate age.  The assembly history of the Milky Way is encoded in
the spatial distributions, kinematics, and chemical composition of the
various distinct Galactic stellar populations. This information can
test cosmological predictions for how galaxies like the Milky Way form
and evolve on small scales that are difficult or impossible to test
elsewhere in the Universe, and provide a critical test of the
small-scale predictions of the $\Lambda$CDM model.


The European Space Agency {\it GAIA} satellite has been successfully
launched and will provide a catalog of parallaxes, proper motions, and
spectrophotometry for a billion point sources down to $V \sim 20$ over
the whole sky. The satellite's RVS spectrograph will supplement those
data with radial velocities for millions of brighter stars, although
the flux limit is still under investigation due to higher than
expected scattered light.  DESI can substantially enhance the science
return from {\it GAIA} by providing radial velocities and
metallicities for stars much fainter than what the {\it GAIA}
spectrograph can provide.  While other projects are planned for
spectroscopic follow-up of {\it GAIA} stars, DESI's higher multiplex,
wide field of view, and extremely rapid reconfiguration give it a
clear advantage. 

The stellar program will put exceptional new constraints on the
distribution of dark matter in the Milky Way, a vital measurement that
links Galactic science, galaxy formation and cosmology.  The Milky Way
gravitational potential can be probed via the rotation of the Milky
Way beyond 15 kiloparsecs, the motions of newly discovered tidal
streams, and the kinematics of bright stars in the distant stellar
halo.  The uncertainty in the Milky Way mass, density profile, and
internal structure currently are critically important systematics in
the interpretation of direct and indirect dark matter searches, and
the measurements possible with the stellar program will substantially
reduce these uncertainties.

Joint metallicity and velocity distribution functions for stars far
beyond the solar neighborhood will reveal the recent assembly
history of the outer disk and vastly improve our understanding of
the structure and formation of the thick disk. The first-ever deep
spectroscopic survey of halo main-sequence turn-off stars to 30
kiloparsecs can be used to reconstruct the history of the Galaxy
in its first two billion years and its interaction with other
galaxies, shedding new light on enigmatic halo substructures like
the Virgo overdensity and Hercules--Aquila cloud. Moreover, a survey
of millions of stars will have huge potential for the discovery
of kinematically and chemically peculiar stars in as-yet unexplored
regions of the Galaxy.


\clearpage
\subsection{Complementarity with Other Surveys}

While DESI's spectroscopic survey will by itself yield incisive
results in cosmology, its power is increased when combined with other
experiments.  DESI's BAO results are directly connected to CMB
measurements via its dependence on the acoustic scale, but additional
information can be obtained by directly cross-correlating the CMB with
the density distribution and redshift space distortions from DESI.
Large imaging surveys, including DES and LSST, will provide vast amounts of
complementary data, allowing increased precision for both cosmological
and neutrino measurements. This combination of imaging and
spectroscopic surveys is particularly powerful for distinguishing dark
energy from modified gravity models for cosmic acceleration.


\subsubsection{Synergies with \textit{Planck} and Future CMB Experiments}

The cross-correlation of Planck and potential future CMB experiments, such as 
Advanced ACTPol and CMB-S4, with DESI enables cosmological
measurements not possible with either individually, and opens up new opportunities to constrain fundamental physics, in the properties of dark energy and gravity discussed in \ref{sec:scienceForecastIntro} and the nature of neutrinos and inflation summarized in \ref{sec:scienceBeyond}.

On large scales, cross-correlating CMB temperature fluctuations
with the galaxy density field measures the Integrated Sachs-Wolfe
effect, probing the time evolution of the gravitational potential
and independently constraining dark energy \citep{2008PhRvD..78d3519H}.
The combination of CMB lensing and the foreground galaxies or quasars
will also improve not only the signal-to-noise of CMB lensing leading
to stronger cosmological constraints on the matter content, but
also our understanding of the foreground tracers in large-scale
structure, as lensing allows a clean measurement of the bias of the
foreground sources.

The combination of CMB lensing and the RSD measurements from DESI
will allow a probe of the two relativistic gravitational potentials
independently (see e.g. \cite{2010Natur.464..256R} for an application
of this test but for the case of gravitational lensing of background
galaxies, not the CMB), testing the GR prediction of their equality
over a wide redshift range \cite{2007PhRvL..99n1302Z}. CMB lensing
and RSD measurements will also provide complementary constraints
on the sum and differences of the neutrino masses, that in combination
could  help infer the neutrino hierarchy.

DESI will provide highly complementary constraints on inflation to
those from {\it Planck} and a number of upcoming CMB small scale
temperature and polarization experiments.  An exciting  realization
in inflationary theory  is that discerning the scale-dependence,
or `shape', of the bispectrum (the 3-point function) could provide
a direct insight into the inflationary mechanism, through how
non-Gaussianity is generated \citep{Fergusson:2009nv,Liguori:2010hx}.
CMB 3-point correlation measurements constrain a wide range of
primordial bispectrum configurations, while DESI will provide more
detailed information about the properties in the squeezed limit, a
regime that could provide characteristic information about the
underlying mechanism driving inflation e.g. whether it is multi-field,
sourced from a non-Bunch Davies vacuum  state, or includes non-trivial
kinetic terms in the inflationary action.

Cross correlating the galaxy velocity field (inferred from the 3D
density distribution) with the CMB will measure the kinetic
Sunyaev-Zeldovich (kSZ) effect at the percent level. These measurements
provide constraints on more exotic deviations from our standard
cosmological models
\cite{2010MNRAS.407L..36Z,2011PhRvL.107d1301Z,2012ApJ...758..130L}.  In
addition, these measurements are astrophysically important since
the kSZ effect is an unbiased probe of electrons and can be used
to inventory the baryons in the Universe \cite{2009arXiv0903.2845H}.

\subsubsection{Synergies of DESI with DES and LSST}

The massive spectroscopic survey provided by DESI will provide a
unique and important complement to direct-imaging science projects
currently being planned.  We focus here on the Dark Energy Survey
(DES) and the Large Synoptic Survey Telescope (LSST), but DESI will
complement other future imaging surveys in similar ways.  Although
both DES and LSST are located in the Southern Hemisphere, their
planned surveys have overlap of a few thousand square degrees with the
baseline DESI survey.  In addition, only some of the cosmological tests
described below rely on overlap between the photometric and
spectroscopic surveys.

DESI can provide critical input into photometric redshifts which
can help control the systematic uncertainty associated with
cosmological measurements from photometric surveys like DES and LSST.
For instance, cross correlation of photometric lensing sources with
spectroscopic galaxy samples enable the reconstruction of the
redshift distribution of the lensing sources
\citep{Newman08,2008PhRvD..78d3519H,2013MNRAS.433.2857M,2013MNRAS.431.3307S}
providing a critical consistency test on the photometric redshifts
used for cosmic shear and/or calibrating the mass of galaxy clusters
for cluster abundance tests.  Likewise, magnification-based lensing
measurements of spectroscopic sources \citep{couponetal13} can provide
a consistency test for shape systematics and/or photometric redshift
systematics in shear-based calibration of cluster masses.

Just as importantly, the combination of photometric and spectroscopic
surveys is significantly more powerful than either set of surveys
alone.  An example is the utility of using galaxy-galaxy lensing,
in which one uses the lensing of background galaxies by galaxies
from the spectroscopic sample to measure the galaxy-mass
cross-correlation of the spectroscopic sample.  On small scales,
this measures the properties of the host dark matter halo, testing
galaxy bias models; on larger scales, it can be used to measure the
mass-mass auto-correlation and hence the amplitude of structure
\citep{2012PhRvD..86h3504Y,2013MNRAS.430..767C}.
Several studies have forecast cosmological constraints from a
combination of DES-like and DESI-like experiments
\citep{cai_bernstein, gaztanaga12, kirk13, deputter13}, and while the
range of assumptions and forecasts varies from work to work, there is
agreement that the combination of DES and DESI/LSST gives substantial
benefits in terms of measured cosmological and non-cosmological
parameters.  This is particularly true within the context of modified
gravity models, where the combination of surveys enables entirely new
types of measurements that are ideally suited for addressing such
questions.  
For instance, recent theoretical work suggests that comparing the
shear field generated by galaxy clusters to the corresponding galaxy
velocity can significantly improve current modified gravity
constraints \citep{lametal12}.


\begin{figure}[!hbt]
\centering
\includegraphics[height=3in]{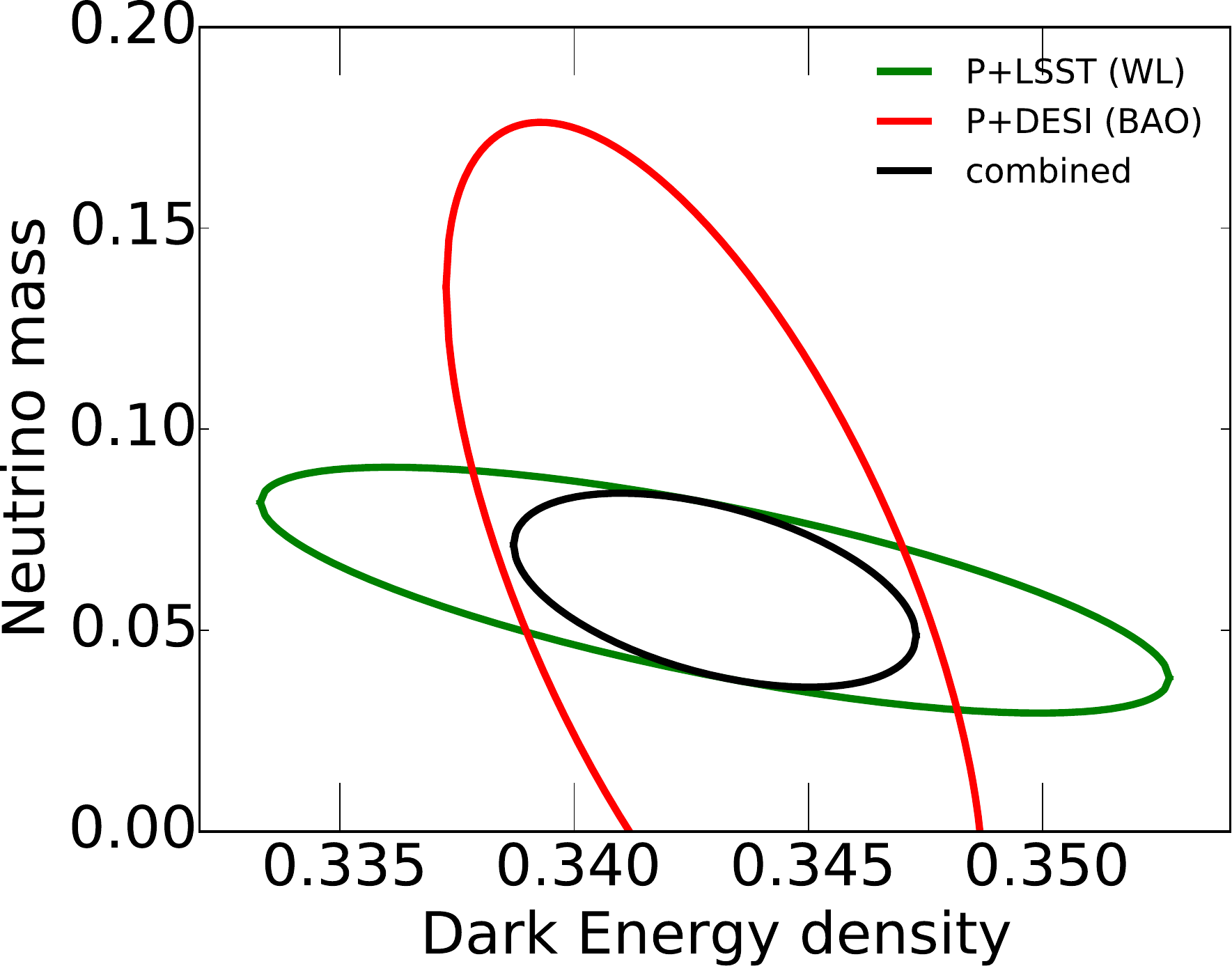} 
\caption{Constraint on the sum of the neutrino masses in eV against the dark 
energy density $\omega_{DE}=\Omega_{DE}h^2$ obtained by combining DESI BAO with
LSST weak lensing, in each case including Planck CMB constraints. 
More powerful constraints are obtained when the full power 
spectrum from DESI is used.  See Table \ref{tab:FisherNeutrinos}.}
\label{fig:omdemnu}
\end{figure}

\begin{figure}[!htb]
\centering
\includegraphics[height=3in]{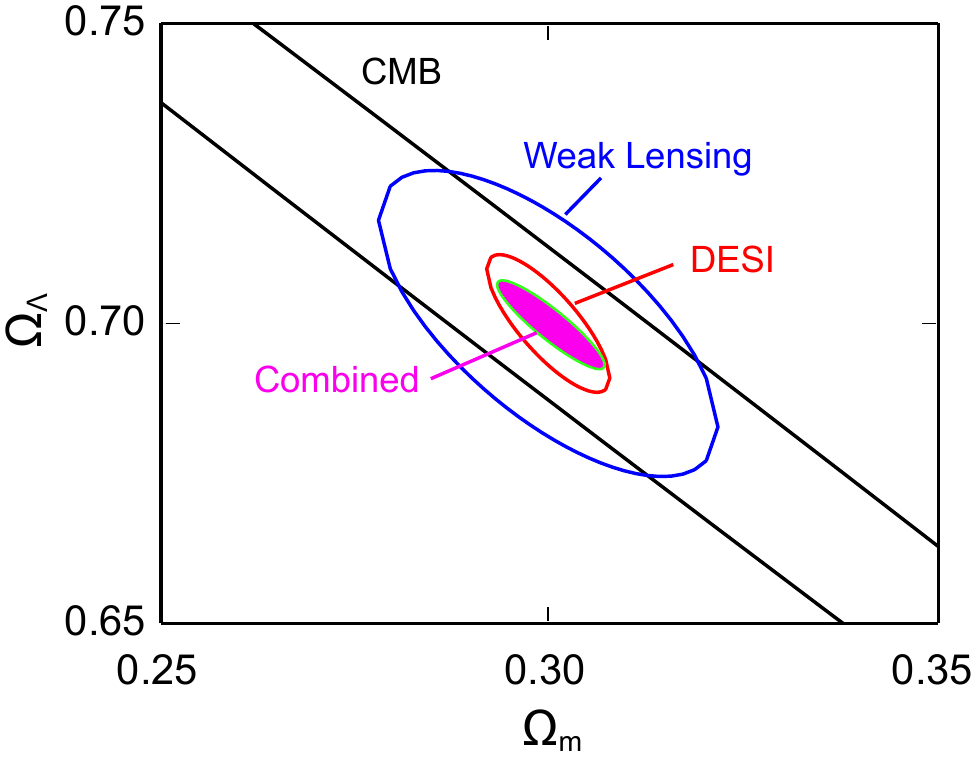} 
\caption{Prospective constraints in the $\Omega_m$--$\Omega_{\Lambda}$ plane  obtained by combining DESI BAO with LSST weak lensing. More powerful constraints are obtained when the full power spectrum from DESI is used.  See Table \ref{tab:FoM}.}
\label{fig:OmmOmde}
\end{figure}

As an example of improvement in another type of constraint that can be
achieved through the combination of DESI with imaging surveys,
Figure~\ref{fig:omdemnu} shows the joint constraint on the sum of the
neutrino masses in eV against the dark energy density
$\omega_{DE}=\Omega_{DE} h^2$ obtained by combining anticipated
results for DESI BAO with LSST weak lensing.  Similarly,
Figure~\ref{fig:OmmOmde} shows prospective constraints in the
$\Omega_m$--$\Omega_{\Lambda}$ plane obtained by combining anticipated
results for DESI BAO with LSST weak lensing (these forecasts assume the surveys
are not overlapping on the sky, although it makes practically no difference 
\cite{2014JCAP...05..023F,2013arXiv1308.6070D}).

Finally, DES and LSST will provide world-leading samples for supernova
cosmology.  The BAO and SNe Ia methods for measuring the cosmic
distance scale are highly complementary: supernovae excel at low
redshifts, where the SNe are brighter and where the BAO is more
limited by cosmic variance due to the small cosmic volume.  The
combination of DESI with ground-based supernovae samples spanning from
$z=0$ to $z\approx 0.8$ will be a powerful view of the
distance-redshift relation and the expansion history of the Universe.
While we have focused on Figure of Merits drawn only from BAO and the
DESI clustering samples, the inclusion of low to intermediate-redshift
supernovae provides a notable improvement to current BAO constraints,
as highlighted in numerous papers, such as
\cite{2014MNRAS.441...24A,2014A&A...568A..22B}.  Essentially one is
using BAO to calibrate the relative distance scale provided by the
SNe.  The redshift overlap of the two methods provides a further
systematic cross-check.  The exquisite precision of DESI at $z>0.6$
will find an excellent partner in the DES and LSST supernova samples.

DESI will directly support the coming decade of supernova cosmology
by providing spectroscopic redshifts for many tens of thousands of
SNe host galaxies.  This will happen both for the faint galaxy
survey out to $z\sim 1$, but also with the BGS at $z<0.4$.  Over a
10-year period, a typical ($L^*$) galaxy has at least a 1\% probability
of having a detectable SN Ia.  This means that the BGS will contain
of order $10^5$ supernova host galaxies, and the LRG sample of more
massive galaxies could produce a comparable number at higher redshift.
While photometric redshifts are planned for the large LSST and DES
supernova samples, spectroscopic redshifts allow more precision,
particularly at low redshift where the uncertainty in the redshift
and resulting luminosity distance overwhelm the intrinsic precision
of the standard candle.  Samples of many tens of thousands of hosts
can only be achieved with multi-object wide-field surveys.  We note
that with DESI there is no need to wait to select the host galaxies
after the explosion: at $z<0.2$, the BGS will include more than
half of all SN Ia host galaxies in the survey footprint.  Having a
pre-existing redshift will also enable better allocation of follow-up
resources for rare transients from surveys such as LSST.

\subsubsection{Synergies of DESI with \textit{Euclid/WFIRST}}

{\it Euclid} is a medium class European Space Agency survey mission designed to
measure Dark Energy \citep{2011arXiv1110.3193L}. Recently, NASA
has become a partner, enabling a group of 40 US astronomers to join the
international consortium. {\it Euclid} will perform a 15,000\,deg$^2$ survey jointly
undertaking visible imaging to measure weak lensing and simultaneous near-
infrared observations split into sequential imaging (for photometric redshift
measurement) and slitless spectroscopy. Two
Deep Fields about 2 magnitudes deeper than the wide survey and covering
around 20\,deg$^2$ each will be also observed, primarily for calibrations of
the wide survey data but also extending the scientific scope of the mission to
faint high redshift galaxies, quasars and AGNs. The spectroscopic survey is
focused on H$\alpha$ emitting galaxies and is most powerful at high
redshifts $1<z<2$. 


The 
timeline for DESI is prior to {\it Euclid} (which is
scheduled to launch in December 2020 for 6 scheduled years of data collection), but even in the era of {\it Euclid}, at
redshifts $z<1$ the combination of LRGs and ELGs that DESI will
observe will remain the world-leading data set for spectroscopically
confirmed galaxies with good redshift measurements. At $z>2$ the DESI
measurements from \lya\ will also remain unique. Euclid may surpass
DESI in the redshift range $1<z<2$ provided the slitless
spectroscopy is as effective as hoped.  DESI could help {\it Euclid}
clustering measurements by providing important information on the
potential confusion of the {\it Euclid} slitless spectroscopy in this
redshift range.  The combination of Euclid space-based weak lensing 
with the large spectroscopic samples from DESI will be a strong 
opportunity for galaxy-galaxy weak lensing, similar to what was discussed
in the DES/LSST context in the previous subsection.  DESI's contribution
of $z<1$ lenses is particularly important in this regard.

{\it WFIRST-AFTA} is an envisaged NASA mission using a 2.4~m diameter
primary mirror satellite being designed to perform a 2000\,deg$^2$
near-infrared survey, including a slitless spectroscopic component
\citep{2013arXiv1305.5422S}. The current narrow/deep {\it WFIRST-AFTA}
concept is highly complementary to the wide/shallow {\it Euclid}
strategy, and will provide deeper, denser galaxy samples.  However,
the smaller area covered compared to either {\it Euclid} or DESI means
that the direct expansion rate and growth rate measurements would be
weaker.

Comparisons of the precision of the BAO measurement projected for
DESI, $Euclid$, and $WFIRST$ are shown in Figure~\ref{fig:binnedR}.

DESI will be highly complementary to the weak lensing surveys to be
performed for {\it Euclid} and {\it WFIRST-AFTA}, providing
spectroscopic galaxy samples at the same redshifts as the matter that
is causing the lensing, thus enabling many innovative analyses from
these combined datasets. DESI will help in the calibration of
photometric redshifts - which are essential for these lensing
experiments - and aid in investigating systematic issues such as
intrinsic alignments.  Likewise, {\it Euclid} and {\it WFIRST-AFTA}
will greatly enhance the legacy value of DESI, providing high
resolution optical and NIR imaging of all DESI targets, greatly
improving the prospects for non-dark energy science, e.g., the
morphology--density relationship at $z>1$.




\clearpage

\section{Target Selection}\label{s4:TargetSelection}
\setcounter{equation}{0}\setcounter{figure}{0}\setcounter{table}{0}




\label{ss4:introduction}

The DESI survey will measure with high precision the baryon acoustic
feature imprinted on the large-scale structure of the Universe, as
well as the distortions of galaxy clustering due to redshift-space
effects.  To achieve these goals, the survey will make spectroscopic
observations of four distinct classes of extragalactic sources -- the
bright galaxy sample (BGS), luminous red galaxies (LRGs), star-forming
emission line galaxies (ELGs), and quasi-stellar objects (QSOs).  Each
of these categories requires a different set of selection techniques
to acquire sufficiently large samples of spectroscopic targets from
photometric data.  To ensure high efficiency and spectroscopic
completeness, we select objects with spectral features expected to
produce a reliable redshift determination or a \lyaf\ measurement
within the DESI wavelength range.


The characteristics of our baseline samples for each of these target
classes are summarized in Table~\ref{tab:TargetReqGalaxyTypes}. This
Table specifies the primary redshift range, the photometric bands for
targeting, the projected areal density (in terms of number of targets,
number of fibers allocated across all pointings accounting for
multiple exposures, and the number of useful redshifts resulting per
square degree), as well as the total number of objects in the desired
class for which redshifts are expected to be obtained for each of these samples.
This table may be compared to Table 1 in the Science Requirements
Document (SRD). The SRD considers both a threshold survey of 9,000~deg$^2$
and a baseline survey of 14,000~deg$^2$.  Throughout this
chapter, we consider only the latter scenario; simulations for reduced
focal planes indicate that we would achieve essentially the same
sample surface densities as for the baseline scenario, so that sample
sizes would simply scale with survey area.
In the following sections, we will describe the basis of these numbers in more detail.

\begin{table}[!b]
\centering
\caption{Summary of the properties for each DESI target class.
The bands listed are for the target selection, where
$g$, $r$, and $z$ are optical photometry and 
$W1$ and $W2$ denote are {\it WISE} infrared photometry.
The exposure densities are increased over the target densities
due to some objects being observed on multiple passes.
The number of good redshifts and baseline sample sizes (in millions)
are for successful redshifts.  
}
\begin{tabularx}{1.0\textwidth}{lcccccc}
\hline
Galaxy type& Redshift &Bands& Targets &Exposures&           Good $z$'s  & Baseline\\
                  & range & used & per deg$^2$ &  per deg$^2$  &  per deg$^2$  &sample\\
\hline
LRG&  0.4--1.0 & $r$,$z$,$W1$&350& 580 &285& 4.0 M\\
ELG&  0.6--1.6 & $g$,$r$,$z$  &{ 2400}&{ 1870} & { 1220} & { 17.1 M}\\
QSO (tracers)& $<2.1$& $g$,$r$,$z$,$W1$,$W2$&170&170&120 & 1.7 M \\
QSO (Ly-$\alpha$)&  $>2.1$& $g$,$r$,$z$,$W1$,$W2$&90&250&50 & 0.7 M \\
\hline
{\bf Total in dark time} & & & {\bf 3010} & {\bf 2870} & {\bf 1675} & {\bf 23.6 M}\\
\hline\hline
BGS &  0.05--0.4 & $r$ & 700 & 700 & 700 & 9.8 M \\
\hline
{\bf Total in bright time} & & & {\bf 700} & {\bf 700} & {\bf 700} & {\bf 9.8 M} \\
\hline\hline
\end{tabularx}
\label{tab:TargetReqGalaxyTypes}
\end{table}

\vspace{0.1in}
\noindent{\it {\bf Summary of Target Samples}}

\vspace{0.1in}
The lowest-redshift sample of DESI targets will be the Bright Galaxy
Sample (BGS).  These galaxies will be observed during the time
when the moon is significantly above the horizon, and the sky
is too bright to allow efficient observation of fainter targets.
Approximately the 10 million brightest galaxies within the DESI
footprint will be observed over the course of the survey,
sampling the redshift range $0.05 < z < 0.4$ at high density.
This sample alone will be ten times larger than the SDSS-I and SDSS-II
``main sample'' of 1 million bright galaxies observed from 1999-2008.

Above redshift $z=0.4$, DESI will observe luminous red galaxies (LRGs).
These luminous, massive galaxies have long since ceased star formation
and therefore exhibit evolved, red composite spectral energy distributions
(SEDs).  The BOSS survey targeted these objects to $z\approx 0.6$ using
SDSS $gri$ colors and measured spectroscopic redshifts using the prominent
4000~\AA~break continuum feature.  While DESI will aim to achieve 
350 LRGs/deg$^2$ over 14,000 square degrees, the BOSS sample of 119 LRGs/deg$^2$
will contribute significantly to our science analyses over the 10,000~deg$^2$
footprint in which it exists; DESI may extend this low-redshift
sample over a larger footprint, but this is not in the current baseline plan.
DESI will target LRGs to $z\approx 1.0$, where they may be most efficiently
selected using the prominent 1.6~\micron~(rest frame) ``bump,'' which causes
a strong correlation between optical/near-infrared (NIR) color and redshift
in this regime.  We will use 3.4~\micron~photometry from the space-based
Wide-Field Infrared Survey Explorer ({\it WISE}) to select LRGs efficiently
in the redshift range of $0.6<z<1.0$. DESI can exploit the 4000~\AA~break
to obtain secure redshifts for LRGs over this full redshift range.

The majority of the spectroscopic redshift measurements for DESI will
come from ELGs at redshifts $0.6<z<1.6$.  These galaxies possess high
star formation rates, and therefore exhibit strong emission lines from
ionized H~II regions around massive stars, as well as SEDs with a
relatively blue continuum, which allows their selection from optical
$grz$-band photometry. The prominent \otwo~$\lambda\lambda3726,29$
doublet in ELG spectra consists of a pair of emission lines separated
in rest-frame wavelength by 2.783~\AA.  This wavelength separation of the doublet
provides a unique signature, allowing definitive line identification
and secure redshift measurements.
The goal of the DESI ELG target selection will be to provide a large sample
of ELGs with sufficient \otwo line flux to obtain a detection and redshift
measurement to $z=1.6$.  

The highest-redshift target sample will consist of QSOs.  We will
measure large-scale structure using QSOs as direct tracers of dark matter
in the redshift range $0.9<z<2.1$.  At higher redshifts, we will utilize the foreground neutral-hydrogen
absorption systems that make up the \lyaf ; DESI spectra cover the Ly-$\alpha$ transition at $\lambda=1216$~\AA\  for objects at $z>2.1$. 
We will use optical photometry combined with{\it  WISE} infrared photometry
in the W1 and W2 bands to select our primary sample of QSOs.
QSOs are $\sim 2$ mag brighter in the near-infrared at all redshifts compared to stars of similar optical magnitude and color, providing a powerful method for discriminating against contaminating stars.
QSOs at $z>2.1$ used for \lyaf\ measurements do not require homogeneous
selection on the sky for cosmological measurements, as we do not rely on
the clustering of the QSOs themselves.  As a result, DESI may exploit optical
variability and additional passbands where available to enhance this sample.  Those $z>2.1$ QSOs which are selected via uniform methods across the sky may also be used to enhance clustering measurements.  
DESI will obtain additional exposures on the confirmed $z>2.1$ quasars
to measure the \lyaf\ to the required S/N.

\vspace{0.1in}
\noindent {\it {\bf Summary of Required Imaging}}
 
\vspace{0.1in}
All DESI target samples will be selected 
using optical $grz$-band photometry from ground-based
telescopes and near-infrared photometry from the {\it WISE}
satellite.  The observations assumed in our baseline targeting plan
are summarized in Table \ref{tab:TargetReqTelescopes}.  This imaging
plan has been developed through a detailed analysis of alternative
telescope/instrument combinations.  
The imaging depths will be at least 24.0, 23.4, 22.5 AB
(5$\sigma$ for an exponential profile $r_3=0.45\arcsec$)
in $g$,$r$,$z$ and 20.0, 19.3 AB (5$\sigma$) in {\it WISE} W1,W2.
All sample magnitude limits quoted in this section are total
(model-like) magnitudes for the BGS and for LRGs and ELGs, or PSF
magnitudes for QSOs.

The optical imaging for the DESI targets will be provided from three
telescopes at two sites, Cerro Tololo and Kitt Peak.  The DECam camera
on the Blanco 4-m telescope will provide $grz$ imaging over 9000~deg$^2$
in the DESI footprint at Dec $\le +34$ deg.
The first 6700~deg$^2$ of this footprint (DECaLS) has been approved as
a 64-night NOAO ``Large Survey'' program during the period August 2014
through July 2017 and is 40\% completed.
An 8-night extension of this program (DECaLS+) has been approved for the
2016A semester to obtain another 800~deg$^2$ in the Northern Galactic Cap.
A proposal to observer the remainder of the DESI footprint in the South
Galactic Sky will be submitted in future semesters.
The Bok 2.3-m telescope is providing $gr$ imaging over
the 5000~deg$^2$ region of the North Galactic Cap (NGC) that lies at Dec $\ge +34$ deg
with the existing 90Prime camera.  The 220 nights necessary for these
observations are guaranteed via an MOU with the University of Arizona
/ Steward Observatory.  Observations were taken in Spring 2015
which identified electronics problems in the camera that were corrected
in September 2015.
The Bok observations re-started in January 2016 and are now 15\% complete.
The Mayall 4-m telescope will provide $z$-band imaging over the same NGC
footprint using the existing MOSAIC-2 camera upgraded with 4
red-sensitive CCDs.
Those observations will be conducted over 220 nights in 2016 and 2017.
The Mayall observations began in February 2016 and are now 15\% complete.
All imaging data are planned to be completed by August 2017,
where the Mayall observations must be complete as that telescope
will being taken off-line for DESI installation.


The {\it WISE} satellite is obtaining
infrared imaging to sufficient depths for DESI target selection over
the full sky.  An initial 13-month survey is being supplemented with
a 3-year extended mission known as {\it NEOWISE} that began 1 December 2013
and will complete in December 2016.  The initial {\it WISE} survey
and first year of {\it NEOWISE} data are publicly available, with
the final two data releases scheduled for March 2016 and March 2017.

The DESI analyses will be performed separately in
each of the three regions of the DESI footprint: the NGC
at DEC $>+34$ deg, the NGC at DEC $<+34$ deg,
and the South Galactic Cap (SGC).  Based on SDSS-III/BOSS experience with separately-calibrated regions,
we expect to analyze these separately and combine
the cosmological constraints downstream.
The DECam and Bok/MOSAIC coverage will have some overlap (at the DEC $\approx +34^\circ$ strip and 
by targeting specific calibration fields like COSMOS, Bo\"otes, and DEEP-2)
in order to tie together calibrations and understand the subtle variations in target selection 
resulting from 
differences in filter+telescope response between the two datasets.


\begin{table}[!t]
\centering
\caption{Summary of telescopes being used for targeting.}
\begin{tabularx}{5.5in}{lcccl}
\hline
Telescope &  Bands   &Area& Location & Status \\
 &  & deg$^2$ & & \\
\hline
Blanco DECam &  $g$,$r$,$z$&   9k& NGC+SGC (Dec~$\le+34$ deg) & Begun 2014B \\
Bok 90Prime& $g$,$r$ & 5k & NGC (Dec~$\ge+34$ deg) & Begun 2015A \\
Mayall MOSAIC-3& $z$ & 5k & NGC (Dec~$\ge+34$ deg) & To begin 2016A\\
WISE-W1 & 3.4 $\mu$m&all sky &all-sky & Completed \\
WISE-W2 & 4.6 $\mu$m&all sky &all-sky & Completed \\
\hline
\end{tabularx}
\label{tab:TargetReqTelescopes}
\end{table}

In the remainder of this Section, we demonstrate that our baseline
optical/infrared color selections can select the targets listed in
Table~\ref{tab:TargetReqGalaxyTypes}, and summarize the key properties
of each sample.   
The accompanying instrument volume of the FDR details the design of the DESI instrument,
which informs a spectral simulator presented in that volume.  The spectral simulator
aids in the design of the targeting strategy (such as magnitude limits),
calculates exposure times, and estimates redshift measurement efficiencies.
Given the expected target densities and exposure times, the overall survey
strategy is developed in Section~\ref{sec:survey}. Included in the survey
strategy is an optimized method to tile the sky that maximizes the area
covered and number of target redshifts obtained, while minimizing the
overall time required for the survey.  The outlines for a strategy for
fiber allocation are given in the accompanying FDR.  This strategy leads
to the values given in Table \ref{tab:TargetReqGalaxyTypes}.

\subsection{Targets: Bright Galaxy Sample}
\label{sec:bgs}
\subsubsection{Overview of the Sample}

The galaxy sample for the BGS will be a flux-limited, $r$-band
selected sample of galaxies. The magnitude limit is determined by the
total amount of observing bright time and the exposure times required
to achieve our desired redshift efficiency. This target selection is,
in essence, a deeper version of the galaxy target selection for the
SDSS main galaxy sample (MGS). We explore the properties of the BGS
target sample through mock catalogs created from numerical
simulations. These mocks have identical properties to the MGS at low
redshift, including the luminosity function, color distribution, and
clustering properties. At higher redshifts, the mock BGS is calibrated
using data from the much smaller areas of the
GAMA ($z\sim 0.3$) and DEEP2 ($z\lesssim 1.0$) surveys.

\subsubsection{Sample Properties}

\begin{figure}[!t]
\centering
\includegraphics[height=3in]{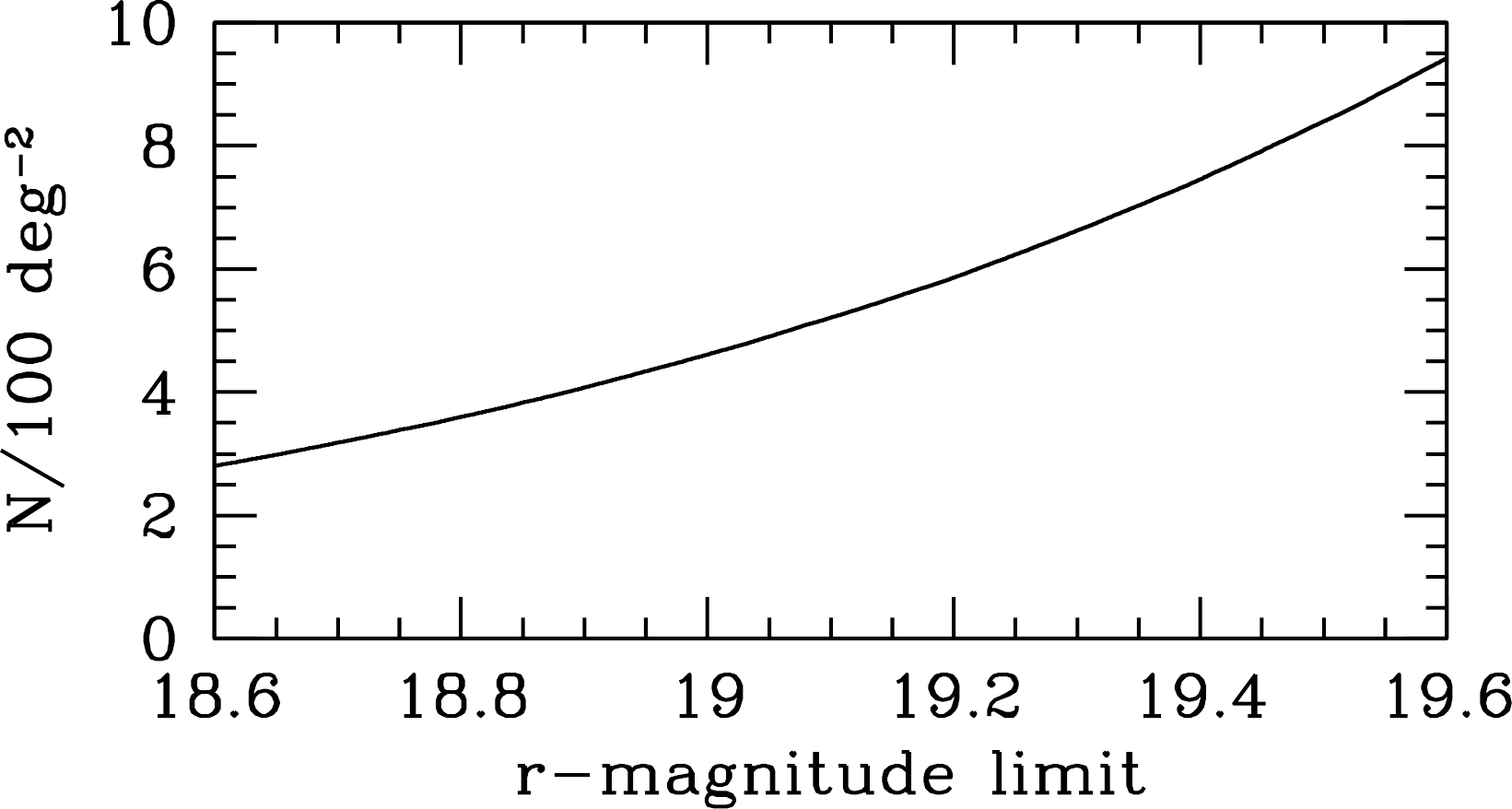}
\caption{ Surface density of BGS targets as a function of $r$-band
  magnitude from a numerical simulation.  This mock is calibrated to match
  low-redshift data from SDSS. \label{fig:bgs_selection_limit}}
\end{figure}

\paragraph{ Surface Density} 
Figure \ref{fig:bgs_selection_limit} shows the surface density of targets
as a function of limiting magnitude.  We expect to have a density of
just over 800 deg$^{-2}$ for an $r$-band limit of 19.5, somewhat higher
than the goal of 700 targets per deg$^{-2}$.

\begin{figure}[!tb]
\centering
\includegraphics[height=4in]{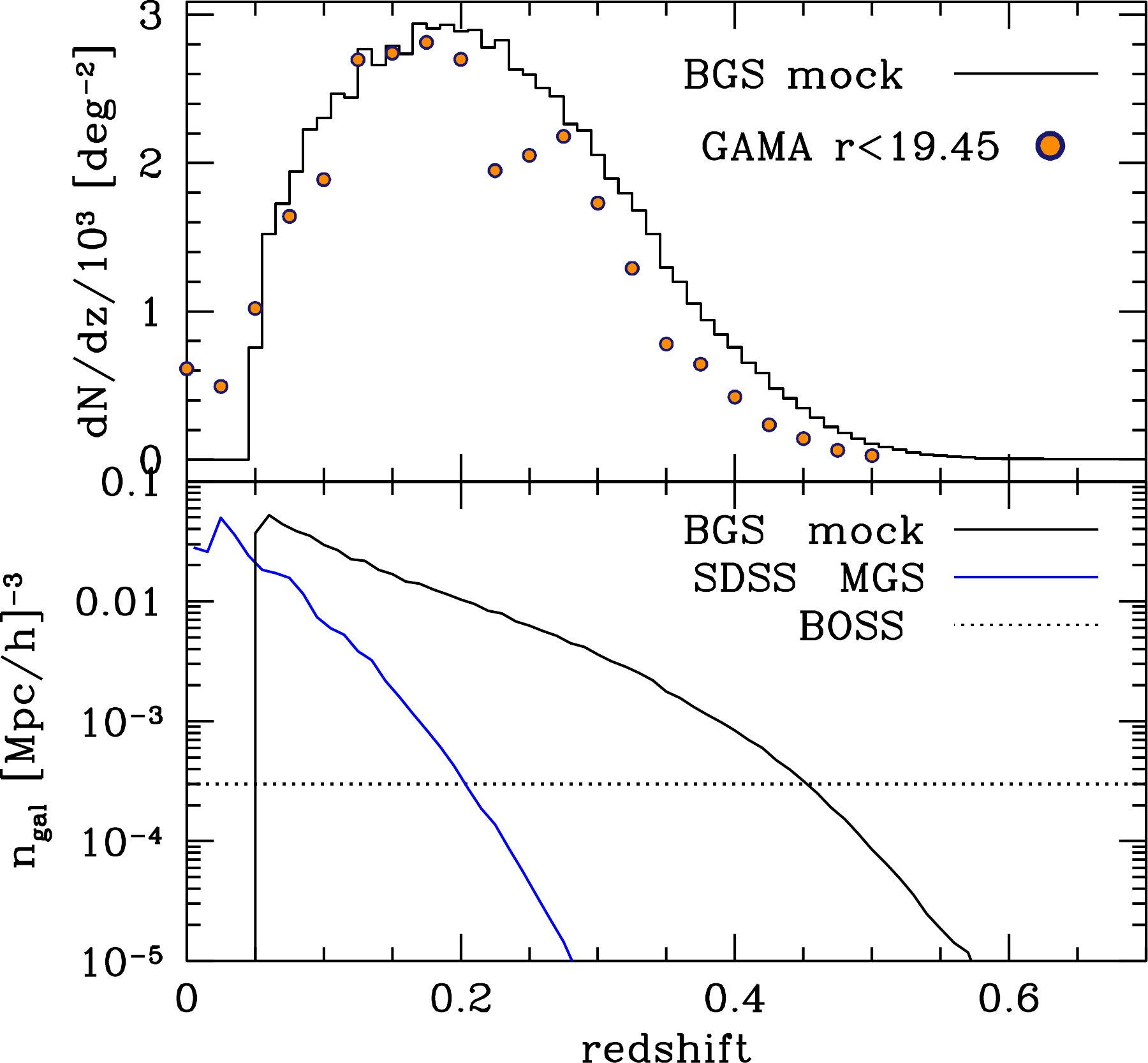}
\caption{ {\it Upper panel}: The redshift distribution of the mock BGS
  sample. The distribution peaks at $z=0.18$ with a median redshift of
  $z=0.204$. {\it Lower panel:} The space density of BGS galaxies as a
  function of redshift. For comparison, the space density of the MGS
  is shown with the blue curve, and the approximate space density of
  the full BOSS LRG sample (LOWZ+CMASS) is shown with the dotted
  line. The space density of the BGS sample is larger than the
  MGS+BOSS samples up to $z\sim 0.4$. \label{fig:bgs_nz}}
\end{figure}

\paragraph{Redshift Distribution}
Figure \ref{fig:bgs_nz} shows the estimated redshift distribution and
space density of galaxies. The upper panel shows the redshift
distribution $dN/dz$ in units of $10^3~{\rm deg^{-2}}$ per unit redshift. The area under the curve
is 800 targets/deg$^{-2}$. The redshift distribution peaks at $z\sim
0.2$, a factor of 2 higher than the MGS, with a tail out past
$z=0.4$. For comparison, results from GAMA at $r<19.45$ are shown with
the filled circles. The lower panel shows the space density of galaxies in units
of comoving $($Mpc$/h)^{-3}$. For reference, the space density of the
MGS is shown, as well as the density of BOSS LOWZ+CMASS objects, which
is roughly constant at $3\times 10^{-4}$ $($Mpc$/h)^{-3}$. The BGS
sample has a significantly higher density than either the MGS or BOSS
out to a redshift of $z=0.4$. At $z=0.3$, the sampling of the density
field is over an over of magnitude higher in the BGS than in the sum
of all SDSS targets.

\begin{figure}[!t]
\centering
\includegraphics[height=3in]{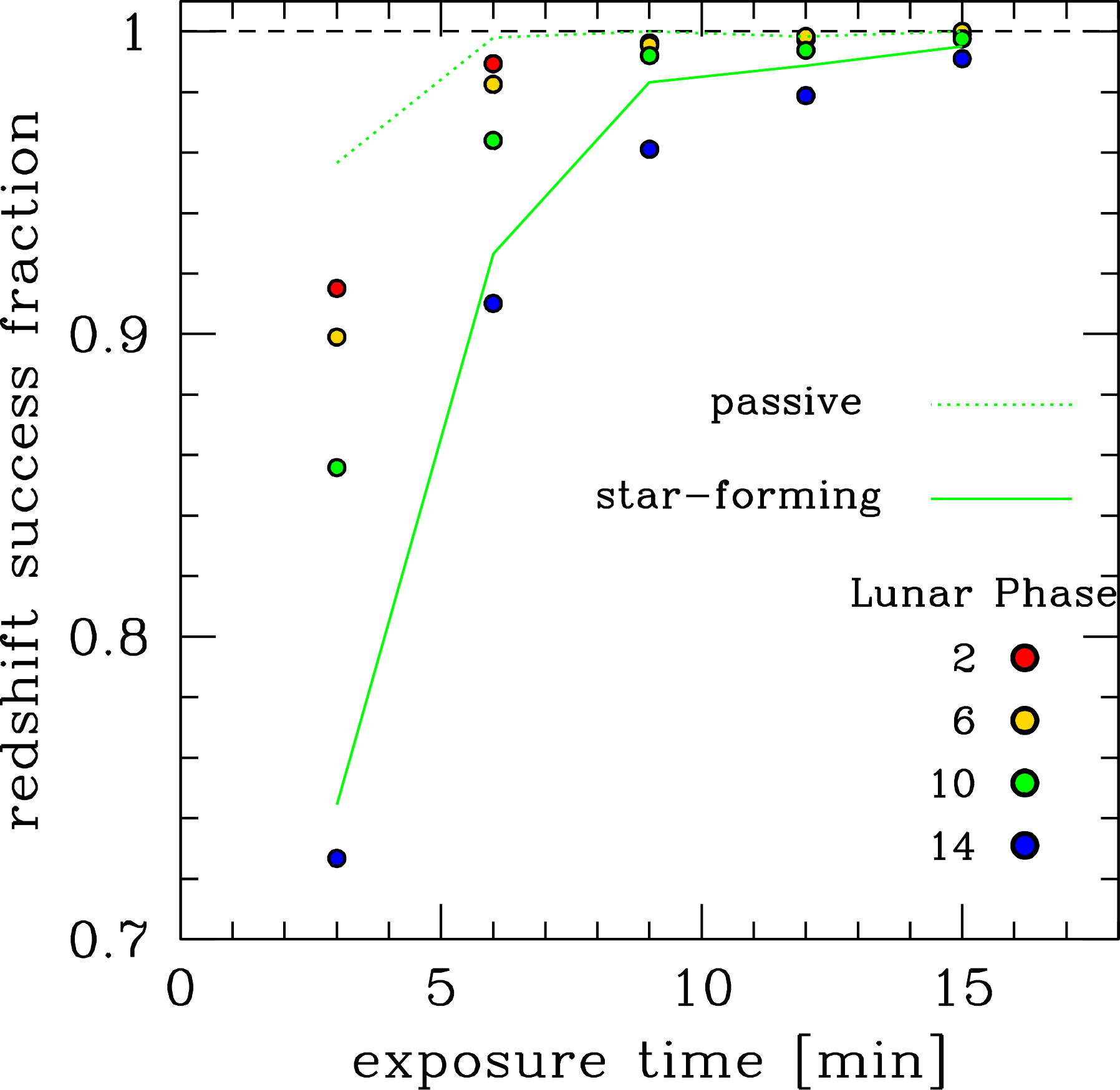}
\caption{ \label{fig:bgs_redshift_efficiency} The redshift success
  rate for BGS-like targets. Test targets are created by `observing'
  MGS galaxies at twice the true redshift of the galaxies. Test
  spectra are created using {\tt desi$\_$quicksim}, incorporating a
  lunar model that incorporates the phase, zenith angle of the moon,
  zenith of the target, and the angle between the target and the
  moon. Results are shown as a function of exposure time and lunar
  phase. The green curves show the results for a 10-day lunar phase
  for passive and star-forming galaxies.  }
\end{figure}

\paragraph {Redshift measurement method}
As a simple flux-limited sample,
the BGS will target both star-forming and
quiescent galaxies.  Redshifts will be obtained from template fits
over the full DESI spectral range, with the significance of the fits
dominated by the emission lines for star-forming galaxies and by
the 4000\AA\ break and Mg absorption features for quiescent galaxies.
Figure \ref{fig:bgs_redshift_efficiency} shows the redshift
efficiency as a function of both exposure time and lunar phase for a
test sample of galaxies. The test sample is constructed by taking
random MGS galaxies and `moving' them further away from the observer
by a factor of 2 in redshift. Because the median redshift of the MGS
is $z\sim 0.1$, this process creates a test sample with the same
median redshift as the BGS sample.  We take into account the change in
the fraction of light from the galaxy that enters the fiber aperture
through redshifting, the change in the angular diameter distance, the
change in the point spread function from SDSS to DESI, and the
different fiber diameters. {\tt desi\_quicksim} is used to create DESI
spectra for each test galaxy at a variety of exposure times and lunar
phases. Redshifts are obtained using the BOSS redshift code {\tt
  zfind}, and compared to the true redshift ($2\times
z_{SDSS}$). Phase, as indicated in the key, is in units of days, with
maximum illumination at 14 days and zero illumination at 0
days. Typical BGS observing conditions will be at 10 days, on
average. At this phase, the overall redshift success rate is 96\% at
$t_{\rm exp}=6$min, increasing to $\gtrsim 99$\% at 9 min. Success
fraction decreases monotonically with increasing moon
illumination. An additional factor in the degree to which the moon
affects observations is the angular separation between the moon and
the target. All results here are for a separation of 60 degrees. 

The results for star-forming and passive galaxies for 10-day phase are
shown as well. Galaxies are classified as star forming or passive by
their $D_n(4000)$ value, with $D_n(4000)>1.5$ being passive. At fixed
observing conditions, the redshift success rate for star-forming
galaxies is lower than for the passive galaxies, indicating that the
4000\AA\ break is more efficient as a redshift indicator given the
spectral noise imparted by the observing conditions. But the redshift
success rate for star-forming galaxies is still $\sim 98\%$ for 9
minute exposure times. 

\begin{figure}[t]
\centering
\includegraphics[width=4in]{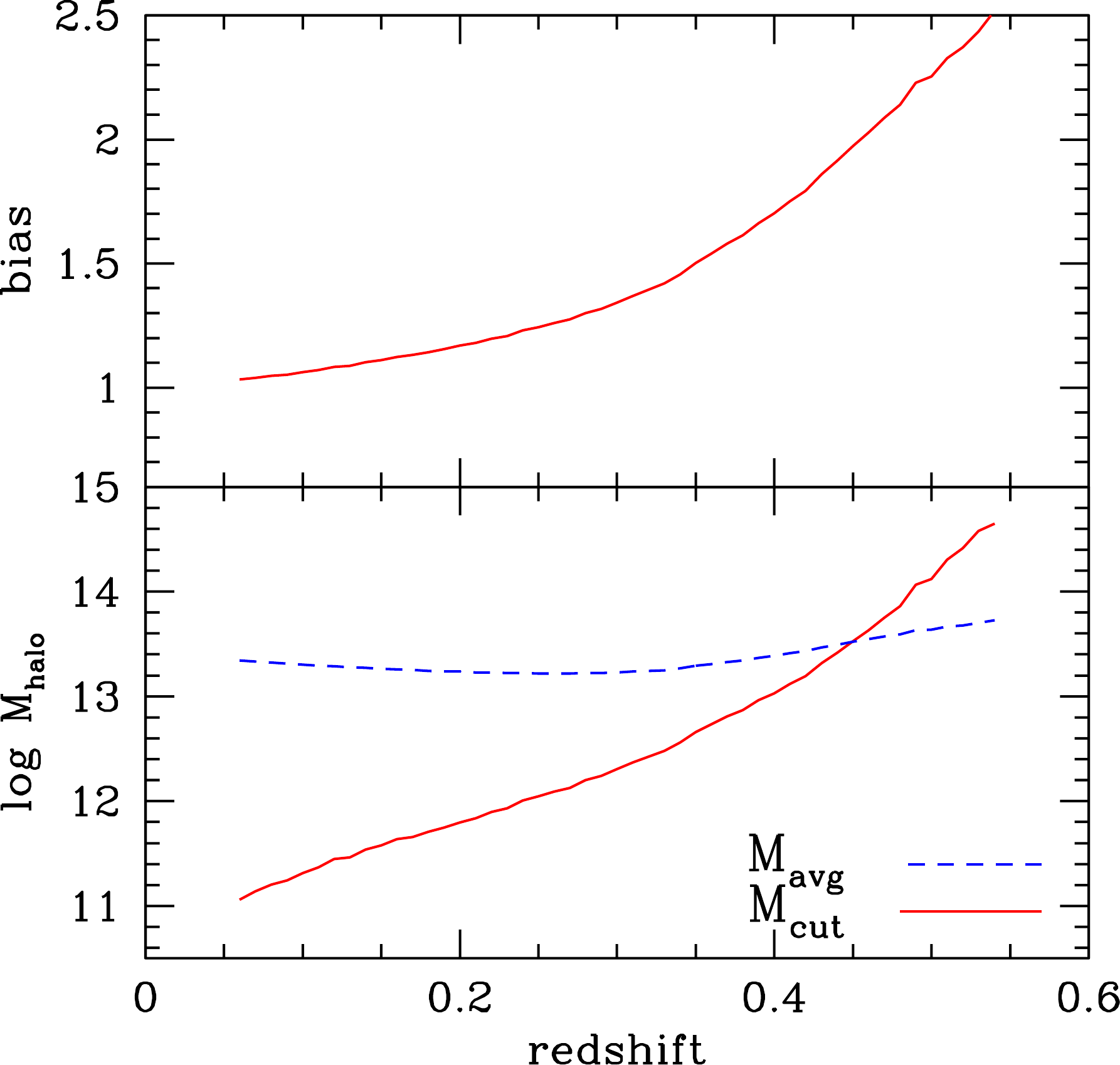}
\caption{ {\it Upper panel}: The bias of the BGS sample as a function
  of redshift. The bias is calculated using the abundance matching
  model and the space density from Figure \ref{fig:bgs_nz}. {\it Lower
    panel:} The halo mass scales probed in the BGS sample. $M_{\rm
    avg}$ is the mean halo mass of the sample. $M_{\rm cut}$ is a
  cutoff mass scale where halos have 50\% probability of containing a
  galaxy in the sample. The scatter in halo mass at fixed luminosity
  increases with luminosity, thus increases with redshift. This causes
  the inversion between $M_{\rm avg}$ and $M_{\rm cut}$ when the
  number density drops below the BOSS value. \label{fig:bgs_bias}}
\end{figure}

\paragraph{Large-scale-structure bias}
Estimating the bias of the BGS sample is straightforward due to its
completeness in magnitude. We use the abundance matching technique
(e.g., \citep{conroy_etal:06}) to match galaxies to halos as a
function of their luminosity. The bias is then estimated by
integrating over the halo mass function, weighted by the number of
galaxies per halo. The upper panel in Figure \ref{fig:bgs_bias} shows
the bias as a function of redshift obtained with this technique. At
low redshift, where the magnitude-limited nature of the survey spans a
wide range of absolute magnitudes, the bias is near unity. As redshift
increases, the bias monotonically increases. This is for two reasons:
for a flux-limited sample, the objects at higher redshift are
intrinsically brighter and therefore have higher clustering amplitude,
and at higher redshift the bias increases because the amplitude of dark
matter clustering is decreasing. 

The bottom panel shows the halo masses probed by the BGS target
selection as a function of redshift. 
$M_{\rm cut}$ is a cutoff mass scale: halos of this mass have a 50\%
probability of having galaxies in the sample. Significantly above
$M_{\rm cut}$, this probability asymptotes to 100\%, but the width of
this transition is reflective of the scatter of halo mass at fixed
luminosity.  This scatter increases with luminosity, which causes
the mean halo mass,
$M_{\rm avg}$, to vary more slowly than $M_{\rm cut}$. For the
brightest galaxies, this scatter is so large that $M_{\rm avg}$ is
actually below $M_{\rm cut}$.

\paragraph{Target selection efficiency}
The dominant loss of targets is due to fiber assignment inefficiencies.
Low-redshift galaxies have higher angular clustering on the sky, which
can lead to more contention for fibers in high density regions.  However,
as described in \S \ref{sec:bts}, the BGS is being observed in 3 layers
to achieve fairly high completeness.

A few percent of galaxies will be lost by deblending errors,
superpositions with bright stars, and other artifacts that typically
affect imaging catalogs.

\paragraph{Areas of risk}
Given the straightforward nature of the target selection, the BGS has
minimal risks. There are two possible sources of low-level risk. As shown
in Figure \ref{fig:bgs_redshift_efficiency}, the redshift efficiency
for star-forming objects lags behind that of passive galaxies at fixed
observing conditions. The majority of these redshift failures lie in
the green valley, in between the main star-forming sequence and the
red sequence. These objects have low star formation rates and thus
weak emission lines, but do not have stellar populations evolved
enough to have strong $D_n(4000)$ values. Dependent on the integration
time and observing conditions, the BGS may be incomplete for
green-valley objects.

Another possible source of incompleteness is low surface brightness
objects, which become more difficult to observe under bright time
conditions.

\subsection{Targets: Luminous Red Galaxies} 



\label{sec:lrg}
\subsubsection{Overview of the Sample}

The lowest-redshift dark-time sample for DESI will come from targeting 350 candidate luminous red galaxies (LRGs) per square degree \cite{Eis01}.  These objects are both high in luminosity and red in rest-frame optical wavelengths due to their high stellar mass and lack of ongoing star formation. They exhibit strong clustering and a relatively
high large-scale-structure bias, which enhances the amplitude of their power spectrum, and hence the BAO signal (\cite{Eis05}, \cite{Ho09}, \cite{Kazin10}).  Because of their strong
4000~\AA~breaks and 
their well-behaved red spectral energy distributions, low-redshift 
LRGs at $z < 0.6$ can be selected efficiently and their
redshifts estimated based on SDSS-depth photometry \cite{Padman06}.
The BOSS survey has targeted 119 LRGs per deg$^{2}$  with $z \lesssim 0.6$ using SDSS imaging.

DESI science analyses will incorporate existing BOSS spectroscopic samples (which cover 10,000 deg$^2$ of the DESI footprint) where available, as well as applying BOSS-like target selection algorithms (in regions not yet covered) to target LRGs at low $z$.  Because the BOSS target selection is well understood and documented in SDSS papers, we will not discuss it further here.  Extending the LRG sample to redshifts $z>0.6$,  where the 4000~\AA~break passes beyond the $r$ band and the optical colors of LRGs overlap with those of red stars, requires different selection techniques, taking advantage of available near-infrared imaging from space.  The remainder of this section will focus on the strategy we will use in that domain.

\subsubsection{Selection Technique for \texorpdfstring{$z>0.6$}{z gt 0.6}  LRGs}

%

\begin{figure}[p]
\centering
\includegraphics[height=2.9in]{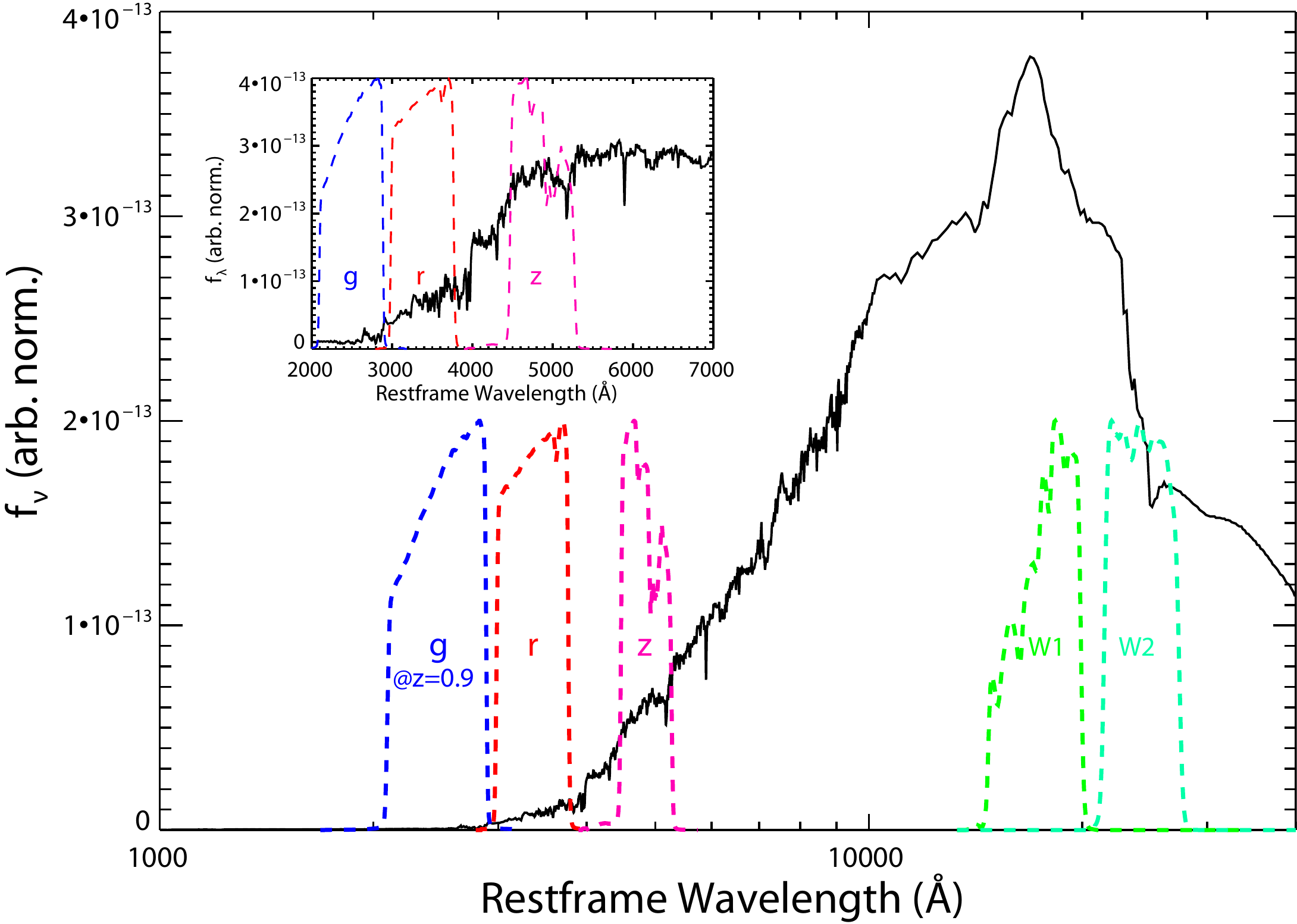}
\caption{A template spectrum based upon observations of the nearby elliptical galaxy NGC 4552, drawn from the work of \cite{Brown14}.  The spectrum $f_\nu$ is plotted as a function of rest-frame wavelength; we overplot the total (telescope + instrument + detector) response curves for DECam $grz$ and {\it WISE} $W1$ and $W2$ imaging at the appropriate rest frame wavelengths for an LRG at $z = 0.9$.  The 1.6 micron bump, the key spectral feature that enables our LRG selection method, corresponds to the peak in this spectrum.  In the inset, we plot flux  $f_\lambda$ over a limited wavelength range in order to illustrate clearly the 4000~\AA~break and the abundance of spectral absorption features in this vicinity, which will be exploited by DESI to measure redshifts for LRGs.  }
\label{fig:lrgspectrum}

\vskip 0.1in

\centering
\includegraphics[height=2.9in]{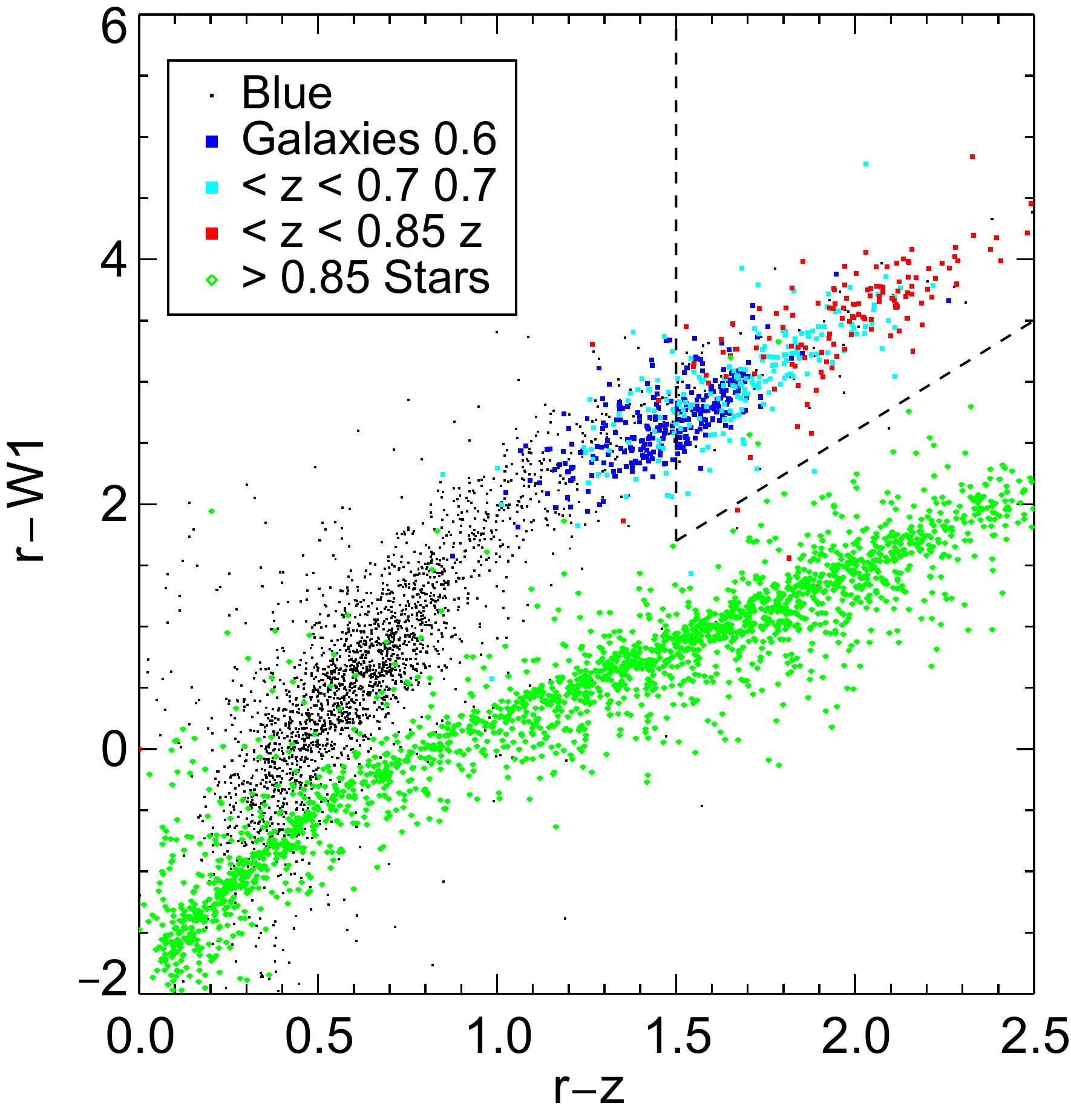}
\caption{An optical/near-infrared color-color diagram for galaxies observed by both DECam and WISE in the COSMOS field, where highly accurate 30-band photometric redshifts are available and used to label points the points shown.  In this and subsequent figures, $r$ indicates DECam $r$-band AB magnitude, $z$ indicates DECam $z_{AB}$, and W1 indicates {\it WISE} 3.4~\micron~AB magnitude.  Galaxies with LRG-like spectral energy distributions also having $z>0.6$ are indicated by points color-coded according to their redshift, whereas small black points indicate blue galaxies at all redshifts.  The dashed lines indicate the borders of our LRG selection box; our baseline sample assumes that objects above and to the right of these lines that also have magnitude $z_{AB}<20.46$ will be targeted by DESI as high-redshift LRGs.}
\label{fig:lrgselection}
\end{figure}

The spectral energy distributions of cool stars exhibit a local maximum around a wavelength of $1.6$~\micron, corresponding to a local minimum in the opacity of H$^{-}$ ions \cite{John88}.  This feature, commonly referred to as the {\textquotedblleft}1.6~\micron~bump{\textquotedblright},
represents the global peak in the flux density ($f_{\nu}$) for stellar populations older than about $ 500$ Myr \cite{Sawicki02}, such as those in LRGs.  In Figure \ref{fig:lrgspectrum} we plot an example LRG template spectrum from \cite{Brown14}, illustrating both the strength of this peak and the depth of the 4000~\AA~break.  
The lowest-wavelength channel in {\it WISE}, the W1 band centered at 3.4~\micron, is nearly optimal for selecting luminous red galaxies; it overlaps the bump at redshift near $z= 1$, so that higher-redshift LRGs will be bright in {\it WISE} photometry but comparatively faint in the optical.
As may be seen in Figure~\ref{fig:lrgselection}, a simple cut in $r$ - $W1$ color can therefore select LRGs effectively, while adding in information on $r-z$ color can help in rejecting non-LRGs.  {\it WISE} data are particularly well suited for this application, as the survey depth was designed specifically for detection of $L_*$ red-sequence galaxies to $z=1$; LRGs are generally significantly brighter than this limit. In addition, we currently apply  an $ i_{SDSS} > 19.9$ cut to emulate rejection
of previous BOSS-like targets.
%

We have tested selection techniques using optical $grz$ catalogs derived from CFHT Legacy Survey \citep{Gwyn08}, SDSS Stripe 82 data, or DECam $grz$ imaging; NIR imaging from {\it WISE}; and redshifts and rest-frame colors derived from DEEP2 spectra \citep{Newman2012} or accurate 30-band COSMOS photometric \citep{Ilbert09} redshifts.  A BOSS ancillary program has obtained roughly $ 10,000$ redshifts of magnitude $z_{SDSS}<20$ LRG candidates selected using SDSS and {\it WISE} photometry with somewhat broader color cuts than DESI will likely use, which provide additional tests of our basic techniques.  

\subsubsection{Sample Properties}

The baseline LRG selection cuts for DESI are shown by the solid lines in Figure \ref{fig:lrgselection}.  This selection, applied to a sample with a total DECam $z$-band magnitude limit of $z_{AB}=20.46$, relies on optical photometry in the $r$ and $z$ bands and  infrared photometry in the {\it WISE} W1 band.  DESI target LRGs will often not be detected in the anticipated $g$ band imaging, but are well above the depth limits in the $r$, $z$, and $W1$ bands, having $r < 23$ and $W1 < 19.5$.

\begin{figure}[!t]
\centering

\includegraphics[height=3in]{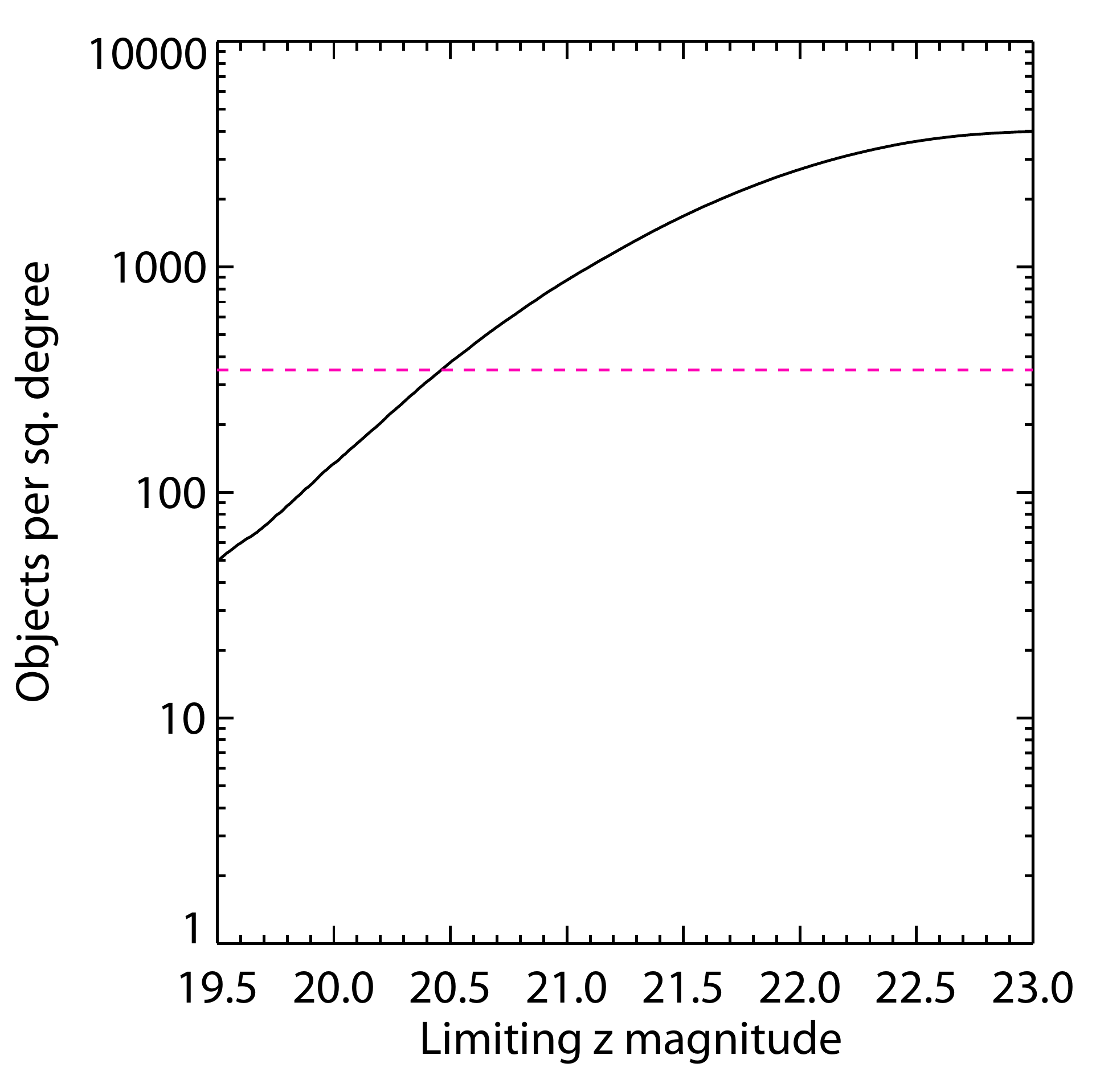}

\caption{Surface densities of targeted candidate $z>0.6$ LRGs as a function of limiting $z$-band magnitude.   We plot here the surface density of objects that lie within the target selection box shown in Figure \ref{fig:lrgselection} as a function of their $z_{AB}$ magnitude, as determined from DECam data in the 35 square degree DECaLS Early Data Release region.  We also indicate our goal density of 350 targets per square degree via the magenta dashed line.  Our baseline LRG sample size is attained at a depth of $z_{AB} < 20.46$.  At this limit, an average of roughly two spectroscopic measurements per LRG will be required to attain secure redshifts for $>98\%$ of targets.}
\label{fig:lrgselectionmagdist}
\end{figure}

This selection is already sufficient to meet all DESI design requirements, though we anticipate further improvements in the future.  The major properties of this sample are:

\paragraph{Surface Density} 
Figure \ref{fig:lrgselectionmagdist} shows the effect of changing the limiting magnitude on the surface density of selected targets using the color cuts shown in Figure \ref{fig:lrgselection}.  Based upon tests with DECam $grz$ data in the Early Data Release field, we find that the baseline sample density of 350 LRG targets/deg$^2$ is achieved when selecting objects down to a magnitude limit $z_{AB}=20.46$. 

Based on the results of the BOSS ancillary {\it WISE} LRG program, we can expect high ($>98$\%) redshift completeness for $z_{AB}<20$ LRGs with one DESI visit, for $z_{SDSS}<20.38$ with two visits, or for $z_{SDSS}<20.57$ with three visits.  For our baseline sample, a mean of two visits per object will thus be required (given the fractions of the sample with $z_{SDSS}<20$ or $z_{SDSS}>20.38$). We note, however, that redshift completeness has been somewhat lower than this for the eBOSS LRG sample, due to a combination of a bright $i$ magnitude limit applied to exclude CMASS galaxies, instrumental issues, and limitations of the BOSS data pipelines when handling low-S/N objects; improvements are underway to address the latter issues.  A more conservative estimate of anticipated completeness based on the eBOSS experience may be 90--95\%; however, adopting these lower completeness numbers would have negligible effect on cosmological constraint forecasts.    



\begin{figure}[!t]
\centering
\includegraphics[height=3in]{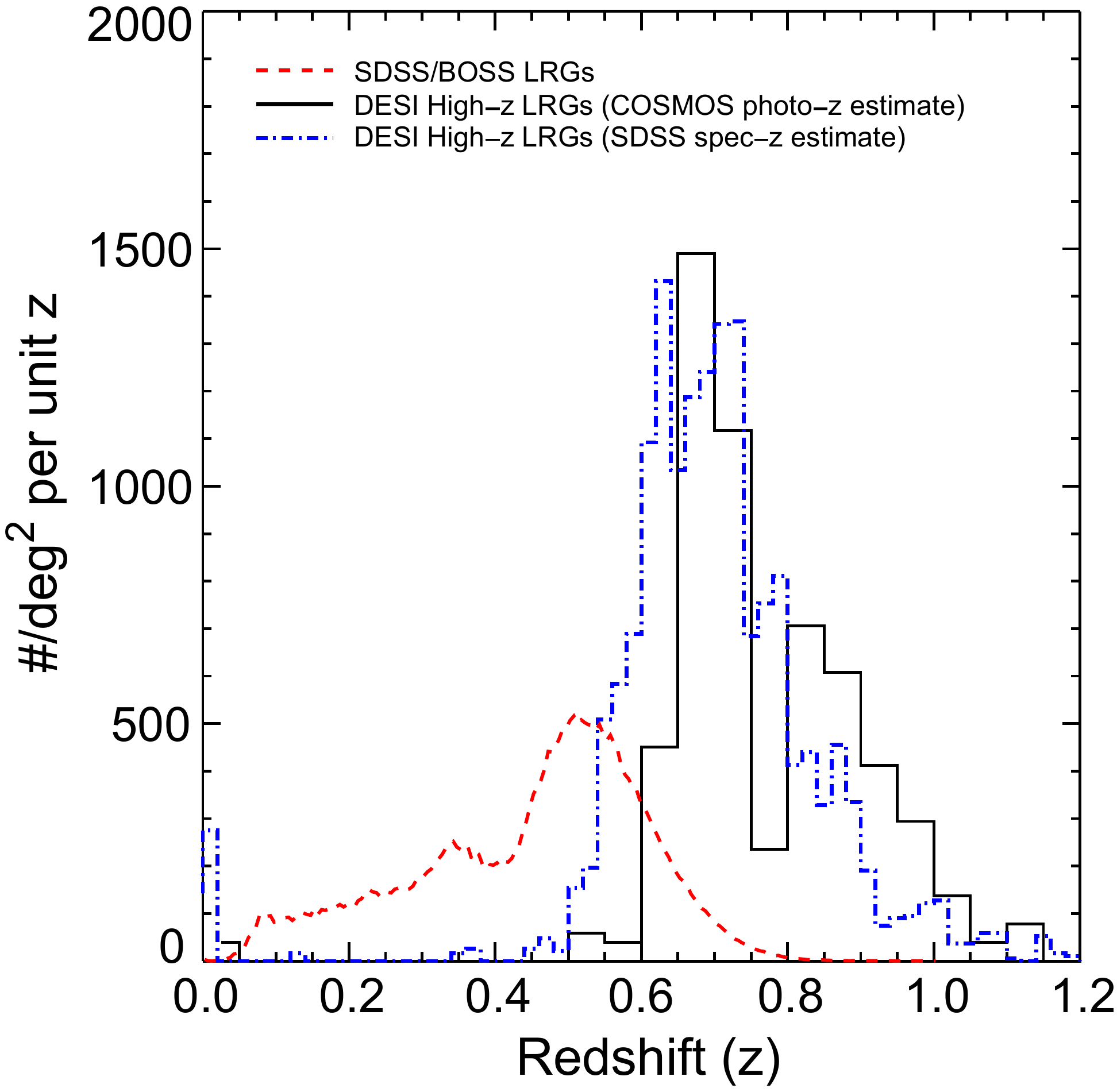}
\caption{DESI LRG redshift distribution for our candidate sample from two studies:
\emph{(black)} Photometric redshift distribution for a sample selected using DECam imaging in the COSMOS field, which has 
full redshift coverage but suffers from high sample variance
(as seen from the feature at $z\approx 0.77$).
\emph{(blue)} Spectroscopic redshift distribution for galaxies
selected using SDSS Stripe 82 photometry and assigned the
redshift of the object with the nearest color from a BOSS ancillary program.
The latter sample has low sample variance, but the high-redshift tail
is suppressed by the lack of redshifts at $20 < z_{\rm SDSS} < 20.46$.
Shown in red is the redshift distribution of low-$z$ LRGs,
many of them already observed by SDSS-I/II and SDSS-III/BOSS,
which will be included in the DESI analysis.
}
\label{fig:wise-selection-nz}
\end{figure}

\paragraph{Redshift Distribution} 
We have estimated the redshift distributions resulting from the DESI baseline target selection (see Figure~\ref{fig:wise-selection-nz}) by using both
COSMOS photometric redshifts and spectroscopic redshifts from our SDSS-III/BOSS ancillary program. 
Specifically, for the latter we applied an SDSS-passband-optimized version of the DESI selection cuts to SDSS Stripe 82 + {\it WISE} photometry, and then assigned the selected galaxies the spectroscopic redshift of the nearest-color object from our BOSS ancillary program.  The larger noise in the SDSS imaging over the ancillary program's footprint causes the redshift assignment to be contaminated by lower-redshift objects, while the high-redshift tail is suppressed by the lack of redshifts at $20 < z_{\rm SDSS} < 20.46$, making the resulting redshift distribution somewhat more weighted toward low redshift than DESI's should be.  In contrast, in the COSMOS field we can use DECam imaging for selection and photo-z's are available to much fainter than $z=20.46$, but due to the small area of the field sample/cosmic variance yields strong fluctuations in the redshift distribution.  Even given these limitations, we find that our sample selection meets or exceeds all requirements for the DESI baseline LRG sample.

As this figure shows, we have a particularly large density of objects  at $z < 0.8$ and will likely down-sample at those redshifts accordingly (e.g., by using a brighter magnitude limit for objects with blue $r-z$ colors).  The apparent magnitude of LRGs is strongly correlated with their redshift,
allowing us to sculpt the LRG redshift distribution efficiently. 

%
%

\paragraph{Redshift measurement method} 
LRGs exhibit a prominent break in their spectral energy distribution around 
4000~\AA\ (rest-frame), associated with multiple strong absorption-line features. 
This feature will be covered by the DESI spectrograph at redshifts up to $z=1.45$.  Our exposure times per target are set to achieve equivalent signal-to-noise at the wavelengths of interest as our BOSS ancillary program targeting $z_{SDSS}<20$, $z>0.6$ LRGs attained in one hour of SDSS exposure time.  We therefore expect to obtain highly-secure redshifts for a comparable fraction of targets ($>98\%$) as in that ancillary program.

\paragraph{Large-scale-structure bias} 
In order to predict the strength of the BAO feature in galaxy clustering measurements, we must assume a value for the ratio of galaxy clustering to dark matter clustering, commonly referred to as the large-scale structure bias.  On large scales this may be approximated as a function of redshift that is independent of scale, $b(z)$.  We can anticipate that the bias for $z>0.6$ luminous red galaxies should be at least as large as that of BOSS LRGs, as only the most extreme objects will be able to assemble a large amount of mass and cease star formation by this earlier epoch.  We therefore assume a bias of the form $b(z) = 1.7 / D(z)$, where $D(z)$ is the growth factor; this matches the value measured by SDSS-I at $z=0.34$ \citep{Eis05} and by SDSS-III at $z=0.57$ \citep{white2011}.
We have extrapolated this trend to $z=1$ for the DESI LRGs.


\paragraph{Target selection efficiency} 
Targets selected as LRGs could fall short in several ways: they could fail to yield redshifts entirely; they could prove to be stars rather than galaxies; they could be outside of the desired redshift range; or they could turn out to be blue (i.e., star forming and less highly biased).  Based on results from the BOSS ancillary program, we expect to obtain redshifts for $>98\%$ of LRGs targets, as described above.  Roughly 2\% of the objects targeted via the baseline selection box (which could be further optimized) are stars.  98\% of the galaxies selected are at $z>0.6$, while 98\% of the galaxies selected prove to have red-sequence rest frame colors.
If we treat all failure modes as independent (the worst-case scenario), this yields a net target selection efficiency of $92\%$; i.e., more than $ 92\%$ of all DESI LRG targets will be luminous red galaxies in the correct redshift range with a secure redshift measurements.

\paragraph{Areas of risk} 
There are few sources of risk in our LRG selection, the most important of which is the possibility that the COSMOS field is unrepresentative of the overall survey and instead contains (due to Poisson statistics or cosmic variance) an unusually large fraction of galaxies with red colors at $z>0.6$. 
%
%
At worst, this would degrade the target selection efficiency to near $ 90\%$.  The second potential source of risk is that the redshift success rate for LRGs is not simply a function of the signal-to-noise ratio, in which case we can not map our BOSS ancillary experience to DESI.
These risks will be reduced with the continuation of the spectroscopic
eBOSS program.

\vskip 0.1in

To summarize, the luminous red galaxy selection methods used for our baseline plan will yield a high-bias sample of about $ 315$ LRGs/deg$^2$ (assuming 90\% efficiency net) from a sample of $350$ targets/deg$^2$; almost all will be galaxies at $z>0.6$.  To be conservative, our projections assume that only 86\% of the targeted LRGs (i.e., 300 per square degree) will in fact be $z>0.6$ luminous red galaxies.  Combined with BOSS LRGs and the Bright Galaxy Sample at lower redshift, this will allow us to measure the BAO scale from $0 < z < 1$.  This sample allows direct comparisons to cosmological results provided by the ELG sample in overlapping redshift ranges, providing a key test for systematic effects.  

\subsection{Targets: Emission Line Galaxies} 
\subsubsection{Overview of the sample}
\label{sec:ELGtargprop}
\large

Emission-line galaxies (ELGs) constitute the largest sample of objects that DESI
will observe.  The galaxies exhibit strong nebular emission lines originating in
the ionized (``\htwo{}'') regions surrounding short-lived but luminous, massive
stars.  ELGs are typically late-type spiral and irregular galaxies, although any
galaxy actively forming new stars at a sufficiently high rate will qualify as an
ELG.  Because of their vigorous ongoing star formation, the integrated
rest-frame colors of ELGs are dominated by massive stars, and hence will
typically be bluer than galaxies with evolved stellar populations such as LRGs.
The optical colors of ELGs at a given redshift will also span a larger range
than LRGs due to the greater diversity of their star formation histories and
dust properties.

\begin{figure}[t]
\centering
\includegraphics[width=4in]{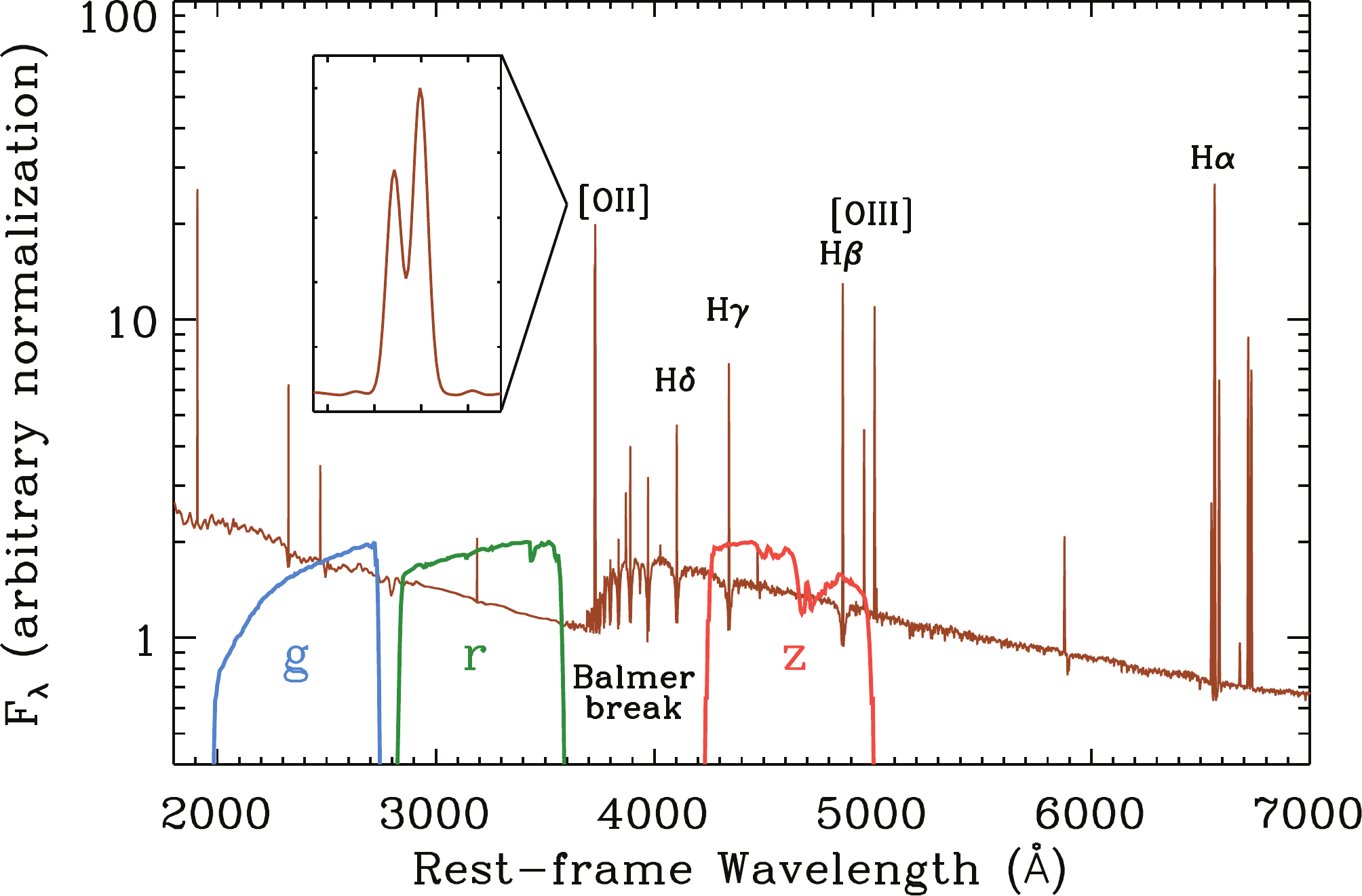}
\caption{Example rest-frame spectrum of an ELG showing the blue
stellar continuum, the prominent Balmer break, and the numerous strong nebular
emission lines.  The inset shows a zoomed-in view of the \otwo doublet, which
DESI is designed to resolve over the full redshift range of interest,
$0.6<z<1.6$.  The figure also shows the portion of the rest-frame spectrum the
DECam $grz$ optical filters would sample for such an object at redshift
$z=1$. \label{fig:ELGspectrum}}
\end{figure}

DESI leverages the fact that the cosmic star formation rate was roughly an order
of magnitude higher at $z\sim1$ than today, which causes galaxies with strong
line-emission to be very common at that epoch \citep{Hopkins06, Zhu09,
rujopakarn10}.  Figure~\ref{fig:ELGspectrum} shows an example rest-frame
spectrum of an ELG, which is characterized by a blue stellar continuum dominated
by massive stars, a Balmer break at $\sim3700$~\AA~ (whose strength depends on
the age of the stellar population), and numerous nebular emission lines, the
most prominent of which are H$\alpha~\lambda6563$, H$\beta~\lambda4861$, the
higher-order Balmer lines, and the forbidden \othree~$\lambda\lambda4959,5007$
and \otwo~$\lambda\lambda3726,3729$ nebular emission-line doublets.  The inset
provides a zoomed-in view of the \otwo{} doublet (assuming an intrinsic
line-width of $70$~km~s$^{-1}$), which the DESI instrument is designed to
resolve over the full redshift range, $0.6<z<1.6$.  By resolving the \otwo{}
doublet, DESI will avoid the ambiguity of lower-resolution spectroscopic
observations, which cannot differentiate between this doublet and other single
emission lines \citep{Comparat13c}.

\subsubsection{Selection Technique for \texorpdfstring{$z>0.6$}{z gt 0.6}
ELGs}\label{sec:ELGselection}

\begin{figure}
\centering
\includegraphics[height=3.0in]{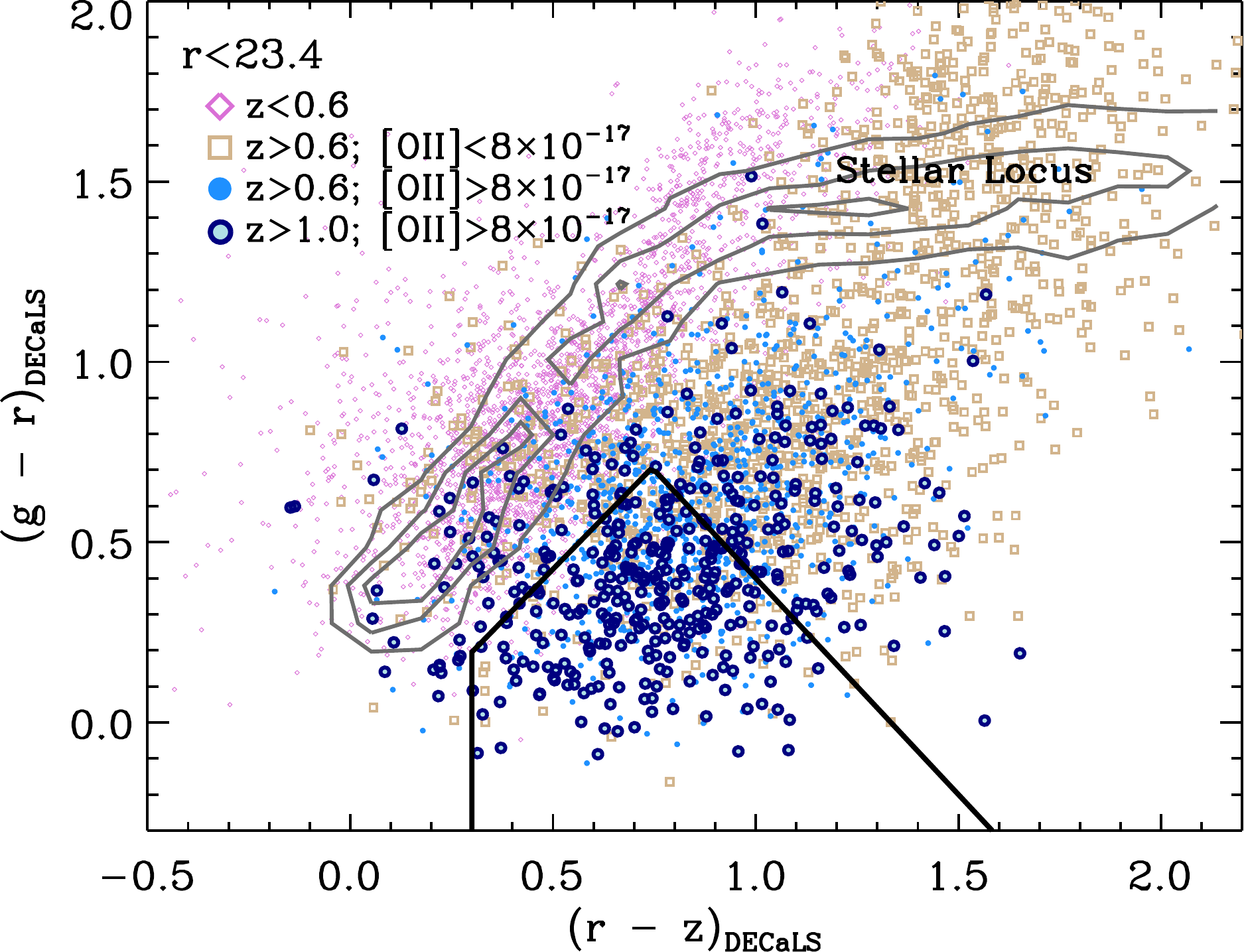}
\caption{Optical $g-r$ vs.~$r-z$ color-color diagram based on spectroscopy from
the DEEP2 Galaxy Redshift Survey, illustrating our preliminary selection for ELGs
at $z>0.6$ with significant \otwo{} emission-line flux.  Although the galaxy
photometry is based on deep CFHTLS imaging \citep{Matthews2013}, the colors have
been transformed and degraded to the expected depth of the DECaLS imaging.  This
plot shows that strong \otwo-emitting galaxies at $z>0.6$ (blue points) are in
general well-separated from both the population of lower-redshift galaxies (pink
diamonds) and from the locus of stars in this color space (grey contours).  The
selection box (thick black polygon) selects those galaxies with
strong \otwo-emission while minimizing
contamination from stars and lower-redshift interlopers. \label{fig:DEEP2grz}}
\end{figure}


The DESI/ELG targeting strategy builds upon the success of the DEEP2
galaxy redshift survey, which used cuts in optical color-color space to
effectively isolate the population of $z\gtrsim0.7$ galaxies for follow-up
high-resolution spectroscopy using the Keck/DEIMOS spectrograph \citep{Davis03,
Newman2012}.  More recently, several SDSS-III/BOSS and SDSS-IV/eBOSS ancillary
programs have confirmed that optical color-selection techniques can be used to
optimally select bright ELGs at $0.6<z<1.7$ \citep{comparat15a, raichoor15a,
Comparat13a}. 



In Figure~\ref{fig:DEEP2grz} we plot the $g-r$ vs $r-z$ color-color diagram for
those galaxies with both highly-secure spectroscopic redshifts and
well-measured \otwo{} emission-line strengths from the DEEP2 survey of the
Extended Groth Strip (EGS) \citep{Newman2012}.  The $grz$ photometry of these
objects is drawn from CFHTLS-Deep observations of this
field \citep{Matthews2013}, transformed and degraded to the anticipated depth of
our DECam imaging (see \S\ref{sec:decam}).  As discussed in the next section, we
expect to achieve a very high redshift success rate for ELGs with
integrated \otwo{} emission-line strengths in excess of approximately
$8\times10^{-17}$~erg~s$^{-1}$~cm$^{-2}$.
This integrated \otwo{} flux
corresponds to a limiting star-formation rate of approximately $1.5$, $5$, and
$15~M_{\odot}$~yr$^{-1}$ at $z\sim0.6$, $1$, and $1.6$, respectively, which lies
below the `knee' of the star formation rate function of galaxies at these
redshifts \citep{Kennicutt98, Kennicutt2012}.

\begin{figure}[!t]
\centering
\includegraphics[height=3in]{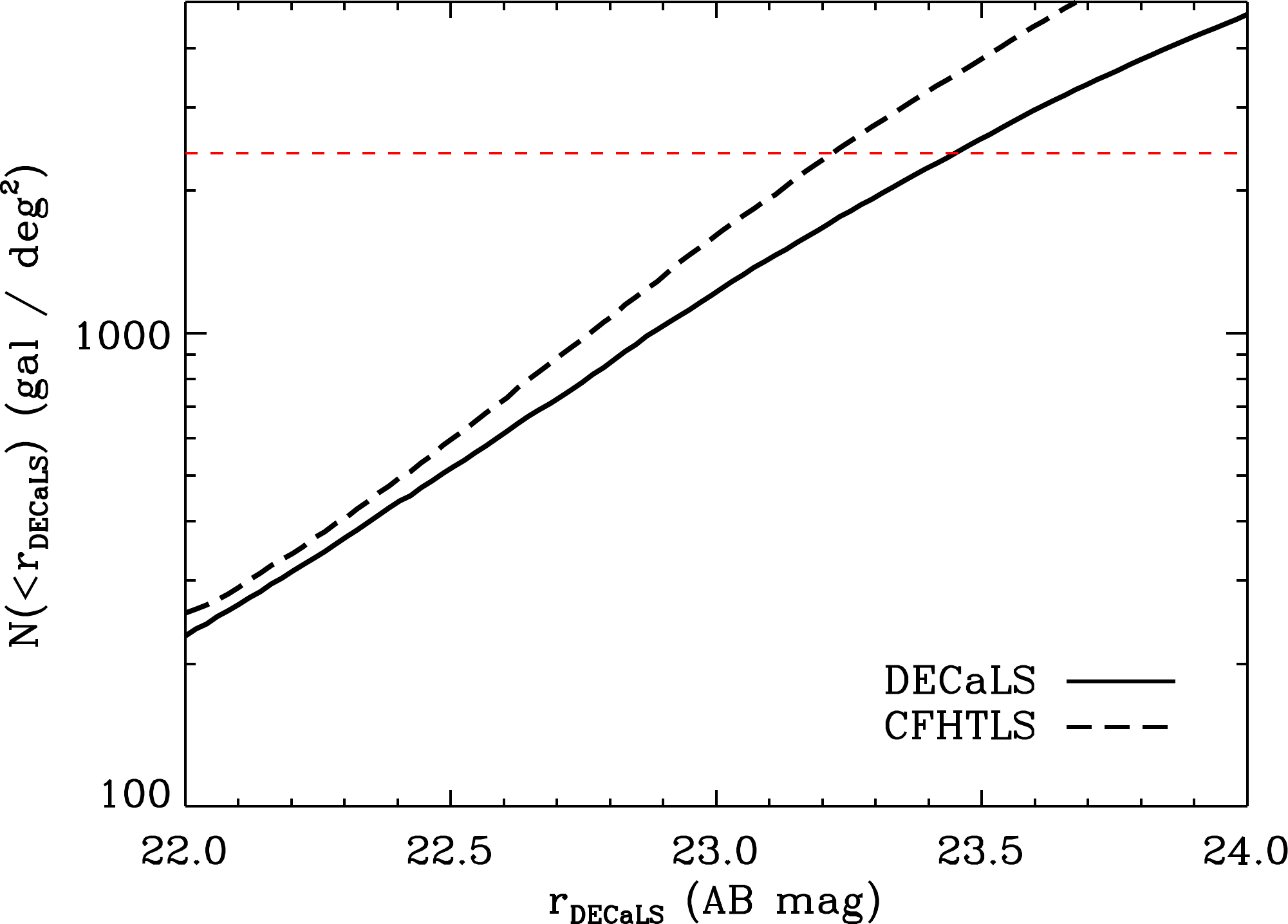}
\caption{Surface density of ELGs as a function of limiting $r$-band magnitude.
The solid black line shows the surface density of objects which lie within the
target selection box shown in Figure~\ref{fig:DEEP2grz} as a function of
$r_{AB}$ magnitude based on a $35$~deg$^{2}$ region of DECaLS observed to the
final survey depth.  For comparison, the dashed line is the set of objects
selected from CFHTLS-Deep photometry \citep{Gwyn08} which has been transformed
and degraded to the anticipated depth of DECaLS.  The horizontal dashed red line
shows our goal density of $2400$~targets~deg$^{-2}$, which is achieved at a
depth of $r_{AB} \lesssim 23.4$.  We note that the differences in the two curves
is most likely due to the scatter in the transformations between the CFHTLS and
DECaLS photometric systems. \label{fig:elgmagdist}}
\end{figure}

Figure~\ref{fig:DEEP2grz} shows that strong \otwo-emitting galaxies at $z>0.6$
(blue points) are well-isolated from the population of lower-redshift galaxies
(pink diamonds) and the stellar locus (grey contours).
The separation between galaxies above and below $z\simeq0.6$ occurs due to the
spectrum blueward of the Balmer break ($\lambda_{\rm rest}\sim3700$~\AA;
cf. Figure~\ref{fig:ELGspectrum}) shifting into the $r$-band filter, which
rapidly reddens the $r-z$ color.  Similarly, at $z\gtrsim1.2$ the Balmer break
moves into the $z$-band filter, causing both the $g-r$ and $r-z$ colors to be
relatively blue at higher redshifts.  The black polygon in
Figure~\ref{fig:DEEP2grz} delineates the target selection box to
isolate the population of strong \otwo-emitting ELGs at $0.6 < z < 1.6$.
By targeting galaxies in this box to a depth of $r_{AB}=23.4$, we strike a
balance between maximizing the number of $z\sim1$ ELGs with significant \otwo
flux while simultaneously minimizing contamination from stars and lower-redshift
galaxies.  ELGs galaxies with the very bluest colors are not included
in the selection box, as their ``flat'' spectra exhibit similar colors
at all redshifts and are therefore difficult to select in our redshift range.

\subsubsection{Sample Properties}\label{sec:ELGsampleprops}


The baseline ELG selection criteria for DESI are based on our analysis of the
DEEP2/EGS survey data, which targeted galaxies more than half a magnitude
fainter and with considerably higher spectroscopic signal-to-noise ratio than
DESI.  Because of this greater depth, we anticipate that any galaxies with
sufficiently strong \otwo{} flux to yield a redshift with DESI also yielded a
successful redshift measurement in DEEP2.  We have also cross-verified our
selection criteria and redshift distributions for ELGs using data from the
$1.3$~deg$^{2}$ COSMOS field \citep{ilbert13b} and from the $0.6$~deg$^{2}$
VVDS-Deep field \citep{le-fevre13b}; both of these samples give consistent
results, within the expected variation due to both sample variance and
systematic differences between the samples.  Our selection, when applied to
imaging with magnitude limits of $g_{AB}=24$, $r_{AB}=23.4$ and $z_{AB}=22.5$
(i.e., the anticipated depth of DECam Legacy imaging), is sufficient to meet all
DESI science requirements (although we do anticipate to refine the sample
selection even further).  The major properties of this sample are as follows. 

\paragraph{Surface Density} 
In Figure~\ref{fig:elgmagdist} we show the surface density of candidate ELGs in
our $grz$ selection box (see Figure~\ref{fig:DEEP2grz}) as a function of the
$r$-band magnitude limit.  At a depth of $r_{AB}\approx23.4$, we achieve our
goal of $2400$~targets per square degree.  As we discuss below, we
conservatively estimate that at least $65\%$ of these will be bona fide ELGs in
the redshift range $0.6<z<1.6$ with a strong enough \otwo{} emission-line
doublet (in tandem with other nebular emission lines available at $z\lesssim1$)
to yield a secure redshift.  Out of this sample, at most 270,000 ELGs over
500-1,500~deg$^{2}$ may be targeted by SDSS-IV/eBOSS, representing a sample that
could be used for further validation of DESI targets.




\begin{figure}[!t]
\centering
\includegraphics[height=3in]{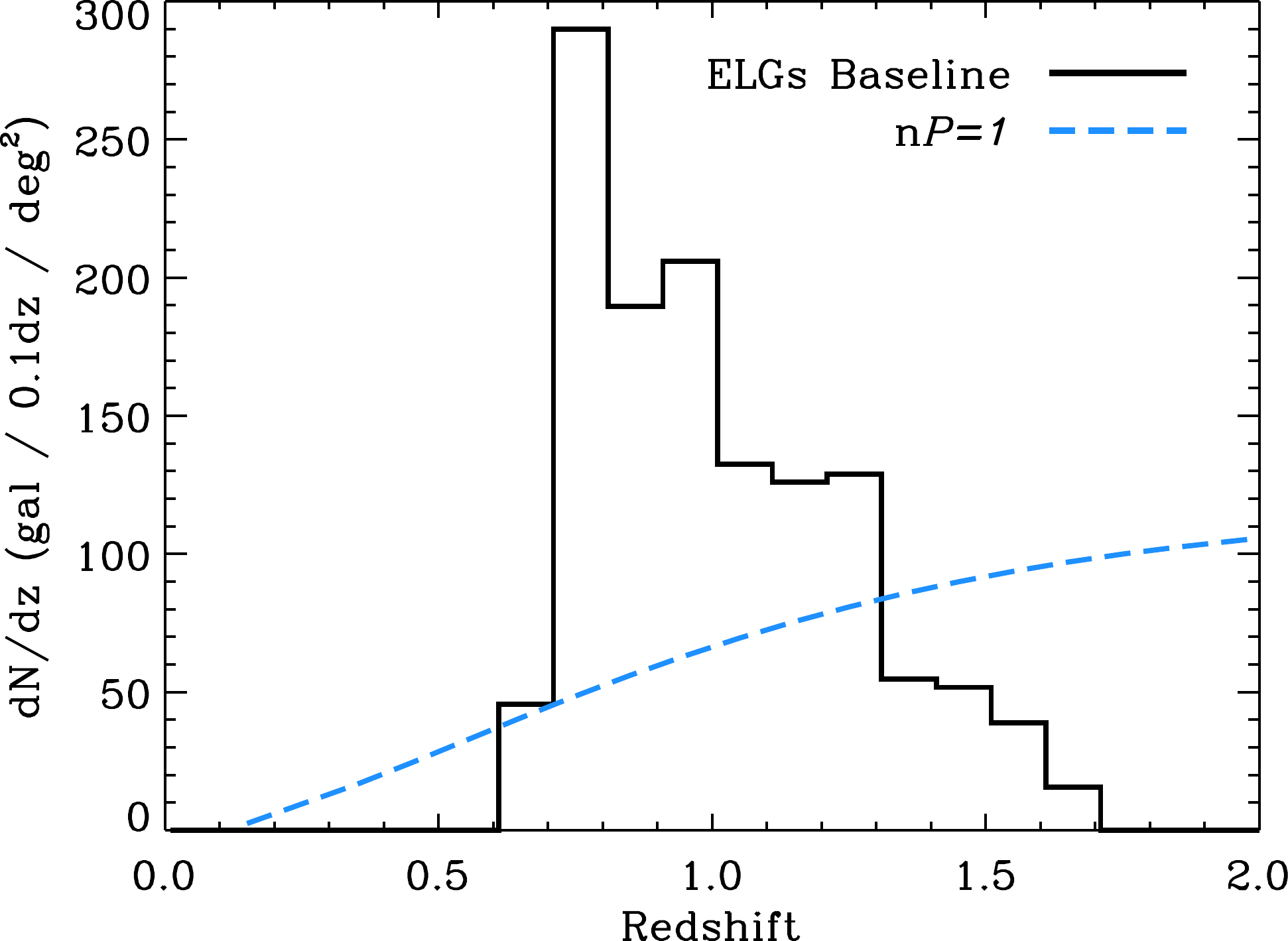}
\caption{Expected redshift distribution of ELG targets based on our
analysis of the DEEP2/EGS survey data (see Figure~\ref{fig:DEEP2grz}).
The overall normalization of the distribution has been fixed to
$1280$~ELGs~deg$^{-2}$ (from a targeted sample of
$2400$~targets~deg$^{-2}$) to reflect conservative estimates of the
overall efficiencies of fiber assignment, target selection, and
redshift measurement.
The ELG redshift distribution drops to a level where shot noise
dominates errors in BAO measurements (i.e., $\bar{n} P < 1$) only at
$z\gtrsim1.3$ (dashed blue line). \label{fig:ELGnz}}
\end{figure}


\paragraph{Redshift Distribution} 
Figure~\ref{fig:ELGnz} shows the anticipated redshift distribution of our
candidate $grz$-selected sample of ELGs, determined based on those DEEP2/EGS
objects which are both selected by our candidate cuts (after transforming to the
DECaLS photometric system and degrading to the expected depth of the survey) and
exhibit sufficient \otwo{} flux for DESI redshift measurements to succeed,
reweighted to account for DEEP2 target selection rates.\footnote{DEEP2 does not
cover \otwo{} at $z<\sim 0.8$ or $z \simgt 1.4$.  We handle this at low redshift
by assigning \otwo{} fluxes to galaxies at slightly higher redshift which have
comparable (rest-frame) color and luminosity.  For $z>1.4$, we plot a power-law
extrapolation of the redshift distribution measured at lower redshift, as DEEP2
would in general not obtain a redshift at all for objects where \otwo{} is past
the red end of the spectrum.  An analysis of COSMOS photometric redshifts for
objects meeting our selection cuts suggests that this extrapolation if anything
underestimates the number of objects at $1.4 < z < 1.6$.}

The ELG sample is designed to have a product of the number density
and the power spectrum, $\bar{n} P$, that exceeds 1 over some scales.
This is shown as the dashed blue line in Figure \ref{fig:ELGnz},
which is the surface density
for which $\bar{n} P = 1$ when evaluated at wave number $k=0.14~\ihMpc$ and
orientation relative to the line-of-sight $\mu=0.6$).  Below this limit, shot
noise will dominate errors in measuring the BAO signal
(cf. \S \ref{sec:scienceForecastBaseline}).  Our candidate ELG selection exceeds
the $\bar{n} P = 1$ curve to redshift $z\sim1.3$.

\begin{figure}[!ht]
\centering
\includegraphics[height=3in]{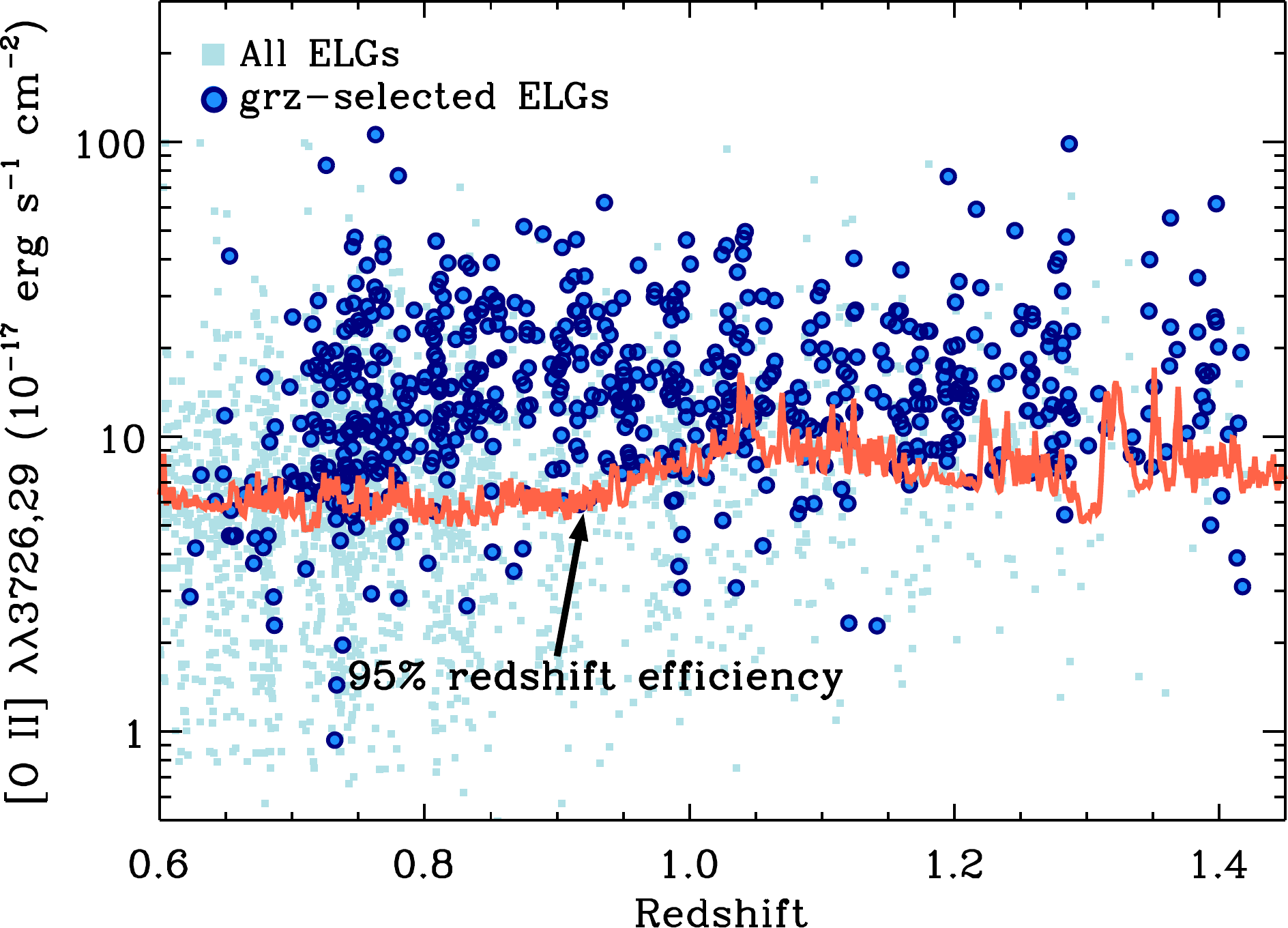}
\caption{\otwo{} flux as a function of redshift for DEEP2/EGS galaxies.  
The light blue squares represent all galaxies in the sample, while the dark blue
points are those objects targeted as DESI ELGs (see
Figure~\ref{fig:DEEP2grz}).
DESI will detect emission lines at $7\sigma$ for the bulk of the
targeted sample, corresponding to those objects above
the 95\% efficiency line in red.
\label{fig:ELGo2flux}}
\end{figure}

\paragraph{Redshift measurement method} 
The adopted $grz$ color-cuts are designed to maximize the selection of galaxies
at $z\approx1$ with significant \otwo emission-line flux.  In
Figure~\ref{fig:ELGo2flux} we plot \otwo flux as a function of redshift using
the DEEP2/EGS sample.  The red curve shows the limiting \otwo{} flux above
which DESI simulations predict we will detect emission lines at
$>7 \sigma$, resulting in secure redshifts.
Galaxies at redshift $z>1.0$ will have the \otwo{} doublet as the
only strong spectroscopic feature, while those at lower redshifts
will show H$\beta$ (at $z<0.5$) and \othree{} (at $z<1.0$).

\paragraph{Large-Scale Structure Bias}
\label{sec:ELGbias} We estimate the linear clustering bias of our sample of ELGs
relative to their dark matter halos using the DEEP2 data.  Employing methods
similar to those of \cite{Coil08} and \cite{Mostek13}, we have measured the
clustering of ELGs at
quasilinear scales of $1-10~h^{-1}$~Mpc in three overlapping redshift bins
centered at $z=0.87$, $1.0$ and $1.2$.  The observed galaxy clustering is constant
within errors at all redshifts, even as the amplitude of matter clustering
increases at lower redshift \cite{Smith03}.  The observations can thus be
described by a galaxy bias which is inversely proportional to the growth factor
of dark matter fluctuations.  Based on our measurements we adopt $b(z) = 0.84 /
D(z)$, where $D(z)$ is the growth factor at redshift $z$ ($D(z)=1$ today). This
increase in the bias with redshift for star-forming galaxies is
consistent with other studies of similar objects at $z$=0.5--2.2 \citep{Geach08,
Blake09, Sumiyoshi09}.

\paragraph{Target selection efficiency} 
Targets selected as ELGs could fall short in several ways: they could entirely
fail to yield a redshift (e.g., if the galaxy is at $z\gtrsim1.63$ then no
strong emission lines will be detected by DESI); they could prove to be
low-redshift galaxies, $z<0.6$; they could be QSOs instead of galaxies (and
hence useful for higher-redshift clustering analyses but likely outside the
redshift range of the ELGs); or they could be stars.  Based on the DEEP2/EGS
sample, we estimate that $\sim 10\%$ of the objects targeted via the baseline
selection criteria are expected to be stars, $\sim 5\%$ will be lower-redshift
interlopers, and $\sim 5\%$ will be at $z\gtrsim1.6$, while contamination from
QSOs is expected to be negligible.  Combining all these factors, the fraction of
ELG targets which are in fact galaxies in the correct redshift range is
approximately $80\%$.  Among these objects, about $85\%$ will have a high
enough \otwo{} flux to securely measure a redshift more than $95\%$ of the time
(see Figure~\ref{fig:ELGo2flux}).  Combining all these factors with the $78\%$
fiber assignment rate expected for an input target density of
$2400$~targets~deg$^{-2}$, we obtain an a final density of
$1220$~ELGs~deg$^{-2}$.




\paragraph{Areas of risk} 
The primary source of risk in our ELG selection is the limitations of the
datasets available for developing and assessing selection algorithms.  DEEP2 is
the only large current survey which resolves the \otwo{} doublet critical for
obtaining secure redshifts at $z>1$; however, due to the $z>0.75$ color cut
applied by DEEP2 in three of four survey fields, it can be used to assess the
low-redshift tail of the ELG selection in only a limited area, the Extended
Groth Strip used for all analyses here.  Because of the limited area, the number
of DEEP2 ELGs within our color box is relatively small, so both Poisson noise
and sample/cosmic variance have a significant effect on our predicted redshift
distributions.  Furthermore, the lack of DEEP2 coverage of \otwo{} at $z>\sim
1.4$ means that our assessments of performance in that regime are subject to
some amount of uncertainty.  Despite these shortcomings, even more assumptions
and extrapolations would be necessary with any other existing dataset.
The consistency of VVDS and
COSMOS results---together with the initial SDSS-IV/eBOSS observations---with the
DEEP2-based predictions builds confidence that these uncertainties are not
substantial.


The second potential source of risk which would cause performance to
fall short of our projections is that the redshift success rate for
DESI ELGs could not simply be a function of signal-to-noise ratio, but
may also depend in more subtle ways upon the instrumental resolution
and the intrinsic galaxy velocity dispersions.  For example, it would
be difficult to directly discriminate between \otwo or another
single-line feature at lower redshift for a population of ELGs with
unusually large velocity dispersions $\sigma_v > 150 \kms$ (though the
rarity of low-luminosity objects with extremely high velocity
dispersions, as would be implied by a false identification, may allow
such cases to be resolved). 




To conclude, the ELG selection methods used for our baseline plan will yield
$2400$~targets~deg$^{-2}$.  From these targets, DESI should securely measure
redshifts for approximately $1220$~ELGs~deg$^{-2}$ in the redshift range $0.6<z
< 1.6$ (see Table~\ref{tab:TargetReqGalaxyTypes}).  This sample will enable
constraints on cosmological parameters over a broad redshift range centered on
$z\approx1$, which can be directly compared to results from the independently
observed samples of LRGs at $z<1$ and quasars at $z>1$.

\subsection{Targets: QSOs} 

\subsubsection{Overview of the sample}
\label{sec:qsotargets}

The highest-redshift coverage of DESI will come from quasars
(a.k.a. quasi-stellar objects, or QSOs),
extremely luminous extragalactic sources associated with active
galactic nuclei. QSOs are fueled by gravitational accretion onto
supermassive black holes at the centers of these galaxies.
The QSO emission can outshine that of the host galaxy by a large factor.
Even in the nearest QSOs, the emitting regions are too small to be resolved,
so QSOs will generally appear as point sources in images. 
These are the brightest population of astrophysical targets with a
useful target density at redshifts $z >1$ where the population
peaks~\cite{Richards09, 2013A&A...551A..29P}.

DESI will use  QSOs as point tracers of the matter clustering
mostly at redshifts lower than 2.1, in addition to using QSOs at higher redshift
as backlights for clustering in the \lyaf.
This enlarges the role of QSOs relative to the BOSS project, which only
selected QSOs at $z>2.15$ for use in the \lyaf, and enhances their role relative 
to eBOSS where QSOs are used in a similar fashion as in DESI although with lower densities. 
DESI will select 170 QSOs per deg$^2$ over its footprint,
of which 50 per ${\rm deg}^2$ will
be at $z>2.1$ and suitable for the \lyaf.

\begin{figure}[!tb]
\centering
\includegraphics[height=2.5in]{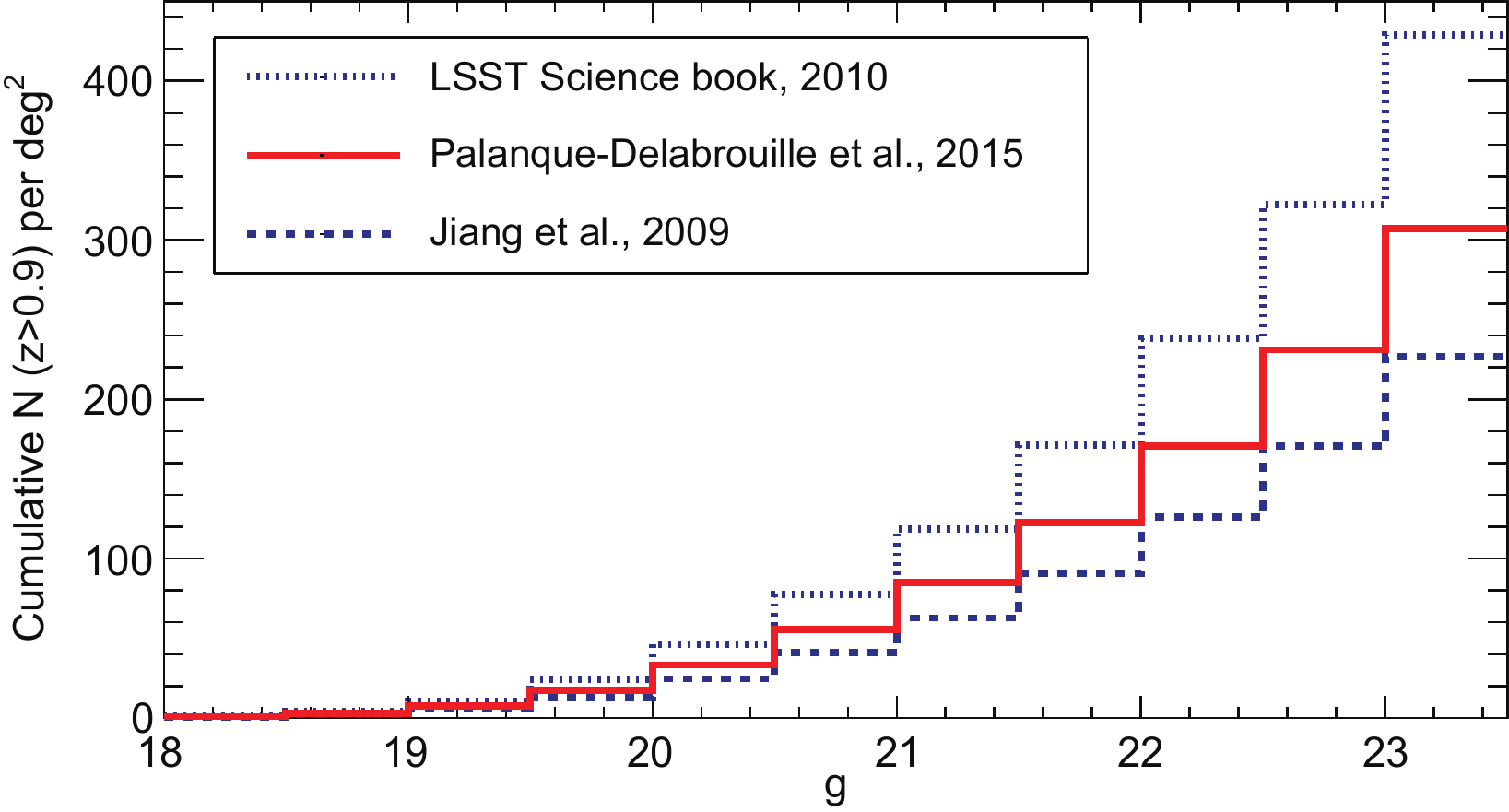}
\caption{Cumulative surface density of quasars (objects per deg$^2$) as a function of $g$ magnitude for $z>0.9$, derived from different estimates of the QSO luminosity function. }
\label{fig:LaFlumFunc}
\end{figure}

DESI pilot programs, \cite{2013A&A...551A..29P}  updated in \cite{Palanque2016LF},
have answered the long-standing uncertainties in the faint end of the QSO
luminosity function.
The surface density for $z>0.9$ QSOs derived from these studies is shown
in Figure~\ref{fig:LaFlumFunc}, along with previous estimates
from \cite{Jiang06} (25\% lower) or from the LSST science book
\cite{Hopkins07b, LSSTSB} (40\% higher).  Brighter than magnitude $g=23.0$ ($r=23.0$ respectively),
we infer that a \emph{complete} QSO sample would contain about 185  (200, resp.) QSOs
per deg$^2$ at $z<2.1$ and about 75  (90, resp.) at $z>2.1$. DESI will target and obtain redshifts for 120 and 50 QSOs per deg$^2$
in the redshift ranges $z<2.1$ and $z>2.1$, respectively. 

Because of  their point-like morphologies and with  {\it photometric} characteristics that mimic faint blue stars in optical wavelengths (Figure~\ref{fig:colorsQSO}, middle plot), QSO selection is challenging. The photometric selection used by BOSS to target \lya QSOs at $z>2.15$ has attained a 42\% targeting efficiency (i.e., fraction of targets that prove to have the desired class and be in the desired redshift range), yielding 17 $z>2.15$ QSOs per deg$^{2}$  down to the SDSS photometric limit of $g<22.1$ \cite{Ross12}. The selection technique for DESI needs to achieve a minimum efficiency of about 65\%; unlike for BOSS, however, QSOs at $z<2.15$ are considered successes. A  baseline scheme for QSO selection that  achieves  our goals for DESI is presented below.  

\subsubsection{Selection Technique}
\label{sec:QSOselectiontechnique}

\begin{figure}[!bt]
\centering
\includegraphics[width=3in]{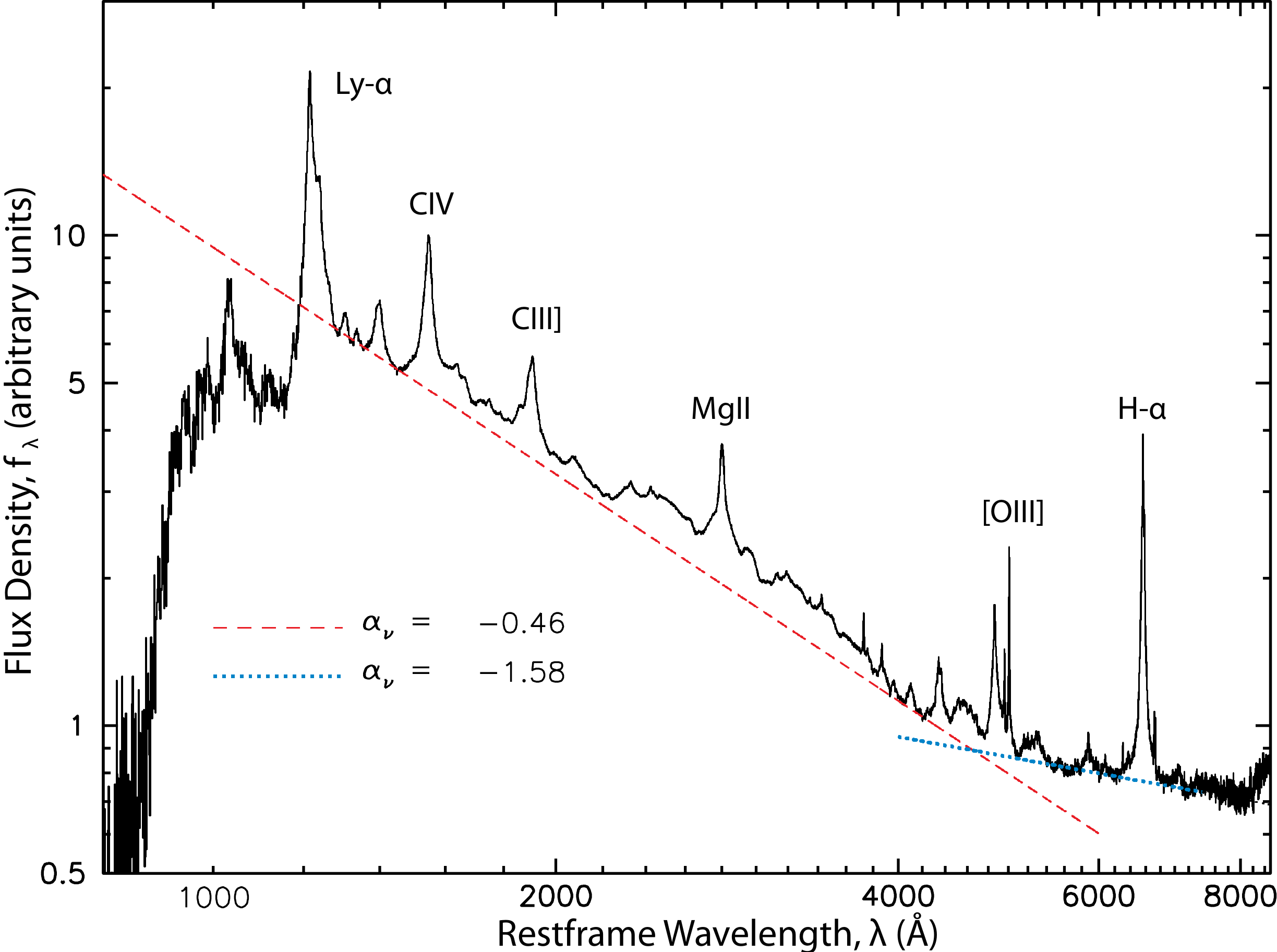}
\caption{QSO spectrum exhibiting  the main emission lines used in their identification. }
\label{fig:QSOspectrum}
\end{figure}

QSOs commonly exhibit hard spectra in the X-ray wavelength regime,
bright \lya\ emission in the rest-frame UV, and a power-law spectrum
behaving as $F_{\nu}\propto\nu^{\alpha}$ with $\alpha<0$ in the mid-infrared bands
\cite{Stern05} (c.f.\ Figure~\ref{fig:QSOspectrum}).
In the mid-optical colors, QSOs at most redshifts are not easily
distinguished from the much more numerous stars.
Successful selection of a highly-complete and pure QSO sample
must make use of either UV or infrared photometry; DESI relies
upon optical and infrared photometry for its baseline selection.

\begin{figure}[!b]
\centering
\includegraphics[width=\textwidth]{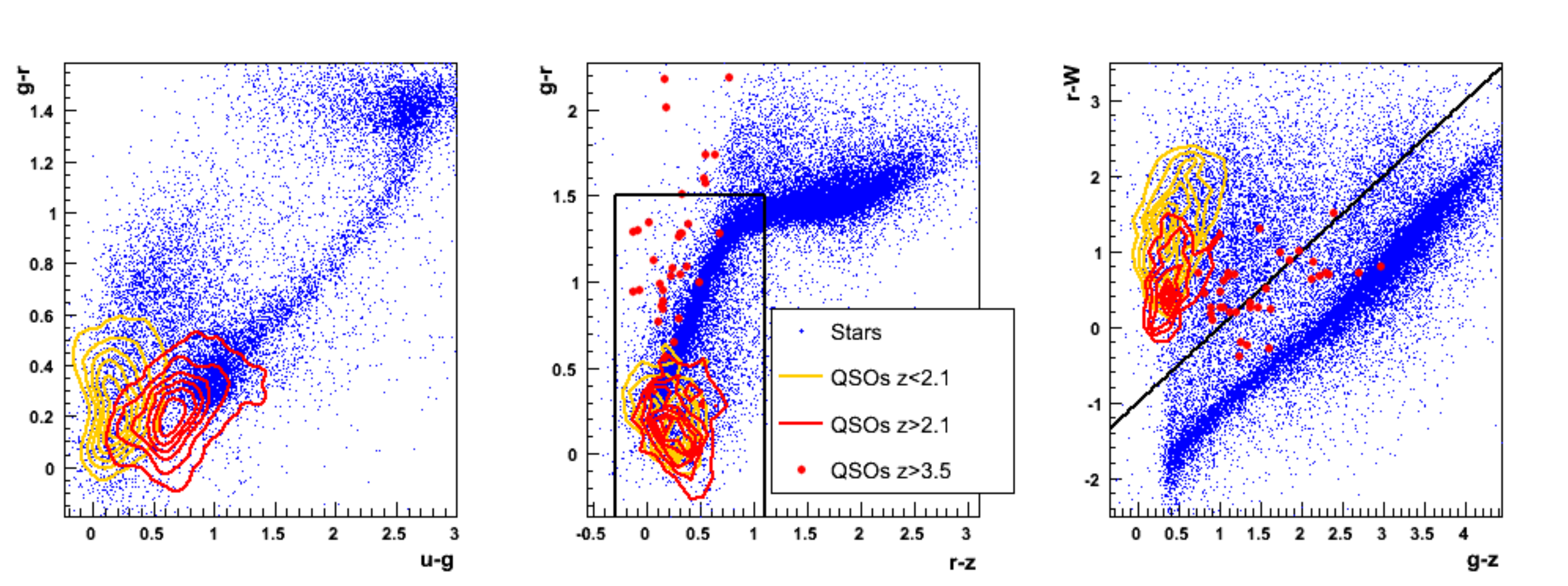}
\caption{Colors in the optical ($ugrz$) or  near-infrared (W is a linear combination of \it{WISE}~W1 and W2 bands)  of  objects photometrically classified as stars  (blue points) or spectroscopically classified as QSOs. Orange  contours indicate the locus of tracer QSOs at $z<2.1$, red contours of \lya QSOs at $z>2.1$, and red dots are for $z>3.5$ QSOs.
Left panel is based on SDSS photometry, middle and right panels on DECaLS-DR2. Black lines mark the boundaries of the selection regions described in the text.
 }
\label{fig:colorsQSO}
\end{figure}

The QSO target selection is a combination of  optical-only 
and  optical+IR selections.
The greatest separation from the stellar locus in the optical comes from $ugr$ colors where the ``UV excess'' in $u-g$ produces bluer colors than those of most stars (Figure \ref{fig:colorsQSO} left). 
In the absence of $u$ band in the baseline imaging plan, 
the bulk of the QSO targets are identified in an optical+IR selection (Figure \ref{fig:colorsQSO} right),
where the excess infrared emission from QSOs results in a clear
segregation from stars with similar optical fluxes. Stellar
SEDs indeed sample the rapidly declining tail of the blackbody spectrum at those wavelengths, 
where QSOs  have a much flatter SED. 
We defined a color selection to depths $r=23.0$ with cuts
in $g-r$ vs. $r-z$ and in $r-$W vs. $g-z$ shown in Figure \ref{fig:colorsQSO},  using DECaLS+{\it WISE} photometry from the DR2 data release.
We restrain the selection to objects with stellar morphology, to avoid an almost 10-fold contamination by  galaxies that otherwise enter our selection region.  
The {\it WISE} data are available on the whole sky, and are
photometered deeper than the public {\it WISE} catalogs using
the Tractor-forced photometry (see section \ref{sec:tractor}).
Although {\it WISE} and optical data are not synchronous, the  
color difference between QSOs and stars is so large that QSO variability
has minimal effect on the color selection.
The {\it WISE} satellite has been reactivated, and will improve
by a factor of two in signal-to-noise prior to DESI.

This baseline QSO target selection was tuned  over the  Stripe82 region where we led DESI pilot surveys 
(ancillary programs in BOSS and eBOSS, complemented by MMT observations)  in order 
 to build catalogs of  spectroscopically identified QSOs at all redshift, which we use as truth tables. 
 These pilot surveys selected highly complete  samples of 
 $g<23$ or $r<23$ QSOs from combined color and variability information (cf. section~\ref{sec:variability}),
using deep SDSS $ugriz$  and {\it WISE} near-infrared  data sets.
 Our baseline selection was  then tested on an independent region of Stripe82. 

 We  also investigated an alternative algorithm based on a machine-learning algorithm called Random Forest. We trained it on all 47000 identified QSOs over the DECaLS-DR2 footprint, and  used, for the star sample, a selection of 80000 unresolved objects in Stripe82, stripped of known QSOs and sources exhibiting QSO-like variations in their light curve.  As for the previous selection, the algorithm relies solely on object colors and is  restrained to unresolved sources with $r<23$. It selects 97\% of the known QSOs recovered by the more traditional color selection, but exhibits a better  performance than the latter, in particular at redshifts above 2.1 or faint magnitudes. 
 
 Considering the completeness of  the color cut or of the Random Forest approach as a function of redshift and magnitude, measured over  truth regions, and applying it to the QSO luminosity function of~\cite{Palanque2016LF}, both selections
result in  over 170 QSOs per deg$^2$, among which over 40 per deg$^2$ (55 per deg$^2$ for the Random Forest) are at $z > 2.1$. The non-QSO targets are stellar contaminants  
(about 80 per deg$^2$ in the color-cut selection, and 60 per deg$^2$ in the Random Forest selection). 

DESI may supplement its high-redshift QSOs with more sophisticated
selection algorithms and other supplementary photometry as it becomes
available.  Time-domain data enable variability selection methods (as
described in Section~\ref{sec:variability}).  UV ($u$-band) data
improve QSO selection, and allow discrimination between low-redshift
and high-redshift QSOs.  Algorithmically, neural-network based
algorithms \cite{Yeche10} and an extreme deconvolution method that models
the distributions of stars and quasars at the flux limit \cite{Bovy11}
have been in use by BOSS where they allowed an increase of up to 20\% in 
selection efficiency over traditional selection algorithms~\cite{2012ApJS..199....3R}. 
They are also  applied, and thus  further tested, in eBOSS.  A combination of these additional data
and algorithms will allow DESI to  target 
 QSOs in excess to those currently planned,
 with a small impact on the overall fiber budget.

The main contaminants to a $grz$+{\it WISE} QSO selection are very
low-redshift star-forming galaxies with strong PAH emission, currently
excluded using a star-galaxy separation based on ground-based
optical imaging; a few high-redshift obscured galaxies, which are rare
at bright optical magnitudes; and faint stars that artificially drift
into the QSO locus because of poor optical photometry. 

\subsubsection{Sample Properties}

Two selections using optical $grz$ and near-infrared data  achieved a performance at the level of our goals
for the DESI sample. Application of additional data and more sophisticated
selection algorithms may be used to boost, in particular, the high-redshift QSO densities.
To be conservative, we consider below the color-cut selection as the baseline DESI QSO selection.  
The major properties of the baseline DESI QSO sample are :

$\bullet$ {\it Surface Density:} 
The  current $grz$+{\it WISE} color-box selection yields a total of 260 targets
per deg$^2$ to a limit $r=23$, of which 
about 140 per deg$^2$ are expected to be QSOs with $z<2.1$ and about 40 per deg$^2$
are QSOs at $z>2.1$, similar to the required densities of
Table~\ref{tab:TargetReqGalaxyTypes}. Based on the QSO luminosity
function of \cite{Palanque2016LF}, this  corresponds to about 60\%
of all QSOs in this magnitude range. The Random Forest selection increases this rate to 67\%, with 55 $z>2.1$ QSOs per deg$^2$.  We anticipate that the deeper {\it WISE}
data expected before the  start of DESI will allow us to further 
 increase the completeness and decrease the stellar
contamination.

\begin{figure}[!tb]
\centering
\includegraphics[height=3in]{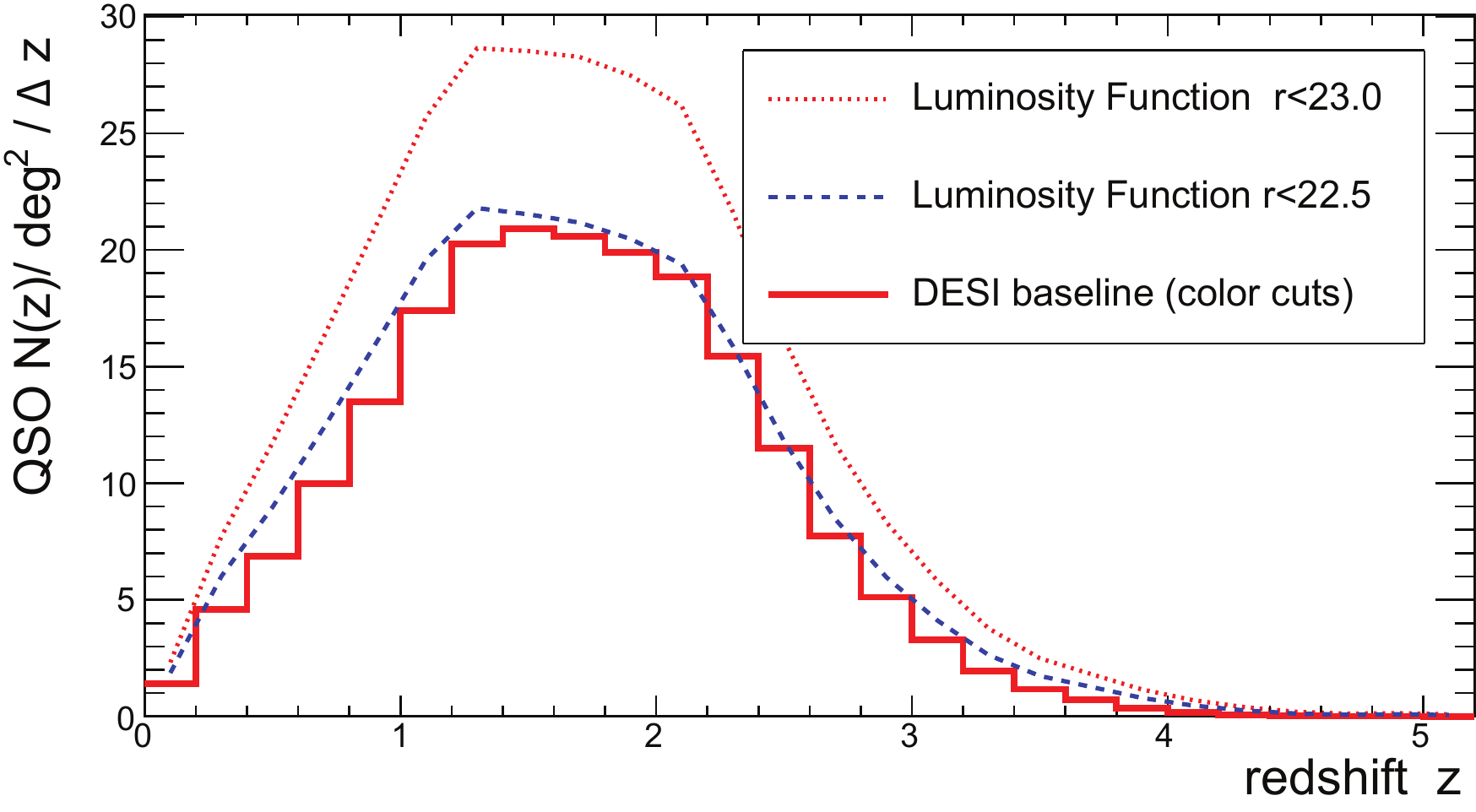}
\caption{Expected distribution of QSO redshifts from DESI (thick red
histogram) using the  targeting efficiency measured for the baseline DECaLS-DR2 selection over  truth regions.
For comparison, we also show the QSO luminosity function to $r<22.5$ (blue dashed line)
and $r<23.0$ (red dotted line). }
\label{fig:QSOnz}
\end{figure}

$\bullet$ {\it Redshift distribution:}
The expected redshift distribution of the QSO sample is illustrated in
Figure~\ref{fig:QSOnz} as the thick red histogram, which is determined by
assuming the QSO completeness for QSOs brighter than $r<23$ measured in the truth region for 
the color-cut selection. For comparison, we show on the same plot 
the QSO luminosity function to $r<22.5$ (blue dashed line)
and $r<23$ (red dotted line). 

$\bullet$ {\it Redshift measurement method:}
The key features contributing to the classification and redshifts
of QSOs are the \lya, CIV, CIII] and MgII emissions
(c.f.\ Figure~\ref{fig:QSOspectrum}).
From our experience with BOSS, eBOSS and MMT pilot programs, 
we estimate that in a single DESI visit we will fail to obtain
redshifts for QSO targets about 10\% of the time, mostly for objects at
$g>22.5$~\cite{2013A&A...551A..29P,Palanque2016LF}.
All QSO targets will be observed once early in the survey.  
Those confirmed to be QSOs at $z>2.1$ will be re-observed in subsequent
passes over the sky in order to obtain higher signal-to-noise spectra
of the \lya.

$\bullet$ {\it Large-scale-structure bias:}
QSO bias has been measured in BOSS via QSO-\lya cross-correlation studies
to be $3.6$ at $z=2.4$ \cite{fontribera13}, in agreement with previous
measurements \cite{Ross09,White12}. For QSOs at lower redshifts, we
project a bias of the form $b(z) = 1.2/D(z)$, where $D(z)$ is the
growth factor.  At $z>2.1$, clustering information is computed from the
transmitted flux in the \lyaf\ and not directly from correlations between
objects; the flux bias of \lya absorbers is estimated to be about -0.2
(it is negative because a larger matter density implies a higher absorption
and thus a lesser transmitted flux) \cite{slosar11}, and is strongly
enhanced along the line of sight by redshift-space distortions.

$\bullet$ {\it Target selection efficiency:} 
From the first pass of targeting over the sky, we expect to identify 170
QSOs per deg$^2$ from a sample of  260 targets per deg$^2$, for a target
selection efficiency (including redshift failures) of 65\%. 
For the subsequent passes, the target selection efficiency will be
near $100\%$, as only objects identified as $z>2.1$ QSOs will be re-observed.
After four passes, the average target selection efficiency is therefore of order 80\%.

\subsubsection{Recent and near-term developments  for QSO target selection}

During 2015, we focused on building large truth tables of QSOs against which to test current and improved selection algorithms. We developed comprehensive selections of quasars using the deep and multi-epoch SDSS photometry in the Southern Equatorial region called Stripe 82, where   variability selections are notably efficient (cf. Sec.~\ref{sec:variability} and \cite{Palanque11,2013A&A...551A..29P}). These pilot programs led, in particular, to a sample of 18,000 spectroscopically-confirmed QSOs over  120~deg$^2$ to an extinction-corrected magnitude $g_c<22.5$, as well as to a smaller but deep sample of $175~{\rm deg}^{-2}$ QSOs to $g_c<23$ over $\sim 10~{\rm deg}^2$. They also allowed us to update the QSO luminosity function and make it more robust at faint magnitudes~\cite{Palanque2016LF}. We are planning further dedicated programs to be run at MMT and AAT  to extend the truth tables to $r_c<23$ as required for DESI.  We also applied for a program at MMT to test  the current target selection algorithms relying solely upon DECaLS+{\it WISE} data, in a field where the {\it WISE} data already have the depth of the final 4-year survey, with the aim of providing the first validation the QSO selection for DESI. 

In parallel,  work has begun on  machine-learning algorithms to take better advantage of the  imaging data available for DESI. In BOSS, the XDQSO algorithm~\cite{Bovy11} led to  nearly 20\% improvement compared to color cuts and we can reasonably expect a similar increase in the yield of the QSO selection for DESI by the Random Forest algorithm that we are focusing on. These developments have  started with  DECaLS optical data and existing WISE infrared data. They will be iterated as additional depth is acquired on WISE.

\subsubsection{Variability Data Improves Selection of High-Redshift QSOs} 
\label{sec:variability}
Time-domain photometric measurements can enhance QSO selection. They allow us 
to exploit the intrinsic variability of QSOs \cite{Schmidt10} to distinguish
them from stars of similar colors.  They therefore complement the
color-selection techniques presented in Sec. \ref{sec:QSOselectiontechnique}.
We have so far used variability information extensively to build truth tables against which to test QSO selection. In a second step, we will use variability to select additional high-redshift
\lya QSOs, for which uniformity of selection across the sky is not required.

Because the accretion region around a quasar is highly compact, its luminosity can vary substantially on timescales ranging from days to years, with a pattern distinct from that seen in variable stars. 
The time variability of astronomical sources can be described using  a measure of the amplitude of the observed magnitude variability $\Delta m$ as a function of the time delay $\Delta t$ between two observations.  This  ``structure function''  is modeled as a power law parameterized in terms of $A$, the mean variation amplitude  on a one-year time scale (in the observer's reference frame) and $\gamma$, the logarithmic slope of the variation amplitude with respect to time: $\Delta m = A (\Delta t)^\gamma$. 

\begin{figure}[!t]
\centering
\includegraphics[width=3in]{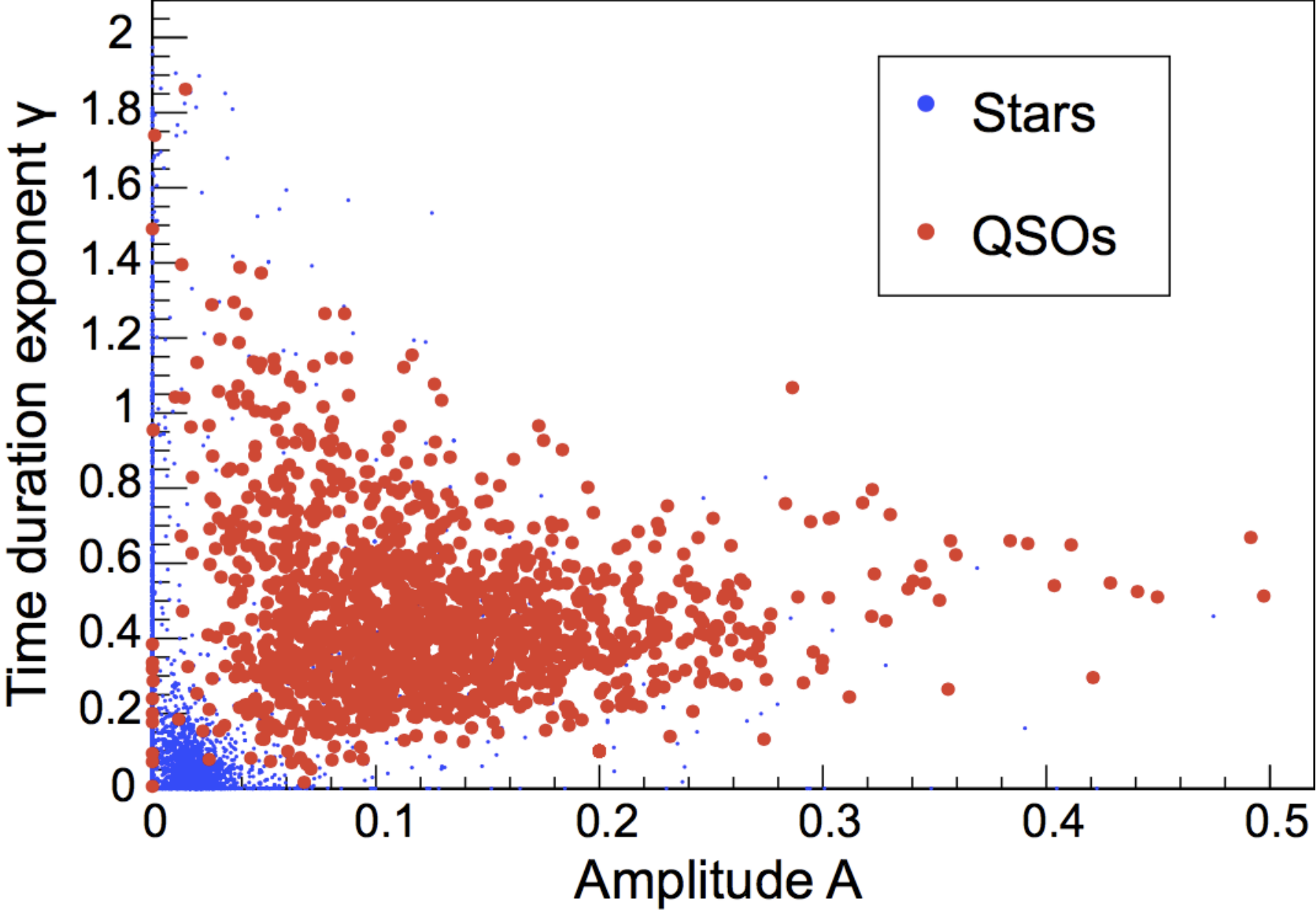}
\hspace{0.25in}
\includegraphics[width=3in]{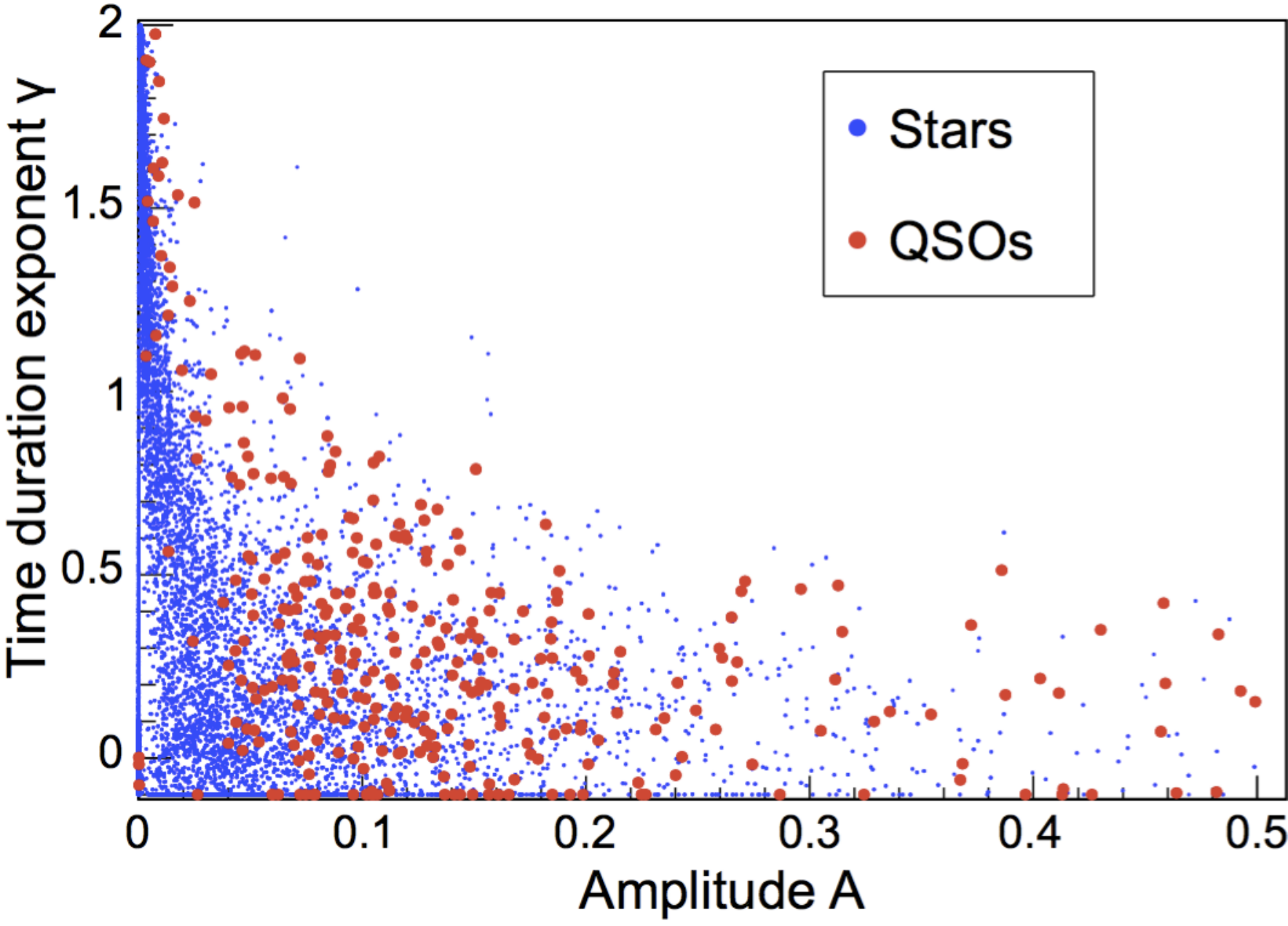}
\caption{
{\it Left panel:} Structure function parameters for 50-epoch $gri$ light curves
from SDSS in Stripe 82 (left), where the parameters are amplitude (A)
and time duration exponent ($\Gamma$).
{\it Right panel:} Structure function parameters for the 6-epoch $R$ light curves
from PTF, where the discriminating power is diminished but still
valuable with fewer epochs and filters.}
\label{fig:Agamma}
\end{figure}

We have tested variability techniques in DESI pilot surveys, both in Stripe 82 \cite{Palanque11} that was the subject of repeated SDSS observations totaling about $50$ epochs, and elsewhere on the sky, where time-domain information was derived from 5-10 epochs of PTF $R$-band data. As illustrated in Figure~\ref{fig:Agamma}, the segregation between QSOs and stars is much reduced with poorer data, but variability remains competitive. 
This technique allowed us to identify 30\% more QSOs in the Stripe 82 field than with time-averaged optical photometry only \cite{Palanque11}, and a combined color and variability selection from CFHT and PTF imaging data in the CFHTLS D3 field allowed us to achieve a record-high surface density of 207 QSOs per ${\rm deg}^2$ to $g=23$. 
The gain relative to the baseline QSO targeting with full
{\it WISE} depth is likely to be less dramatic, and will be evaluated
as those data become available. Even with only four epochs of {\it WISE}  data (two stacks per year,  two years of observations), preliminary tests indicate that variability information from {\it WISE} can be used to reduce the contamination of the target sample by about 10\%. The full 4-year survey  {\it WISE} will allow 8-epoch light curves, and  further gain over current estimates. 

Imaging surveys that could provide useful time-domain information for
variability selection include the  PTF and follow-on iPTF surveys
(in the deepest areas of their footprint), 
the DES survey or the {\it WISE} survey.  Variability is not assumed in our baseline
targeting plan, but it is expected to be valuable for selecting the
\lya QSOs wherever coverage exists.

\subsection{Calibration Targets} 
Target selection is also responsible for providing lists of standard stars
for flux calibration, and lists of blank sky locations to be used for modeling
the sky.

Main-sequence F stars will be used as the primary spectrophotometric standard stars.
These stars are well-described by stellar atmosphere models, making them
ideal targets for spectrophotometric calibration at optical wavelengths.
A stellar template of appropriate temperature, surface gravity and
metallicity will be derived for each star and used to derive the spectral response
including the time-varying atmospheric absorption bands.

The selection will be similar to the color-magnitude selection of BOSS
to identify low-metallicity targets through a selection in ($u-g$),($g-r$),
($r-i$), and ($i-z$) colors.
The restrictive BOSS selection yields 10 stars per deg$^2$;
to obtain a larger number of potential targets using the new $grz$ photometry,
DESI will broaden this selection and include higher metallicity standard stars.
With Gaia spectrophotometry of F stars that span a range of metallicity,
and upcoming data from SDSS-IV/eBOSS in which a broader selection is applied,
we plan to evaluate the value of a mix of lower and
higher metallicity F stars to serve as flux
calibration standards for DESI.  Finally, we will perform a cross-calibration
of low-metallicity and higher metallicity F-stars during the commissioning
stages of DESI, thus providing validation of the standard star selection.

Blank sky locations will be determined as part of the object detection
algorithms applied to the input imaging, ensuring that there are no detectable
sources within the fiber diameter in any of the input bands.
These will be provided at a density
such that every fiber (when possible) will have the option of a blank sky
if it isn't otherwise assigned to a science target.

\subsection{Baseline Imaging Datasets}
\label{sec:imagingDataSets}

\begin{figure}[!t]
\includegraphics[width=\textwidth]{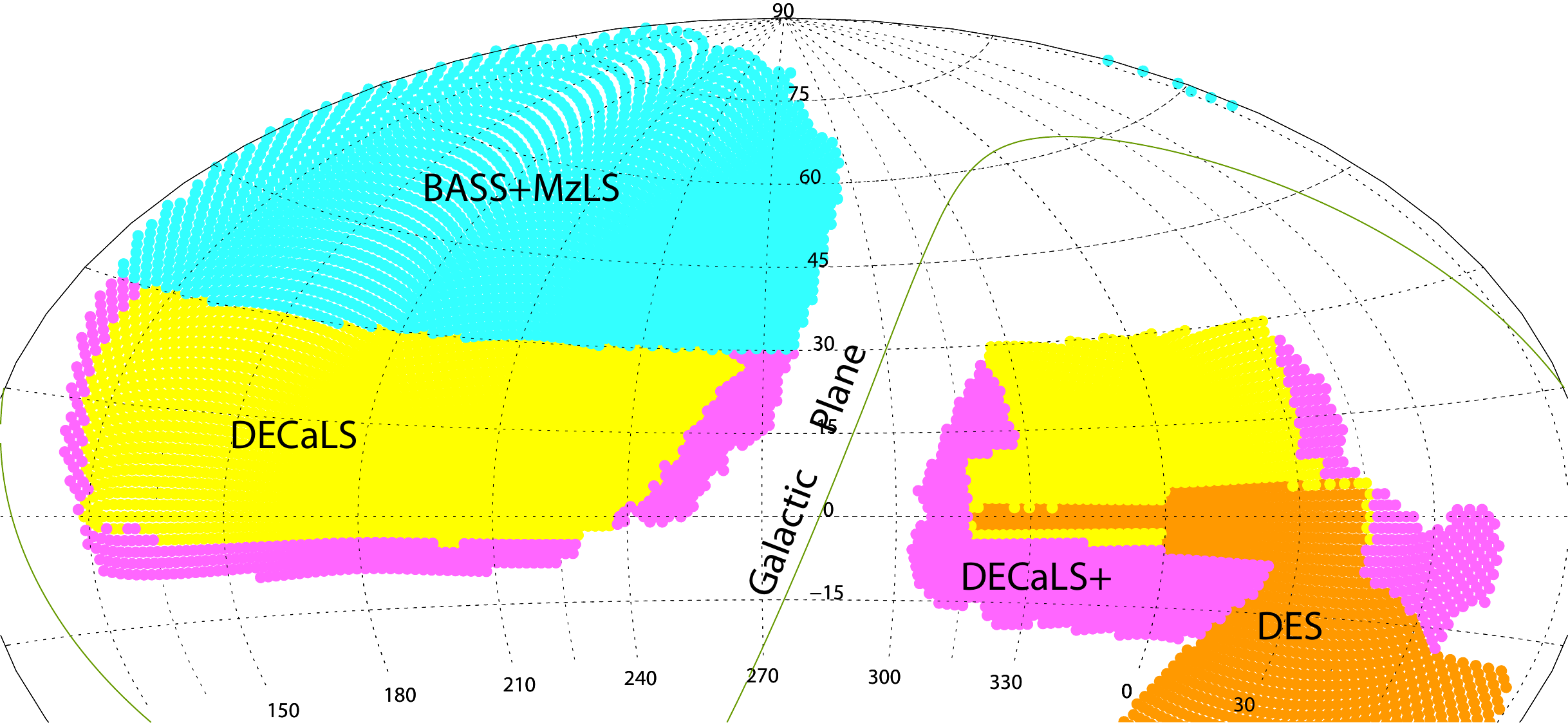}
\caption{The primary imaging surveys that will result in targeting data
for the DESI project. The footprint at DEC~$\le +34^\circ$ will be
covered using the Dark Energy Camera (DECam) on the Blanco 4m telescope
at Cerro Tololo Inter-American Observatory. The Dark Energy Camera Legacy
Survey (DECaLS, in yellow), the Dark Energy Survey (DES, in orange), and 
the extended DECaLS in the North Galactic Cap (DECaLS+, in purple on left)
are underway.  A proposal for the remaining extended DECaLS
in the South Galactic Cap (DECaLS+, in purple on right) will be submitted.
Imaging of the North Galactic Cap region at 
DEC~$\ge +34^\circ$ (cyan) will be covered with the 
90Prime camera at the Bok 2.3-m telescope in $g-$ and $r-$bands
(BASS: the Beijing-Arizona Sky Survey) and with the upgraded MOSAIC-3 camera
on the Mayall 4m telescope in $z$-band (MzLS: the MOSAIC
z-band Legacy Survey). Both the Bok and Mayall telescopes are located
on Kitt Peak National Observatory.
\label{fig:allsurveys}}
\end{figure}

The samples described above can be selected given highly-uniform
optical imaging data in the $g$, $r$, and $z$ bands, as well as all-sky
imaging from the {\it WISE} satellite.  The same imaging data for
selected science targets will be used to identify calibration targets
(standard stars and sky fibers).
A combination of three telescopes will be used to provide the baseline targeting 
data for DESI: the Blanco 4m telescope at Cerro Tololo, the Bok 90-inch and the 
Mayall 4m telescope at Kitt Peak. The footprints of the primary surveys using these telescopes that will deliver the targeting data are shown in Figure~\ref{fig:allsurveys} and the next three subsections discuss these surveys and their current status in more detail. The status of the {\it WISE} data is presented in \S~\ref{sec:wise}.

\subsubsection{Blanco/DECam Surveys (DEC$\le$34$^\circ$)} 
\label{sec:decam}

The Dark Energy Camera (DECam) on the Blanco 4m telescope, located at the Cerro Tololo Inter-American Observatory, will provide the optical
imaging for targeting over 2/3 of the DESI footprint, covering both the North and South Galactic Cap regions at Dec $\le 34^\circ$.  Due to the
combination of large field of view and high sensitivity from 400-1000~nm, 
DECam is the most efficient option for obtaining photometry in the
$g$, $r$, and $z$ bands.  


DECam can reach the required depths for DESI targets in modest total
exposure times of 100, 100 and 200 sec in $g$, $r$, $z$ in median
conditions.  These data reach required 5$\sigma$ depths
of $g$=24.0, $r$=23.4 and $z$=22.5 for an ELG galaxy with half-light
radius of 0.45 arcsec.  For a 3-dither observing strategy, accounting
for weather loss, DECam is capable of imaging 9000 deg$^2$ of the DESI
footprint to this depth in 81 scheduled nights.  These depth estimates have
been vetted with $grz$ photometry in the COSMOS field in Spring 2013
(Section~\ref{sec:ELGtargprop}).

\begin{figure}[!b]
\centering
\includegraphics[width=0.48\textwidth]{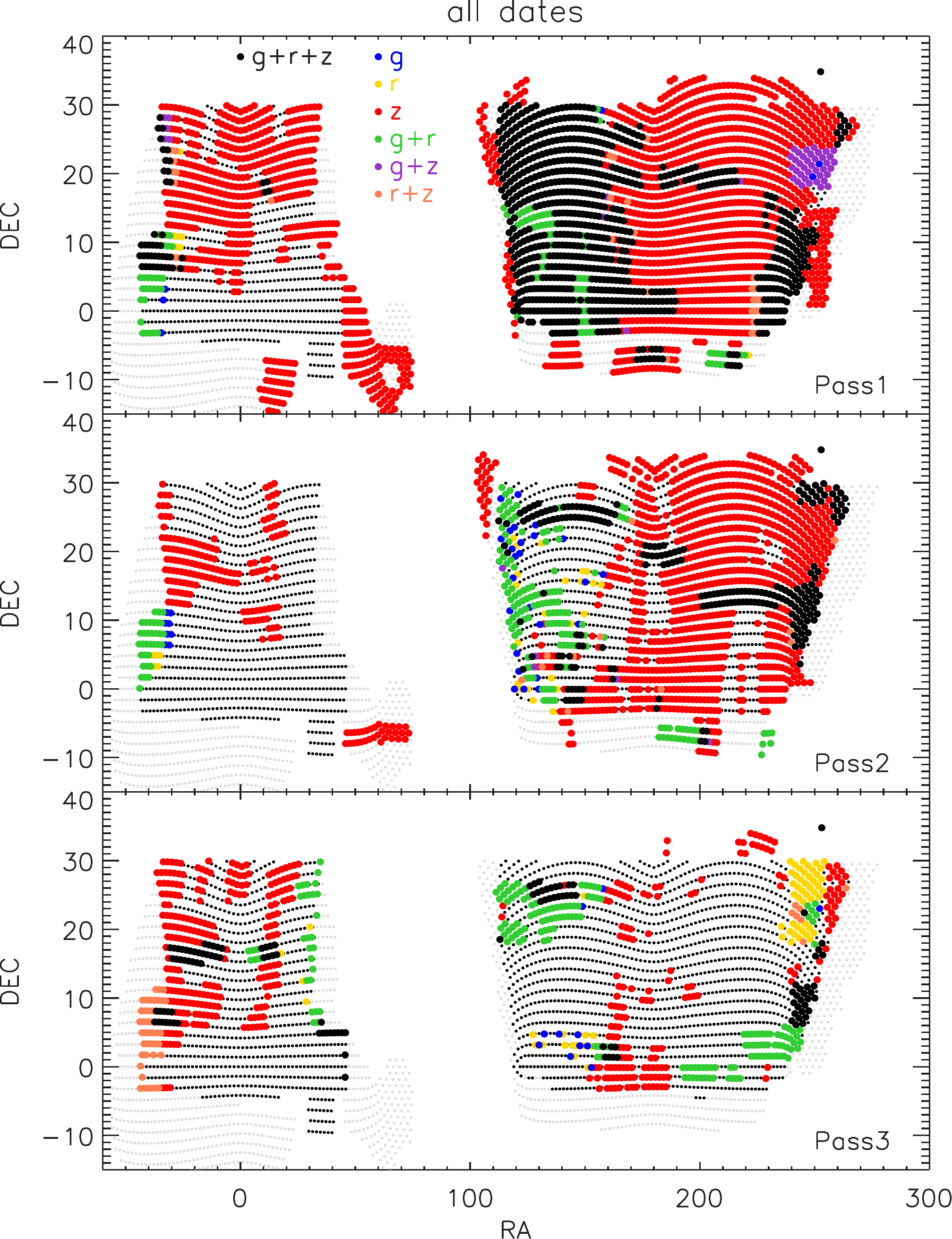}
\caption{{\it Left panel:} Coverage map of the DECaLS survey through
February 2016.
The coverage in the $g$, $r$ and $z$ filters is indicated by
the color as blue ($g$-only), yellow ($r$-only), green ($g$+$r$),
purple ($g$+$z$), orange ($r$+$z$) or black ($g$+$r$+$z$).
Each panel represents one of the 3 passes, where pass 1 is observed
in the best weather conditions.
\label{fig:DECcoverage}}
\end{figure}

A public survey, ``The DECam Legacy Survey of the SDSS Equatorial
Sky'' (DESI collaborators D. Schlegel and A. Dey are PIs), has been
approved to obtain optical imaging to the required depth over 6200
deg$^2$.  This ``DECaLS'' survey has been allocated 64
nights spread out over 3 years (2014A to 2017B semesters)
as part of the NOAO Large Surveys program.
The survey began in August 2014 and has thus far had 30 scheduled
nights and 6 Director's discretionary nights (near full moon),
during which 23\% of the $g+r$ and 49\%
of the $z$ imaging has been completed. The current coverage is shown in
Figure~\ref{fig:DECcoverage}.

The DECaLS program is making use of other DECam data
within the DESI footprint as those data become public.
The most significant of these other data sets is from
the Dark Energy Survey, which includes a 500~deg$^2$
contiguous area in the South Galactic Cap.
DECaLS is explicitly not re-imaging that area,
and making use of those raw data as the proprietary period
expires 12 months after the date of observation.

DECaLS will cover $\approx$2/3 of the planned DESI footprint
at Dec $\le 34^\circ$.
The DECaLS team successfully applied for an 8-night extension
(DECaLS+) that will obtain imaging for the
remaining 800~deg$^2$ in the North Galactic Cap.
An additional 500 deg$^2$ in the
South Galactic Cap is being observed by the Dark Energy Survey,
with those raw data publicly available 12 months after the date
of observation.
A proposal to observer the remainder of the DESI footprint in the South
Galactic Sky will be submitted in future semesters.
Data from these programs are treated the same and reprocessed uniformly
to ensure consistency for DESI target selection.

The DECam data have been reduced to calibrated images at NOAO
and catalogs constructed using the Tractor algorithm (see \S~\ref{sec:tractor}).
These catalogs have been used for the DESI target selection tests
described elsewhere in this chapter.

\subsubsection{Bok/90Prime Survey (DEC$\ge$34$^\circ$)}
\label{sec:bok}

The NGC footprint at Dec~$\ge +34$ deg will be observed by the Bok 2.3-m 
telescope in two optical bands ($g$ and $r$) for DESI targeting.
The Bok Telescope, owned and operated by the University
of Arizona,  is located on Kitt Peak,
adjacent to the Mayall Telescope. 
The 90Prime instrument is a prime focus 8k$\times$8k
CCD imager, with four University of Arizona ITL 4k$\times$4k CCDs
that have been thinned and UV optimized with peak QE of 95\% at
4000\AA\ \citep{90Prime}.  These CCDs were installed
in 2009 and have been operating routinely since then.  90Prime
delivers a 1.12 deg field of view, with 0.45$''$ pixels, and 94\%
filling factor. Typical delivered image quality at the telescope
is 1.5$''$.  The $g$ and $r$-band survey over 5000 deg$^2$ is projected
to require
180 nights of scheduled telescope time for average weather.
The throughput and performance
in these bands were demonstrated with data in September 2013.

The BASS survey tiles the sky in three passes, similar to
the DECaLS survey strategy.  At least one of these passes
will be observed in photometric conditions (P1) and seeing
conditions better than 1.7 arcsec.

The Bok survey (known as the Beijing-Arizona Sky Survey; Zhou Xu
and Xiaohui Fan, PIs; see {\tt http://batc.bao.ac.cn/BASS}) 
was awarded 56 nights in Spring 2015 and 100 nights in each
of Spring 2016 and 2017.
The Bok survey will target 5500~deg$^2$ in the NGC, including 500~deg$^2$ of overlap 
with the region covered by the DECam surveys in order to understand and 
correct for any systematic biases in the target selection.
The existing Bok $g$-band filter is well-matched to the DECam
$g$-band filter.
The existing Bok $r$-band filter had a significantly different
bandpass as compared to the DECam $r$-band filter, therefore
we acquired a new $r$-band filter from Asahi that was delivered
in April 2015.

The BASS survey began observations in Spring 2015.
A number of instrument
control software updates, new flexure maps, and new observing tools
were implemented that greatly improve the pointing accuracy,
focusing of the telescope, and observing efficiency.
15\% of the $g$-band and 2\% of the $r$-band tiles were observed
in that semester.  It was discovered that those data suffered
from defective electronics in the read-out system that introduced
A/D errors, gain variations and non-linearities.  Those electronics
were replaced in September 2015 followed by a recommissioning of
the system in Fall 2015.

BASS has been scheduled for the 100 darkest nights in the 2016A semester (January-June),
and expects to schedule a comparable number of nights in 2017A.
Through February 17, 2016, the survey has completed 10, 10 and 0\%
of the pass 1, 2, 3 tiles in $g$-band, and 14, 13, 5\% of the
tiles in $r$-band (see Figure~\ref{fig:Mzcoverage}).
The raw and calibrated images will be publicly served through the
NOAO Science Archive.
These data will be included in the Legacy Survey catalogs
beginning with Data Release 3 in 2016.


\subsubsection{Mayall/MOSAIC Survey (DEC$\ge$34$^\circ$)}
\label{sec:mayall}

The Mayall $z$-band Legacy Survey (MzLS) will image the 
DEC~$\ge +34^\circ$ region of the DESI North Galactic footprint.
It will use the MOSAIC-3 camera at the prime focus of the 4-meter
Mayall telescope at Kitt Peak National Observatory.
MzLS will be scheduled for
230 nights during semesters 2016A and 2017A through an agreement
between the National Science Foundation and the Department of Energy.
116 of these nights have been scheduled in the 2016A semester,
with a survey start on February 2, 2016.
The imaging camera has undergone a major upgrade in 2015 to improve
its $z$-band efficiency.
The KPNO 4m telescope control system and the imaging camera
software have been upgraded for improved operational efficiency.
NOAO has purchased a new $z$-band filter to match the DECam filter
bandpass and to thereby minimize any differences between the DECam
and MOSAIC $z$ surveys. 

The MOSAIC-3 camera is a new version of the prime focus imaging
system.  This upgrade has made use of the dewar from the MOSAIC-2
camera at CTIO and the MOSAIC-1.1 mechanical system and guider
from KPNO.
Yale University designed and built a new cold plate for the dewar
which it populated with four super-thick (00$\mu$m-thick) fully-depleted
4096$^2$ pixel CCDs with the same 15-micron pitch.
The readout system consists of four DESI controllers, one
for each CCD that simultaneously reads the four quadrants of each device.
These controllers were modified to synchronize to a single clock.
The dewar was delivered to NOAO in September 2015 where it
was integrated with the MOSAIC-1.1 mechanical enclosure,
shutter, filter wheel and acquisition and guider system.
This upgraded camera, christened MOSAIC-3, saw first light in
October 2015 and underwent further on-sky commissioning runs in 
November and December 2015.  The $z$-band efficiency has
been measured to be improved by 60\% as compared to the MOSAIC-1.1 camera.

\begin{figure}[!tb]
\centering
\includegraphics[width=0.47\textwidth]{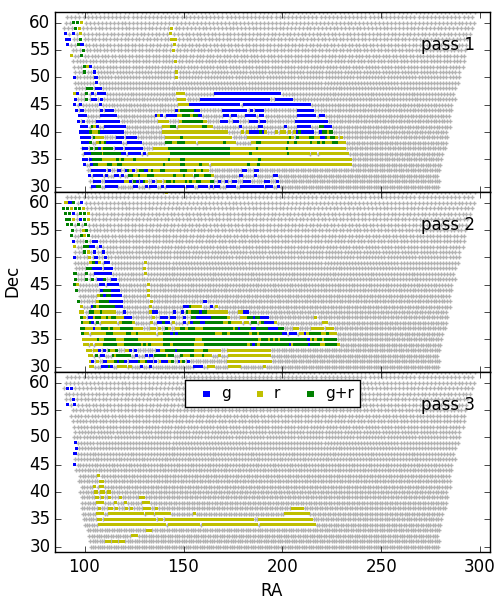}
\includegraphics[width=0.49\textwidth]{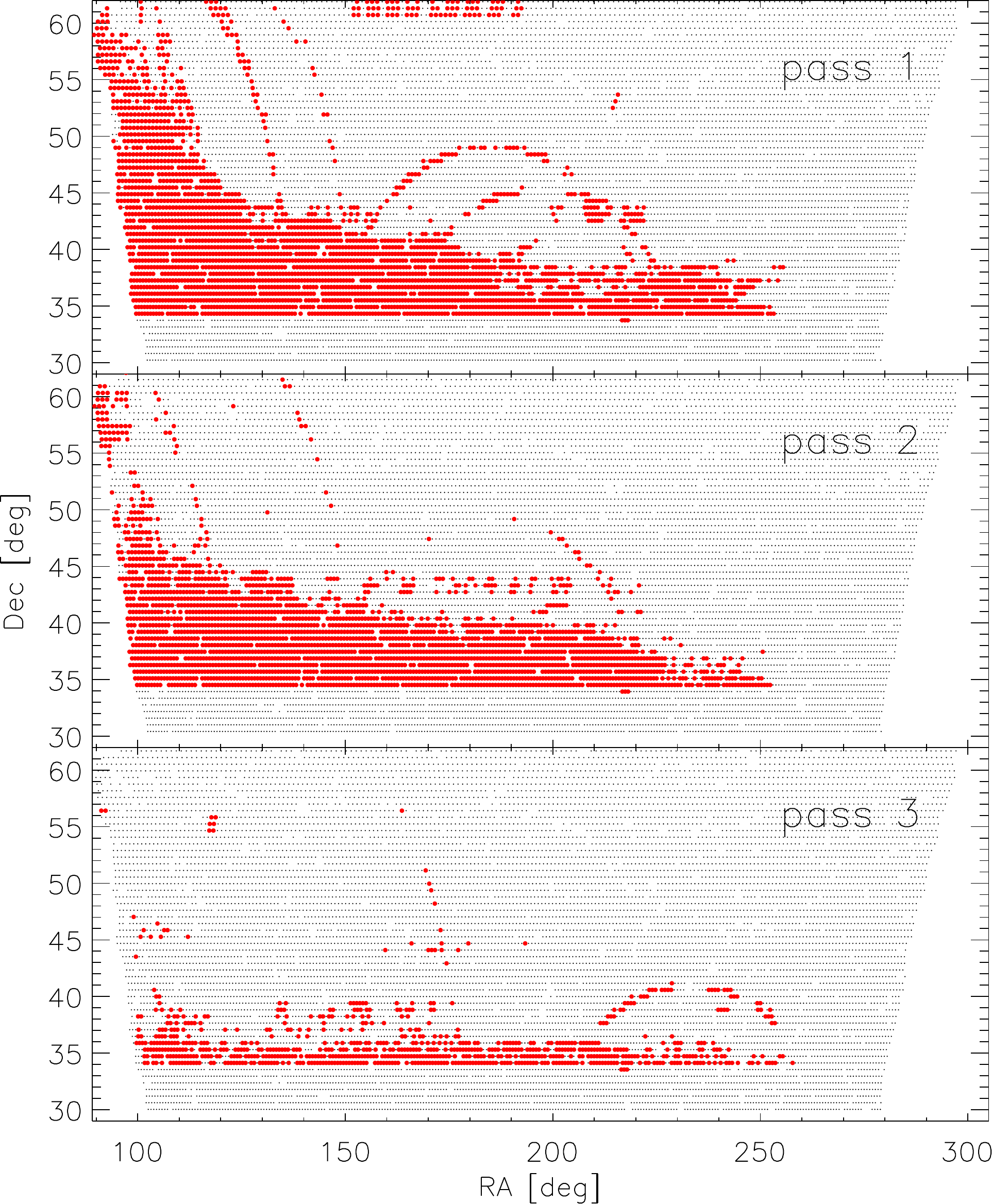}
\caption{
{\it Left panel:} Coverage map of the Bok/BASS survey based on data
collected though March 6, 2016, and excluding data prior to the electronics
fixes in September 2015.
The coverage in the $g$ and $r$ filters is indicated by the color
as blue ($g$-only), yellow ($r$-only) or green ($g$+$r$).
Each panel represents one of the 3 passes, where pass 1 is observed
in the best weather conditions.
{\it Right panel:} Coverage map of the MzLS $z$-band survey based on data
collected though March 6, 2016.
The coverage is indicated in each of the three passes.
\label{fig:Mzcoverage}}
\end{figure}

The MzLS survey tiles the sky in three passes, similar to
the DECaLS survey strategy.  At least one of these passes
will be observed in photometric conditions (P1) and seeing
conditions better than 1.3 arcsec.
Through March 8, 2016, the survey has completed 23, 19 and 8\%
of its pass 1, 2 and 3 tiles (see Figure~\ref{fig:Mzcoverage}).

The 
The MOSAIC $z$-band survey project will be run similarly to the DECaLS survey, with 
the initial processing being done using the NOAO pipeline and calibration and 
catalog construction being carried out at LBNL/NERSC. The raw and pipeline-processed 
images are public as they are available, typically at the end of
each lunar cycle, through the NOAO Science Archive.
These data will be included in the Legacy Survey catalogs
beginning with Data Release 3 in 2016.

\subsubsection{$WISE$ All-Sky Survey} 
\label{sec:wise}

Infrared imaging from the Wide-field Infrared Survey Explorer ({\it
WISE}) satellite are critical to the DESI targeting algorithm for LRGs
and QSOs.  During its primary 7-month mission from through August
2010, {\it WISE} conducted an all-sky survey in four bands
centered at 3.4, 4.6, 12 and 22~$\mu$m (known as W1, W2, W3 and W4)
\cite{WISE}  99.99\% of the
sky was imaged at least 8 times, while regions near the ecliptic poles
were observed more than 100 times.  Following the primary 4-band
mission, {\it WISE} continued survey operations in the three shortest
bands for 2 months, then the two shortest bands for an additional 4
months for a total of a 13-month mission that completed in September 2011.
NASA re-activated the satellite in Fall 2013 and is continuing two-band
survey observations for an additional 3 years starting December 1, 2013,
as the {\it NEOWISE} project.  The first {\it NEOWISE} data release occurred on
March 25, 2015, the second release will be March 23, 2016, and the final
release will be in March 2017.

DESI target selection utilizes the two shortest-wavelength bands
at 3.4 (W1) and 4.6~$\mu$m (W2).  Photometry in these bands is measured
using the \emph{the Tractor} algorithm (see Section~\ref{sec:tractor})
measured on the re-stacked {\it WISE} and {\it NEOWISE} Level 1
imaging that retains the intrinsic resolution of the data and are
appropriate for preserving the available signal-to-noise
\citep{2014AJ....147..108L}.
Data Release 1 of the Legacy Survey (DECaLS and {\it WISE})
made use of the initial 13-month data set, reaching
5-$\sigma$ limiting magnitudes of 20.0 and 19.3 AB mag in W1 and W2.
Data Release 2 made use of approximately twice as much {\it WISE}
data with the first year of {\it NEOWISE}.
The final Legacy Survey catalogs will use the full {\it WISE}
and {\it NEOWISE} data sets, reaching 0.7 mag fainter than
the Legacy Survey Data Release 1 or the {\it WISE} All-Sky Data Release.


\subsection{Additional Imaging Data} 

Additional imaging data, if available, can supplement the target
selection data and may be used, in particular, to improve the
selection of the high-redshift \lyaf~QSO sample.  This is because
the \lyaf\ analysis is based on the clustering of absorption systems
along the line of sight, and therefore does not require a spatially
uniform QSO sample.
%
%
As a result, the QSO target selection can utilize datasets that may
not be uniform (in depth, bandpass, or time sampling) over the DESI
footprint.  In this section, we summarize the key datasets that may
contribute to this effort, if they prove to be available.  These
data sets are not assumed to be available for our baseline target
selection plans, but rather should improve the efficiency of targeting
higher-redshift ($z>2.1$) QSOs beyond the baseline targeting strategy
presented above.


\subsubsection{SDSS}
The Sloan Digital Sky Survey \cite{SDSS09} has obtained multi-band
($ugriz$) photometry (in photometric conditions) over a 10,000 deg$^2$ extragalactic footprint
in the North Galactic and South Galactic Caps.
The Northern Cap and four stripes in the Southern Cap were imaged
in 1998-2004.  The bulk of the Southern Cap was imaged in 2008-2009,
and the SDSS camera was then retired from service in December 2009.
The median 5$\sigma$ magnitude depths for the SDSS $ugriz$ bands are
22.15, 23.13, 22.70, 22.20, and 20.71, respectively, but with substantial
variation in depth from seeing.
SDSS may provide a reference photometric point for variability selection of
high-redshift QSOs, allowing variability over long time baselines to be measured.

\subsubsection{PanSTARRS-1}

The PanSTARRS-1 (PS1) $3\pi$ survey \cite{PS1web} is a transient-sensitive survey designed to 
observe 30,000 deg$^{2}$ of sky over 12 epochs in each of the five $grizy$ filters.
The multi-band photometry generated from the co-added exposures reaches depths
that are comparable to SDSS in $gr$ and potentially deeper in $iz$.
These depths would potentially be adequate for the DESI BGS and LRG samples,
but not the ELG or QSO samples.
The PS1 survey completed observations in 2013.
The PS1 time-domain photometry may be useful for enhancing the selection
of \lya~QSOs at the brighter magnitudes. The DECaLS survey is currently using
a bright star catalog from PS1
to provide initial photometric and astrometric calibration across its footprint.
The PS1 co-added imaging and catalogs are not available
as of March 2016.

\subsubsection{PTF, iPTF, and ZTF}

The Palomar Transient Factory (PTF) \cite{PTF09} was a photometric
survey designed to find transients via repeated imaging over 20,000
deg$^2$ in the Northern Hemisphere.  In February 2013, the next phase
of the program, iPTF (intermediate PTF) began.  Both have used the
CFH12K camera on the 1.2~m Oschin Telescope at Palomar Observatory,
which covers 7.2 deg$^2$ of sky in a single pointing with a pixel
scale of 1.01~arcsec.

Four years of survey operations have so far yielded a total of 5,000
deg$^2$ in $R$-band and 1,000 deg$^2$ in $g$-band to useful depths for
QSO selection based on variability.  LBNL is a partner in the PTF and iPTF
collaborations, and DESI has access to these data.

An upgraded Zwicky Transient Factory (ZTF) has been funded through
an NSF Mid-Scale Innovations Program in Astronomical Sciences.
ZTF will utilize the same telescope with a new 46 square-degree imager,
beginning operations in 2017.
The ZTF survey will cover the entire sky
at declinations Dec $>-20$ deg, including the full DESI footprint.
ZTF will operate with a $g$-band similar to the DECam and Bok
$g$-band, an $R$-band (Mould-$R$) that is broader, and potentially an
$i$-band.  These data, which will be available to DESI collaboration 
for the purposes of target selection, are expected to eventually achieve
the DESI targeting depths in $g$ and $R$ bands, but likely not before
the start of DESI spectroscopic operations.  The time sampling of ZTF
is planned to be highly non-uniform over the DESI footprint, with
different areas of sky covered in different years.  Therefore, ZTF is
not viable for the baseline DESI target selection, but PTF, iPTF and
ZTF may be used to supplement the high-redshift QSO selection for
DESI.

\subsubsection{CFHT}
\label{par:cfht}
The Canada-France-Hawaii Telescope (CFHT) is a 3.6--m meter telescope
on Mauna Kea, Hawaii.  CFHT is a joint facility of the National
Research Council of Canada, the Centre National de la Recherche
Scientifique of France, and the University of Hawaii.  The CFHT prime focus
imager MegaCam, a very efficient instrument for imaging large areas of sky, 
consists of 36 2k$\times$4k e2v CCDs, 
covering a field of view of 0.97 deg$^2$ with a pixel scale
of 0.185~arcsec per pixel.  MegaCam started operations in 2003 and has
conducted a number of large imaging surveys, the largest being the
CFHT Legacy Survey covering 155 deg$^2$.


The CFHT community is in discussions with the Euclid consortium and
may play a role in providing $ugri$ imaging data over the northern
Euclid footprint. However, no plan is currently in place. There is
an ongoing $u$-band survey (``CFHT-Luau: The CFHT Legacy Survey for
the u-band all-sky universe"; A. McConnachie and R. Ibata, PIs)
aimed at providing imaging over 4000~deg$^2$ of the high-Galactic-latitude
northern sky, approximately split between the North and South
Galactic caps.  CFHT-Luau will complete in the 2016B semester (with
data becoming public 1 year after observation).  $(u-g)$ color
selection is an efficient discriminator between low-redshift and
high-redshift QSOs. Hence, the CFHT data may be used to supplement
the high-redshift \lyaf~QSO selection in DESI, especially in
combination with variability data.

%
%

\subsubsection{SCUSS}
\label{par:scuss}
The South Galactic Cap U-band Sky Survey \cite{SCUSS} is a survey of
4000 deg$^2$ in the South Galactic Cap using the 90Prime instrument on
the Bok 2.3-m telescope.  The survey was a joint project among the
Chinese Academy of Sciences, its National Astronomical Observatories
unit, and Steward Observatory \footnote{SCUSS survey
  http://batc.bao.ac.cn/Uband/}.  The survey was conducted between
September 2010 and October 2014 with typical exposure times of 5
minutes per field.  The limiting magnitude reached by the data is
$u\sim 23$ mag ($5\sigma$ point source), with some variation due to
varying seeing conditions.  These data may be used to supplement the
high-redshift \lyaf~QSO selection in DESI, especially in combination
with variability data.

\subsection{The Tractor Photometry for Target Selection}
\label{sec:tractor}

The DESI target selection combines photometry from optical imaging and
from {\it WISE}.  DESI Imaging Scientist Dustin Lang has developed \emph{the Tractor}
forward-modeling approach to perform source extraction on pixel-level data \cite{Lang14}.
\footnote{https://github.com/dstndstn/tractor}
This is a statistically rigorous approach to fitting the
differing PSF and pixel sampling of these data, which is particularly important as the optical
data have a typical PSF of $\approx 1$ arcsec and the {\it WISE} PSF is
$\approx 6$ arcsec.

\emph{The Tractor} takes as input the individual images from
multiple exposures in multiple bands, with different seeing in each.
A simultaneous fit is performed for sources to the pixel-level data
of all images.
Thus, if a source is determined
to be a point source, it is photometered as a point source in every band
and every exposure.
If it is found to be a morphologically extended source, then the same
light profile is consistently fit in all images.
This produces object fluxes
and colors that are consistently-measured across the wide-area
imaging surveys input to DESI target selection

\begin{figure} [!p]
\centering
\includegraphics[width=0.49\textwidth]{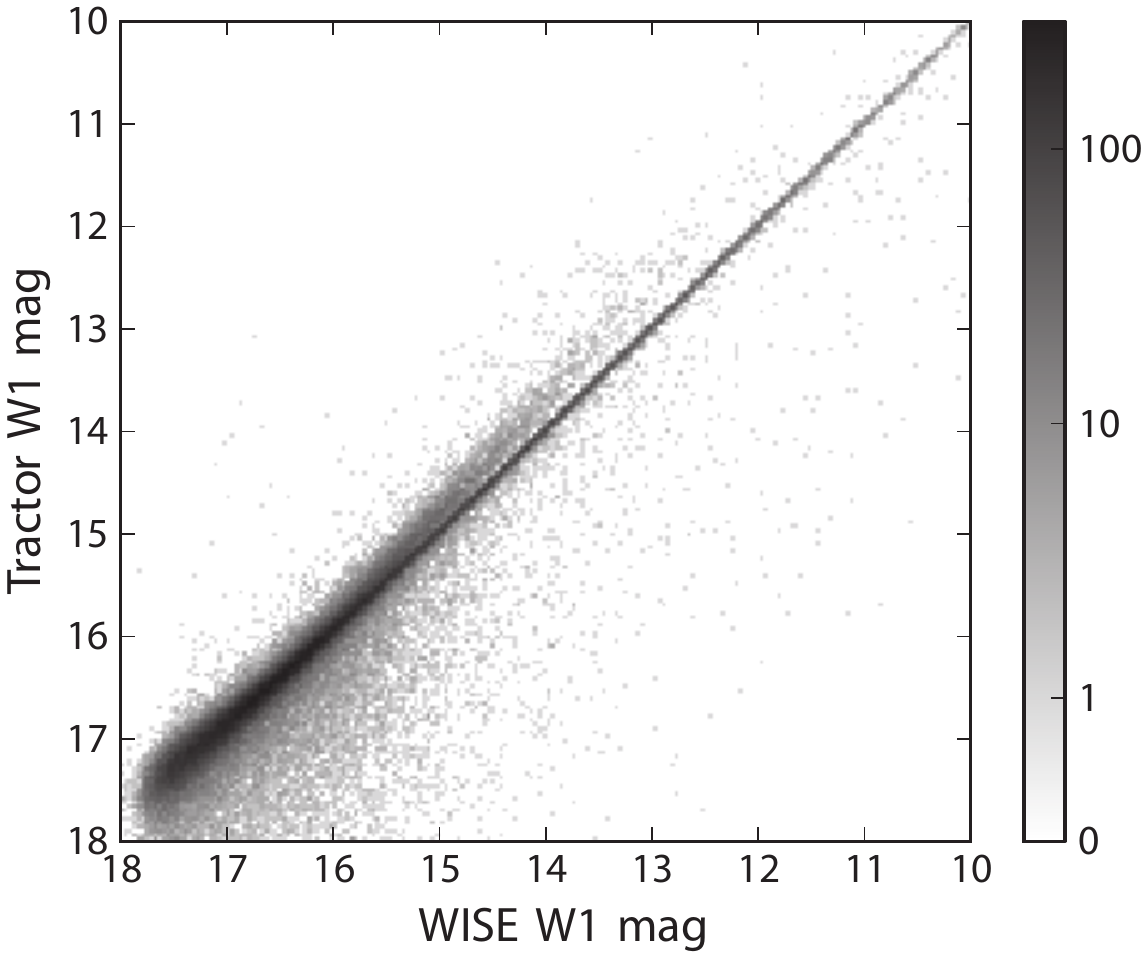}
\includegraphics[width=0.49\textwidth]{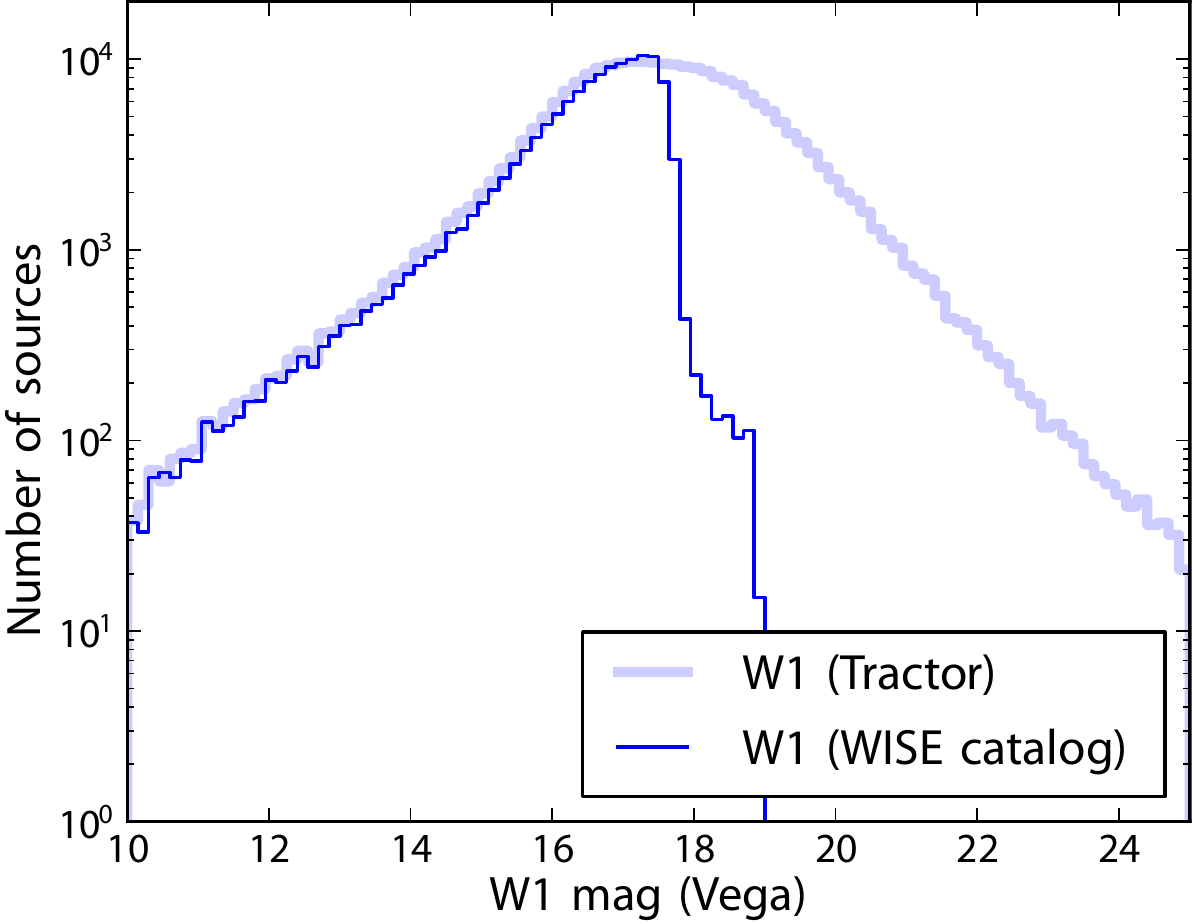}
\caption{Forced photometry results from \emph{the Tractor} code, using
  information from SDSS detections and light profiles to measure the
  flux from objects in the {\it WISE} images to below the {\it
    WISE} detection limit.  {\it Left panel:} The results agree for bright
  objects that are detected in the {\it WISE} catalog.  The widening
  locus below W1$\sim$14 is due to our photometry treating larger
  objects as truly extended, in contrast to the point-source-only
  assumptions in the public {\it WISE} catalog.  {\it Right panel:} A
  demonstration of the increased depth made possible from using
  \emph{the Tractor}.  By using optical imaging from SDSS to detect
  objects, photometry is measured for objects that are well
  below the {\it WISE} detection limit. \label{fig:tractor-bright}}
\vspace{0.2in}
\centering
\includegraphics[width=0.49\textwidth]{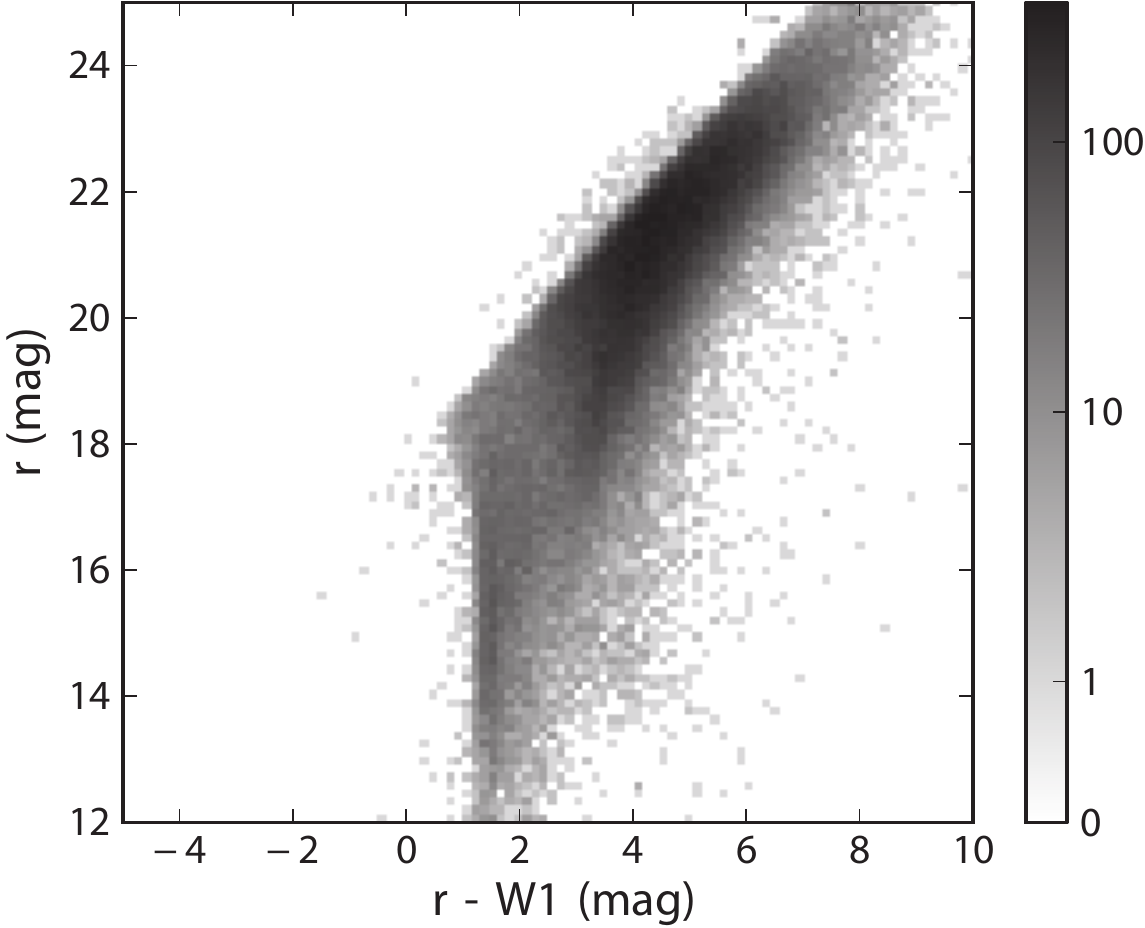}
\includegraphics[width=0.49\textwidth]{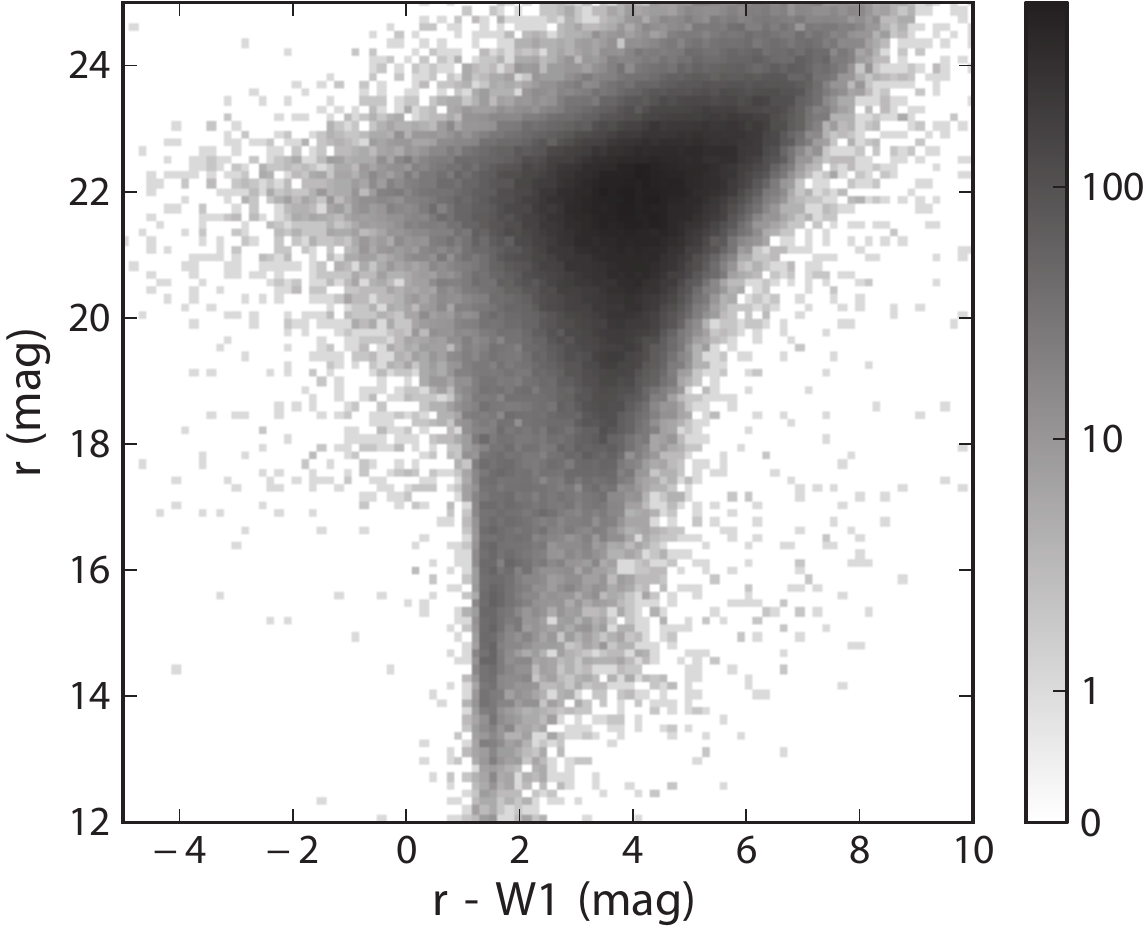}
\caption{Forced photometry results from \emph{the Tractor} code,
  contrasted with traditional ``catalog-matching''.  {\it Left:}
  Color-magnitude diagram from matching SDSS to {\it WISE} catalogs.
  Many objects below the {\it WISE} catalog detection limits are
  lost. {\it Right:} Results from forced photometry of the {\it WISE}
  images based on SDSS detections.  No matching is required, and
  objects that would be detected in {\it WISE} at only few-sigma
  significance can readily provide flux
  measurements.\label{fig:tractor-cmd}}
\end{figure}

For bright objects that were cleanly detected by {\it WISE} alone, we
find our pixel-level measurements to be consistent with catalog-level measurements
(see Figure \ref{fig:tractor-bright}).  However, we are also able to
measure the fluxes of significantly fainter objects, as well as to
study collections of objects that are blended in the {\it WISE}
imaging but that are resolved in the optical images.  Figure
\ref{fig:tractor-cmd} compares a traditional 
optical-infrared color-magnitude diagram, based on matching sources between
catalogs at different wavelengths, to the results of our
{\it WISE} forced photometry, which requires no such matching.
This demonstrates how \emph{The Tractor} increases the
color-space information available to DESI targeting.

In general, {\em The Tractor} improves target selection for all
DESI classes by allowing
information from low signal-to-noise measurements to be utilized.
\emph{The Tractor} is particularly important for QSO targeting.
Up to 15\% of QSO spectra exhibit broad absorption lines that
potentially reduce the measured flux in broadband imaging. High-redshift
(Ly-$\alpha$) QSOs will drop out of some imaging bands completely. 
Finally, the $5\sigma$ optical limit at the extremes of DESI targeting corresponds to
a $< 5\sigma$ limit in {\it WISE} for QSOs (c.f.\ Sec.\ \ref{sec:QSOselectiontechnique}).
\emph{The Tractor} successfully differentiates between the QSOs that
are detected in {\it WISE}, and the QSOs that in general are not detected
(c.f.\ Figure \ref{fig:colorsQSO}), whereas
traditional ``catalog-matching'' approaches would not be successful.

Target selection of
LRGs and QSOs for the SDSS-IV/eBOSS, which began observations in July 2014, 
utilized \emph{The Tractor}.
For eBOSS targets, \emph{the Tractor} was applied to obtain forced photometry based
upon galaxy profiles measured by the SDSS imaging pipeline. Those profiles were
convolved with the {\it WISE} point-spread function, and then a linear fit
was performed on the full set of {\it WISE} imaging data.
The result was a set of
flux estimates for all SDSS objects, constructed so that the sum of
flux-weighted profiles best matched the {\it WISE} images. DESI will make use of 
this same fitting approach, using optical images from surveys being conducted with 
the DECam, Bok and Mayall telescopes (c.f.\  Sec.\ \ref{sec:imagingDataSets}) 
in place of the SDSS images. 

\begin{figure}[!t]
\centering
\includegraphics[width=6.5in]{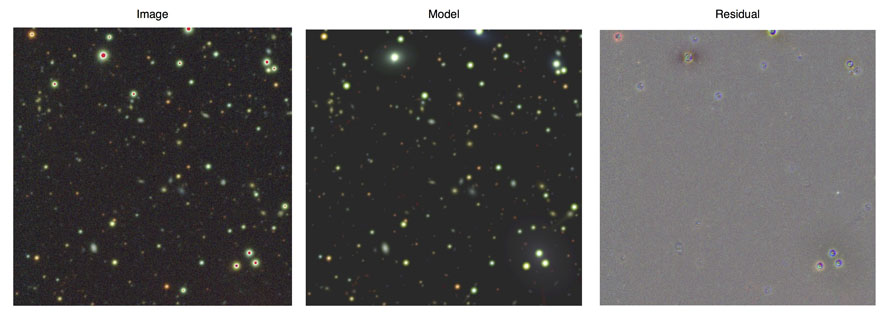}
\caption{An example ``brick" covering $0.25\times0.25$~deg$^2$ from
the DECaLS survey. From left to right, the panels show the actual
$grz$ imaging data, the rendered model based on \emph{the Tractor} catalog
of the region, and the residual map. \emph{The Tractor} catalog represents
an inference-based model of the sky that best fits the observed
data.\label{fig:DECaLStrac}}
\end{figure}

\emph{The Tractor} has already been applied to the DECam survey imaging that will 
be used as part of DESI target selection. This survey, which is known as DECaLS, attained 
its second release (DR2) of imaging early in 2016. \emph{Tractor} catalogs based on this 
DR2 data are publicly available\footnote{\url{http://legacysurvey.org}}.
DECaLS DR2 comprises all $grz$ imaging conducted with DECam prior to June 2015 
that lies within the DESI footprint. This includes both imaging conducted specifically 
for DECaLS and public raw imaging re-extracted using \emph{the Tractor}. The 
co-added images and \emph{Tractor} catalogs are  presented in ``bricks" of 
approximate size 0.25$^\circ \times 0.25^\circ$ (see Figure~\ref{fig:DECaLStrac}) 
and DECaLS DR2 contains approximately 260 million unique sources spread 
over 97{,}554 bricks.

In total, DECaLS DR2 contains about 2000\,${\rm deg}^2$ of imaging in both 
$g$- and $r$-band and roughly 5300\,${\rm deg}^2$ in $z$-band only.
1800\,${\rm deg}^2$ has been observed in all three optical filters. DECaLS is 
on schedule to observe its projected 6200\,${\rm deg}^2$ of imaging over 
3 years (c.f.\ Sec.\ \ref{sec:decam}). Based on formal errors from 
\emph{The Tractor}, the median 5$\sigma$ point source depths 
for areas in DECaLS DR2 with full coverage in each band are 
$g=24.65$, $r=23.61$, $z=22.84$, meeting the depth requirements for DESI
target selection.
{\it WISE} fluxes based on forced photometry using \emph{the Tractor}
are available for all sources extracted as part of DECaLS DR2.

Catalogs generated by \emph{the Tractor} will be vetted
for DESI target selection using a series of image validation tests. Catalogs of galaxies are expected to be
generated in a manner that is model-independent across all bands and that
should achieve a $5\sigma$, extinction-corrected depth of $g$=24.0, $r$=23.4 and $z$=22.5. 
90\% of the DESI footprint requires full-depth imaging, but 95\% (98\%) 
must be within 0.3 (0.6) magnitudes of full-depth. The photometric system 
produced by \emph{the Tractor} must be uniform and stable, with $< 1$\% systematic
errors (RMS) in $g$- and $r$-band, $< 2$\% in $z$-band, and $< 2$\% from 
morphological mis-classifications. The $z$-band image quality must exceed 1.3\,\arcsec\
in at least one pass everywhere in the DESI footprint. The systematic and random 
errors in astrometry must be less than 30\,mas and 90\,mas RMS, respectively. 
In order to facilitate these imaging tests, which are ongoing, \emph{The Tractor} catalogs will ultimately 
include source positions, fluxes, shape parameters, and morphological quantities that can be used to 
discriminate extended sources from point-sources, together with errors on these quantities.

\clearpage

\section{Survey Design}\label{s5:surveydesign}
\setcounter{equation}{0}\setcounter{figure}{0}\setcounter{table}{0}
\label{sec:survey}

\subsection{Introduction}
The DESI instrument will make largest spectroscopic survey to date.  The design of the survey is optimized by selecting a footprint that is as large as possible from the Mayall telescope while 
staying clear of the Milky Way.  The survey strategy will establish the order in which the observations will be made.  The strategy will be modified in detail by atmospheric conditions, but the overall plan will be established to optimize the best science results for both the
complete survey and results from intermediate years.



\subsection{Survey Footprint}
\label{sec:footprint}
The DESI survey footprint is defined to be 14,000 square degrees
that can be observed spectroscopically from Kitt Peak.
This footprint will be one contiguous region selected from the North Galactic Cap (NGC)
and one contiguous region in the South Galactic Cap (SGC).
The instrumented area of the
focal plane is 7.50 square degrees.
14,000 square degrees can be covered nearly completely with little overlap using 
2,000 tiles, where each tile represents one DESI observation.
We refer to the full 2000-tile coverage of the footprint as a ``layer''.
Five layers with altogether 10,000 tiles covers each coordinate of the footprint with an average of 5.24 fibers.
The DESI footprint is formally defined as any position on the sky
within 1.605 deg of any of these selected tile centers.

The DESI spectroscopic survey will primarily select targets from
catalogs derived from imaging with the Blanco/DECam camera, the Bok/90Prime camera,
the Mayall/MOSAIC-2 camera and the Wide-field Infrared Survey Explorer (\emph{WISE}).
Although WISE imaging covers the entire sky, the imaging from
DECam, the Bok Telescope, and the Mayall telescope impose an external constraint on the DESI footprint,
as targets must be selected from large contiguous regions imaged
with the same instruments.
The Bok and Mayall will provide targeting in the NGC at Dec $> +30$ deg.
The Blanco will provide targeting in both the NGC and SGC at Dec $< +30$ deg.
An area of approximately 800 sq.\ deg.\ in the SGC
at Dec $> +30$ (and $-32<b<-15$) is ``orphaned'' and excluded
from the DESI survey as it would be a small area observed with
a different camera.

\begin{figure}[!tb]
\centering
\includegraphics[height=4in]{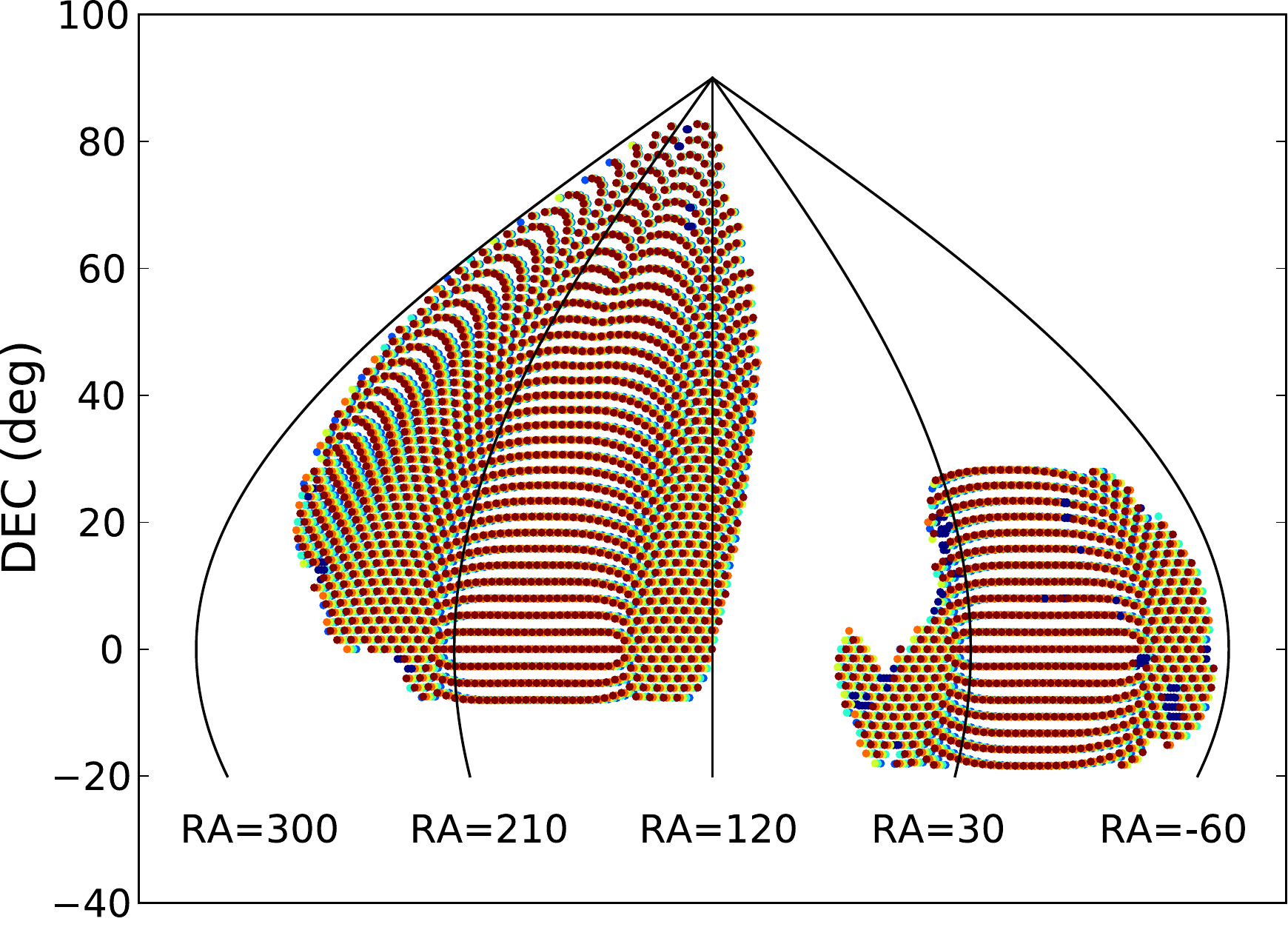}
\caption{Tile centers for the DESI footprint in an equal-area projection.
Declination limits are imposed at $-8.2 <$ Dec in the NGC (left),
and $-18.4 <$ Dec $< +30$ in the SGC (right).
Approximately $1\%$ of tiles have exposure factors larger than 2.5 (shown in blue), but
are included to avoid unwanted holes in the footprint.
The five layers are shown in separate (but overlapping) colors.
The spots indicate the centers of focal plane positions and overlap between layers.
The symbols do not represent the size of the area in the sky
subtended by the focal plane.
Every location inside the footprint is within reach of a fiber, on average, 5.24 times during the survey.
\label{fig:desi-tiles}}
\end{figure}

The footprint is constrained, as well, by the need to avoid regions that
would require long exposures due to airmass or dust, by weather patterns at
Kitt Peak, and by regions of high stellar density.  The resulting footprint
is shown in Figure~\ref{fig:desi-tiles}.

\subsection{Field Centers}
\label{sec:tiling_fiber_assignment}
We refer to ``tiling'' as the process by which field centers are assigned in a manner to cover the
footprint with optimal coverage of each coordinate on the sky.
The single-layer tiling of the sky mentioned in Section~\ref{sec:footprint} is a preliminary
solution that is achieved using the
icosahedral tiling \citep{Hardin2001} with 5762 tile centers distributed
on the full sphere\footnote{\url{http://neilsloane.com/icosahedral.codes/}}.
This tiling is very-well matched to the DESI focal plane size.
The first layer rotates the above tiling solution by 90 deg in RA.
This rotation conveniently puts rows of tile centers
along lines of approximately constant declination at the north and
south boundaries of the DESI survey.  Each of additional layers 2 through 5
have an additional rotation of the tile centers by 1.08 deg in RA.
This gives large dithers on most of the sky (except at the pole, which is not in
the DESI footprint), thus filling the gaps in the focal plane
with subsequent visits.
Non-uniformity in coverage could artificially introduce structure
in the targeting of LSS-tracers;  alternative tilings based
on the same first layer but with subsequent layers obtained with more
disparate rotations will be further studied for possible improvements
to the uniformity.

A descoped instrument has been considered which would conduct the
DESI survey over 9000 square degrees rather than 14,000 square degrees.
This descoped instrument would populate only six of the 10 wedges
on the focal plane with 3000 instead of 5000 fibers.
The populated wedges are best arranged
in a ``Pacman'' format.
A different tiling solution is necessary for covering a smaller survey
area with the same mean coverage per survey area.
First, 240 tile centers are placed
on the celestial equator uniformly separated in RA.  Stripes
of tiles are then placed on lines of constant celestial latitude
spaced every 2.765 deg.  At each stripe, the number of tiles is reduced
by the factor cos(Dec) from the 240 placed on the celestial equator.
This results in a tiling solution with similar uniformity and coverage statistics
as the baseline survey, with 4\% more tiles than would be necessary under
the assumption of a simple scaling with focal plane area.

\begin{figure}[!h]
\centering
\includegraphics[width=0.9\textwidth]{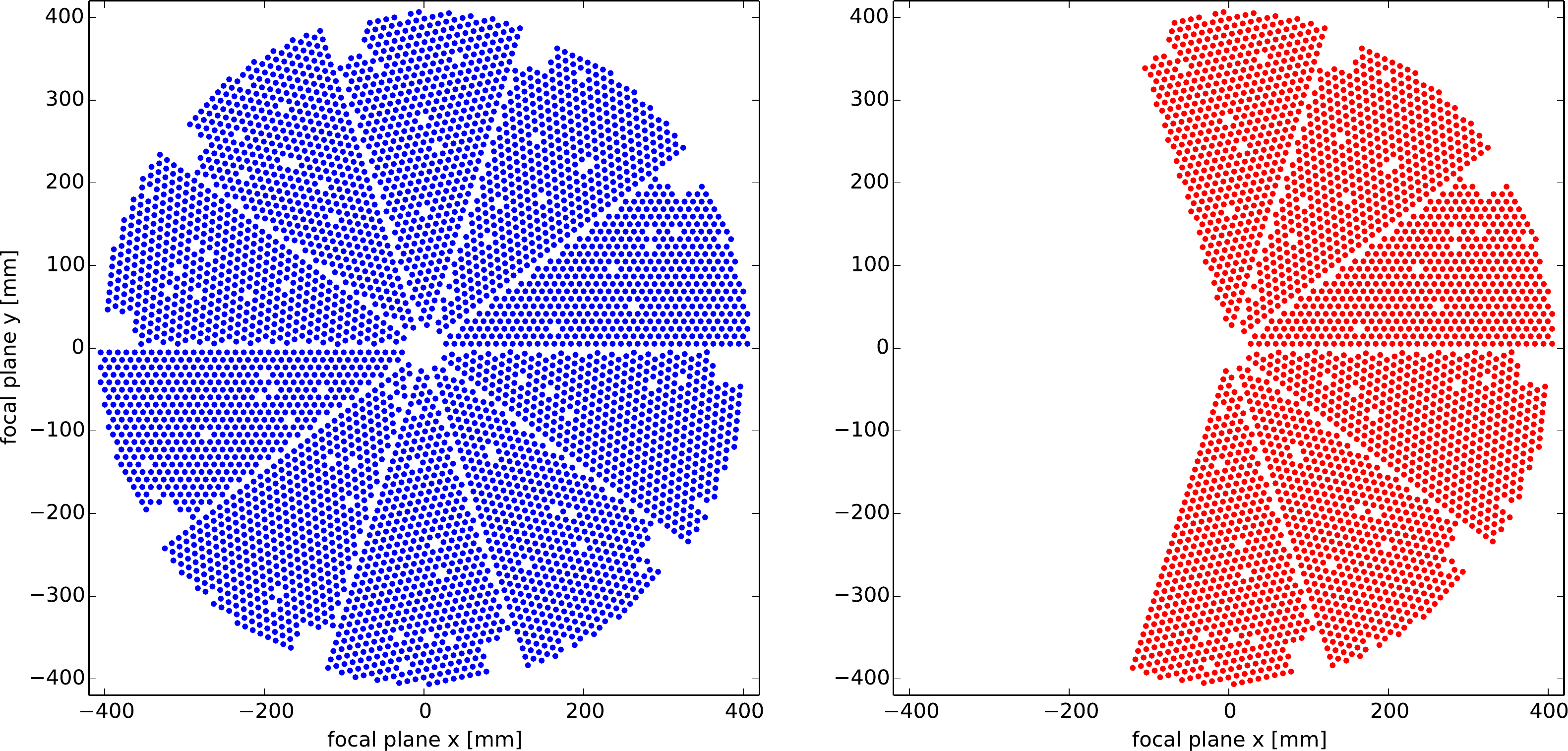}
\caption{
{\it Left:  }Fiber positioner locations for the full DESI instrument.
{\it Right:  }The locations for the
reduced instrument ``pacman'' configuration.  ``Missing''
positioner locations are for the guide-focus arrays (square regions)
and fiducial markers for the fiber view camera.
\label{fig:fiber_assignment_focalplane}}
\end{figure}

\begin{figure}[!h]
\begin{center}
\includegraphics[height=2.5in]{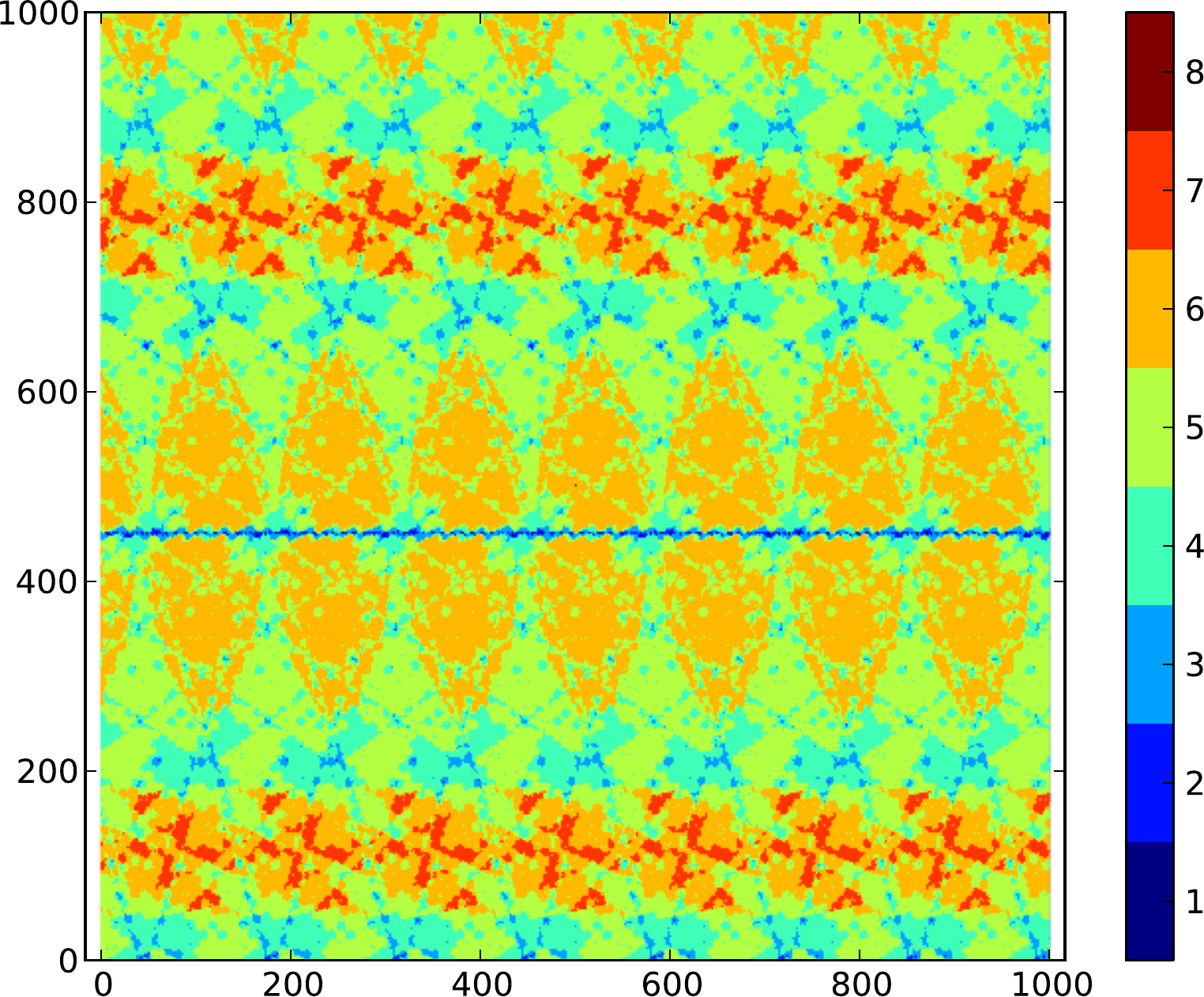}
\includegraphics[height=2.5in]{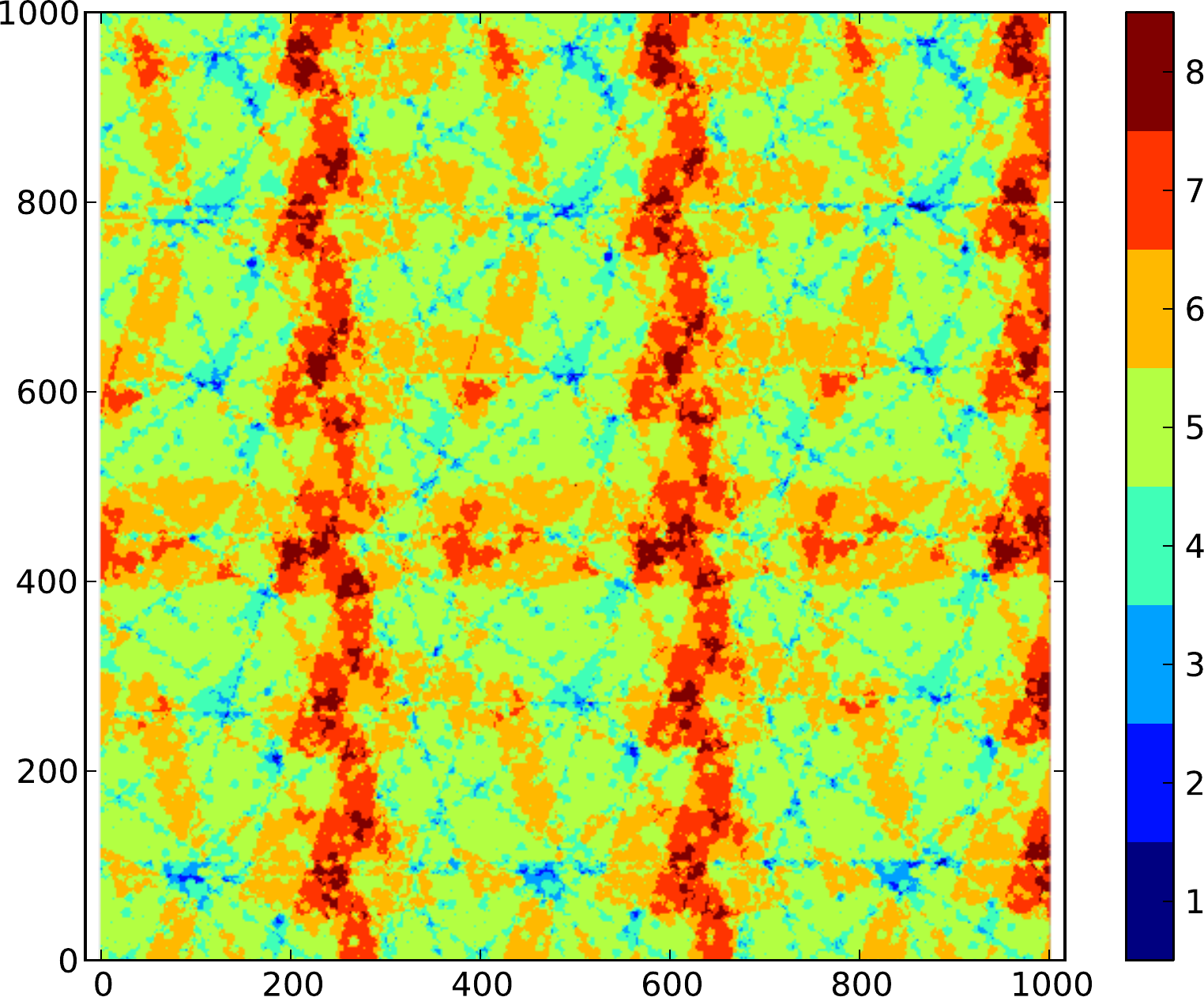}
\caption{Coverage pattern on the sky 
after five layers over a 4 degree by 4 degree patch.  This is shown for a region away from the edges of the footprint.  
{\it Left:  }The fully-populated focal plane with 5000 fibers.
{\it Right:  }The reduced focal plane with 3000 fibers.
}
\label{fig:full}
\end{center}
\end{figure}

The pattern of fiber positioners in the focal plane is  shown in Figure~\ref{fig:fiber_assignment_focalplane}.
Combining this with the tiling gives a purely geometric measure of the coverage
for each position within the DESI footprint.
The distribution of this coverage
is shown in Figure~\ref{fig:full} and in Table \ref{tab:coverage_fractions}.
The average coverage is about 5.1 fibers available per coordinate, with only 3.5\% of the footprint having
a coverage of less than 3 fibers.  The mean relative to the value of 5.24 reported earlier is slightly
reduced due to increase of edge effects over the smaller area tested.  The edges of the footprint have the least coverage.
The results of a similar study for the reduced ``Pacman'' focal plane
are shown in the right hand panel of Figure~\ref{fig:full}.

\begin{table}[!h]
\begin{center}
\caption{The fraction of the footprint covered by 1, 2, 3, ... 8 fibers.
The first row shows the results using all five layers
in the baseline survey.  The second row shows the results using just the four layers that include LRG and quasar targets.
The mean is slightly decreased and the rms slightly increased by edge effects. }\label{tab:coverage_fractions}
\begin{tabular}{lrrrrrrrrrr}\hline
Coverage&1&2&3&4&5&6&7&8&Mean&RMS\\ \hline
All five layers &0.016& 0.019 & 0.040 & 0.155 & 0.424 & 0.279 & 0.055 & 0.009&5.06  & 1.175  \\ 
Four layers only&0.021 & 0.037 & 0.152 & 0.482 & 0.462 & 0.040 & 0.004 & 0.000 & 4.04  & 1.012\\ \hline
\end{tabular}

\end{center}
\end{table}


\subsection{Observation Strategy}
\label{sec:longterm_obsplan}

\subsubsection{Sequence of Observations}\label{par:observingsequence}

The placement of field centers presented in Section~\ref{sec:tiling_fiber_assignment} is designed to cover
the footprint in five independent tilings.
Given the 1940 hours of scheduled time each year, roughly 20\% (390 hours)
will occur under conditions when the moon is above the horizon and the remainder under dark conditions.
Each year, 20\% of the fields in a full independent tiling of 2000 field centers
will be observed using the scheduled time in grey conditions.
This layer is planned to include only ELG targets because their spectral features
are predominantly found at redder wavelengths and redshift success rates are less
susceptible to increased sky background from the moon.
On the other hand, the darkest 80\% of the scheduled time (1550 hours) will be used to observe the
QSO and LRG targets at highest priority, leaving the remaining science fibers for ELG targets.

There remains additional freedom to determine the order in which the tiles over the
four dark time layers are observed.
Full simulations of the program will be used to determine the optimal approach.
The simulations will factor in seeing, transparency and weather variations
for each exposure via Monte Carlo simulations to predict the quality of spectra and the variations in
final survey areal coverage.  Each exposure will be tuned to a grid of targets parameterized by magnitude and
redshift using an exposure time calculator that approximates the sensitivity of the instrument.
Weather conditions will be mocked using monthly statistics at Kitt Peak and the results will be used to
determine likely redshift success rates over all target classes.
A description of these simulations is found in the document DESI-1658.
The approach that optimizes intermediate and final cosmology results will
be chosen.

We provide a baseline strategy in the accompanying document on long-term strategy.  This program
is assembled without consideration of weather and other variables.
The survey is designed to get an early complete sample over 10\% of the footprint as early as possible.
The survey also provides distinct milestones for data products and cosmological analysis at the end of each year.

Finally, we will investigate the target strategies, exposure depths, quality of data, and expected
spectroscopic completeness during a phase of 
survey validation.  Survey validation
will occur during the
end of commissioning before the full survey begins.  The baseline program for this phase of the project is
presented in the accompanying document on survey validation.

\subsubsection{Exposure Times and Margin}

Over five years, DESI is projected to observe 14,000 sq.\ deg.\ of the footprint presented
in Section~\ref{sec:footprint}.
The exact subset of this footprint to be observed will be contiguous regions in
each of the NGC and the SGC that best fit the expected allocation of time.
We have simulated the choice of final tile centers and the average exposure times
according to an observing schedule of 1940 hours of dark and grey time
per year as defined in the Site Alternatives study (DESI-311).
The simulation includes a two minute overhead between fields and
variations in exposure time for each field due to airmass and Galactic dust extinction.
All exposure times are split into two separate exposures
(with one minute of read time).  This split limits the number of cosmic rays in an
individual exposure, and also effectively maximize the S/N in variable sky conditions.
The split is not assumed in the baseline for spectroscopic depth and completeness, so the
time per field is larger in these simulations than in the baseline design.
The accumulated S/N will be measured by the exposure time calculator (see the DESI Performance Studies in the Instrument FDR).
We project that 57\% of the scheduled time will deliver usable data,
where ``usable data'' is assumed in conditions when the dome is open
and seeing is better than 1.5 arcsec.
Although DESI will observe when the seeing is worse than 1.5 arcsec,
those data have been ignored in these estimates of survey duration.

We simulate the full suite of observations accounting for airmass and Galactic dust extinction
by choosing an hour angle for each field that maximizes the overall survey depth while
fitting into the allocated time.
Exposure times are estimated for each field to produce uniform depth in
dust-extinction and atmosphere-extinction corrected spectra.
In preliminary estimates, we assume the same dependence of S/N on airmass
as was measured with BOSS, and degradation in S/N due to Galactic extinction
for the sky-noise-limited case of the faintest targets.
In future iterations, we will include a more sophisticated
interpretation of redshift success rate for representative targets,
thus accounting for the wavelength-dependent S/N estimates of each target class.
For the 14,000 sq.\ deg.\ footprint observed with 10,000 tiles,
we find an average exposure time of 1800 seconds.  Scaling this to an observation
taken at zenith with no Galactic dust extinction (as shown in the figure in the DESI Performance Studies in the Instrument FDR)
produces an equivalent exposure time of 1226 seconds.
In other words, each exposure will have a S/N equivalent to a 1226 second
exposure taken at zenith, under photometric conditions, median sky brightness
and median seeing.
As explained in Simulations Section
of the Instrument FDR this fiducial exposure time of 1226 seconds
allows the 1000 second exposures that are predicted to produce
the required redshift success rates for each DESI target class.
This projection leaves a 22\% margin in exposure time for worse-than-projected weather,
throughput performance, instrument downtime, or other factors that could slow the pace of the survey.

Similarly, we have estimated the average and fiducial exposure times for the reduced focal plane of the DESI
KPP survey.  The ``Pacman'' tiling of Section~\ref{sec:footprint} leads
to an average exposure time of 1700 seconds for 10,600 tile centers covering 9,000 sq.\ deg.
Even though the average exposure time is somewhat lower than the 14,000 sq.\ deg.\ survey,
the fiducial exposure time of 1270 seconds is actually larger because the average field
in a 9,000 sq.\ deg.\ program lies at lower airmass and lower Galactic extinction
than the average field in a 14,000 sq.\ deg.\ program.
The projected margin for the 9,000 sq.\ deg.\ KPP survey is 27\%.

\subsection{The Bright Galaxy and Milky Way Surveys}
\label{sec:bts}

\subsubsection{Introduction}

A portion of DESI operations will be affected by increased sky
brightness from the moon, so as to make conditions unsuitable for
observing the targets above $z=0.6$.  DESI expects to observe in the
darkest 21 nights of the month, but some of those nights are affected
in part by moon, adding up to about 440 hours per year of time.
Assuming the same average weather statistics used in planning for dark
time, we expect 250 hours per year on average of open-dome bright
time. During this time, the DESI collaboration will conduct a survey
of bright galaxies which will increase performance for the cosmology
goals.  This Bright Galaxy Survey (BGS) will be the primary
bright-time survey program.  In addition, the density of fibers in the
DESI focal plane will enable a simultaneous survey of Milky Way Stars
(The Milky Way Survey; MWS) during bright time. The MWS will target
some of the oldest stars in the Galaxy with the goal of understanding
the mass distribution, formation and evolution of the Galaxy.  We
refer to these two combined programs as the Bright Time Survey (BTS).

\subsubsection{Survey Footprint}

The Bright Time Survey will use the same 14,000 square degree
footprint as the dark time project.  This will enable the BTS to
benefit from the optimization of the dark time footprint for
observability.   The BGS targets will be selected from the same
imaging data as the dark time targets.  The MWS will use Gaia
photometry and proper motions for target selection. The Gaia survey
is all-sky, and so covers the DESI 14,000 square degree footprint. 

\subsubsection{Field Centers}

The BTS will use the same tiling pattern as the DESI Key Project, but
with only 3 layers totaling 6000 tiles.  There are roughly 1400
galaxies per square degree to $r = 20.0$.  With 4500 science fibers
per tile, the BTS will place about 27 million fibers, more than the
$\sim20$ million BGS targets.  However, the presence of clustering and Poisson
fluctuations of bright galaxies implies that we must incur these extra
layers if we want to achieve a higher completeness. For fiber
assignment, BGS targets are divided into two priorities; brighter
targets with $r<19.5$ ($\sim 800$ deg$^{-2}$) receive high priority,
while fainter $19.5<r<20.0$ ($\sim 600$ deg$^{-2}$) targets receive
secondary priority.  Preliminary simulations of DESI fiber assignments
using this priority scheme yields 3-layer completeness values of 92\%
for the bright sample and 77\% for the faint sample, for an overall
fiber completeness of 86\%, or roughly 17M targets.

There are about 600 stars with effective temperatures higher than 4700
K per square degree to $r = 18$ at Galactic latitude greater than 40
degrees  from the equator. The DESI
focal plane is 7.5 square degrees, so at each layer there will be many
fibers available for the MWS.  

\subsubsection{Observation Strategy}

Completing 6000 tiles in the 1250 hours of available open-dome time
indicates an average time of 12.5 minutes per tile.  Survey
simulations accounting for the increased exposure time required as a
function of airmass and extinction indicate that we would have 400
seconds available for a reference exposure at unit airmass and zero
extinction.  We are planning for a 300 second reference exposure,
therefore leaving a 33\% margin.  Our spectral simulations (\S
\ref{sec:bgs} and Figure~\ref{fig:bgs_redshift_efficiency}) indicate
that a 5-minute exposure will yield a redshift success of 97\% for
galaxies down to $r=19.5$, and 92\% for the fainter $19.5<r<20.0$
sample.


For stars, 5 minute exposures will result in spectra with S/N per
Angstrom of 14 at $\lambda >$ 650 nm for stars of $r$ magnitude
16.5-18, depending on their spectral energy distributions.  Spectra of
that quality are sufficient to yield radial velocity and chemical
abundance information.

Because the BTS needs only three layers, it will be possible to
combine multiple exposures for fainter objects for higher S/N.  For
example, it would be possible to re-expose many of the 5\% of BGS
targets that fail to achieve a redshift in the first two layers.  We
will perform fiber assignment simulations that combine the MWS and BGS
samples to determine the optimal way to assign fibers that accounts
for galaxy clustering and the variation in stellar density across the
footprint, and which achieves maximum redshift and radial velocity
completeness for faint targets.

With this basic strategy we expect to obtain spectra of roughly 10
million galaxies in the BGS and 10 million stars in the MWS.  More
simulations of the BTS are required to determine how to prioritize sky
coverage versus completeness to enable early science. The BTS
simulations will use the same survey simulation code as the dark time
program, adapting it as required to account for scheduling around
lunar phase and separation angle between the field and the moon.

\clearpage

\section*{Acknowledgements}
\addcontentsline{toc}{section}{Acknowledgements}
This research is supported by the Director, Office of Science, Office of High Energy Physics of the U.S. Department of Energy under Contract No. 
DE–AC02–05CH1123, and by the National Energy Research Scientific Computing Center, a DOE Office of Science User Facility under the same 
contract; additional support for DESI is provided by the U.S. National Science Foundation, Division of Astronomical Sciences under Contract No. 
AST-0950945 to the National Optical Astronomy Observatory; the Science and Technologies Facilities Council of the United Kingdom; the Gordon 
and Betty Moore Foundation; the Heising-Simons Foundation; the National Council of Science and Technology of Mexico, and by the DESI 
Member Institutions: Aix-Marseille University;  Argonne National Laboratory; Barcelona Regional Participation Group; Brookhaven National Laboratory; 
Boston University; Carnegie Mellon University; CEA-IRFU, Saclay; China Participation Group; Cornell University; Durham University;  École Polytechnique 
Fédérale de Lausanne; Eidgenössische Technische Hochschule, Zürich;  Fermi National Accelerator Laboratory;  Granada-Madrid-Tenerife Regional 
Participation Group; Harvard University; Korea Astronomy and Space Science Institute; Korea Institute for Advanced Study; Institute of Cosmological  
Sciences, University of Barcelona; Lawrence Berkeley National Laboratory; Laboratoire de Physique Nucléaire et de Hautes Energies; Mexico Regional 
Participation Group; National Optical Astronomy Observatory; Siena College; SLAC National Accelerator Laboratory;  Southern Methodist University; 
Swinburne University; The Ohio State University; Universidad de los Andes; University of Arizona; University of California, Berkeley; University of California, 
rvine; University of California, Santa Cruz; University College London; University of Michigan at Ann Arbor; University of Pennsylvania; University of Pittsburgh; 
University of Portsmouth; University of Queensland; University of Toronto; University of Utah; UK Regional Participation Group; Yale University. T
he authors are honored to be permitted to conduct astronomical research on Iolkam Du’ag (Kitt Peak), a mountain with particular significance to the 
Tohono O’odham Nation.  For more information, visit desi.lbl.gov.

%
%

\clearpage

\addcontentsline{toc}{section}{References}
\printbibliography

\noindent$\rule{4in}{0.15mm}$\\
Author Institutions
\begin{small}
\begin{enumerate}
  \setlength{\itemsep}{1pt}
  \setlength{\parskip}{0pt}
  \setlength{\parsep}{0pt}

\item [$^{1}$]2137 Frederick Reines Hall, Irvine, CA 92697, USA
\item [$^{2}$]Aix Marseille Univ, CNRS, LAM,  13388 Marseille, France
\item [$^{3}$]Aix Marseille Univ, CNRS, OHP, 04870 Saint-Michel-l'Observatoire, France
\item [$^{4}$]Aix Marseille Universit\'{e}, CNRS/IN2P3, CPPM UMR 7346, 13288, Marseille, France
\item [$^{5}$]Alphabet Inc., 1650 Charleston Rd. Mountain View, CA 94043, USA
\item [$^{6}$]AMNH, Department of Astrophysics, American Museum of Natural History, New York, NY 10024, USA
\item [$^{7}$]APC, Universit\'{e} Paris Diderot-Paris 7, CNRS/IN2P3, CEA, Observatoire de Paris, 10, rue Alice Domon \& Léonie Duquet, Paris, France
\item [$^{8}$]Argonne National Laboratory, High-Energy Physics Division, 9700 S. Cass Avenue, Argonne, IL 60439, USA
\item [$^{9}$]Astronomy Department, Yale University, P.O. Box 208101 New Haven, CT 06520-8101, USA
\item [$^{10}$]Brookhaven National Laboratory, Upton NY 11973, USA
\item [$^{11}$]Carreterra M\'{e}xico-Toluca S/N, La Marquesa, Ocoyoacac, Edo. de M\'{e}xico C.P. 52750,  M\'{e}xico
\item [$^{12}$]CEA Saclay, IRFU  F-91191 Gif-sur-Yvette, France
\item [$^{13}$]Center for Cosmology and AstroParticle Physics, The Ohio State University, 191 West Woodruff Avenue, Columbus, OH 43210, USA
\item [$^{14}$]Centre for Advanced Instrumentation, Department of Physics, Durham University, South Road, Durham, DH1 3LE, UK
\item [$^{15}$]Centre for Astrophysics \& Supercomputing, Swinburne University of Technology, P.O. Box 218, Hawthorn, VIC 3122, Australia
\item [$^{16}$]Centre for Extragalactic Astronomy, Department of Physics, Durham University, South Road, Durham, DH1 3LE, UK
\item [$^{17}$]Centre for Theoretical Cosmology, Department of Applied Mathematics and Theoretical Physics, Wilberforce Road, Cambridge CB3 0WA, UK
\item [$^{18}$]Cerro Tololo Inter-American Observatory (CTIO), Colina El Pino s/n, Casilla 603, La Serena, Chile
\item [$^{19}$]CIEMAT, Avenida Complutense 40, E-28040 Madrid, Spain
\item [$^{20}$]Clippinger Laboratories, Room 333, Ohio University, Athens, OH 45701, USA
\item [$^{21}$]Departamento de F\'{i}sica, Universidad de Guanajuato - DCI, C.P. 37150, Leon, Guanajuato, M\'{e}xico
\item [$^{22}$]Departamento de F\'isica, Universidad de los Andes, Cra. 1 No. 18A-10, Edificio Ip, Bogot\'{a}, Colombia 
\item [$^{23}$]Department of Astronomy \& Astrophysics, University of Toronto, 50 St.~George Street, Toronto, ON, Canada M5S 3H4
\item [$^{24}$]Department of Astronomy and Astrophysics, University of California, Santa Cruz, 1156 High Street, Santa Cruz, CA 95065, USA
\item [$^{25}$]Department of Astronomy and Space Science, Sejong University, Seoul 143-747, Republic of Korea
\item [$^{26}$]Department of Astronomy, The Ohio State University, 4055 McPherson Laboratory, 140 W 18th Avenue, Columbus, OH 43210, USA
\item [$^{27}$]Department of Astronomy, University of California, Berkeley, CA 94720-3411, USA
\item [$^{28}$]Department of Astronomy, University of Michigan, 1085 S. University Avenue, Ann Arbor, MI 48109-1107, USA
\item [$^{29}$]Department of Astronomy, Yale University, Steinbach Hall, 52 Hillhouse Avenue, New Haven, CT 06511, USA
\item [$^{30}$]Department of Physics \& Astronomy and Pittsburgh Particle Physics, Astrophysics, and Cosmology Center (PITT PACC), University of Pittsburgh, Pittsburgh, PA 15260, USA
\item [$^{31}$]Department of Physics \& Astronomy, Ohio University, Athens, OH 45701, USA
\item [$^{32}$]Department of Physics \& Astronomy, University  of Wyoming, 1000 E. University, Dept.~3905, Laramie, WY 82071, USA
\item [$^{33}$]Department of Physics \& Astronomy, University College London, Gower Street, London, WC1E 6BT, UK
\item [$^{34}$]Department of Physics and Astronomy, Siena College, 515 Loudon Road, Loudonville, NY 12211, USA
\item [$^{35}$]Department of Physics and Astronomy, The University of Utah, 115 South 1400 East, Salt Lake City, UT 84112, USA
\item [$^{36}$]Department of Physics and Astronomy, University College London, 3rd Floor, 132 Hampstead Road, London, NW1 2PS, UK
\item [$^{37}$]Department of Physics and Astronomy, University of California, 4129 Frederick Reines Hall, Irvine, CA 92697, USA
\item [$^{38}$]Department of Physics and Center for Cosmology and Particle Physics, New York University, New York, NY 10003, USA
\item [$^{39}$]Department of Physics and JINA Center for the Evolution of the Elements, University of Notre Dame, Notre Dame, IN 46556, USA
\item [$^{40}$]Department of Physics and Michigan Center for Theoretical Physics, University of Michigan, Ann Arbor, MI 48109, USA
\item [$^{41}$]Department of Physics, Carnegie Mellon University, 5000 Forbes Avenue, Pittsburgh, PA 15213, USA
\item [$^{42}$]Department of Physics, Harvard University, 17 Oxford Street, Cambridge, MA 02138, USA
\item [$^{43}$]Department of Physics, Kansas State University, 116 Cardwell Hall, Manhattan, KS 66506, USA
\item [$^{44}$]Department of Physics, Southern Methodist University, 3215 Daniel Avenue, Dallas, TX 75275, USA
\item [$^{45}$]Department of Physics, The Ohio State University, 191 West Woodruff Avenue, Columbus, OH 43210, USA
\item [$^{46}$]Department of Physics, University of Arizona, 1118 E. Fourth Street, PO Box 210081, Tucson, AZ 85721, USA
\item [$^{47}$]Department of Physics, University of California, Berkeley, 366 LeConte Hall MC 7300, Berkeley, CA 94720-7300, USA
\item [$^{48}$]Department of Physics, University of Michigan, 450 Church St., Ann Arbor, MI 48109, USA
\item [$^{49}$]Department of Physics, University of Warwick, Gibbet Hill Road, Coventry, CV4 7AL, UK
\item [$^{50}$]Ecole Polytechnique F\'{e}d\'{e}rale de Lausanne, CH-1015 Lausanne, Switzerland
\item [$^{51}$]European Space Astronomy Centre (ESAC), 38205 Villanueva de la Ca\~{n}ada, Madrid, Spain
\item [$^{52}$]Fermi National Accelerator Laboratory, PO Box 500, Batavia, IL 60510, USA
\item [$^{53}$]Harvard-Smithsonian Center for Astrophysics, Harvard University, 60 Garden Street, Cambridge, MA 02138, USA
\item [$^{54}$]HCTLab Research Group, Escuela Politecnica Superior, Universidad Aut\'{o}noma de Madrid, C/Francisco Tomas y Valiente 11, 38049, Spain
\item [$^{55}$]Instituci\'{o} Catalana de Recerca i Estudis Avan\c{c}ats (ICREA), Pg.~de Llu\'{i}s Companys 23, 08010 Barcelona, Spain
\item [$^{56}$]Institut de C\`{i}encies de l'Espai, IEEC-CSIC, Campus UAB, Carrer de Can Magrans s/n, 08913 Bellaterra, Barcelona, Spain
\item [$^{57}$]Institut de Fisica d’Altes Energies (IFAE), The Barcelona Institute of Science and Technology, Campus UAB, 08193 Bellaterra Barcelona, Spain
\item [$^{58}$]Institute for Astronomy, ETH Z\"{u}rich, Wolfgang-Pauli-Strasse 27, CH-8093 Z\"{u}rich, Switzerland
\item [$^{59}$]Institute for Astronomy, University of Edinburgh, Royal Observatory, Edinburgh EH9 3HJ, UK
\item [$^{60}$]Institute for Computational Cosmology, Department of Physics, Durham University, South Road, Durham DH1 3LE, UK
\item [$^{61}$]Institute of Astronomy, University of Cambridge, Madingley Road, Cambridge, CB3 0HA, UK
\item [$^{62}$]Institute of Cosmology \& Gravitation, University of Portsmouth, Dennis Sciama Building, Portsmouth PO1 3FX, UK
\item [$^{63}$]Instituto de Astrofisica de Andaluc\'{i}a, Glorieta de la Astronom\'{i}a, s/n, E-18008 Granada, Spain
\item [$^{64}$]Instituto de Astrof\'{i}sica de Canarias, C/ Vía L\'{a}ctea, s/n, 38205 San Crist\'{o}bal de La Laguna, Santa Cruz de Tenerife, Spain
\item [$^{65}$]Instituto de Astronomia, Universidad Nacional Aut\'{o}noma de M\'{e}xico, Apartado Postal 70–264, 04510 M\'{e}xico D.F., M\'{e}xico
\item [$^{66}$]Instituto de C\`{i}encias del Cosmoc, (ICCUB) Universidad de Barcelona (IEEC-UB), Mart\'{i} i Franqu\`{e}s 1, E08028 Barcelona
\item [$^{67}$]Instituto de F\'{i}sica Te\'{o}rica (IFT) UAM/CSIC, Universidad Aut\'{o}noma de Madrid, Cantoblanco, E-28049, Madrid, Spain
\item [$^{68}$]Instituto de F\'{i}isica, Universidad Nacional Aut\'{o}noma de M\'{e}xico, Cd. M\'{e}xico C.P. 04510
\item [$^{69}$]Kavli Institute for Astronomy and Astrophysics at Peking University, PKU, 5 Yiheyuan Road, Haidian District, Beijing 100871, P.R. China
\item [$^{70}$]Kavli Institute for Cosmology, Cambridge, University of Cambridge, Madingley Road, Cambridge CB3 0HA, UK
\item [$^{71}$]Kavli Institute for Particle Astrophysics and Cosmology and SLAC National Accelerator Laboratory, Menlo Park, CA 94305, USA
\item [$^{72}$]Key Laboratory of Optical Astronomy, National Astronomical Observatories, Chinese Academy of Sciences, Beijing 100012, P.R. China
\item [$^{73}$]Korea Astronomy and Space Science Institute, 776, Daedeokdae-ro, Yuseong-gu, Daejeon 34055, Republic of Korea
\item [$^{74}$]Laboratoire d’Astrophysique, Ecole Polytechnique F\'{e}d\'{e}rale de Lausanne (EPFL), Observatoire de Sauverny, CH-1290 Versoix, Switzerland
\item [$^{75}$]Laborat\'{o}rio Interinstitucional de e-Astronomia, Rua Gal. Jose Cristino 77, Rio de Janeiro, RJ 20921-400, Brazil
\item [$^{76}$]Lawrence Berkeley National Laboratory, 1 Cyclotron Road, Berkeley, CA 94720, USA
\item [$^{77}$]Lawrence Livermore National Laboratory, P.O. Box 808 L-211, Livermore, CA 94551, USA
\item [$^{78}$]Ludwig-Maximilians University Munich, University Observatory, Scheinerstr.~1, 81679 Munich, Germany
\item [$^{79}$]McWilliams Center for Cosmology, Carnegie Mellon University, 5000 Forbes Avenue, Pittsburgh, PA 15213, USA
\item [$^{80}$]National Astronomical Observatories, Chinese Academy of Sciences, A20 Datun Rd. 100012, Beijing, P.R. China
\item [$^{81}$]National Optical Astronomy Observatory, 950 N. Cherry Avenue, Tucson, AZ 85719, USA
\item [$^{82}$]Observatorio Nacional, R. Gal. Jose Cristino 77, Rio de Janeiro, RJ 20921-400, Brazil
\item [$^{83}$]Physics Department, Stanford University, Stanford, CA 93405, USA
\item [$^{84}$]Physics Department, Yale University, P.O. Box 208120, New Haven, CT 06511, USA
\item [$^{85}$]Physics Dept., Boston University, 590 Commonwealth Avenue, Boston, MA 02215, USA
\item [$^{86}$]School of Mathematics and Physics, University of Queensland, 4101, Australia
\item [$^{87}$]School of Physics, Korea Institute for Advanced Study, 85 Hoegiro, Dongdaemun-Gu, Seoul 02455, Republic of Korea
\item [$^{88}$]Sorbonne Universit\'{e}s, UPMC Université Paris 06, Universit\'{e} Paris-Diderot, CNRS-IN2P3 LPNHE 4 Place Jussieu, F-75252, Paris Cedex 05, France
\item [$^{89}$]Space Sciences Laboratory, University of California, Berkeley, 7 Gauss Way, Berkeley, CA  94720, USA
\item [$^{90}$]Steward Observatory, University of Arizona, 933 N. Cherry Avenue, Tucson, AZ 85721, USA
\item [$^{91}$]SUPA, School of Physics and Astronomy, University of St Andrews, St Andrews, KY16 9SS, UK
\item [$^{92}$]University of California Observatories, 1156 High Street, Sana Cruz, CA 95065, USA
\item [$^{93}$]University of Science and Technology, Daejeon 34113, Republic of Korea

\end{enumerate}
\end{small}

\end{document}